\documentclass[11pt]{article}
\pdfoutput=1  
\usepackage{%
  abstract,
  amsmath,
  amssymb,
  amsthm,
  array,
  color,
  comment,
  extarrows,
  latexsym,
  mathrsfs,
  mathtools,
  multirow,
  nccmath,
  rotating,
  setspace,
  slashed,
  textcomp,
  titlesec,
}
\usepackage[sorting=none,backend=biber,sortcites,citestyle=numeric-comp]{biblatex}
\DeclareFieldFormat[misc]{title}{%
  {\mkbibquote{#1}}
}

\usepackage[pdfusetitle]{hyperref}
\hypersetup{
  linktocpage=true,
  breaklinks=true,
  pdfauthor={Arif Er and Meng-Chwan Tan},
}

\makeatletter
\renewcommand{\sectionautorefname}{$\S$\@gobble}
\renewcommand{\subsectionautorefname}{$\S$\@gobble}
\renewcommand{\subsubsectionautorefname}{$\S$\@gobble}
\makeatother

\usepackage[multiple]{footmisc}
\usepackage[auth-lg,affil-it]{authblk}

\numberwithin{equation}{section}

\titleformat*{\section}{\bfseries\large}

\usepackage[a4paper,
textwidth=165mm,
tmargin=25mm,
bmargin=40mm,
]{geometry}

\allowdisplaybreaks[2]

\DeclareFontFamily{U}{MnSymbolC}{}
\DeclareSymbolFont{MnSyC}{U}{MnSymbolC}{m}{n}
\DeclareFontShape{U}{MnSymbolC}{m}{n}{
  <-6>  MnSymbolC5
  <6-7>  MnSymbolC6
  <7-8>  MnSymbolC7
  <8-9>  MnSymbolC8
  <9-10> MnSymbolC9
  <10-12> MnSymbolC10
  <12->   MnSymbolC12
}{}
\DeclareMathSymbol{\intprod}{\mathbin}{MnSyC}{'270}  

\usepackage{graphicx}
\usepackage{tikz}
\usepackage{subcaption}
\usepackage{adjustbox}
\usetikzlibrary{arrows,automata,positioning,calc,bending}

\newcommand{\dv}[2]{\frac{d #1}{d #2}}
\newcommand{\Dv}[2]{\frac{D #1}{D #2}}
\newcommand{\pdv}[2]{\frac{\partial #1}{\partial #2}}
\newcommand{\longto}{\longrightarrow}

\newcommand{\subtitle}[1]{\bigskip\noindent\textit{#1}\vspace*{0.5em}}

\newcommand{\C}{\mathbb{C}}

\newcommand{\R}{\mathbb{R}}

\newcommand{\Tr}{\mathrm{Tr}}

\renewcommand{\Re}{\text{Re}}
\renewcommand{\Im}{\text{Im}}
\usepackage[%
toc,%
sort=def,%
nonumberlist,%
]{glossaries}

\makeglossaries

\newglossaryentry{spin7-structure}{
  name={\ensuremath{\phi}},
  description = {The $\text{Spin}(7)$-structure four-form},
}

\newglossaryentry{metric}{
  name={\ensuremath{g_X}},
  description={The metric on the space $X$},
}

\newglossaryentry{frak-space}{
  name={\ensuremath{\mathfrak{A}_D}},
  description={The space of all fields on a closed $D$-manifold},
}

\newglossaryentry{map-space}{
  name={\ensuremath{\mathcal{M}(\Sigma, X)}},
  description={The space of all smooth maps from $\Sigma$ to $X$},
}

\newglossaryentry{x-soliton}{
  name={\ensuremath{X^{\theta}}-soliton},
  description={A $\theta$-deformed soliton defined on a sigma model with target $X$},
}

\newglossaryentry{x-sheet}{
  name={\ensuremath{X^{\theta}}-sheet},
  description={A $\theta$-deformed sheet defined on a sigma model with target $X$},
}

\newglossaryentry{x-threebrane}{
  name={\ensuremath{X^{\theta}}-threebrane},
  description={A $\theta$-deformed threebrane defined on a sigma model with target $X$},
}

\newglossaryentry{x-soliton-string}{
  name={\ensuremath{X^{\theta}}-soliton string},
  description={A $\theta$-deformed soliton string defined on an LG model with target $X$ corresponding to an $X^{\theta}$-soliton},
}

\newglossaryentry{x-soliton-membrane}{
  name={\ensuremath{X^{\theta}}-soliton membrane},
  description={A $\theta$-deformed soliton membrane defined on an LG model with target $X$ corresponding to an $X^{\theta}$-sheet},
}

\newglossaryentry{x-soliton-threebrane}{
  name={\ensuremath{X^{\theta}}-soliton threebrane},
  description={A $\theta$-deformed soliton threebrane defined on an LG model with target $X$ corresponding to an $X^{\theta}$-threebrane},
}

\newglossaryentry{config}{
  name={\ensuremath{\mathcal{E}^{*}_{\text{XY}}(\theta)}},
  description={A $\theta$-deformed XY configuration on a closed manifold},
}

\newglossaryentry{A-soliton}{
  name={\ensuremath{\Gamma^{IJ}(t, \theta, \mathfrak{A}_D)}},
  description={An $\mathfrak{A}_D^{\theta}$-soliton along the $t$-direction with endpoints corresponding to $\mathcal{E}^I_{\text{XY}}(\theta)$ and $\mathcal{E}^J_{\text{XY}}(\theta)$},
}

\newglossaryentry{A-sheet}{
  name={\ensuremath{\Sigma^{IJ, KL}(\tau, t, \theta, \mathfrak{A}_D)}},
  description={An $\mathfrak{A}_D^{\theta}$-sheet along the $(t, \tau)$-directions with
  (i) edges being $\Gamma^{IJ}(t, \theta, \mathfrak{A}_D)$ and $\Gamma^{KL}(t, \theta, \mathfrak{A}_D)$,
  and (ii) vertices corresponding to $\mathcal{E}^I_{\text{XY}}(\theta)$, $\mathcal{E}^J_{\text{XY}}(\theta)$, $\mathcal{E}^K_{\text{XY}}(\theta)$, and $\mathcal{E}^L_{\text{XY}}(\theta)$},
}

\newglossaryentry{A-threebrane}{
  name={\ensuremath{\Upsilon^{\{IJ, KL\}, \{MN, PQ\}}(\xi, \tau, t, \theta, \mathfrak{A}_D)}},
  description={An $\mathfrak{A}_D^{\theta}$-threebrane along the $(t, \tau, \xi)$-directions with
  (i) faces being $\Sigma^{IJ, KL}(\tau, t, \theta, \mathfrak{A}_D)$ and $\Sigma^{MN, PQ}(\tau, t, \theta, \mathfrak{A}_D)$,
  (ii) edges being $\Gamma^{IJ}(t, \theta, \mathfrak{A}_D)$, $\Gamma^{KL}(t, \theta, \mathfrak{A}_D)$, $\Gamma^{MN}(t, \theta, \mathfrak{A}_D)$, and $\Gamma^{PQ}(t, \theta, \mathfrak{A}_D)$,
  and (iii) vertices corresponding to $\mathcal{E}^I_{\text{XY}}(\theta)$, $\mathcal{E}^J_{\text{XY}}(\theta)$, $\mathcal{E}^K_{\text{XY}}(\theta)$, $\mathcal{E}^L_{\text{XY}}(\theta)$, $\mathcal{E}^M_{\text{XY}}(\theta)$, $\mathcal{E}^N_{\text{XY}}(\theta)$, $\mathcal{E}^P_{\text{XY}}(\theta)$, and $\mathcal{E}^Q_{\text{XY}}(\theta)$},
}

\newglossaryentry{M-path}{
  name={\ensuremath{\gamma^{IJ}(\theta, \mathfrak{A}_D)}},
  description={A $\theta$-deformed non-constant path in $\mathcal{M}(\R_t, \mathfrak{A}_D)$, and corresponds to $\Gamma^{IJ}(t, \theta, \mathfrak{A}_D)$},
}

\newglossaryentry{M-path-soliton}{
  name={\ensuremath{\sigma^{IJ, KL}(\tau, \theta, \mathfrak{A}_D)}},
  description={An $\mathcal{M}^{\theta}(\R_t, \mathfrak{A}_D)$-soliton along the $\tau$-direction with endpoints being $\gamma^{IJ}(\theta, \mathfrak{A}_D)$ and $\gamma^{KL}(\theta, \mathfrak{A}_D)$, and corresponds to $\Sigma^{IJ, KL}(\tau, t, \theta, \mathfrak{A}_D)$},
}

\newglossaryentry{M-double-path}{
  name={\ensuremath{\varsigma^{IJ, KL}(\theta, \mathfrak{A}_D)}},
  description={A $\theta$-deformed non-constant double path in $\mathcal{M}(\R_t \times \R_{\tau}, \mathfrak{A}_D)$, and corresponds to $\Sigma^{IJ, KL}(\tau, t, \theta, \mathfrak{A}_D)$},
}

\newglossaryentry{M-double-path-soliton}{
  name={\ensuremath{\Xi^{\{IJ, KL\}, \{MN, PQ\}}(\xi, \theta, \mathfrak{A}_D)}},
  description={An $\mathcal{M}^{\theta}(\R_t \times \R_{\tau}, \mathfrak{A}_D)$-soliton along the $\xi$-direction with endpoints being $\varsigma^{IJ, KL}(\theta, \mathfrak{A}_D)$ and $\varsigma^{MN, PQ}(\theta, \mathfrak{A}_D)$, and corresponds to $\Upsilon^{\{IJ, KL\}, \{MN, PQ\}}(\xi, \tau, t, \theta, \mathfrak{A}_D)$},
}

\newglossaryentry{LG-M-path-string}{
  name={\ensuremath{p^{IJ, KL}_{\text{XY}}(\theta)}},
  description={An LG $\mathcal{M}^{\theta}(\R_t, \mathfrak{A}_D)$-soliton string corresponding to $\sigma^{IJ, KL}(\tau, \theta, \mathfrak{A}_D)$.
  It can be interpreted as an intersection of thimbles corresponding to $\gamma^{IJ}(\theta, \mathfrak{A}_D)$ and $\gamma^{KL}(\theta, \mathfrak{A}_D)$, which, in turn, correspond to $\mathfrak{A}_D^{\theta}$-solitons with endpoints corresponding to $\mathcal{E}^{*}_{\text{XY}}(\theta)$},
}

\newglossaryentry{LG-A-string}{
  name={\ensuremath{q^{IJ}_{\text{XY}}(\theta)}},
  description={An LG $\mathfrak{A}_D^{\theta}$-soliton string corresponding to $\Gamma^{IJ}(t, \theta, \mathfrak{A}_D)$.
  It can be interpreted as an intersection of thimbles corresponding to $\mathcal{E}^I_{\text{XY}}$ and $\mathcal{E}^J_{\text{XY}}$},
}

\newglossaryentry{LG-A-membrane}{
  name={\ensuremath{\mathfrak{P}^{IJ, KL}_{\text{XY}}(\theta)}},
  description={An LG $\mathfrak{A}_D^{\theta}$-soliton membrane corresponding to $\Sigma^{IJ, KL}(\tau, t, \theta, \mathfrak{A}_D)$.
  It can be interpreted as an intersection of thimble-intersections $q^{IJ}_{\text{XY}}(\theta)$ and $q^{KL}_{\text{XY}}(\theta)$},
}

\newglossaryentry{LG-M-double-path-string}{
  name={\ensuremath{\mathfrak{p}^{\{IJ, KL\}, \{MN, PQ\}}_{\text{XY}}(\theta)}},
  description={An LG $\mathcal{M}^{\theta}(\C, \mathfrak{A}_D)$-soliton string corresponding to $\Xi^{\{IJ, KL\}, \{MN, PQ\}}(\xi, \theta, \mathfrak{A}_D)$.
  It can be interpreted as an intersection of thimbles corresponding to $\varsigma^{IJ, KL}(\theta, \mathfrak{A}_D)$ and $\varsigma^{MN, PQ}(\theta, \mathfrak{A}_D)$, which, in turn, correspond to $\mathfrak{A}_D^{\theta}$-sheets with vertices corresponding to $\mathcal{E}^{*}_{\text{XY}}(\theta)$},
}

\newglossaryentry{LG-A-threebrane}{
  name={\ensuremath{\mathfrak{Q}^{\{IJ, KL\}, \{MN, PQ\}}_{\text{XY}}(\theta)}},
  description={An LG $\mathfrak{A}_D^{\theta}$-soliton threebrane corresponding to $\Upsilon^{\{IJ, KL\}, \{MN, PQ\}}(\xi, \tau, t, \theta, \mathfrak{A}_D)$.
  It can be interpreted as an intersection of an ``intersecting pair of thimble-intersections'' $\mathfrak{P}^{IJ, KL}_{\text{XY}}(\theta)$ and $\mathfrak{P}^{MN, PQ}_{\text{XY}}(\theta)$},
}

\newglossaryentry{scattering-m-r-strings}{
  name={\ensuremath{\mu^n_{\mathfrak{A}_D}}},
  description={A tree-level scattering amplitude of $n$ incoming LG $\mathcal{M}^{\theta}(\R_t, \mathfrak{A}_D)$-soliton strings and a single outgoing LG $\mathcal{M}^{\theta}(\R_t, \mathfrak{A}_D)$-soliton string},
}

\newglossaryentry{scattering-a-membranes}{
  name={\ensuremath{\varPi^n_{\mathfrak{A}_D}}},
  description={A tree-level scattering amplitude of $n$ incoming LG $\mathfrak{A}_D^{\theta}$-soliton membranes and a single outgoing LG $\mathfrak{A}_D^{\theta}$-soliton membrane},
}

\newglossaryentry{scattering-m-c-strings}{
  name={\ensuremath{\mho^n_{\mathfrak{A}_D}}},
  description={A tree-level scattering amplitude of $n$ incoming LG $\mathcal{M}^{\theta}(\C, \mathfrak{A}_D)$-soliton strings and a single LG $\mathcal{M}^{\theta}(\C, \mathfrak{A}_D)$-soliton string},
}

\newglossaryentry{scattering-a-threebranes}{
  name={\ensuremath{\varDelta^n_{\mathfrak{A}_D}}},
  description={A tree-level scattering amplitude of $n$ incoming LG $\mathfrak{A}_D^{\theta}$-soliton threebranes and a single LG $\mathfrak{A}_D^{\theta}$-soliton threebranes},
}

\title{%
  Topological Gauge Theories with Sixteen Supercharges:
  \texorpdfstring{\bigskip \\}{}
  Higher \texorpdfstring{$A_{\infty}$}{A-infinity}-categorification of Floer Homologies
  \texorpdfstring{\vspace{2.0cm}}{}
}%

\author{Arif Er\thanks{
    Email: \href{mailto:arif.er@u.nus.edu}{arif.er@u.nus.edu}
  }
}%

\author{Meng-Chwan Tan\thanks{
    Email:\href{mailto:mctan@nus.edu.sg}{mctan@nus.edu.sg}
  }
}%

\affil{%
  Department of Physics \\ \medskip%
  National University of Singapore \\ \medskip%
  2 Science Drive 3, Singapore 117551%
}%
\date{}

\begin{document}
\addtolength{\baselineskip}{1.5mm}

\maketitle
\pagenumbering{gobble} 

\begin{abstract}
  \normalsize \singlespacing \noindent%
  This work is a sequel to~\cite{er-2024-topol-gauge-theor}, and a third and final installment of the program initiated in~\cite{er-2023-topol-n}.
  We show how, via a 3d gauged Landau-Ginzburg model interpretation of certain topologically-twisted 5d $\mathcal{N} = 2$ and 8d $\mathcal{N} = 1$ gauge theories, one can derive novel Fueter type $A_{\infty}$-2-categories that 2-categorify the 3d-Haydys-Witten, Haydys-Witten, and holomorphic Donaldson-Thomas Floer homology of two, four, and five-manifolds, respectively.
  Via a 2d gauged Landau-Ginzburg model interpretation of the aforementioned twisted gauge theories, these Fueter type $A_{\infty}$-2-categories can be shown to be equivalent to corresponding Fukaya-Seidel type $A_{\infty}$-categories.
  In the 8d case, one can also derive higher $A_{\infty}$-categories, such as a novel Cauchy-Riemann-Fueter type $A_{\infty}$-3-category that 3-categorifies the Haydys-Witten Floer homology of four-manifolds via a 4d gauged Landau-Ginzburg model interpretation of the theory.
  Together with previous results from~\cite{er-2024-topol-gauge-theor} and~\cite{er-2023-topol-n}, our work furnishes purely physical proofs and generalizations of the mathematical conjectures by Bousseau~\cite{bousseau-2024-holom-floer}, Doan-Rezchikov~\cite{doan-2022-holom-floer}, and Cao \cite{cao-2016-gauge-theor}.
\end{abstract}

\clearpage
\pagenumbering{arabic} 

\tableofcontents

\section{Introduction, Summary and Acknowledgements}
\vspace{0.4cm} \setlength{\parskip}{5pt}

\subtitle{Introduction}

The program of studying the $A_{\infty}$-categorical aspects of topologically-twisted gauge theories with sixteen supercharges began in \cite{er-2023-topol-n}, where we physically realized, from a 5d $\mathcal{N} = 2$ gauge theory, Haydys' construction of a \emph{gauge-theoretic} Fukaya-Seidel (FS) type $A_{\infty}$-category of three-manifolds categorifying a Floer homology of three-manifolds \cite{haydys-2015-fukay-seidel}.
In a sequel paper \cite{er-2024-topol-gauge-theor}, we were able to physically realize, from an 8d $\mathcal{N} = 1$ gauge theory, FS type $A_{\infty}$-categories of six, five, and four-manifolds categorifying Floer homologies of the said manifolds.

In the previous two works, we saw that to configurations on a $D$-manifold, $M_D$, one can associate a Floer homology of $M_D$ 0-category realized by the partition function of a gauge theory on $M_D \times \R$, which, in turn, can be categorified into an FS type $A_\infty$-category of $M_D$ 1-category realized by the partition function of a gauge theory on $M_D \times \R^2$.

This systematic process hints at the possibility of a higher categorification of Floer homologies simply by studying gauge theories with an increasing number of $\R$ lines. As a final installment to this program, we will do just this.

To this end, we will study, on various five and Spin$(7)$-manifolds\footnote{%
  A Spin$(7)$-manifold is an eight-manifold with Spin$(7)$ holonomy \cite[Prop. 10.5.3]{joyce-2000-compac-manif}, \cite{donaldson-1996-gauge}.
  \label{ft:spin7 manifold definition}
} with $\R^3$ submanifolds within, the aforementioned 5d $\mathcal{N} = 2$ and 8d $\mathcal{N} = 1$ topologically-twisted gauge theories whose BPS equations are the Haydys-Witten (HW) and Spin$(7)$ instanton equation, respectively.

The computational techniques we employ in this paper are mainly that of standard Kaluza-Klein reduction;
recasting gauge theories as supersymmetric quantum mechanics as pioneered in \cite{blau-1993-topol-gauge};
the physical realization of Floer homologies via supersymmetric quantum mechanics of gauge theories with sixteen supercharges as elucidated in \cite{ong-2023-vafa-witten-theor, er-2023-topol-n};
and the physical realization of FS type $A_{\infty}$-categories via a soliton string theory as elucidated in \cite{er-2023-topol-n, er-2024-topol-gauge-theor}.

Let us now give a brief plan and summary of the paper.

\subtitle{A Brief Plan and Summary of the Paper}


In \autoref{sec:topo theories}, we discuss certain topologically-twisted gauge theories with sixteen supercharges, where the gauge group $G$ is taken to be a real, simple, compact Lie group.
In particular, we discuss (i) the HW theory, i.e., a 5d $\mathcal{N} = 2$ theory on $M_5 = M_4 \times \R$ with an HW twist; and (ii) the Spin$(7)$ theory, i.e., an 8d $\mathcal{N} = 1$ theory on a Spin$(7)$-manifold\footnote{%
  See~\autoref{ft:spin7 manifold definition}.
} with a ``trivial'' twist.


In \autoref{sec:m2 x r3}, we consider HW theory where $M_4 = M_2 \times \R^2$, where $M_2$ is a closed and compact Riemann surface, and it is equivalently  recast as (i) a 3d gauged Landau-Ginzburg (LG) model on $\R^3$ with target the space $\mathfrak{A}_2$ of irreducible $(\mathscr{A}, C)$ fields on $M_2$, where $\mathscr{A} \in \Omega^1(M_2, \text{ad}(G_{\C}))$ and $C \in \Omega^0(M_2, \text{ad}(G))$ are a holomorphic gauge connection and real scalar, with $G_{\C}$ being the corresponding complex Lie group; or
(ii) a 2d gauged LG model on $\R^2$ with target the path space $\mathcal{M}(\R, \mathfrak{A}_2)$ of maps from $\R$ to $\mathfrak{A}_2$, which, in turn, can be recast as a 1d SQM theory  in the double path space $\mathcal{M}(\R^2, \mathfrak{A}_2)$ of double paths from $\R^2$ to $\mathfrak{A}_2$.

From the SQM and its critical points that can be interpreted as LG $\mathcal{M}^{\theta}(\R, \mathfrak{A}_2)$-solitons in the 2d gauged LG model, we obtain \eqref{eq:m2 x r3:2d lg:floer hom as morphism}:
\begin{equation}
  \label{summary:eq:m2 x r3:2d lg:floer hom as morphism}
  \boxed{
    \text{Hom} \left(
      \gamma^{IJ}(\mathfrak{A}_2), \gamma^{KL}(\mathfrak{A}_2)
    \right)_{\pm}
    \Longleftrightarrow
    \text{HF}^G_{d_m} \left(
      p^{IJ, KL}_{\text{BF}, \pm}
    \right)
  }
\end{equation}
Here, $\text{HF}^G_{d_m} (p^{IJ, KL}_{\text{BF}, \pm})$ is a Floer homology class, of degree $d_m$, generated by $p^{IJ, KL}_{\text{BF}, \pm}$, the intersection points of left and right thimbles representing LG $\mathcal{M}^{\theta}(\R, \mathfrak{A}_2)$-solitons that can be described as morphisms $\text{Hom}(\gamma^{IJ}(\mathfrak{A}_2), \gamma^{KL}(\mathfrak{A}_2))_{\pm}$ whose endpoints $\gamma^{**}(\mathfrak{A}_2)$ correspond to $\mathfrak{A}_2^{\theta}$-solitons.
Furthermore, via the aforementioned equivalent description of HW theory as a 2d gauged LG model, we can interpret the normalized 5d partition function as a sum over tree-level scattering amplitudes of LG $\mathcal{M}^{\theta}(\R, \mathfrak{A}_2)$-soliton strings given by the composition map of morphisms \eqref{eq:m2 x r3:2d lg:composition maps}:
\begin{equation}
  \label{summary:eq:m2 x r3:fs-cat composition maps}
  \boxed{
    \mu^{\mathfrak{N}_l}_{\mathfrak{A}_2}: \bigotimes_{i = 1}^{\mathfrak{N}_l}
    \text{Hom} \left(
      \gamma^{I_{2i - 1} I_{2i}}(\mathfrak{A}_2), \gamma^{I_{2(i + 1) - 1} I_{2(i + 1)}}(\mathfrak{A}_2)
    \right)_-
    \longto
    \text{Hom} \left(
      \gamma^{I_1 I_2}(\mathfrak{A}_2), \gamma^{I_{2\mathfrak{N}_l + 1} I_{2\mathfrak{N}_l + 2}}(\mathfrak{A}_2)
    \right)_+
  }
\end{equation}
where $\text{Hom}(\gamma^{**}(\mathfrak{A}_2), \gamma^{**}(\mathfrak{A}_2))_-$ and $\text{Hom}(\gamma^{**}(\mathfrak{A}_2), \gamma^{**}(\mathfrak{A}_2))_+$ represent incoming and outgoing scattering LG $\mathcal{M}^{\theta}(\R, \mathfrak{A}_2)$-soliton strings, as shown in \autoref{fig:m2 x r3:mu-d maps}.

The SQM and its critical points can also be interpreted as LG $\mathfrak{A}_2^{\theta}$-sheets in the 3d gauged LG model, from which we obtain \eqref{eq:m2 x r3:3d lg:2-morphism}:
\begin{equation}
  \label{summary:eq:m2 x r3:3d lg:2-morphism}
  \boxed{
    \text{Hom} \left(
      \Gamma^{IJ}(\tau, \mathfrak{A}_2),
      \Gamma^{KL}(\tau, \mathfrak{A}_2)
    \right)_\pm
    \Longleftrightarrow
    \text{Hom} \left(
      \text{Hom}(\mathcal{E}^I_{\text{BF}}, \mathcal{E}^J_{\text{BF}}),
      \text{Hom}(\mathcal{E}^K_{\text{BF}}, \mathcal{E}^L_{\text{BF}})
    \right)_\pm
    \Longleftrightarrow
    \text{HF}^G_{d_m} \left(
      \mathfrak{P}^{IJ, KL}_{\text{BF}, \pm}
    \right)
  }
\end{equation}
Here, $\text{HF}^G_{d_m}(\mathfrak{P}^{IJ, KL}_{\text{BF}\pm})$ is a Floer homology class, of degree $d_m$, generated by $\mathfrak{P}^{IJ, KL}_{\text{BF}, \pm}$, the intersection points of thimble-intersections representing LG $\mathfrak{A}_2^{\theta}$-sheets that can be described as
(i) 1-morphisms $\text{Hom} (\Gamma^{IJ}(\tau, \mathfrak{A}_2), \Gamma^{KL}(\tau, \mathfrak{A}_2) )_\pm$ whose edges $\Gamma^{**}(\tau, \mathfrak{A}_2)$ correspond to $\mathfrak{A}_2^{\theta}$-solitons, or
(ii) 2-morphisms $\text{Hom} (\text{Hom}(\mathcal{E}^I_{\text{BF}}, \mathcal{E}^J_{\text{BF}}), \text{Hom}(\mathcal{E}^K_{\text{BF}}, \mathcal{E}^L_{\text{BF}}) )_\pm$ whose vertices $\mathcal{E}^*_{\text{BF}}$ correspond to ($\theta$-deformed) $G_{\C}$-BF configurations on $M_2$ which generate a 3d-HW Floer homology.
Furthermore, via the aforementioned equivalent description of HW theory as a 3d gauged LG model, we can also interpret the normalized 5d partition function as a sum over tree-level scattering amplitudes of LG $\mathfrak{A}_2^{\theta}$-soliton membranes given by the composition map of 2-morphisms \eqref{eq:m2 x r3:fueter composition maps}:
\begin{equation}
  \label{summary:eq:m2 x r3:fueter composition maps}
  \boxed{
    \begin{aligned}
      \varPi^{\mathfrak{N}_l}_{\mathfrak{A}_2}: \bigotimes_{i = 1}^{\mathfrak{N}_l}
      & \text{Hom} \left(
        \text{Hom} \left( \mathcal{E}^{I_{2i - 1}}_{\text{BF}}, \mathcal{E}^{I_{2i}}_{\text{BF}} \right),
        \text{Hom} \left( \mathcal{E}^{I_{2(i + 1) - 1}}_{\text{BF}}, \mathcal{E}^{I_{2(i + 1)}}_{\text{BF}} \right)
        \right)_-
      \\
      &\longto
        \text{Hom} \left(
        \text{Hom} \left( \mathcal{E}^{I_1}_{\text{BF}}, \mathcal{E}^{I_2}_{\text{BF}} \right),
        \text{Hom} \left( \mathcal{E}^{I_{2 \mathfrak{N}_l + 1}}_{\text{BF}}, \mathcal{E}^{I_{2 \mathfrak{N}_l + 2}}_{\text{BF}} \right)
        \right)_+
    \end{aligned}
  }
\end{equation}
where $\text{Hom}\Big( \text{Hom}(\mathcal{E}^{*}_{\text{BF}}, \mathcal{E}^{*}_{\text{BF}}), \text{Hom}(\mathcal{E}^{*}_{\text{BF}}, \mathcal{E}^{*}_{\text{BF}}) \Big)_-$ and $\text{Hom}\Big( \text{Hom}(\mathcal{E}^{*}_{\text{BF}}, \mathcal{E}^{*}_{\text{BF}}), \text{Hom}(\mathcal{E}^{*}_{\text{BF}}, \mathcal{E}^{*}_{\text{BF}}) \Big)_+$ represent incoming and outgoing scattering LG $\mathfrak{A}_2^{\theta}$-membranes, as shown in \autoref{fig:m2 x r3:fueter composition maps}.

Note that \eqref{summary:eq:m2 x r3:2d lg:floer hom as morphism} and \eqref{summary:eq:m2 x r3:fs-cat composition maps} underlie a \emph{novel} FS type $A_{\infty}$-category of $\mathfrak{A}_2^{\theta}$-solitons, while \eqref{summary:eq:m2 x r3:3d lg:2-morphism} and \eqref{summary:eq:m2 x r3:fueter composition maps} underlie a \emph{novel} Fueter type $A_{\infty}$-2-category which 2-categorifies the 3d-HW Floer homology of $M_2$! Moreover, these categories ought to be equivalent!


In \autoref{sec:cy2 x s x r3}, we consider Spin$(7)$ theory on $CY_2 \times S^1 \times \R^3$, where $CY_2$ is a closed and compact Calabi-Yau twofold, and it is equivalently recast as (i) a 3d gauged LG model on $\R^3$ with target the space $\mathfrak{A}_5$ of irreducible $(\mathcal{A}, C)$ fields on $CY_2 \times S^1$, where $\mathcal{A} \in \Omega^{(1, 0)}(CY_2, \text{ad}(G)) \otimes \Omega^0(S^1 \times \R^3, \text{ad}(G))$ and $C \in \Omega^1(S^1, \text{ad}(G)) \otimes \Omega^0(CY_2 \times \R^3, \text{ad}(G))$; or (ii) a 2d gauged LG model on $\R^2$ with target the path space $\mathcal{M}(\R, \mathfrak{A}_5)$ of maps from $\R$ to $\mathfrak{A}_5$, which, in turn, can be recast as a 1d SQM theory  in the double path space $\mathcal{M}(\R^2, \mathfrak{A}_5)$ of double paths from $\R^2$ to $\mathfrak{A}_5$.

From the SQM and its critical points that can be interpreted as LG $\mathcal{M}^{\theta}(\R, \mathfrak{A}_5)$-solitons in the 2d gauged LG model, we obtain \eqref{eq:cy2 x s x r3:2d lg:floer hom as morphism}:
\begin{equation}
  \label{summary:eq:cy2 x s x r3:2d lg:floer hom as morphism}
  \boxed{
    \text{Hom} \left(
      \gamma^{IJ}(\mathfrak{A}_5), \gamma^{KL}(\mathfrak{A}_5)
    \right)_{\pm}
    \Longleftrightarrow
    \text{HF}^G_{d_u} \left(
      p^{IJ, KL}_{\text{HW}, \pm}
    \right)
  }
\end{equation}
Here, $\text{HF}^G_{d_u} (p^{IJ, KL}_{\text{HW}, \pm})$ is a Floer homology class, of degree $d_u$, generated by $p^{IJ, KL}_{\text{HW}, \pm}$, the intersection points of left and right thimbles representing LG $\mathcal{M}^{\theta}(\R, \mathfrak{A}_5)$-solitons that can be described as morphisms $\text{Hom}(\gamma^{IJ}(\mathfrak{A}_5), \gamma^{KL}(\mathfrak{A}_5))_{\pm}$ whose endpoints $\gamma^{**}(\mathfrak{A}_5)$ correspond to $\mathfrak{A}_5^{\theta}$-solitons.
Furthermore, via the aforementioned equivalent description of Spin$(7)$ theory as a 2d gauged LG model, we can interpret the normalized 8d partition function as a sum over tree-level scattering amplitudes of LG $\mathcal{M}^{\theta}(\R, \mathfrak{A}_5)$-soliton strings given by the composition map of morphisms \eqref{eq:cy2 x s x r3:2d lg:composition maps}:
\begin{equation}
  \label{summary:eq:cy2 x s x r3:fs-cat composition maps}
  \boxed{
    \mu^{\mathfrak{N}_m}_{\mathfrak{A}_5}: \bigotimes_{i = 1}^{\mathfrak{N}_m}
    \text{Hom} \left(
      \gamma^{I_{2i - 1} I_{2i}}(\mathfrak{A}_5), \gamma^{I_{2(i + 1) - 1} I_{2(i + 1)}}(\mathfrak{A}_5)
    \right)_-
    \longto
    \text{Hom} \left(
      \gamma^{I_1 I_2}(\mathfrak{A}_5), \gamma^{I_{2\mathfrak{N}_m + 1} I_{2\mathfrak{N}_m + 2}}(\mathfrak{A}_5)
    \right)_+
  }
\end{equation}
where $\text{Hom}(\gamma^{**}(\mathfrak{A}_5), \gamma^{**}(\mathfrak{A}_5))_-$ and $\text{Hom}(\gamma^{**}(\mathfrak{A}_5), \gamma^{**}(\mathfrak{A}_5))_+$ represent incoming and outgoing scattering LG $\mathcal{M}^{\theta}(\R, \mathfrak{A}_5)$-soliton strings.

The SQM and its critical points can also be interpreted as LG $\mathfrak{A}_5^{\theta}$-sheets in the 3d gauged LG model, from which we obtain \eqref{eq:cy2 x s x r3:3d lg:2-morphism}:
\begin{equation}
  \label{summary:eq:cy2 x s x r3:3d lg:2-morphism}
  \boxed{
    \text{Hom} \left(
      \Gamma^{IJ}(t, \mathfrak{A}_5),
      \Gamma^{KL}(t, \mathfrak{A}_5)
    \right)_\pm
    \Longleftrightarrow
    \text{Hom} \left(
      \text{Hom}(\mathcal{E}^I_{\text{HW}}, \mathcal{E}^J_{\text{HW}}),
      \text{Hom}(\mathcal{E}^K_{\text{HW}}, \mathcal{E}^L_{\text{HW}})
    \right)_\pm
    \Longleftrightarrow
    \text{HF}^G_{d_u} \left(
      \mathfrak{P}^{IJ, KL}_{\text{HW}, \pm}
    \right)
  }
\end{equation}
Here, $\text{HF}^G_{d_u}(\mathfrak{P}^{IJ, KL}_{\text{HW}\pm})$ is a Floer homology class, of degree $d_u$, generated by $\mathfrak{P}^{IJ, KL}_{\text{HW}, \pm}$, the intersection points of thimble-intersections representing LG $\mathfrak{A}_5^{\theta}$-sheets that can be described as
(i) 1-morphisms $\text{Hom} (\Gamma^{IJ}(t, \mathfrak{A}_5), \Gamma^{KL}(t, \mathfrak{A}_5) )_\pm$ whose edges $\Gamma^{**}(t, \mathfrak{A}_5)$ correspond to $\mathfrak{A}_5^{\theta}$-solitons, or
(ii) 2-morphisms $\text{Hom} (\text{Hom}(\mathcal{E}^I_{\text{HW}}, \mathcal{E}^J_{\text{HW}}), \text{Hom}(\mathcal{E}^K_{\text{HW}}, \mathcal{E}^L_{\text{HW}}) )_\pm$ whose vertices $\mathcal{E}^*_{\text{HW}}$ correspond to ($\theta$-deformed) HW configurations on $CY_2 \times S^1$ which generate a holomorphic Donaldson-Thomas (DT) Floer homology.
Furthermore, via the aforementioned equivalent description of Spin$(7)$ theory as a 3d gauged LG model, we can also interpret the normalized 8d partition function as a sum over tree-level scattering amplitudes of LG $\mathfrak{A}_5^{\theta}$-soliton membranes given by the composition map of 2-morphisms \eqref{eq:cy2 x s x r3:fueter composition maps}:
\begin{equation}
  \label{summary:eq:cy2 x s x r3:fueter composition maps}
  \boxed{
    \begin{aligned}
      \varPi^{\mathfrak{N}_m}_{\mathfrak{A}_5}: \bigotimes_{i = 1}^{\mathfrak{N}_m}
      & \text{Hom} \left(
        \text{Hom} \left( \mathcal{E}^{I_{2i - 1}}_{\text{HW}}, \mathcal{E}^{I_{2i}}_{\text{HW}} \right),
        \text{Hom} \left( \mathcal{E}^{I_{2(i + 1) - 1}}_{\text{HW}}, \mathcal{E}^{I_{2(i + 1)}}_{\text{HW}} \right)
        \right)_-
      \\
      &\longto
        \text{Hom} \left(
        \text{Hom} \left( \mathcal{E}^{I_1}_{\text{HW}}, \mathcal{E}^{I_2}_{\text{HW}} \right),
        \text{Hom} \left( \mathcal{E}^{I_{2 \mathfrak{N}_m + 1}}_{\text{HW}}, \mathcal{E}^{I_{2 \mathfrak{N}_m + 2}}_{\text{HW}} \right)
        \right)_+
    \end{aligned}
  }
\end{equation}
where $\text{Hom}\Big( \text{Hom}(\mathcal{E}^{*}_{\text{HW}}, \mathcal{E}^{*}_{\text{HW}}), \text{Hom}(\mathcal{E}^{*}_{\text{HW}}, \mathcal{E}^{*}_{\text{HW}}) \Big)_-$ and $\text{Hom}\Big( \text{Hom}(\mathcal{E}^{*}_{\text{HW}}, \mathcal{E}^{*}_{\text{HW}}), \text{Hom}(\mathcal{E}^{*}_{\text{HW}}, \mathcal{E}^{*}_{\text{HW}}) \Big)_+$ represent incoming and outgoing scattering LG $\mathfrak{A}_5^{\theta}$-membranes.

Note that \eqref{summary:eq:cy2 x s x r3:2d lg:floer hom as morphism} and \eqref{summary:eq:cy2 x s x r3:fs-cat composition maps} underlie a \emph{novel} FS type $A_{\infty}$-category of $\mathfrak{A}_5^{\theta}$-solitons, while \eqref{summary:eq:cy2 x s x r3:3d lg:2-morphism} and \eqref{summary:eq:cy2 x s x r3:fueter composition maps} underlie a \emph{novel} Fueter type $A_{\infty}$-2-category which 2-categorifies the holomorphic DT Floer homology of $CY_2 \times S^1$! Moreover, these categories ought to be equivalent!


In \autoref{sec:cy2 x r3}, we perform a Kaluza-Klein dimensional reduction of Spin$(7)$ theory on $CY_2 \times S^1 \times \R^3$ along the $S^1$ circle by shrinking it to be infinitesimally small. The resulting 7d-Spin$(7)$ theory on $CY_2 \times \R^3$ is equivalently recast as (i) a 3d gauged LG model on $\R^3$ with target the space $\mathfrak{A}_4$ of irreducible $(\mathcal{A}, C)$ fields on $CY_2$, where $\mathcal{A} \in \Omega^{(1, 0)}(CY_2, \text{ad}(G)) \otimes \Omega^0(\R^3, \text{ad}(G))$ and $C \in \Omega^0(CY_2 \times \R^3, \text{ad}(G))$;
or (ii) a 2d gauged LG model on $\R^2$ with target the path space $\mathcal{M}(\R, \mathfrak{A}_4)$ of maps from $\R$ to $\mathfrak{A}_4$, which, in turn, can be recast as a 1d SQM theory  in the double path space $\mathcal{M}(\R^2, \mathfrak{A}_4)$ of double paths from $\R^2$ to $\mathfrak{A}_4$.

From the SQM and its critical points that can be interpreted as LG $\mathcal{M}^{\theta}(\R, \mathfrak{A}_4)$-solitons in the 2d gauged LG model, we obtain \eqref{eq:cy2 x r3:2d lg:floer hom as morphism}:
\begin{equation}
  \label{summary:eq:cy2 x r3:2d lg:floer hom as morphism}
  \boxed{
    \text{Hom} \left(
      \gamma^{IJ}(\mathfrak{A}_4), \gamma^{KL}(\mathfrak{A}_4)
    \right)_{\pm}
    \Longleftrightarrow
    \text{HF}^G_{d_v} \left(
      p^{IJ, KL}_{\text{VW}, \pm}
    \right)
  }
\end{equation}
Here, $\text{HF}^G_{d_v} (p^{IJ, KL}_{\text{VW}, \pm})$ is a Floer homology class, of degree $d_v$, generated by $p^{IJ, KL}_{\text{VW}, \pm}$, the intersection points of left and right thimbles representing LG $\mathcal{M}^{\theta}(\R, \mathfrak{A}_4)$-solitons that can be described as morphisms $\text{Hom}(\gamma^{IJ}(\mathfrak{A}_4), \gamma^{KL}(\mathfrak{A}_4))_{\pm}$ whose endpoints $\gamma^{**}(\mathfrak{A}_4)$ correspond to $\mathfrak{A}_4^{\theta}$-solitons.
Furthermore, via the aforementioned equivalent description of 7d-Spin$(7)$ theory as a 2d gauged LG model, we can interpret the normalized 7d partition function as a sum over tree-level scattering amplitudes of LG $\mathcal{M}^{\theta}(\R, \mathfrak{A}_4)$-soliton strings given by the composition map of morphisms \eqref{eq:cy2 x r3:2d lg:composition maps}:
\begin{equation}
  \label{summary:eq:cy2 x r3:fs-cat composition maps}
  \boxed{
    \mu^{\mathfrak{N}_n}_{\mathfrak{A}_4}: \bigotimes_{i = 1}^{\mathfrak{N}_n}
    \text{Hom} \left(
      \gamma^{I_{2i - 1} I_{2i}}(\mathfrak{A}_4), \gamma^{I_{2(i + 1) - 1} I_{2(i + 1)}}(\mathfrak{A}_4)
    \right)_-
    \longto
    \text{Hom} \left(
      \gamma^{I_1 I_2}(\mathfrak{A}_4), \gamma^{I_{2\mathfrak{N}_n + 1} I_{2\mathfrak{N}_n + 2}}(\mathfrak{A}_4)
    \right)_+
  }
\end{equation}
where $\text{Hom}(\gamma^{**}(\mathfrak{A}_4), \gamma^{**}(\mathfrak{A}_4))_-$ and $\text{Hom}(\gamma^{**}(\mathfrak{A}_4), \gamma^{**}(\mathfrak{A}_4))_+$ represent incoming and outgoing scattering LG $\mathcal{M}^{\theta}(\R, \mathfrak{A}_4)$-soliton strings.

The SQM and its critical points can also be interpreted as LG $\mathfrak{A}_4^{\theta}$-sheets in the 3d gauged LG model, from which we obtain \eqref{eq:cy2 x r3:3d lg:2-morphism}:
\begin{equation}
  \label{summary:eq:cy2 x r3:3d lg:2-morphism}
  \boxed{
    \text{Hom} \left(
      \Gamma^{IJ}(t, \mathfrak{A}_4),
      \Gamma^{KL}(t, \mathfrak{A}_4)
    \right)_\pm
    \Longleftrightarrow
    \text{Hom} \left(
      \text{Hom}(\mathcal{E}^I_{\text{VW}}, \mathcal{E}^J_{\text{VW}}),
      \text{Hom}(\mathcal{E}^K_{\text{VW}}, \mathcal{E}^L_{\text{VW}})
    \right)_\pm
    \Longleftrightarrow
    \text{HF}^G_{d_v} \left(
      \mathfrak{P}^{IJ, KL}_{\text{VW}, \pm}
    \right)
  }
\end{equation}
Here, $\text{HF}^G_{d_v}(\mathfrak{P}^{IJ, KL}_{\text{VW}\pm})$ is a Floer homology class, of degree $d_v$, generated by $\mathfrak{P}^{IJ, KL}_{\text{VW}, \pm}$, the intersection points of thimble-intersections representing LG $\mathfrak{A}_4^{\theta}$-sheets that can be described as
(i) 1-morphisms $\text{Hom} (\Gamma^{IJ}(t, \mathfrak{A}_4), \Gamma^{KL}(t, \mathfrak{A}_4) )_\pm$ whose edges $\Gamma^{**}(t, \mathfrak{A}_4)$ correspond to $\mathfrak{A}_4^{\theta}$-solitons, or
(ii) 2-morphisms $\text{Hom} (\text{Hom}(\mathcal{E}^I_{\text{VW}}, \mathcal{E}^J_{\text{VW}}), \text{Hom}(\mathcal{E}^K_{\text{VW}}, \mathcal{E}^L_{\text{VW}}) )_\pm$ whose vertices $\mathcal{E}^*_{\text{VW}}$ correspond to ($\theta$-deformed) VW configurations on $CY_2$ which generate a HW Floer homology.
Furthermore, via the aforementioned equivalent description of 7d-Spin$(7)$ theory as a 3d gauged LG model, we can also interpret the normalized 7d partition function as a sum over tree-level scattering amplitudes of LG $\mathfrak{A}_4^{\theta}$-soliton membranes given by the composition map of 2-morphisms \eqref{eq:cy2 x r3:fueter composition maps}:
\begin{equation}
  \label{summary:eq:cy2 x r3:fueter composition maps}
  \boxed{
    \begin{aligned}
      \varPi^{\mathfrak{N}_n}_{\mathfrak{A}_4}: \bigotimes_{i = 1}^{\mathfrak{N}_n}
      & \text{Hom} \left(
        \text{Hom} \left( \mathcal{E}^{I_{2i - 1}}_{\text{VW}}, \mathcal{E}^{I_{2i}}_{\text{VW}} \right),
        \text{Hom} \left( \mathcal{E}^{I_{2(i + 1) - 1}}_{\text{VW}}, \mathcal{E}^{I_{2(i + 1)}}_{\text{VW}} \right)
        \right)_-
      \\
      &\longto
        \text{Hom} \left(
        \text{Hom} \left( \mathcal{E}^{I_1}_{\text{VW}}, \mathcal{E}^{I_2}_{\text{VW}} \right),
        \text{Hom} \left( \mathcal{E}^{I_{2 \mathfrak{N}_n + 1}}_{\text{VW}}, \mathcal{E}^{I_{2 \mathfrak{N}_n + 2}}_{\text{VW}} \right)
        \right)_+
    \end{aligned}
  }
\end{equation}
where $\text{Hom}\Big( \text{Hom}(\mathcal{E}^{*}_{\text{VW}}, \mathcal{E}^{*}_{\text{VW}}), \text{Hom}(\mathcal{E}^{*}_{\text{VW}}, \mathcal{E}^{*}_{\text{VW}}) \Big)_-$ and $\text{Hom}\Big( \text{Hom}(\mathcal{E}^{*}_{\text{VW}}, \mathcal{E}^{*}_{\text{VW}}), \text{Hom}(\mathcal{E}^{*}_{\text{VW}}, \mathcal{E}^{*}_{\text{VW}}) \Big)_+$ represent incoming and outgoing scattering LG $\mathfrak{A}_4^{\theta}$-membranes.

Note that \eqref{summary:eq:cy2 x r3:2d lg:floer hom as morphism} and \eqref{summary:eq:cy2 x r3:fs-cat composition maps} underlie a \emph{novel} FS type $A_{\infty}$-category of $\mathfrak{A}_4^{\theta}$-solitons, while \eqref{summary:eq:cy2 x r3:3d lg:2-morphism} and \eqref{summary:eq:cy2 x r3:fueter composition maps} underlie a \emph{novel} Fueter type $A_{\infty}$-2-category which 2-categorifies the HW Floer homology of $CY_2$! Moreover, these categories ought to be equivalent!


In \autoref{sec:cy2 x r4}, we consider Spin$(7)$ theory on $CY_2 \times \R^4$, and equivalently recast it as (i) a 4d gauged LG model on $\R^4$ with target the space $\mathfrak{A}_4$ of irreducible $\mathcal{A}$ fields on $CY_2$, where $\mathcal{A} \in \Omega^{(1, 0)}(CY_2, \text{ad}(G)) \otimes \Omega^0(\R^4, \text{ad}(G))$;
or (ii) a 2d gauged LG model on $\R^2$ with target the complex path space $\mathcal{M}(\C, \mathfrak{A}_4)$ of maps from $\C$ to $\mathfrak{A}_4$, which, in turn, can be recast as a 1d SQM theory in the triple path space $\mathcal{M}(\R^3, \mathfrak{A}_4)$ of triple paths from $\R^3$ to $\mathfrak{A}_4$.

From the SQM and its critical points that can be interpreted as LG $\mathcal{M}^{\theta}(\C, \mathfrak{A}_4)$-solitons in the 2d gauged LG model, we obtain \eqref{eq:cy2 x r4:2d lg:floer hom as morphism}:
\begin{equation}
  \label{summary:eq:cy2 x r4:2d lg:floer hom as morphism}
  \boxed{
    \text{Hom} \left(
      \varsigma^{IJ, KL}(\mathfrak{A}_4),
      \varsigma^{MN, PQ}(\mathfrak{A}_4)
    \right)_{\pm}
    \Longleftrightarrow
    \text{HF}^G_{d_v} \left(
      \mathfrak{p}^{\{IJ, KL\}, \{MN, PQ\}}_{\text{VW}, \pm}
    \right)
  }
\end{equation}
Here, $\text{HF}^G_{d_v} (\mathfrak{p}^{\{IJ, KL\}, \{MN, PQ\}}_{\text{VW}, \pm})$ is a Floer homology class, of degree $d_v$, generated by $\mathfrak{p}^{\{IJ, KL\}, \{MN, PQ\}}_{\text{VW}, \pm}$, the intersection points of left and right thimbles representing LG $\mathcal{M}^{\theta}(\C, \mathfrak{A}_4)$-solitons that can be described as morphisms $\text{Hom}(\varsigma^{IJ, KL}(\mathfrak{A}_4), \varsigma^{MN, PQ}(\mathfrak{A}_4))_{\pm}$ whose endpoints $\varsigma^{**, **}(\mathfrak{A}_4)$ correspond to $\mathfrak{A}_4^{\theta}$-sheets.
Furthermore, via the aforementioned equivalent description of Spin$(7)$ theory as a 2d gauged LG model, we can interpret the normalized 8d partition function as a sum over tree-level scattering amplitudes of LG $\mathcal{M}^{\theta}(\C, \mathfrak{A}_4)$-soliton strings given by the composition map of morphisms \eqref{eq:cy2 x r4:2d lg:composition maps}:
\begin{equation}
  \label{summary:eq:cy2 x r4:fs-cat composition maps}
  \boxed{
    \begin{aligned}
      \mho^{\mathfrak{K}_n}_{\mathfrak{A}_4} : \bigotimes_{i = 1}^{\mathfrak{K}_n}
      & \text{Hom} \left(
        \varsigma^{\{I_{4i - 3} I_{4i - 2}\}, \{I_{4i - 1} I_{4i}\}}(\mathfrak{A}_4),
        \varsigma^{\{I_{4(i + 1) - 3} I_{4(i + 1) - 2}\}, \{I_{4(i + 1) - 1} I_{4(i + 1)}\}}(\mathfrak{A}_4)
      \right)_-
      \\
      & \longto
      \text{Hom} \left(
        \varsigma^{\{I_1 I_2\}, \{I_3 I_4 \}}(\mathfrak{A}_4),
        \varsigma^{\{I_{4\mathfrak{K}_n + 1} I_{4\mathfrak{K}_n + 2}\}, \{I_{4\mathfrak{K}_n + 3} I_{4\mathfrak{K}_n + 4}\}}(\mathfrak{A}_4)
      \right)_+
    \end{aligned}
  }
\end{equation}
where $\text{Hom}(\varsigma^{**, **}(\mathfrak{A}_4), \varsigma^{**, **}(\mathfrak{A}_4))_-$ and $\text{Hom}(\varsigma^{**, **}(\mathfrak{A}_4), \varsigma^{**, **}(\mathfrak{A}_4))_+$ represent incoming and outgoing scattering LG $\mathcal{M}^{\theta}(\C, \mathfrak{A}_4)$-soliton strings.

The SQM and its critical points can also be interpreted as LG $\mathfrak{A}_4^{\theta}$-threebranes in the 4d gauged LG model, from which we obtain \eqref{eq:cy2 x r4:4d-lg:3-morphism}:
\begin{equation}
  \label{summary:eq:cy2 x r4:4d lg:3-morphism}
  \boxed{
    \begin{gathered}
      \text{Hom} \left[
        \Sigma^{IJ, KL}(\tau, t, \mathfrak{A}_4),
        \Sigma^{MN, PQ}(\tau, t, \mathfrak{A}_4)
      \right]_{\pm}
      \\
      \Updownarrow
      \\
      \text{Hom} \left[
        \text{Hom} \left(
          \Gamma^{IJ}(t, \mathfrak{A}_4),
          \Gamma^{KL}(t, \mathfrak{A}_4)
        \right),
        \text{Hom} \left(
          \Gamma^{MN}(t, \mathfrak{A}_4),
          \Gamma^{PQ}(t, \mathfrak{A}_4)
        \right)
      \right]_{\pm}
      \\
      \Updownarrow
      \\
      \text{Hom} \left[
        \text{Hom} \Big(
          \text{Hom} \left(
            \mathcal{E}^I_{\text{VW}},
            \mathcal{E}^J_{\text{VW}}
          \right),
          \text{Hom} \left(
            \mathcal{E}^K_{\text{VW}},
            \mathcal{E}^L_{\text{VW}}
          \right)
        \Big),
        \text{Hom} \Big(
          \text{Hom} \left(
            \mathcal{E}^M_{\text{VW}},
            \mathcal{E}^N_{\text{VW}}
          \right),
          \text{Hom} \big(
            \mathcal{E}^P_{\text{VW}},
            \mathcal{E}^Q_{\text{VW}}
          \big)
        \Big)
      \right]_{\pm}
      \\
      \Updownarrow
      \\
      \text{HF}^G_{d_v} \left(
        \mathfrak{Q}^{\{IJ, KL\}, \{MN, PQ\}}_{\text{VW}, \pm}
      \right)
    \end{gathered}
  }
\end{equation}
Here, $\text{HF}^G_{d_v}(\mathfrak{Q}^{\{IJ, KL\}, \{MN, PQ\}}_{\text{VW}, \pm})$ is a Floer homology class, of degree $d_v$, generated by $\mathfrak{Q}^{\{IJ, KL\}, \{MN, PQ\}}_{\text{VW}, \pm}$, the intersection points of ``intersecting pairs of thimble-intersections'' representing LG $\mathfrak{A}_4^{\theta}$-threebranes that can be described as
(i) 1-morphisms $\text{Hom}[ \Sigma^{IJ, KL}(\tau, t, \mathfrak{A}_4), \Sigma^{MN, PQ}(\tau, t, \mathfrak{A}_4)]_{\pm}$ whose faces correspond to $\mathfrak{A}_4^{\theta}$-sheets,
(ii) 2-morphisms $\text{Hom}\Big[ \text{Hom} \Big(\Gamma^{IJ}(t, \mathfrak{A}_4), \Gamma^{KL}(t, \mathfrak{A}_4) \Big), \text{Hom} \Big(\Gamma^{IJ}(t, \mathfrak{A}_4), \Gamma^{KL}(t, \mathfrak{A}_4) \Big) \Big]_\pm$ whose edges $\Gamma^{**}(t, \mathfrak{A}_4)$ correspond to $\mathfrak{A}_4^{\theta}$-solitons, or
(iii) 3-morphisms $\text{Hom}\Big[ \text{Hom}\Big( \text{Hom}(\mathcal{E}^I_{\text{VW}}, \mathcal{E}^J_{\text{VW}}),$ \\ $\text{Hom}(\mathcal{E}^K_{\text{VW}}, \mathcal{E}^L_{\text{VW}}) \Big), \text{Hom}\Big( \text{Hom}(\mathcal{E}^M_{\text{VW}}, \mathcal{E}^N_{\text{VW}}), \text{Hom}(\mathcal{E}^P_{\text{VW}}, \mathcal{E}^Q_{\text{VW}}) \Big) \Big]_\pm$ whose vertices $\mathcal{E}^*_{\text{VW}}$ correspond to ($\theta$-deformed) VW configurations on $CY_2$ which generate a HW Floer homology.
Furthermore, via the aforementioned equivalent description of Spin$(7)$ theory as a 4d gauged LG model, we can also interpret the normalized 8d partition function as a sum over tree-level scattering amplitudes of LG $\mathfrak{A}_4^{\theta}$-soliton threebranes given by the composition map of 3-morphisms \eqref{eq:cy2 x r4:4d lg:crf composition maps}:
\begin{equation}
  \label{summary:eq:cy2 x r4:4d lg:crf composition maps}
  \boxed{
    \begin{aligned}
      \varDelta^{\mathfrak{K}_n}_{\mathfrak{A}_4}
      : \bigotimes_{i = 1}^{\mathfrak{K}_n}
      & \text{Hom} \bigg[
        \text{Hom} \left(
        \text{Hom} \left(
        \mathcal{E}^{I_{4i-3}}_{\text{VW}},
        \mathcal{E}^{I_{4i-2}}_{\text{VW}}
        \right),
        \text{Hom} \left(
        \mathcal{E}^{I_{4i-1}}_{\text{VW}},
        \mathcal{E}^{I_{4i}}_{\text{VW}}
        \right)
        \right),
      \\
      & \qquad \quad
        \text{Hom} \left(
        \mathcal{E}^{I_{4(i + 1)-3}}_{\text{VW}},
        \mathcal{E}^{I_{4(i + 1)-2}}_{\text{VW}}
        \right),
        \text{Hom} \left(
        \mathcal{E}^{I_{4(i + 1)-1}}_{\text{VW}},
        \mathcal{E}^{I_{4(i + 1)}}_{\text{VW}}
        \right)
        \bigg]_-
      \\
      & \longrightarrow
        \text{Hom} \bigg[
        \text{Hom} \left(
        \text{Hom} \left(
        \mathcal{E}^{I_1}_{\text{VW}},
        \mathcal{E}^{I_2}_{\text{VW}}
        \right),
        \text{Hom} \left(
        \mathcal{E}^{I_3}_{\text{VW}},
        \mathcal{E}^{I_4}_{\text{VW}}
        \right)
        \right),
      \\
      & \qquad \qquad \quad
        \text{Hom} \left(
        \mathcal{E}^{I_{4\mathfrak{K}_n + 1}}_{\text{VW}},
        \mathcal{E}^{I_{4\mathfrak{K}_n + 2}}_{\text{VW}}
        \right),
        \text{Hom} \left(
        \mathcal{E}^{I_{4\mathfrak{K}_n + 3}}_{\text{VW}},
        \mathcal{E}^{I_{4\mathfrak{K}_n + 4}}_{\text{VW}}
        \right)
        \bigg]_+
    \end{aligned}
  }
\end{equation}
where $\text{Hom}\Big[ \text{Hom} \Big( \text{Hom}(\mathcal{E}^{*}_{\text{VW}}, \mathcal{E}^{*}_{\text{VW}}), \text{Hom}(\mathcal{E}^{*}_{\text{VW}}, \mathcal{E}^{*}_{\text{VW}}) \Big), \text{Hom}\Big( \text{Hom}(\mathcal{E}^{*}_{\text{VW}}, \mathcal{E}^{*}_{\text{VW}}), \text{Hom}(\mathcal{E}^{*}_{\text{VW}}, \mathcal{E}^{*}_{\text{VW}}) \Big) \Big]_-$ and $\text{Hom}\Big[ \text{Hom} \Big( \text{Hom}(\mathcal{E}^{*}_{\text{VW}}, \mathcal{E}^{*}_{\text{VW}}), \text{Hom}(\mathcal{E}^{*}_{\text{VW}}, \mathcal{E}^{*}_{\text{VW}}) \Big), \text{Hom}\Big( \text{Hom}(\mathcal{E}^{*}_{\text{VW}}, \mathcal{E}^{*}_{\text{VW}}), \text{Hom}(\mathcal{E}^{*}_{\text{VW}}, \mathcal{E}^{*}_{\text{VW}}) \Big) \Big]_+$ represent incoming and outgoing scattering LG $\mathfrak{A}_4^{\theta}$-threebranes.

Note that \eqref{summary:eq:cy2 x r4:2d lg:floer hom as morphism} and \eqref{summary:eq:cy2 x r4:fs-cat composition maps} underlie a \emph{novel} FS type $A_{\infty}$-category of $\mathfrak{A}_4^{\theta}$-sheets, while \eqref{summary:eq:cy2 x r4:4d lg:3-morphism} and \eqref{summary:eq:cy2 x r4:4d lg:crf composition maps} underlie a \emph{novel} Cauchy-Riemann-Fueter type $A_{\infty}$-3-category which 3-categorifies the HW Floer homology of $CY_2$!
Moreover, these categories ought to be equivalent!


In \autoref{sec:proofs}, we find, in \autoref{sec:proofs:atiyah-floer}, that we can interpret~\cite[eqn.~(9.43)]{er-2023-topol-n} and \cite[eqn~(12.15)]{er-2024-topol-gauge-theor} as Bousseau-Doan-Rezchikov's Fueter 2-category.
This will then allow us to prove and generalize Bousseau's mathematical conjectures in \cite[Conjectures 2.14 and 2.15]{bousseau-2024-holom-floer}.
In \autoref{sec:proofs:equivalences}, using the results of \autoref{sec:proofs:atiyah-floer}, we prove the mathematical conjectures by Bousseau \cite[Conjecture 2.10]{bousseau-2024-holom-floer} and Doan-Rezchikov \cite[Conjecture 1.5]{doan-2022-holom-floer}.
And finally, via the results of \autoref{sec:m2 x r3}--\autoref{sec:cy2 x r4}, we will arrive at a gauge-theoretic generalization of the latter and Cao's mathematical conjecture in \cite[Proposal 1.2.2 and Remark 1.2.3]{cao-2016-gauge-theor}!

\subtitle{Acknowledgements}

We would like to thank R.P. Thomas and D. Joyce for useful discussions.
We would also like to thank the ATMP referee for suggesting improvements to our paper.
This work is supported in part by the MOE AcRF Tier 1 grant R-144-000-470-114.

\section{The Topologically-twisted Gauge Theories}
\label{sec:topo theories}

In this section, we will introduce the various topologically-twisted gauge theories that we will use in the rest of this paper, where the gauge group is taken to be a real, simple, and compact Lie group $G$.

\subsection{Haydys-Witten Theory: A Topologically-twisted 5d \texorpdfstring{$\mathcal{N} = 2$}{N = 2} Gauge Theory}
\label{sec:topo theory:hw}

The first topologically-twisted gauge theory that we will introduce, is a topologically-twisted 5d $\mathcal{N} = 2$ gauge theory on a five manifold $M_5 = M_4 \times \R$ with a Haydys-Witten (HW) twist,\footnote{%
  We refer the reader to~\cite{er-2023-topol-n, anderson-2013-five-dimen, elliott-2022-taxon-twist} for a detailed description of the twist.
  \label{ft:hw twist}
}
whose BPS equations are the HW equations.
We shall henceforth refer to this theory as HW theory.

The bosonic field content of HW theory consists of scalars $\upsilon, \bar{\upsilon}\in \Omega^0 (M_4 \times \R, \text{ad}(G))$, gauge connections $A_{\mu} \in \Omega^1 (M_4, \text{ad}(G)) \otimes \Omega^0(\R, \text{ad}(G))$ and $A_t \in \Omega^0(M_4, \text{ad}(G)) \otimes \Omega^1(\R, \text{ad}(G))$, and a self-dual two-form $B_{\mu\nu}\in \Omega^{2,+}(M_4, \text{ad}(G)) \otimes \Omega^0(\R, \text{ad}(G))$.
The fermionic field content consists of scalars $\eta, \tilde{\eta}\in \Omega^0 (M_4 \times \R, \text{ad}(G))$, one-forms $\psi_{\mu}, \tilde{\psi}_{\mu} \in \Omega^1(M_4, \text{ad}(G)) \otimes \Omega^0(\R, \text{ad}(G))$, and self-dual two-forms $\chi_{\mu\nu}, \tilde{\chi}_{\mu\nu}\in\Omega^{2, +}(M_4, \text{ad}(G)) \otimes \Omega^0(\R, \text{ad}(G))$.
Here, $\mu \in \{1, \dots 4\}$ are the indices of $M_4$, $t$ is the coordinate along $\R$, and $\text{ad}(G)$ is the adjoint bundle of the underlying principal $G$-bundle.

As the supersymmetry generators transform in the same representation as the fermions above, the HW twist will result in two, linearly-dependent, \emph{scalar} supersymmetry generators $\mathcal{Q}$ and $\bar{\mathcal{Q}}$.
Thus, we can just choose $\mathcal{Q}$ to define the twisted theory. Its supersymmetry transformations of the twisted fields work out to be~\cite{er-2023-topol-n}
\begin{equation}
 \label{eq:hw susy variations}
  \begin{aligned}
    \delta \upsilon
    & = 0
      \, ,
    & \qquad
     \delta \eta
    & = -2i D_t \upsilon
      \, , \\
    \delta \bar{\upsilon}
    & = -i \tilde{\eta}
      \, ,
    & \qquad
     \delta \tilde{\eta}
    & = - 2[\upsilon, \bar{\upsilon}]
      \, , \\
    \delta A_{\mu}
    & = i \psi_{\mu}
      \, ,
    & \qquad
     \delta \psi_{\mu}
    & = -2 D_{\mu}\upsilon
      \, , \\
    \delta A_t
    & = \eta
      \, ,
    & \qquad
     \delta \tilde{\psi}_{\mu}
    & = i \left( F_{t\mu} + D^{\nu}B_{\nu\mu} \right)
      \, , \\
    \delta B_{\mu\nu}
    & = 2 {\chi}_{\mu\nu}
      \, ,
    & \qquad
     \delta \chi_{\mu\nu}
    & = -i [B_{\mu\nu}, \upsilon]
      \, , \\
    & &
    \delta \tilde{\chi}_{\mu\nu}
    &= - \left(
        F^+_{\mu\nu}
        - \frac{1}{4}[B_{\mu\rho}, B_{\nu}^{\rho}]
        - \frac{1}{2}D_t B_{\mu\nu}
      \right)
      \, ,
  \end{aligned}
\end{equation}
where $\delta$ denotes a $\mathcal{Q}$-variation; $F^+_{\mu\nu} = \frac{1}{2} (F_{\mu\nu} + \frac{1}{2} \epsilon_{\mu\nu\rho\lambda} F^{\rho\lambda})$ is the self-dual component of $F_{\mu\nu}$; self-duality of the self-dual fields $\Phi = \{B, \chi, \tilde{\chi}\}$ means that $\Phi_{\mu\nu} = \frac{1}{2} \epsilon_{\mu\nu\rho\lambda} \Phi^{\rho\lambda}$;
and $[B_{\mu\rho}, B_{\nu}^{\rho}] \equiv g^{\rho\varrho}[B_{\mu\rho}, B_{\nu\varrho}]$.

With auxiliary fields, a straightforward calculation will show that $\mathcal{Q}$ is nilpotent up to gauge transformations generated by $\upsilon$.
As we wish to study the theory whereby the relevant moduli space is well-behaved (which corresponds to no reducible connections), we shall consider the case where $\upsilon$ has no zero-modes.

Setting to zero the $\mathcal{Q}$-variations of the fermions in~\eqref{eq:hw susy variations}, we obtain the BPS equations of HW theory as\footnote{%
  As we are only considering the case where $\upsilon$ has no zero-modes, we can take it to be zero in the variations of the fermions.
  \label{ft:hw ignore variations}
}
\begin{equation}
  \label{eq:hw eqns}
  \begin{aligned}
    F_{t\mu} + D^{\nu}B_{\nu\mu}
    &= 0
      \, , \\
    F^+_{\mu\nu}
    - \frac{1}{4} [B_{\mu\rho}, B_{\nu}^{\rho}]
    - \frac{1}{2}D_t B_{\mu\nu}
    &= 0
      \, . \\
  \end{aligned}
\end{equation}
These are the HW equations on $M_4 \times \R$.
Configurations of $(A_\mu, B_{\mu\nu})$ satisfying~\eqref{eq:hw eqns} constitute a moduli space $\mathcal{M}_{\text{HW}}$ that the HW theory path integral will localize onto.

Finally, the $\mathcal{Q}$-exact topological action of HW theory is given as~\cite{er-2023-topol-n, anderson-2013-five-dimen}
\begin{equation}
  \label{eq:hw action}
  \begin{aligned}
    S_{\text{HW}}
    = \frac{1}{e^2}
    & \int_{M_4 \times \R} dt d^4x \, \Tr \bigg(
      \frac{1}{2} \left|
        F^+_{\mu\nu}
        - \frac{1}{4} [B_{\mu\rho}, B^{\rho}_{\nu}]
        - \frac{1}{2} D_t B_{\mu\nu}
      \right|^2
      + \frac{1}{2} \left| F_{t\mu} + D^{\nu} B_{\nu\mu} \right|^2
    \\
    & + 2 D_{\mu} \upsilon D^{\mu} \bar{\upsilon}
      + 2 D_t \upsilon D^t \bar{\upsilon}
      - 2 [\upsilon, \bar{\upsilon}]^2
      + \frac{1}{2} [B_{\mu\nu}, \upsilon] [B^{\mu\nu}, \bar{\upsilon}]
    \\
    & + \frac{1}{2} B^{\mu\nu} \{\tilde{\eta}, \chi_{\mu\nu}\}
      - \frac{1}{2} B^{\mu\nu} \{\eta, \tilde{\chi}_{\mu\nu}\}
      - B^{\mu\nu} \{\psi_{\mu}, \tilde{\psi}_{\nu}\}
      - B^{\mu\nu} \{\tilde{\chi}_{\mu\rho}, \chi^{\rho}_{\nu}\}
    \\
    & - i \tilde{\eta} D_{\mu} \psi^{\mu}
      - i \eta D_{\mu} \tilde{\psi}^{\mu}
      - 2 i \tilde{\psi}_{\mu} D_{\nu} \chi^{\mu\nu}
      - 2 i \psi_{\mu} D_{\nu} \tilde{\chi}^{\mu\nu}
      - \tilde{\eta} D_t \eta
      - \tilde{\psi}_{\mu} D_t \psi^{\mu}
      - \tilde{\chi}_{\mu\nu} D_t \chi^{\mu\nu}
    \\
    & - i \upsilon \{\tilde{\eta}, \tilde{\eta}\}
      - i \bar{\upsilon} \{\eta, \eta\}
      + i \upsilon \{\tilde{\psi}_{\mu}, \tilde{\psi}^{\mu}\}
      + i \bar{\upsilon} \{\psi_{\mu}, \psi^{\mu}\}
      - i \upsilon \{\tilde{\chi}_{\mu\nu}, \tilde{\chi}^{\mu\nu}\}
      - i \bar{\upsilon} \{\chi_{\mu\nu}, \chi^{\mu\nu}\}
      \bigg)
      \, .
  \end{aligned}
\end{equation}

\subsection{\texorpdfstring{Spin$(7)$}{Spin(7)} Theory: A Topologically-twisted 8d \texorpdfstring{$\mathcal{N} = 1$}{N = 1} Gauge Theory}
\label{sec:topo theories:spin7}

The next topologically-twisted gauge theory that we will introduce, is a topologically-twisted 8d $\mathcal{N} = 1$ gauge theory on a Spin$(7)$-manifold\footnote{See~\autoref{ft:spin7 manifold definition}.} with a ``trivial'' twist,\footnote{%
  We refer the reader to~\cite{acharya-1997-higher-dimen, elliott-2022-taxon-twist} for a detailed description of the twist.
  \label{fn:acharya twist reference}
} whose BPS equation is the Spin$(7)$ instanton equation.
We shall henceforth refer to this theory as Spin$(7)$ theory.

The bosonic field content of Spin$(7)$ theory consists of complex scalars $\varphi, \lambda \in \Omega^0(\text{Spin}(7), \text{ad}(G))$ and a gauge connection $A_\mu \in \Omega^1(\text{Spin}(7), \text{ad}(G))$.
The fermionic field content consists of a scalar $\eta \in \Omega^0(\text{Spin}(7), \text{ad}(G))$, a one-form $\psi \in \Omega^1(\text{Spin}(7), \text{ad}(G))$, and a self-dual two-form $\chi \in \Omega^{2+}(\text{Spin}(7), \text{ad}(G))$.
Here, $\mu \in \{0, \dots, 7\}$ are the indices of the Spin$(7)$ manifold.

As the supersymmetry generators transform in the same representation as the fermions above, the ``trivial'' twist will result in a single, \emph{scalar} supersymmetry generator $\widetilde{\mathcal{Q}}$.
Its supersymmetry transformations of the twisted fields work out to be~\cite{er-2024-topol-gauge-theor}
\begin{equation}
  \label{eq:spin7 susy variations}
  \begin{aligned}
    \tilde{\delta} \varphi
    &= 0
      \, ,
    &\qquad
    \tilde{\delta} \eta
    &= \frac{i}{2} [\varphi, \lambda]
      \, , \\
    \tilde{\delta} \lambda
    &= - 2 \eta
      \, ,
    &\qquad
    \tilde{\delta} \chi_{\mu\nu}
    &= i F^+_{\mu\nu}
      \, , \\
    \tilde{\delta} A_\mu
    &= - \psi_\mu
      \, ,
    &\qquad
    \tilde{\delta} \psi_\mu
    &= - i D_\mu \varphi
      \, ,
  \end{aligned}
\end{equation}
where $\tilde{\delta}$ denotes a $\widetilde{\mathcal{Q}}$-variation;
$F^+_{\mu\nu} = \frac{1}{2} (F_{\mu\nu} + \frac{1}{2} \phi_{\mu\nu\rho\pi} F^{\rho\pi})$ is the self-dual component of $F_{\mu\nu}$, with $\phi$ being the closed Hodge self-dual Spin$(7)$ structure;
and self-duality of $\Phi \in \{\chi, F^+\}$ means $\Phi_{\mu\nu} = \frac{1}{6} \phi_{\mu\nu\rho\pi} \Phi^{\rho\pi}$.
Our choice of $\phi$ will be that in~\cite{er-2024-topol-gauge-theor}, where its nonvanishing components, denoted as $[\mu\nu\rho\pi] \equiv \phi_{\mu\nu\rho\pi}$, are
\begin{equation}
  \label{eq:spin7 structure}
  \begin{aligned}
    \relax  
    [0145] = [0167] = [2345] = [2367]
    &= [0246] = [1357] = [0123] = [4567] = 1
      \, , \\
    [0257] = [1346] = [0347]
    &= [0356] = [1247] = [1256] = -1
      \, .
  \end{aligned}
\end{equation}

With auxiliary fields, a straightforward calculation will show that $\widetilde{\mathcal{Q}}$ is nilpotent up to gauge transformations generated by $\varphi$.
As we wish to study the theory whereby the relevant moduli space is well-behaved (which corresponds to no reducible connections), we shall consider the case where $\varphi$ has no zero-modes.

Setting to zero the $\widetilde{\mathcal{Q}}$-variations of the fermions in~\eqref{eq:spin7 susy variations}, we obtain the BPS equation of Spin$(7)$ theory as\footnote{%
  As we are only considering the case where $\varphi$ has no zero-modes, we can take it to be zero in the variations of the fermions.
  \label{ft:spin7 ignore variations}
}
\begin{equation}
  \label{eq:spin7 inst eq}
  F^+_{\mu\nu} = 0 \, .
\end{equation}
This is the Spin$(7)$ instanton equation.
Configurations of $A_{\mu}$ satisfying~\eqref{eq:spin7 inst eq} constitute the Spin$(7)$ instanton moduli space $\mathcal{M}_{\text{Spin}(7)}$ that the Spin$(7)$ theory path integral will localize onto.

Finally, the $\mathcal{Q}$-exact topological action of Spin$(7)$ theory is given as~\cite{er-2024-topol-gauge-theor, acharya-1997-higher-dimen}
\begin{equation}
  \label{eq:spin7 action}
  \begin{aligned}
    S_{\text{Spin}(7)} = \frac{1}{e^2} \int_{\text{Spin}(7)} d^8x \,
    \Tr \bigg(
    & \left| F^+_{\mu\nu} \right|^2
      - \frac{1}{2} D_\mu \varphi D^{\mu} \lambda
      - \frac{1}{8} [\varphi, \lambda]^2
      - i \eta D^\mu \psi_\mu
      + 2i D_\mu \psi_\nu \chi^{\mu\nu} \\
    & - \frac{i}{2} \varphi \{\eta, \eta\}
      - \frac{i}{4} \varphi \left\lbrace \chi_{\mu\nu}, \chi^{\mu\nu} \right\rbrace
      - \frac{i}{2} \lambda \left\lbrace \psi_\mu, \psi^\mu \right\rbrace \bigg)
      \, .
  \end{aligned}
\end{equation}

\section{A Fueter type \texorpdfstring{$A_\infty$}{A-infinity}-2-category of Two-Manifolds}
\label{sec:m2 x r3}

In this section, we will study HW theory on $M_5 = M_2 \times \R^3$, with $M_2$ being a closed and compact Riemann surface.
We will recast it as a 3d gauged Landau-Ginzburg (LG) model on $\R^3$, a 2d gauged LG model on $\R^2$, or a 1d LG SQM.
Following the approach in \cite[$\S$9]{er-2023-topol-n}, we will, via the 5d HW partition function and its equivalent 2d gauged LG model, be able to physically realize a novel Fukaya-Seidel (FS) type $A_{\infty}$-category of solitons whose endpoints correspond to $G_{\C}$-BF configurations on $M_2$ that generate a  3d-HW Floer homology.
Similarly, via the 5d HW partition function and its equivalent 3d gauged LG model, we will be able to also physically realize a novel Fueter type $A_{\infty}$-2-category that 2-categorifies the 3d-HW Floer homology of $M_2$.

\subsection{HW Theory on \texorpdfstring{$M_2 \times \R^3$}{M2 x R3} as a 3d Model on \texorpdfstring{$\R^3$}{R3}, 2d Model on \texorpdfstring{$\R^2$}{R2}, or 1d SQM}
\label{sec:m2 x r3:theory}

Recall from \autoref{sec:topo theory:hw} that our HW theory is defined on a five-manifold $M_5 = M_4 \times \R$.
Let us take $M_4 = M_2 \times \R^2$, and relabel the $(x^1, x^2)$ coordinates of this $\R^2$ as $(\tau, \xi)$ for later convenience.
Doing so, both self-dual two-form fields $B$ and $\chi$ can each be interpreted as a combination of a scalar and a one-form on $M_2$.
In particular, from the three linearly independent components of $B_{\mu\nu} \in \Omega^{2, +}(M_2 \times \R^2) \otimes \Omega^0(\R)$, we get $B_\alpha \in \Omega^1(M_2) \otimes \Omega^0(\R^3)$ and $C \in \Omega^0(M_2 \times \R^3)$, where $\alpha = \{3, 4\}$ are the indices on $M_2$.
Similarly, from the three linearly-independent components of $\chi_{\mu\nu} \in \Omega^{2, +}(M_2 \times \R^2) \otimes \Omega^0(\R)$, we get $\chi_\alpha \in \Omega^1(M_2) \otimes \Omega^0(\R^3)$ and $\varkappa \in \Omega^0(M_2 \times \R^3)$.

Subtracting $\mathcal{Q}$-exact and topological terms from the action, \eqref{eq:hw action} becomes
\begin{equation}
  \label{eq:hw action:m2 x r3}
  S_{\text{HW}, M_2 \times \R^3}
  = \frac{1}{e^2} \int_{M_2 \times \R^3} dt d\tau d\xi dx^2 \, \Tr \big(
  L_A + L_B + L_C
  \big)
  \, ,
\end{equation}
where
\begin{equation}
  \label{eq:hw action:m2 x r3:La}
  \begin{aligned}
    L_A
    =
    & \frac{1}{2} \left|
        F_{t\tau}
        - D_{\xi} C
        - D^{\alpha} B_{\alpha}
      \right|^2
      + \frac{1}{2} \left|
        F_{t\xi}
        + D_{\tau} C
        - \epsilon_{\alpha\beta} D^{\alpha} B^{\beta}
      \right|^2
    \\
    & + \frac{1}{2} \left|
        F_{\tau\xi}
        - D_t C
        + \frac{1}{2} \epsilon_{\alpha\beta} \left(
          F^{\alpha\beta} - [B^{\alpha}, B^{\beta}]
        \right)
      \right|^2
    \\
    & + \frac{1}{2} \left|
        D_t A_{\alpha} - \partial_{\alpha} A_t
        + D_{\tau} B_{\alpha}
        + \epsilon_{\alpha\beta} \left(
          D_{\xi} B^{\beta}
          - D^{\beta} C
        \right)
      \right|^2
    \\
    & + \frac{1}{2} \left|
        D_{\tau} A_{\alpha} - \partial_{\alpha} A_{\tau}
        - D_t B_{\alpha}
        - \epsilon_{\alpha\beta} \left( D_{\xi} A^{\beta} - \partial^{\beta} A_{\xi} - [C, B^{\beta}] \right)
      \right|^2
    \\
    & + 2 \left| D_{\alpha} \upsilon \right|^2
      + 2 \left| D_{\xi} \upsilon \right|^2
      + 2 \left| D_{\tau} \upsilon \right|^2
      \, ,
  \end{aligned}
\end{equation}
\begin{equation}
  \label{eq:hw action:m2 x r3:Lb}
  \begin{aligned}
    L_B
    =
    & - i \tilde{\eta} D_{\alpha} \psi^{\alpha}
      - i \tilde{\eta} D_{\xi} \psi^{\xi}
      - i \tilde{\eta} D_{\tau} \psi^{\tau}
      - i \eta D_{\alpha} \tilde{\psi}^{\alpha}
      - i \eta D_{\xi} \tilde{\psi}^{\xi}
      - i \eta D_{\tau} \tilde{\psi}^{\tau}
    \\
    & - 2i \left(
      \tilde{\psi}_{\tau} D_{\alpha} - \tilde{\psi}_{\alpha} D_{\tau}
      \right) \chi^{\alpha}
      - 2i \left(
      \tilde{\psi}_{\tau} D_{\xi} - \tilde{\psi}_{\xi} D_{\tau}
      \right) \varkappa
    \\
    & - 2i \left(
        \psi_{\tau} D_{\alpha} - \psi_{\alpha} D_{\tau}
      \right) \tilde{\chi}^{\alpha}
      - 2i \left(
        \psi_{\tau} D_{\xi} - \psi_{\xi} D_{\tau}
      \right) \tilde{\varkappa}
    \\
    & - 2 \epsilon^{\alpha\beta} \left(
        \tilde{\psi}_{\tau} D_{\alpha} \chi_{\beta}
        + \psi_{\tau} D_{\alpha} \tilde{\chi}_{\beta}
      + \tilde{\psi}_{\alpha} D_{\beta} \varkappa
      + \psi_{\alpha} D_{\beta} \tilde{\varkappa}
      - \tilde{\psi}_{\alpha} D_{\xi} \chi_{\beta}
      - \psi_{\alpha} D_{\chi} \tilde{\chi}_{\beta}
      \right)
    \\
    & - \tilde{\psi}_{\tau} D_t \psi^{\tau}
      - \tilde{\psi}_{\xi} D_t \psi^{\xi}
      - \tilde{\psi}_{\alpha} D_t \psi^{\alpha}
      - 2 \tilde{\varkappa} D_t \varkappa
      - 2 \tilde{\chi}_{\alpha} D_t \chi^{\alpha}
      \, ,
  \end{aligned}
\end{equation}
and
\begin{equation}
  \label{eq:hw action:m2 x r3:Lc}
  \begin{aligned}
    L_C
    =
    & i \upsilon \{\tilde{\psi}_{\tau}, \tilde{\psi}^{\tau}\}
      + i \upsilon \{\tilde{\psi}_{\xi}, \tilde{\psi}^{\xi}\}
      + i \upsilon \{\tilde{\psi}_{\alpha}, \tilde{\psi}^{\alpha}\}
      - 2i \upsilon \{\tilde{\varkappa}, \tilde{\varkappa}\}
      - 2i \upsilon \{\tilde{\chi}_{\alpha}, \tilde{\chi}^{\alpha}\}
    \\
    & + i \bar{\upsilon} \{\psi_{\tau}, \psi^{\tau}\}
      + i \bar{\upsilon} \{\psi_{\xi}, \psi^{\xi}\}
      + i \bar{\upsilon} \{\psi_{\alpha}, \psi^{\alpha}\}
      - 2i \bar{\upsilon} \{\varkappa, \varkappa\}
      - 2i \bar{\upsilon} \{\chi_{\alpha}, \chi^{\alpha}\}
      \, .
  \end{aligned}
\end{equation}

The conditions on the bosons that minimize the action \eqref{eq:hw action:m2 x r3} are easily identified by setting to zero the expression within the squared terms in \eqref{eq:hw action:m2 x r3:La}, i.e.,\footnote{%
  We can ignore the terms with $\upsilon$, because they have no zero-modes as explained in \autoref{ft:hw ignore variations}.
  \label{fn:ignore upsilon term}
}
\begin{equation}
  \label{eq:hw bps eqns:m2 x r3}
  \begin{aligned}
    F_{t\tau} - D_{\xi} C
    &= D^{\alpha} B_{\alpha}
      \, ,
    \\
    F_{t\xi} + D_{\tau} C
    &= \epsilon_{\alpha\beta} D^{\alpha} B^{\beta}
      \, ,
    \\
    F_{\tau\xi} - D_t C
    &= - \frac{1}{2} \epsilon_{\alpha\beta} \left(
      F^{\alpha\beta} - [B^{\alpha}, B^{\beta}]
      \right)
      \, ,
    \\
    D_t A_{\alpha} - \partial_{\alpha} A_t + D_{\tau} B_{\alpha} + \epsilon_{\alpha\beta} D_{\xi} B^{\beta}
    &= \epsilon_{\alpha\beta} D^{\beta} C
      \, ,
    \\
    D_{\tau} A_{\alpha} - \partial_{\alpha} A_{\tau}
    - D_t B_{\alpha}
    - \epsilon_{\alpha\beta} ( D_{\xi} A^{\beta} - \partial^{\beta} A_{\xi})
    &= - \epsilon_{\alpha\beta} [C, B^{\beta}]
      \, .
  \end{aligned}
\end{equation}

\subtitle{HW Theory as a 3d Model}

We now want to recast HW theory as a 3d model on $\R^3$.
To this end, first note that by (i) introducing $G_{\C}$ connections via $\mathscr{A} = A + i B \in \Omega^1(M_2, \text{ad}(G_{\C})) \otimes \Omega^0(\R^3, \text{ad}(G_\C))$ and $\bar{\mathscr{A}} = A - i B \in \Omega^1(M_2, \text{ad}(G^{*}_{\C})) \otimes \Omega^0(\R^3, \text{ad}(G^*_\C))$, where $G_{\C}$ is the complexification of $G$ and $G^{*}_{\C}$ is its complex conjugate, and (ii) using $w = x^3 + j x^4$ and $\bar{w} = x^3 - j x^4$ as the complex coordinates for $M_2$, we can express \eqref{eq:hw bps eqns:m2 x r3} as\footnote{%
  Trivially vanishing terms $g^{\alpha\beta} F_{\alpha\beta} = 0 = g^{\alpha\beta} [B_{\alpha}, B_{\beta}]$ are used to arrive at the RHS of the first equation.
  \label{ft:derivation of hat-calF}
}
\begin{equation}
  \label{eq:hw bps eqns:m2 x r3:gc}
  \begin{aligned}
    D_\xi C
    - F_{t\tau}
    &= - 2 i \hat{\mathscr{F}}_{w \bar{w}} g^{w \bar{w}}
      \, ,
    \\
    D_t C - i D_{\tau} C
    - (F_{\tau \xi} + i F_{t \xi})
    &= - 2 j \mathscr{F}_{w \bar{w}}
      \, ,
    \\
    D_{\xi} \mathscr{A}_w
    - k D_t \mathscr{A}_w
    + j D_{\tau} \mathscr{A}_w
    - \partial_w (A_{\xi} - k A_t + j A_{\tau})
    &= i \mathscr{D}_w C
      \, ,
  \end{aligned}
\end{equation}
where $g^{w\bar{w}}$ is the corresponding inverse metric on $M_2$;
$\hat{\mathscr{F}}_{w \bar{w}} = \partial_{w} \bar{\mathscr{A}}_{\bar{w}} - \partial_{\bar{w}} \mathscr{A}_{w} + [\mathscr{A}_{w}, \bar{\mathscr{A}}_{\bar{w}}]$ is a $(1, 1)$-form field strength in $\mathscr{A}$ and $\bar{\mathscr{A}}$;
$\mathscr{F}_{w \bar{w}} = \partial_{w} \mathscr{A}_{\bar{w}} - \partial_{\bar{w}} \mathscr{A}_w + [\mathscr{A}_w, \mathscr{A}_{\bar{w}}]$ is a $(1, 1)$-form field strength in $\mathscr{A}$;
$\mathscr{D}_{w} = \partial_{w} + [\mathscr{A}_{w}, \cdot]$ is the covariant derivative in $\mathscr{A}$;
and $(i, j, k)$ are unit imaginary numbers satisfying the quaternionic relations, i.e., $i^2 = j^2 = k^2 = -1$ and $ij = k$.\footnote{%
  One can understand $(i, j, k)$ to imply the existence of three complex structures $(I, J, K)$ defining the three different ways of complexifying $\R^3$ into $\R \times \C$.
  In particular, $I$ would be the complex structure of the $(t, \tau)$-subplane, $J$ would be the complex structure of the $(\tau, \xi)$-subplane, and $K$ would be the complex structure of the $(\xi, t)$-subplane.
  Also, them satisfying the quaternionic relations comes from the fact that the complexification of $\R^3$ should maintain its orientation.
  \label{ft:complex structures of r3}
}

Next, note that we are physically free to rotate the $(t, \tau)$-subplane of $\R^3$ about the origin by an angle $\theta$, whence~\eqref{eq:hw bps eqns:m2 x r3:gc} becomes
\begin{equation}
  \label{eq:hw bps eqns:m2 x r3:rotated}
  \begin{aligned}
    D_\xi C
    - F_{t\tau}
    &= - 2 i \hat{\mathscr{F}}_{w \bar{w}} g^{w \bar{w}}
      \, ,
    \\
    D_t C - i D_{\tau} C
    + F_{\xi \tau} + i F_{\xi t}
    &= - 2 j_{\theta} \mathscr{F}_{w \bar{w}}
      \, ,
    \\
    D_{\xi} \mathscr{A}_w
    - k_{\theta} D_t \mathscr{A}_w
    + j_{\theta} D_{\tau} \mathscr{A}_w
    - \partial_w (A_{\xi} - k_{\theta} A_t + j_{\theta} A_{\tau})
    &= i \mathscr{D}_w C
      \, ,
  \end{aligned}
\end{equation}
where $j_{\theta} \coloneqq e^{i\theta/2} j e^{-i\theta/2} \equiv j \cos \theta + k \sin \theta$, and $k_{\theta} = i j_{\theta}$.
This allows us to write the action for HW theory on $M_2 \times \R^3$ as
\begin{equation}
  \label{eq:m2 x r3:hw action:rotated}
  \begin{aligned}
    S_{\text{HW}, M_2 \times \R^3}
    = \frac{1}{e^2} \int_{\R^3} dt d\tau d\xi \int_{M_2} |dw|^2 \, \Tr
    \bigg(
    & \left| D_{\xi} C
      - F_{t\tau}
      + p
      \right|^2
      + \left| D_t C - i D_{\tau} C
      + (F_{\xi \tau} + i F_{\xi t})
      + q
      \right|^2
    \\
    & + \left|
      D_{\xi} \mathscr{A}_w
      - k_{\theta} D_t \mathscr{A}_w
      + j_{\theta} D_{\tau} \mathscr{A}_w
      + r_w
      \right|^2
      + \dots
      \bigg)
      \, ,
  \end{aligned}
\end{equation}
where the ``$\dots$'' contains the fermion terms in~\eqref{eq:hw action:m2 x r3}, and
\begin{equation}
  \label{eq:m2 x r3:hw action:rotated:components}
  \begin{aligned}
    p
    &= 2i \hat{\mathscr{F}}_{w \bar{w}} g^{w \bar{w}}
      \, ,
    \\
    q
    &= 2j_{\theta} \mathscr{F}_{w \bar{w}}
      \, ,
    \\
    r_w
    &= - \partial_w (A_{\xi} - k_{\theta} A_t + j_{\theta} A_{\tau})
      - i \mathscr{D}_w C
    \, .
  \end{aligned}
\end{equation}

Lastly, after suitable rescalings, we can recast~\eqref{eq:m2 x r3:hw action:rotated} as a 3d model, where the action is\footnote{%
  To arrive at the following expression, we have (i) employed Stokes' theorem and the fact that $M_2$ has no boundary to omit terms with $\partial_w A_{\{t, \tau, \xi\}}$ as they will vanish when integrated over $M_2$, and (ii) integrated out the scalar field $\mathfrak{h}_2(p) = 2i \hat{\mathscr{F}}_{a \bar{a}} g^{a \bar{a}}$ corresponding to the scalar $p$, whose contribution to the action is $|\mathfrak{h}_2(p)|^2$.
  \label{ft:stokes theorem for m2 x r3:3d model}
}
\begin{equation}
  \label{eq:m2 x r3:3d model action}
  \begin{aligned}
    S_{\text{3d}, \mathfrak{A}_2}
    &= \frac{1}{e^2} \int_{\R^3} dt d\tau d\xi \bigg(
      \left| D_{\xi} C^a - F_{t\tau} \right|^2
      + \left| D_t C^a - i D_{\tau} C^a
      + F_{\xi \tau} + i F_{\xi t}
      + q^a
      \right|^2
    \\
    & \qquad \qquad \qquad \qquad
      + \left|
      D_{\xi} \mathscr{A}^a
      - k_{\theta} D_t \mathscr{A}^a
      + j_{\theta} D_{\tau} \mathscr{A}^a
      + r^a
      \right|^2
      + \dots
      \bigg)
    \\
    &= \frac{1}{e^2} \int_{\R^2} d\tau d\xi \int_{\R} dt \bigg(
      \left| D_{\xi} C^a - D_t A_\tau + P \right|^2
      + \left| D_t C^a
      + i F_{\xi t} + D_{\xi} A_{\tau}
      + Q^a + q^a \right|^2
    \\
    & \qquad \qquad \qquad \qquad \qquad
      + \left| D_{\xi} \mathscr{A}^a
      - k_{\theta} D_t \mathscr{A}^a
      + R^a + r^a \right|^2
      + \dots
      \bigg)
      \, .
  \end{aligned}
\end{equation}
Here, $(\mathscr{A}^a, C^a)$ and $a$ are coordinates and indices on the space $\mathfrak{A}_2$ of irreducible $(\mathscr{A}_w, C)$ fields on $M_2$;\footnote{%
  Since we will ultimately consider only gauge-inequivalent configurations, $\mathfrak{A}_2$ is more precisely the space of irreducible $(\mathscr{A}_w, C)$ fields on $M_2$ modulo gauge equivalence.
  Similar such spaces to appear in later sections should also be understood as spaces of fields modulo gauge equivalence.
  \label{ft:modulo gauge inequivalence}
}
\begin{equation}
  \label{eq:m2 x r3:3d model action:components}
  \begin{gathered}
    P
    = \partial_{\tau} A_t
    \, ,
    \qquad
    Q^a
    = - i D_{\tau} C^a
    - \partial_{\tau} A_{\xi}
    \, ,
    \qquad
    R^a
    = j_{\theta} D_{\tau} \mathscr{A}^a
    \, ,
    \\
    q^a
    = j_{\theta} \mathscr{F}^a
    \, ,
    \qquad
    r^a
    = - i (\mathscr{D} C)^a
    \, ,
  \end{gathered}
\end{equation}
with $(q^a, r^a)$ corresponding to $(q, r_w)$ in~\eqref{eq:m2 x r3:hw action:rotated:components}; and $(\mathscr{F}^a, (\mathscr{D}C)^a)$ are scalars in $\mathfrak{A}_2$ corresponding to $(\mathscr{F}_{w \bar{w}}, \mathscr{D}_w C)$ in the underlying 5d theory.

In other words, HW theory on $M_2 \times \R^3$ can be regarded as 3d gauged sigma model along the $(t, \tau, \xi)$-directions with target space $\mathfrak{A}_2$ and action~\eqref{eq:m2 x r3:3d model action}.

\subtitle{HW Theory as a 2d Model}

From~\eqref{eq:m2 x r3:3d model action}, one can see that we can, after suitable rescalings, also recast the 3d model action as the following 2d model action\footnote{%
  To arrive at the following expression, we have employed Stokes' theorem and the fact that the finite-energy 3d gauge fields $A_{\{t, \tau, \xi\}}$ would vanish at $\tau \rightarrow \pm \infty$.
  \label{ft:stokes theorem for m2 x r3:2d model}
}
\begin{equation}
  \label{eq:m2 x r3:2d model action}
  \begin{aligned}
    S_{\text{2d}, \mathcal{M}(\R_{\tau}, \mathfrak{A}_2)}
    = \frac{1}{e^2} \int_{\R^2} dt d\xi \bigg(
    & \left| D_{\xi} C^m
      - D_t (\tilde{A}_{\tau})^m
      \right|^2
      + \left| D_t C^m
      + i F_{\xi t}
      + D_{\xi} (\tilde{A}_{\tau})^m
      + Q^m
      + q^m
      \right|^2
    \\
    & + \left| D_{\xi} \mathscr{A}^m
      - k_{\theta} D_t \mathscr{A}^m
      + R^m + r^m \right|^2
      + \dots
      \bigg)
      \, .
  \end{aligned}
\end{equation}
Here, $(\mathscr{A}^m, C^m, (\tilde{A}_{\tau})^m)$ and $m$ are coordinates and indices on the path space $\mathcal{M}(\R_{\tau}, \mathfrak{A}_{2})$ of smooth paths from $\R_{\tau}$ to $\mathfrak{A}_2$, and
\begin{equation}
  \label{eq:m2 x r3:2d model action:components}
  \begin{aligned}
    Q^m
    &= - i (\tilde{D}_{\tau} C)^m
    \, ,
    &\qquad
    R^m
    &= j_{\theta} (\tilde{D}_{\tau} \mathscr{A})^m
    \, ,
    \\
    q^m
    &= j_{\theta} \mathscr{F}^m
    \, ,
    &\qquad
    r^m
    &= - i (\mathscr{D} C)^m
    \, ,
  \end{aligned}
\end{equation}
corresponding to~\eqref{eq:m2 x r3:3d model action:components}, with $(\tilde{A}_{\tau}, \tilde{D}_{\tau})$ in $\mathcal{M}(\R_{\tau}, \mathfrak{A}_2)$ corresponding to $(A_{\tau}, D_{\tau})$ in the underlying 3d model.

In other words, HW theory on $M_2 \times \R^3$ can also be regarded as a 2d gauged sigma model along the $(t, \xi)$-directions with target space $\mathcal{M}(\R_{\tau}, \mathfrak{A}_2)$ and action~\eqref{eq:m2 x r3:2d model action}.

\subtitle{HW Theory as a 1d SQM}

Singling out $\xi$ as the direction in ``time'', the equivalent SQM action can be obtained from~\eqref{eq:m2 x r3:2d model action} after suitable rescalings as\footnote{%
  In the resulting SQM, as $A_\xi$ is a non-dynamical field, it will be integrated out to furnish the Christoffel connections for the fermions in the SQM \cite{er-2023-topol-n}.
  We have again applied Stokes' theorem and the fact that the fields corresponding to finite-energy gauge fields $A_{\{t, \tau\}}$ would vanish at $t \rightarrow \pm \infty$, to arrive at the following expression.
  \label{ft:stokes theorem for m2 x r3:sqm}
}
\begin{equation}
  \label{eq:m2 x r3:sqm action}
  \begin{aligned}
    S_{\text{SQM}, \mathcal{M}(\R_t, \mathcal{M}(\R_{\tau}, \mathfrak{A}_2))}
    = \frac{1}{e^2} \int d\xi \Bigg(
    & \left| \partial_\xi C^u
      + g_{\mathcal{M}(\R_t, \mathcal{M}(\R_{\tau}, \mathfrak{A}_2))}^{uv} \pdv{h_2}{C^v}
      \right|^2
      + \left| \partial_{\xi} \breve{A}^u
      + g_{\mathcal{M}(\R_t, \mathcal{M}(\R_{\tau}, \mathfrak{A}_2))}^{uv} \pdv{h_2}{\breve{A}^v}
      \right|^2
    \\
    & + \left| \partial_\xi \mathscr{A}^u
      + g_{\mathcal{M}(\R_t, \mathcal{M}(\R_{\tau}, \mathfrak{A}_2))}^{uv} \pdv{h_2}{\mathscr{A}^v}
      \right|^2
      + \dots
      \Bigg)
      \, ,
  \end{aligned}
\end{equation}
where $(\mathscr{A}^u, C^u, \breve{A}^u)$ and $(u, v)$ are coordinates on the path space $\mathcal{M}(\R_t, \mathcal{M}(\R_{\tau}, \mathfrak{A}_2))$ of smooth maps from $\R_t$ to $\mathcal{M}(\R_{\tau}, \mathfrak{A}_2)$ with $\breve{A}^u \coloneqq (\breve{A}_{\tau} + i \breve{A}_t)^u$ in $\mathcal{M}(\R_t, \mathcal{M}(\R_{\tau}, \mathfrak{A}_2))$ corresponding to $(\tilde{A}_{\tau})^m + i A_t$ in the underlying 2d model, and to $A_{\tau} + i A_t$ in the underlying 3d model;
$g_{\mathcal{M}(\R_t, \mathcal{M}(\R_{\tau}, \mathfrak{A}_2))}$ is the metric of $\mathcal{M}(\R_t, \mathcal{M}(\R_{\tau}, \mathfrak{A}_2))$;
and $h_2(\mathscr{A}, C, \breve{A})$ is the SQM potential function.
Note also that we can interpret $\mathcal{M}(\R_t, \mathcal{M}(\R_{\tau}, \mathfrak{A}_2))$ as the double path space $\mathcal{M}(\R^2, \mathfrak{A}_2)$ of smooth maps from $\R^2$ to $\mathfrak{A}_2$.

In short, HW theory on $M_2 \times \R^3$ can also be regarded as a 1d SQM along the $\xi$-direction in $\mathcal{M}(\R^2, \mathfrak{A}_2)$ whose action is~\eqref{eq:m2 x r3:sqm action}.

\subsection{Non-constant Double Paths, Sheets, Solitons, and the 3d-HW Floer Homology of \texorpdfstring{$M_2$}{M2}}
\label{sec:m2 x r3:gc-bf}

\subtitle{$\theta$-deformed, Non-constant Double Paths in the SQM}

Applying the squaring argument~\cite{blau-1993-topol-gauge} to~\eqref{eq:m2 x r3:sqm action}, we find that the path integral of the equivalent SQM will localize onto configurations that set the LHS and RHS of the expression within the squared terms therein \emph{simultaneously} to zero, i.e., the SQM path integral localizes onto $\xi$-invariant critical points of $h_2(\mathscr{A}, C, \breve{A})$ that obey
\begin{equation}
  \label{eq:m2 x r3:sqm:crit pts}
  \begin{aligned}
    0
    &= - ( [\breve{A}_t, \breve{A}_{\tau}] )^u
      \, ,
    \\
    ( \breve{\partial}_t C - i \breve{\partial}_{\tau} C )^u
    &= - ( [\breve{A}_t - i \breve{A}_{\tau}, C] )^u
      - j_{\theta} \mathscr{F}^u
      \, ,
    \\
    (k_{\theta} \breve{\partial}_t \mathscr{A} - j_{\theta} \breve{\partial}_{\tau} \mathscr{A})^u
    &= - ( [ k_{\theta} \breve{A}_t - j_{\theta} \breve{A}_{\tau}, \mathscr{A}] )^u
      - i (\mathscr{D} C)^u
      \, .
  \end{aligned}
\end{equation}
These are \emph{$\xi$-invariant, $\theta$-deformed}, non-constant double paths in $\mathcal{M}(\R_t, \mathcal{M}(\R_{\tau}, \mathfrak{A}_2))$.

\subtitle{$\mathcal{M}^{\theta}(\R_{\tau}, \mathfrak{A}_2)$-solitons in the 2d Gauged Model}

By comparing~\eqref{eq:m2 x r3:sqm action} with \eqref{eq:m2 x r3:2d model action}, we find that such $\xi$-invariant, $\theta$-deformed, non-constant double paths in the SQM defined by~\eqref{eq:m2 x r3:sqm:crit pts}, will correspond, in the 2d gauged sigma model with target space $\mathcal{M}(\R_{\tau}, \mathfrak{A}_2)$, to configurations defined by
\begin{equation}
  \label{eq:m2 x r3:m-soliton eqns:components}
  \begin{aligned}
    \partial_t (\tilde{A}_{\tau})^m
    &= - [A_t, (\tilde{A}_{\tau})^m]
      + [A_{\xi}, C^m]
      \, ,
    \\
    \partial_t C^m - i \partial_t A_{\xi}
    &= - [A_t, C^m - i A_{\xi}]
      - [A_{\xi}, (\tilde{A}_{\tau})^m]
      - Q^m
      - q^m
      \, ,
    \\
    k_{\theta} \partial_t \mathscr{A}^m
    &= - [k_{\theta} A_t, \mathscr{A}^m]
      + [A_{\xi}, \mathscr{A}^m]
      + R^m
      + r^m
      \, .
  \end{aligned}
\end{equation}
Via~\eqref{eq:m2 x r3:2d model action:components}, we can write~\eqref{eq:m2 x r3:m-soliton eqns:components} as
\begin{equation}
  \label{eq:m2 x r3:m-soliton eqns}
  \begin{aligned}
    \partial_t (\tilde{A}_{\tau})^m
    &= - [A_t, (\tilde{A}_{\tau})^m]
      + [A_{\xi}, C^m]
      \, ,
    \\
    \partial_t C^m - i \partial_t A_{\xi}
    &= - [A_t, C^m - i A_{\xi}]
      + i (\tilde{D}_{\tau} C)^m
      + [(\tilde{A}_{\tau})^m, A_{\xi}]
      - j_{\theta} \mathscr{F}^m
      \, ,
    \\
    k_{\theta} \partial_t \mathscr{A}^m
    &= [A_{\xi} - k_{\theta} A_t, \mathscr{A}^m]
      + j_{\theta} (\tilde{D}_{\tau} \mathscr{A})^m
      - i (\mathscr{D} C)^m
      \, .
  \end{aligned}
\end{equation}
These are $\xi$-invariant, $\theta$-deformed solitons along the $\tau$-direction in the 2d gauged sigma model with target space $\mathcal{M}(\R_{\tau}, \mathfrak{A}_2)$.
We shall henceforth refer to such solitons as $\mathcal{M}^{\theta}(\R_{\tau}, \mathfrak{A}_2)$-solitons.

\subtitle{$\mathfrak{A}_2^{\theta}$-sheets in the 3d Gauged Model}

By further comparing \eqref{eq:m2 x r3:2d model action} with~\eqref{eq:m2 x r3:3d model action}, we find that such $\mathcal{M}^{\theta}(\R_{\tau}, \mathfrak{A}_2)$-solitons in the 2d gauged sigma model defined by~\eqref{eq:m2 x r3:m-soliton eqns}, will correspond, in the 3d gauged sigma model with target space $\mathfrak{A}_2$, to configurations defined by
\begin{equation}
  \label{eq:m2 x r3:sheet eqns:components}
  \begin{aligned}
    \partial_t A_{\tau} - \partial_{\tau} A_t
    &= - [A_t, A_{\tau}]
      + [A_{\xi}, C^a]
      \, ,
    \\
    \partial_t C^a - i \partial_{\tau} C^a
    - i \partial_t A_{\xi} - \partial_{\tau} A_{\xi}
    &= - [A_t - i A_{\tau}, C^a - i A_{\xi}]
      - q^a
      \, ,
    \\
    k_{\theta} \partial_t \mathscr{A}^a - j_{\theta} \partial_{\tau} \mathscr{A}^a
    &= [A_{\xi} - k_{\theta} A_t + j_{\theta} A_{\tau}, \mathscr{A}^a]
      + r^a
      \, .
  \end{aligned}
\end{equation}
Via \eqref{eq:m2 x r3:3d model action:components}, we can write~\eqref{eq:m2 x r3:sheet eqns:components} as
\begin{equation}
  \label{eq:m2 x r3:sheet eqns}
  \begin{aligned}
    \partial_t A_{\tau} - \partial_{\tau} A_t
    &= - [A_t, A_{\tau}]
      + [A_{\xi}, C^a]
      \, ,
    \\
    \partial_t C^a - i \partial_{\tau} C^a
    - i \partial_t A_{\xi} - \partial_{\tau} A_{\xi}
    &= - [A_t - i A_{\tau}, C^a - i A_{\xi}]
      - j_{\theta} \mathscr{F}^a
      \, ,
    \\
    k_{\theta} \partial_t \mathscr{A}^a - j_{\theta} \partial_{\tau} \mathscr{A}^a
    &= [A_{\xi} - k_{\theta} A_t + j_{\theta} A_{\tau}, \mathscr{A}^a]
      - i (\mathscr{D} C)^a
      \, .
  \end{aligned}
\end{equation}
These are $\xi$-invariant, $\theta$-deformed sheets along the $(\tau, t)$-directions in the 3d gauged sigma model with target space $\mathfrak{A}_2$, which also satisfy the condition
\begin{equation}
  \label{eq:m2 x r3:sheet eqns:aux conds}
  2 i \hat{\mathscr{F}}_{a \bar{a}} g^{a \bar{a}} = 0
  \, ,
\end{equation}
where $2i \hat{\mathscr{F}}_{a \bar{a}} g^{a \bar{a}} = \mathfrak{h}_2(p)$ is the auxiliary scalar field defined in~\autoref{ft:stokes theorem for m2 x r3:3d model}.

We shall henceforth refer to such $\xi$-invariant, $\theta$-deformed sheets in the 3d gauged sigma model with target space $\mathfrak{A}_2$, defined by~\eqref{eq:m2 x r3:sheet eqns} and~\eqref{eq:m2 x r3:sheet eqns:aux conds}, as $\mathfrak{A}_2^{\theta}$-sheets.

\subtitle{$\xi$-independent, $\theta$-deformed HW Configurations in HW Theory}

In turn, by comparing~\eqref{eq:m2 x r3:3d model action} with~\eqref{eq:m2 x r3:hw action:rotated}, we find that the 3d configurations defined by \eqref{eq:m2 x r3:sheet eqns}, will correspond, in HW theory, to 5d configurations defined by
\begin{equation}
  \label{eq:m2 x r3:hw configs:components}
  \begin{aligned}
    \partial_t A_{\tau} - \partial_{\tau} A_t
    &= - [A_t, A_{\tau}]
      + [A_{\xi}, C]
      \, ,
    \\
    \partial_t C - i \partial_{\tau} C
    - i \partial_t A_{\xi} - \partial_{\tau} A_{\xi}
    &= - [A_t - i A_{\tau}, C - i A_{\xi}]
      - q
      \, ,
    \\
    k_{\theta} \partial_t \mathscr{A}_w - j_{\theta} \partial_{\tau} \mathscr{A}_w
    &= [A_{\xi} - k_{\theta} A_t + j_{\theta} A_{\tau}, \mathscr{A}_w]
      + r_w.
  \end{aligned}
\end{equation}
Via~\eqref{eq:m2 x r3:hw action:rotated:components}, we can write~\eqref{eq:m2 x r3:hw configs:components} as
\begin{equation}
  \label{eq:m2 x r3:hw configs}
  \begin{aligned}
    \partial_{\tau} A_t - \partial_t A_{\tau}
    &= - [A_{\tau}, A_t]
      + [A_{\xi}, C]
      \, ,
    \\
    \partial_{\tau} C - i \partial_t C
    - i \partial_{\tau} A_{\xi} - \partial_t A_{\xi}
    &= - [A_{\tau} - i A_t, C - i A_{\xi}]
      - 2 j_{\theta} \mathscr{F}_{w \bar{w}}
      \, ,
    \\
    k_{\theta} \partial_{\tau} \mathscr{A}_w - j_{\theta} \partial_t \mathscr{A}_w
    &= - D_w (A_{\xi} - k_{\theta} A_{\tau} + j_{\theta} A_t)
      - i \mathscr{D}_w C
      \, .
    \end{aligned}
\end{equation}
These are $\xi$-independent, $\theta$-deformed HW configurations on $M_2 \times \R^3$ which also satisfy the condition
\begin{equation}
  \label{eq:m2 x r3:hw configs:aux conds}
  2i \hat{\mathscr{F}}_{w \bar{w}} g^{w \bar{w}} = 0
  \, .
\end{equation}

\subtitle{HW Configurations, $\mathfrak{A}_2^{\theta}$-sheets, $\mathcal{M}^{\theta}(\R_{\tau}, \mathfrak{A}_2)$-solitons, and Non-constant Double Paths}

In short, these \emph{$\xi$-independent, $\theta$-deformed} HW configurations on $M_2 \times \R^3$ that are defined by~\eqref{eq:m2 x r3:hw configs} and~\eqref{eq:m2 x r3:hw configs:aux conds}, will correspond to the $\mathfrak{A}_2^{\theta}$-sheets defined by~\eqref{eq:m2 x r3:sheet eqns} and~\eqref{eq:m2 x r3:sheet eqns:aux conds}, which, in turn, will correspond to the $\mathcal{M}^{\theta}(\R_{\tau}, \mathfrak{A}_2)$-solitons defined by~\eqref{eq:m2 x r3:m-soliton eqns}, which, in turn, will correspond to the $\xi$-invariant, $\theta$-deformed, non-constant double paths in $\mathcal{M}(\R^2, \mathfrak{A}_2)$ defined by~\eqref{eq:m2 x r3:sqm:crit pts}.

\subtitle{$\mathcal{M}^{\theta}(\R_{\tau}, \mathfrak{A}_2)$-soliton Endpoints Corresponding to Non-constant Paths}

Consider now the fixed endpoints of the $\mathcal{M}^{\theta}(\R_{\tau}, \mathfrak{A}_2)$-solitons at $t = \pm \infty$, where we also expect the fields in the 2d gauged sigma model corresponding to the finite-energy 3d gauge fields $A_\xi, A_{\tau}, A_t$ to decay to zero.
They are given by~\eqref{eq:m2 x r3:m-soliton eqns} with $\partial_t C^m = 0 = \partial_t \mathscr{A}^m$ and $A_\xi, A_t, (\tilde{A}_{\tau})^m \rightarrow 0$, i.e,
\begin{equation}
  \label{eq:m2 x r3:m-soliton:endpts}
  \begin{aligned}
    i (\tilde{\partial}_{\tau} C)^m
    &= j_{\theta} \mathscr{F}^m
      \, ,
    \\
    j_{\theta} (\tilde{\partial}_{\tau} \mathscr{A})^m
    &= i (\mathscr{D} C)^m
      \, .
  \end{aligned}
\end{equation}
These are $(\xi, t)$-invariant, $\theta$-deformed, non-constant paths in $\mathcal{M}(\R_{\tau}, \mathfrak{A}_2)$.

\subtitle{$\mathfrak{A}_2^{\theta}$-sheet Edges Corresponding to $\mathfrak{A}_{2}^{\theta}$-solitons in the 3d Gauged Model}

In turn, \eqref{eq:m2 x r3:m-soliton:endpts} will correspond, in the 3d gauged sigma model, to the fixed edges of the $\mathfrak{A}_2^{\theta}$-sheets at $t = \pm \infty$, i.e., $(\xi, t)$-invariant, $\theta$-deformed configurations that obey
\begin{equation}
  \label{eq:m2 x r3:soliton}
  \begin{aligned}
    i \partial_{\tau} C^a
    &= j_{\theta} \mathscr{F}^{a}
      \, ,
    \\
    j_{\theta} \partial_{\tau} \mathscr{A}^a
    &= i (\mathscr{D} C)^a
      \, .
  \end{aligned}
\end{equation}
These are $(\xi, t)$-invariant, $\theta$-deformed solitons along the $\tau$-direction in the 3d gauged sigma model, which also satisfy the condition~\eqref{eq:m2 x r3:sheet eqns:aux conds}.
Notice that~\eqref{eq:m2 x r3:soliton} can also be obtained from~\eqref{eq:m2 x r3:sheet eqns} with $\partial_t C^a = 0 = \partial_t \mathscr{A}^a$ and $A_\xi, A_t, A_\tau \rightarrow 0$.

We shall henceforth refer to such $(\xi, t)$-invariant, $\theta$-deformed solitons in the 3d gauged sigma model with target space $\mathfrak{A}_2$, defined by~\eqref{eq:m2 x r3:soliton} and~\eqref{eq:m2 x r3:sheet eqns:aux conds}, as $\mathfrak{A}_2^{\theta}$-solitons.

\subtitle{$\mathfrak{A}^{\theta}_2$-soliton Endpoints or $\mathfrak{A}_2^{\theta}$-sheet Vertices Corresponding to $\theta$-deformed $G_{\C}$-BF Configurations on $M_2$}

Consider now (i) the fixed endpoints of the $\mathfrak{A}_2^{\theta}$-solitons at $\tau = \pm \infty$, or equivalently (ii) the vertices of the $\mathfrak{A}_2^{\theta}$-sheets at $\tau, t = \pm \infty$.
They are given by (i)~\eqref{eq:m2 x r3:soliton} and~\eqref{eq:m2 x r3:sheet eqns:aux conds} with $\partial_{\tau} C^a = 0 = \partial_{\tau} \mathscr{A}^a$, or equivalently (ii)~\eqref{eq:m2 x r3:sheet eqns} and~\eqref{eq:m2 x r3:sheet eqns:aux conds} with $\partial_{\{t, \tau\}} C^a = 0 = \partial_{\{t, \tau\}} \mathscr{A}^a$ and $A_{\xi}, A_{\tau}, A_t \rightarrow 0$, i.e.,
\begin{equation}
  \label{eq:m2 x r3:soliton:endpts}
  j_{\theta} \mathscr{F}^a
  = 0
  \, ,
  \qquad
  i (\mathscr{D} C)^a
  = 0
  \, ,
  \qquad
  i \hat{\mathscr{F}}_{a \bar{a}} g^{a \bar{a}}
  = 0
  \, .
\end{equation}
In turn, they will correspond, in HW theory, to $(\xi, \tau, t)$-independent, $\theta$-deformed configurations that obey
\begin{equation}
  \label{eq:m2 x r3:soliton:endpts:hw}
  j_{\theta} \mathscr{F}_{w \bar{w}}
  = 0
  \, ,
  \qquad
  i \mathscr{D}_w C
  = 0
  \, ,
  \qquad
  i \hat{\mathscr{F}}_{w \bar{w}} g^{w \bar{w}}
  = 0
  \, .
\end{equation}
Notice that~\eqref{eq:m2 x r3:soliton:endpts:hw} can also be obtained from~\eqref{eq:m2 x r3:hw configs} and~\eqref{eq:m2 x r3:hw configs:aux conds} with $\partial_{\{t, \tau\}} C = 0 = \partial_{\{t, \tau\}} \mathscr{A}_{w}$ and $A_\xi, A_\tau, A_t \rightarrow 0$.

At $\theta = 0$ or $\pi$,~\eqref{eq:m2 x r3:soliton:endpts:hw} can be written as
\begin{equation}
  \label{eq:m2 x r3:gc-bf eqns}
  \mathscr{F}_{w \bar{w}}
  = 0
  \, ,
  \qquad
  \mathscr{D}_w C
  = 0
  \, ,
  \qquad
  \hat{\mathscr{F}}_{w \bar{w}} g^{w \bar{w}}
  = 0
  \, .
\end{equation}
This is a \emph{reduced} version of the $G_{\C}$-BF equations on $M_2$ -- whilst the regular $G_{\C}$-BF equations consist of connections and scalars valued in $\text{ad}(G_{\C})$ and $\text{ad}(G_{\C}^{*})$, the scalar that we have in~\eqref{eq:m2 x r3:gc-bf eqns} is valued only in $\text{ad}(G)$.
Nonetheless, configurations spanning the space of solutions to these equations shall, in the rest of this section, be referred to as $G_{\C}$-BF configurations on $M_2$.
One thing to note at this point is that $G_{\C}$-BF configurations on $M_2$ are known to generate the 3d-HW Floer homology of $M_2$~\cite[$\S$5]{er-2023-topol-n}.

In other words, the $(\xi, \tau, t)$-independent, $\theta$-deformed HW configurations corresponding to the endpoints of the $\mathfrak{A}_2^{\theta}$-solitons (or equivalently, to the vertices of the $\mathfrak{A}_2^{\theta}$-sheets) are $\theta$-deformed $G_{\C}$-BF configurations on $M_2$.
We will also assume choices of $M_2$ and $G$ whereby such configurations are isolated and non-degenerate.\footnote{%
  In \cite[footnote 8]{er-2023-topol-n}, we (i) considered choices of $M_2$ and $G$ such that $\text{dim}(G) (1 + b_2^+) = 4kh$, where $b_2^+$ is the second Betti number of $M_2 \times T^2$, $k$ is the instanton number on $M_2 \times T^2$, and $h$ is the dual Coxeter number of $G$, and (ii) added physically-inconsequential $\mathcal{Q}$-exact perturbations of the action, to ensure that $G_{\C}$-BF configurations on $M_3$  are (i) isolated and (ii) non-degenerate at $\theta = 0$. Therefore, at $\theta = 0$, the endpoints of the $\mathfrak{A}_2^{\theta}$-solitons will be isolated and non-degenerate. As the physical theory is symmetric under a variation of $\theta$, this observation of the endpoints of the $\mathfrak{A}_2^{\theta}$-solitons (or equivalently, the vertices of the $\mathfrak{A}_2^{\theta}$-sheets) will continue to hold true for any value of $\theta$.
  Hence, this presumption that the moduli space of $\theta$-deformed $G_{\C}$-BF configurations on $M_2$ will be made of isolated and non-degenerate points, is justified.
  We would like to thank R.P. Thomas for discussions on this point.
  \label{ft:isolation of gc-bf}
}

\subtitle{Non-constant Double Paths, $\mathcal{M}^{\theta}(\R_{\tau}, \mathfrak{A}_2)$-solitons, $\mathfrak{A}_2^{\theta}$-solitons, and $\mathfrak{A}_2^{\theta}$-sheets}

In short, from the equivalent 1d SQM of HW theory on $M_2 \times \R^3$, the theory localizes onto $\xi$-invariant, $\theta$-deformed, non-constant double paths in $\mathcal{M}(\R^2, \mathfrak{A}_2)$, which, in turn, will correspond to $\mathcal{M}^{\theta}(\R_{\tau}, \mathfrak{A}_2)$-solitons in the 2d gauged sigma model whose endpoints are $(\xi, \tau)$-invariant, $\theta$-deformed, non-constant paths in $\mathcal{M}(\R_{\tau}, \mathfrak{A}_2)$.
In the 3d gauged sigma model, these $\mathcal{M}^{\theta}(\R_{\tau}, \mathfrak{A}_2)$-solitons will correspond to $\mathfrak{A}_2^{\theta}$-sheets, whose edges are $\mathfrak{A}_2^{\theta}$-solitons, and whose vertices will correspond to $\theta$-deformed $G_{\C}$-BF configurations on $M_2$ that generate the 3d-HW Floer homology of $M_2$.

\subsection{The 2d Model and Open Strings, the 3d Model and Open Membranes}
\label{sec:m2 x r3:2d-3d model}

\subtitle{Flow Lines of the SQM as BPS Worldsheets of the 2d Model}

The classical trajectories or flow lines of the equivalent SQM are governed by the gradient flow equations (defined by setting to zero the expression within the squared terms in~\eqref{eq:m2 x r3:sqm action}), i.e.,
\begin{equation}
  \label{eq:m2 x r3:sqm flow}
  \begin{gathered}
    \dv{C^u}{\xi}
    = - g_{\mathcal{M}(\R_t, \mathcal{M}(\R_{\tau}, \mathfrak{A}_2))}^{uv} \pdv{h_2}{C^v}
    \, ,
    \qquad
    \dv{\breve{A}^u}{\xi}
    = - g_{\mathcal{M}(\R_t, \mathcal{M}(\R_{\tau}, \mathfrak{A}_2))}^{uv} \pdv{h_2}{\breve{A}^v}
    \, ,
    \\
    \dv{\mathscr{A}^u}{\xi}
    = - g_{\mathcal{M}(\R_t, \mathcal{M}(\R_{\tau}, \mathfrak{A}_2))}^{uv} \pdv{h_2}{\mathscr{A}^v}
    \, ,
  \end{gathered}
\end{equation}
and they go from one $\xi$-invariant critical point of $h_2(\mathscr{A}, C, \breve{A})$ to another in $\mathcal{M}(\R_t, \mathcal{M}(\R_{\tau}, \mathfrak{A}_2))$.
In the 2d gauged sigma model with target space $\mathcal{M}(\R_{\tau}, \mathfrak{A}_2)$, these flow lines will correspond to worldsheets that have, at $\xi = \pm \infty$, $\mathcal{M}^{\theta}(\R_{\tau}, \mathfrak{A}_2)$-solitons.\footnote{%
  The $\mathcal{M}^{\theta}(\R_{\tau}, \mathfrak{A}_2)$-soliton can translate in the $\xi$-direction due to its ``center of mass'' motion, and because it is $\xi$-invariant, it is effectively degenerate.
  This reflects the fact that generically, each critical point of $h_2$ is degenerate and does not correspond to a point, but to a real line $\R_t$ in $\mathcal{M}(\R_t, \mathcal{M}(\R_{\tau}, \mathfrak{A}_2))$.
  Nonetheless, one can perturb $h_2$ via the addition of physically-inconsequential $\mathcal{Q}$-exact terms to the SQM action, and collapse the degeneracy such that the critical points really correspond to points in $\mathcal{M}(\R_t, \mathcal{M}(\R_{\tau}, \mathfrak{A}_2))$.
  This is equivalent to factoring out the center of mass degree of freedom of the $\mathcal{M}^{\theta}(\R_{\tau}, \mathfrak{A}_2)$-soliton, and fixing it at $\xi = \pm \infty$.
  \label{ft:fixing m-A2-soliton centre of mass dof}
}
These solitons shall be denoted as $\sigma_{\pm}(t, \theta, \mathfrak{A}_2)$, and are defined by~\eqref{eq:m2 x r3:m-soliton eqns} with $A_\xi, A_t, (\tilde{A}_{\tau})^m \rightarrow 0$, i.e.,
\begin{equation}
  \label{eq:m2 x r3:m-soliton:eqns:no gauge}
  \begin{aligned}
    \dv{C^m}{t}
    &= i (\tilde{\partial}_{\tau} C)^m
      - j_{\theta} \mathscr{F}^m
      \, ,
    \\
    k_{\theta} \dv{\mathscr{A}^m}{t}
    &= j_{\theta} (\tilde{\partial}_{\tau} \mathscr{A})^m
      - i (\mathscr{D} C)^m
      \, .
  \end{aligned}
\end{equation}
Their endpoints $\sigma(\pm \infty, \theta, \mathfrak{A}_2)$ at $t = \pm \infty$ are defined by
\begin{equation}
  \label{eq:m2 x r3:m-soliton:endpts:no gauge}
  i (\tilde{\partial}_{\tau} C)^m
  = j_{\theta} \mathscr{F}^m
  \, ,
  \qquad
  j_{\theta} (\tilde{\partial}_{\tau} \mathscr{A})^m
  = i (\mathscr{D} C)^m
  \, ,
\end{equation}
which is simply~\eqref{eq:m2 x r3:m-soliton:eqns:no gauge} with $d_t C^m = 0 = d_t \mathscr{A}^m$.

Note that the flow lines are governed by the gradient flow equations, which are actually the BPS equations of the 1d SQM.
This means that the worldsheets that they will correspond to are governed by the BPS equations of the equivalent 2d gauged sigma model with target space $\mathcal{M}(\R_{\tau}, \mathfrak{A}_2)$ (defined by setting to zero the expression within the squared terms in~\eqref{eq:m2 x r3:2d model action}), i.e.,
\begin{equation}
  \label{eq:m2 x r3:worldsheet:eqn:components}
  \begin{aligned}
    D_{\xi} C^m - D_t (\tilde{A}_{\tau})^m
    &= 0
      \, ,
    \\
    D_t C^m
    + i F_{\xi t}
    + D_{\xi} (\tilde{A}_{\tau})^m
    - i (\tilde{D}_{\tau}C)^m
    + j_{\theta} \mathscr{F}^m
    &= 0
      \, ,
    \\
    D_{\xi} \mathscr{A}^m
    - k_{\theta} D_t \mathscr{A}^m
    + j_{\theta} (\tilde{D}_{\tau} \mathscr{A})^m
    - i (\mathscr{D} C)^m
    &= 0
      \, ,
  \end{aligned}
\end{equation}
or more explicitly,
\begin{equation}
  \label{eq:m2 x r3:worldsheet:eqn}
  \begin{aligned}
    \Dv{C^m}{\xi} - \Dv{(\tilde{A}_{\tau})^m}{t}
    &= 0
      \, ,
    \\
    \Dv{(\tilde{A}_{\tau})^m}{\xi}
    + \Dv{C^m}{t}
    + i \dv{A_t}{\xi} - i \dv{A_{\xi}}{t}
    &= - i [A_{\xi}, A_t]
      + i (\tilde{D}_{\tau}C)^m
      - j_{\theta} \mathscr{F}^m
      \, ,
    \\
    \Dv{\mathscr{A}^m}{\xi}
    - k_{\theta} \Dv{\mathscr{A}^m}{t}
    &= - j_{\theta} (\tilde{D}_{\tau} \mathscr{A})^m
      + i (\mathscr{D} C)^m
      \, .
  \end{aligned}
\end{equation}
In \cite{er-2023-topol-n, er-2024-topol-gauge-theor}, such worldsheets corresponding to the classical trajectories of 2d gauged sigma models were coined as BPS worldsheets.
We shall do the same here.

Therefore, the BPS worldsheets of the 2d gauged sigma model with target space $\mathcal{M}(\R_{\tau}, \mathfrak{A}_2)$ will be defined by~\eqref{eq:m2 x r3:worldsheet:eqn}.

\subtitle{Flow Lines of the SQM as BPS Worldvolumes of the 3d Model}

In the 3d gauged sigma model with target space $\mathfrak{A}_2$, the SQM flow lines in \eqref{eq:m2 x r3:sqm flow} will correspond to worldvolumes that have, at $\xi = \pm \infty$, $\mathfrak{A}_2^{\theta}$-sheets.\footnote{%
  Just like the $\mathcal{M}^{\theta}(\R_{\tau}, \mathfrak{A}_2)$-solitons, the $\mathfrak{A}_2^{\theta}$-sheets can translate in the $\xi$-direction due to its ``center of mass motion''.
  They can, however, also be fixed at $\xi = \pm \infty$ by adding physically-inconsequential $\mathcal{Q}$-exact terms to the SQM action (as explained in~\autoref{ft:fixing m-A2-soliton centre of mass dof}).
  \label{ft:fixing A2-sheet centre of mass dof}
}
These sheets shall be denoted as $\Sigma_{\pm}(t, \tau, \theta, \mathfrak{A}_2)$, and are defined by \eqref{eq:m2 x r3:sheet eqns} with finite-energy 3d gauge fields $A_{\xi}, A_{\tau}, A_t \rightarrow 0$, i.e.,
\begin{equation}
  \label{eq:m2 x r3:sheet:eqns:no gauge}
  \begin{aligned}
    \dv{C^a}{t} - i \dv{C^a}{\tau}
    &= - j_{\theta} \mathscr{F}^a
      \, ,
    \\
    k_{\theta} \dv{\mathscr{A}^a}{t} - j_{\theta} \dv{\mathscr{A}^a}{\tau}
    &= - i (\mathscr{D} C)^a
      \, .
  \end{aligned}
\end{equation}
Their vertices $\Sigma(\pm \infty, \pm \infty, \theta, \mathfrak{A}_2)$ and $\Sigma(\pm \infty, \mp \infty, \theta, \mathfrak{A}_2)$ at $(t, \tau) = (\pm \infty, \pm \infty)$ or $(\pm \infty, \mp \infty)$ are defined by
\begin{equation}
  \label{eq:m2 x r3:sheet:endpts:no gauge}
  j_{\theta} \mathscr{F}^a
  = 0
  \, ,
  \qquad
  i (\mathscr{D} C)^a
  = 0
  \, ,
\end{equation}
which is simply~\eqref{eq:m2 x r3:sheet:eqns:no gauge} with $d_{\{t, \tau\}} C^a = 0 =  d_{\{t, \tau\}} \mathscr{A}^a$.

These worldvolumes, that the SQM flow lines correspond to, are governed by the BPS equations of the equivalent 3d gauged sigma model with target space $\mathfrak{A}_2$ (defined by setting to zero the expression within the squared terms in~\eqref{eq:m2 x r3:3d model action}), i.e.,
\begin{equation}
  \label{eq:m2 x r3:worldvolume:eqn:components}
  \begin{aligned}
    D_{\xi} C^a - F_{t\tau}
    &= 0
      \, ,
    \\
    D_t C^a - i D_{\tau} C^a
    + F_{\xi \tau} + i F_{\xi t}
    + j_{\theta} \mathscr{F}^a
    &= 0
      \, ,
    \\
    D_{\xi} \mathscr{A}^a
    - k_{\theta} D_t \mathscr{A}^a
    + j_{\theta} D_{\tau} \mathscr{A}^a
    - i (\mathscr{D}C)^a
    &= 0
      \, ,
  \end{aligned}
\end{equation}
or more explicitly,
\begin{equation}
  \label{eq:m2 x r3:worldvolume:eqn}
  \begin{aligned}
    \Dv{C^a}{\xi}
    - \dv{A_{\tau}}{t} + \dv{A_t}{\tau}
    &= [A_t, A_{\tau}]
      \, ,
    \\
    \Dv{C^a}{t} - i \Dv{C^a}{\tau}
    + \dv{A_{\tau}}{\xi} - \dv{A_{\xi}}{\tau}
    + i \dv{A_{t}}{\xi} - i \dv{A_{\xi}}{t}
    &= - [A_{\xi}, A_{\tau}]
      - i [A_{\xi}, A_t]
      - j_{\theta} \mathscr{F}^a
      \, ,
    \\
    \Dv{\mathscr{A}^a}{\xi}
    - k_{\theta} \Dv{\mathscr{A}^a}{t}
    + j_{\theta} \Dv{\mathscr{A}^a}{\tau}
    &=
      i (\mathscr{D}C)^a
      \, .
  \end{aligned}
\end{equation}
Just as how the worldsheets that correspond to the classical trajectories of our 2d gauged sigma model are called BPS \emph{worldsheets}, the worldvolumes that correspond to the classical trajectories of our 3d gauged sigma model shall be called BPS \emph{worldvolumes}.

Therefore, the BPS worldvolumes of the 3d gauged sigma model with target space $\mathfrak{A}_2$ will be defined by~\eqref{eq:m2 x r3:worldvolume:eqn}.

\subtitle{BPS Worldsheets with Boundaries Labeled by Non-constant Paths in $\mathcal{M}(R_t, \mathfrak{A}_2)$}

In the 2d gauged sigma model, the boundaries of the BPS worldsheets are traced out by the endpoints of the $\mathcal{M}^{\theta}(\R_{\tau}, \mathfrak{A}_2)$-solitons as they propagate in $\xi$.
As we have seen at the end of~\autoref{sec:m2 x r3:gc-bf}, these endpoints will correspond to $(\xi, t)$-invariant, $\theta$-deformed, non-constant paths in $\mathcal{M}(\R_{\tau}, \mathfrak{A}_2)$ that we shall, at $\xi = \pm \infty$, denote as $\gamma_{\pm}(\theta, \mathfrak{A}_2)$.
In the equivalent 3d gauged sigma model with target space $\mathfrak{A}_2$, $\gamma_{\pm}(\theta, \mathfrak{A}_2)$ will correspond to $\mathfrak{A}_2^{\theta}$-solitons that we shall denote as $\Gamma_{\pm}(\tau, \theta, \mathfrak{A}_2)$, whose endpoints, in turn, will correspond to $\theta$-deformed $G_{\C}$-BF configurations on $M_2$.

At $\theta$ = 0, if there are $l \geq 4$ such undeformed configurations $\{ \mathcal{E}^1_{\text{BF}}(0), \mathcal{E}^2_{\text{BF}}(0), \dots, \mathcal{E}^l_{\text{BF}}(0)\}$, we can further specify $\Gamma_{\pm}(\tau, 0, \mathfrak{A}_2)$ as $\Gamma^{IJ}_{\pm}(\tau, 0, \mathfrak{A}_2)$, where $I, J \in \{1, 2, \dots, l\}$ indicates that its endpoints, given by $\Gamma^I(- \infty, 0, \mathfrak{A}_2)$ and $\Gamma^J(+ \infty, 0, \mathfrak{A}_2)$, would correspond to the configurations $\mathcal{E}^I_{\text{BF}}(0)$ and $\mathcal{E}^J_{\text{BF}}(0)$, respectively.
As the physical theory is symmetric under a variation of $\theta$, we would continue to have `$l$' such configurations at any value of $\theta$.
In other words, we can also further specify any $\mathfrak{A}_2^{\theta}$-soliton at $\xi = \pm \infty$ as $\Gamma^{IJ}_{\pm}(\tau, \theta, \mathfrak{A}_2)$, where its endpoints, given by $\Gamma^{I}(- \infty, \theta, \mathfrak{A}_2)$ and $\Gamma^{J}(+ \infty, \theta, \mathfrak{A}_2)$, would correspond to $\mathcal{E}^I_{\text{BF}}(\theta)$ and $\mathcal{E}^J_{\text{BF}}(\theta)$, respectively, with the $\mathcal{E}^{*}_{\text{BF}}(\theta)$'s being $l$ number of $\theta$-deformed $G_{\C}$-BF configurations on $M_2$.
Consequently, in the 2d gauged sigma model, we can also further specify $\gamma_{\pm}(\theta, \mathfrak{A}_2)$  as $\gamma^{IJ}_{\pm}(\theta, \mathfrak{A}_2)$, where the latter will correspond to a $\Gamma^{IJ}_{\pm}(\tau, \theta, \mathfrak{A}_2)$ $\mathfrak{A}_2^{\theta}$-soliton in the equivalent 3d gauged sigma model.

Since the endpoints of an $\mathcal{M}^{\theta}(\R_{\tau}, \mathfrak{A}_2)$-soliton are now denoted as $\gamma^{**}_{\pm}(\theta, \mathfrak{A}_2)$, we can also denote any $\mathcal{M}^{\theta}(\R_{\tau}, \mathfrak{A}_2)$-soliton at $\xi = \pm \infty$ as $\sigma_{\pm}^{IJ,KL}(t, \theta, \mathfrak{A}_2)$, where its left and right endpoints, given by $\sigma^{IJ}(- \infty, \theta, \mathfrak{A}_2)$ and $\sigma^{KL}(+ \infty, \theta, \mathfrak{A}_2)$, would be $\gamma^{IJ}(\theta, \mathfrak{A}_2)$ and $\gamma^{KL}(\theta, \mathfrak{A}_{2})$, respectively.

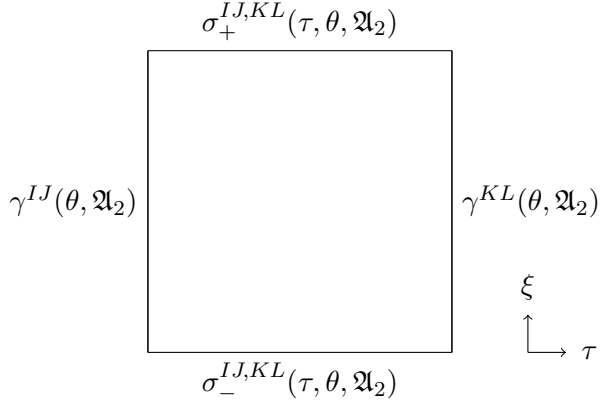
\begin{figure}
  \centering
  \begin{tikzpicture}
    \coordinate (lt) at (0,4) {};
    \coordinate (rt) at (4,4) {}
    edge node[pos=0.5, above] {$\sigma^{IJ, KL}_+(t, \theta, \mathfrak{A}_2)$}
    (lt) {};
    \coordinate (lb) at (0,0) {}
    edge node[pos=0.5, left] {$\gamma^{IJ}(\theta, \mathfrak{A}_2)$}
    (lt) {};
    \coordinate (rb) at (4,0) {}
    edge node[pos=0.5, below] {$\sigma^{IJ, KL}_-(t, \theta, \mathfrak{A}_2)$}
    (lb) {}
    edge node[pos=0.5, right] {$\gamma^{KL}(\theta, \mathfrak{A}_2)$}
    (rt) {};
    \draw (lb) -- (lt);
    \draw (rb) -- (rt);
    \coordinate (co) at (5,0);
    \coordinate (cx) at (5.5,0);
    \node at (cx) [right=2pt of cx] {$t$};
    \coordinate (cy) at (5,0.5);
    \node at (cy) [above=2pt of cy] {$\xi$};
    \draw[->] (co) -- (cx);
    \draw[->] (co) -- (cy);
  \end{tikzpicture}
  \caption[]{BPS worldsheet with $\mathcal{M}^{\theta}(\R_{\tau}, \mathfrak{A}_2)$-solitons $\sigma^{IJ, KL}_\pm(t, \theta, \mathfrak{A}_2)$ and boundaries labeled by $\gamma^{IJ}(\theta, \mathfrak{A}_2)$ and $\gamma^{KL}(\theta, \mathfrak{A}_2)$.
  }
  \label{fig:m2 x r3:bps worldsheet}
\end{figure}

As the $\gamma^{**}(\theta, \mathfrak{A}_2)$'s are $\xi$-invariant and therefore, have the same values for all $\xi$, we have BPS worldsheets of the kind shown in \autoref{fig:m2 x r3:bps worldsheet}.

\subtitle{The 2d Model on $\R^2$ and an Open String Theory in $\mathcal{M}(\R_{\tau}, \mathfrak{A}_2)$}

Hence, one can understand the 2d gauged sigma model on $\R^2$ with target space $\mathcal{M}(\R_{\tau}, \mathfrak{A}_2)$ to define an open string theory in $\mathcal{M}(\R_{\tau}, \mathfrak{A}_2)$, with \emph{effective} worldsheet and boundaries shown in~\autoref{fig:m2 x r3:bps worldsheet}, where $\xi$ and $t$ are the temporal and spatial directions, respectively.

\subtitle{BPS Worldvolumes with Boundaries Labeled by $\mathfrak{A}_2^{\theta}$-solitons, and Edges Labeled by $G_{\C}$-BF Configurations on $M_2$}

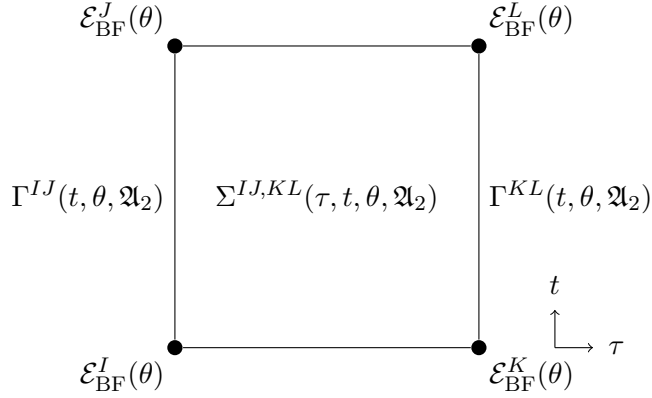
\begin{figure}
  \centering
  \begin{tikzpicture}[%
    auto,%
    every edge/.style={draw},%
    relation/.style={scale=1, sloped, anchor=center, align=center,%
      color=black},%
    vertRelation/.style={scale=1, anchor=center, align=center},%
    dot/.style={circle, fill, minimum size=2*\radius, node contents={},%
      inner sep=0pt},%
    ]
    \let\radius\undefined
    \newlength{\radius}
    \setlength{\radius}{1mm}
    \node at (2, 2) {$\Sigma^{IJ, KL}(t, \tau, \theta, \mathfrak{A}_2)$};
    \node (lt) at (0,4) [dot];
    \node (rt) at (4,4) [dot];
    \node (lb) at (0,0) [dot];
    \node (rb) at (4,0) [dot];
    \node at (lb) [below left]  {$\mathcal{E}^I_{\text{BF}}(\theta)$};
    \node at (lt) [above left]  {$\mathcal{E}^J_{\text{BF}}(\theta)$};
    \node at (rb) [below right]  {$\mathcal{E}^K_{\text{BF}}(\theta)$};
    \node at (rt) [above right]  {$\mathcal{E}^L_{\text{BF}}(\theta)$};
    \draw
    (lb)
    edge node[pos=0.5, left] {$\Gamma^{IJ}(\tau, \theta, \mathfrak{A}_2)$}
    (lt)
    (lt) -- (rt)
    (rt)
    edge node[pos=0.5, right] {$\Gamma^{KL}(\tau, \theta, \mathfrak{A}_2)$}
    (rb)
    (rb) -- (lb)
    ;
    \coordinate (co) at (5,0);
    \coordinate (cx) at (5.5,0);
    \node at (cx) [right=2pt of cx] {$t$};
    \coordinate (cy) at (5,0.5);
    \node at (cy) [above=2pt of cy] {$\tau$};
    \draw[->] (co) -- (cx);
    \draw[->] (co) -- (cy);
  \end{tikzpicture}
  \caption[]{$\mathfrak{A}_2^{\theta}$-sheet $\Sigma^{IJ, KL}(t, \tau, \theta, \mathfrak{A}_2)$ with edges being $\mathfrak{A}_2^{\theta}$-solitons $\Gamma^{IJ}(\tau, \theta, \mathfrak{A}_2)$ and $\Gamma^{KL}(\tau, \theta, \mathfrak{A}_2)$, and vertices corresponding to $\mathcal{E}^I_{\text{BF}}(\theta)$, $\mathcal{E}^J_{\text{BF}}(\theta)$, $\mathcal{E}^K_{\text{BF}}(\theta)$, and $\mathcal{E}^L_{\text{BF}}(\theta)$.
  }
  \label{fig:m2 x r3:frakA-sheet}
\end{figure}

The boundaries and edges of the BPS worldvolumes are traced out by the edges and vertices of the $\mathfrak{A}_2^{\theta}$-sheets, respectively, as they propagate in $\xi$.
As we have seen at the end of~\autoref{sec:m2 x r3:gc-bf}, these edges and vertices would correspond to $\mathfrak{A}_2^{\theta}$-solitons and $\theta$-deformed $G_{\C}$-BF configurations on $M_2$, respectively.
This means that we can further specify any $\mathfrak{A}_2^{\theta}$-sheet at $\xi = \pm \infty$ as $\Sigma_{\pm}^{IJ, KL}(t, \tau, \theta, \mathfrak{A}_2)$, where (i) its left and right edges, given by $\Sigma^{IJ}(- \infty, \tau, \theta, \mathfrak{A}_2)$ and $\Sigma^{KL}(+ \infty, \tau, \theta, \mathfrak{A}_2)$, would correspond to the $\mathfrak{A}_2^{\theta}$-solitons $\Gamma^{IJ}(\tau, \theta, \mathfrak{A}_2)$ and $\Gamma^{KL}(\tau, \theta, \mathfrak{A}_2)$, respectively, and (ii) its four vertices, given by $\Sigma^{I}(- \infty, - \infty, \theta, \mathfrak{A}_2)$, $\Sigma^{J}(- \infty, + \infty, \theta, \mathfrak{A}_2)$, $\Sigma^{K}(+ \infty, - \infty, \theta, \mathfrak{A}_2)$, and $\Sigma^{L}(+ \infty, + \infty, \theta, \mathfrak{A}_2)$, would correspond to $\mathcal{E}^I_{\text{BF}}(\theta)$, $\mathcal{E}^J_{\text{BF}}(\theta)$, $\mathcal{E}^K_{\text{BF}}(\theta)$, and $\mathcal{E}^L_{\text{BF}}(\theta)$, respectively, as shown in \autoref{fig:m2 x r3:frakA-sheet}.

\begin{figure}
  \centering
  \begin{tikzpicture}[%
    auto,%
    every edge/.style={draw},%
    relation/.style={scale=1, sloped, anchor=center, align=center,%
      color=black},%
    vertRelation/.style={scale=1, anchor=center, align=center},%
    dot/.style={circle, fill, minimum size=2*\radius, node contents={},%
      inner sep=0pt},%
    ]
    \let\radius\undefined
    \newlength{\radius}
    \setlength{\radius}{1mm}
    \node (tll) at (-6,4) [dot];
    \node (trl) at (-4,6) [dot];
    \node (brl) at (-4,2) [dot];
    \node (bll) at (-6,0) [dot];
    \draw (bll) edge (tll)
    (tll)--(trl)
    (trl) edge[dashed] (brl)
    (brl) edge[dashed] (bll);
    \node[rotate=45, xslant=0.8, scale=0.9] at (-5,3) {$\Sigma^{IJ, KL}_-(t, \tau, \theta, \mathfrak{A}_2)$};
    \node (tlr) at (4,4) [dot];
    \node (trr) at (6,6) [dot];
    \node (brr) at (6,2) [dot];
    \node (blr) at (4,0) [dot];
    \draw
    (blr) -- (tlr)
    (tlr) -- (trr)
    (trr) edge (brr)
    (brr)--(blr);
    \node[rotate=45, xslant=0.8, scale=0.9] at (5,3.1) {$\Sigma^{IJ, KL}_+(t, \tau, \theta, \mathfrak{A}_2)$};
    \draw
    (bll) edge node[relation, below] {$\mathcal{E}^I_{\text{BF}}(\theta)$} (blr)
    (tll) edge node[relation, above] {$\mathcal{E}^J_{\text{BF}}(\theta)$} (tlr)
    (brl) edge[dashed] node[relation, below] {$\mathcal{E}^K_{\text{BF}}(\theta)$} (brr)
    (trl) edge node[relation, above] {$\mathcal{E}^L_{\text{BF}}(\theta)$} (trr)
    ;
    \node (fn) at ($(bll) + (-2, 0)$) {$\Gamma^{IJ}(\tau, \theta, \mathfrak{A}_2)$} ;
    \node (ff) at ($(bll) + (2, 1)$) {};
    \draw [->]
    (fn) edge [out=350, in=260] (ff);
    \node (bn) at ($(trr) + (2, -0.5)$) {$\Gamma^{KL}(\tau, \theta, \mathfrak{A}_2)$};
    \node (bf) at ($(trr) + (-2, -1)$) {};
    \draw[->]
    (bn) edge [out=170, in=80 , dashed] (bf);
    \coordinate (cco) at (7, 1);
    \coordinate (ccx) at (7.5, 1);
    \coordinate (ccy) at (7.25,1.25);
    \coordinate (ccz) at (7, 1.5);
    \node at (ccx) [right=2pt of ccx] {$\xi$};
    \node at (ccy) [above right=2pt of ccy] {$t$};
    \node at (ccz) [above=2pt of ccz] {$\tau$};
    \draw[->] (cco) -- (ccx);
    \draw[->] (cco) -- (ccy);
    \draw[->] (cco) -- (ccz);
  \end{tikzpicture}
  \caption{%
    BPS worldvolume with $\mathfrak{A}_2^{\theta}$-sheets $\Sigma^{IJ, KL}_\pm(t, \tau, \theta, \mathfrak{A}_2)$, boundaries labeled by $\mathfrak{A}_2^{\theta}$-solitons $\Gamma^{IJ}(\tau, \theta, \mathfrak{A}_2)$ and $\Gamma^{KL}(\tau, \theta, \mathfrak{A}_2)$, and edges labeled by $\mathcal{E}^I_{\text{BF}}(\theta)$, $\mathcal{E}^J_{\text{BF}}(\theta)$, $\mathcal{E}^K_{\text{BF}}(\theta)$, and $\mathcal{E}^L_{\text{BF}}(\theta)$.}
  \label{fig:m2 x r3:bps worldvolume}
\end{figure}
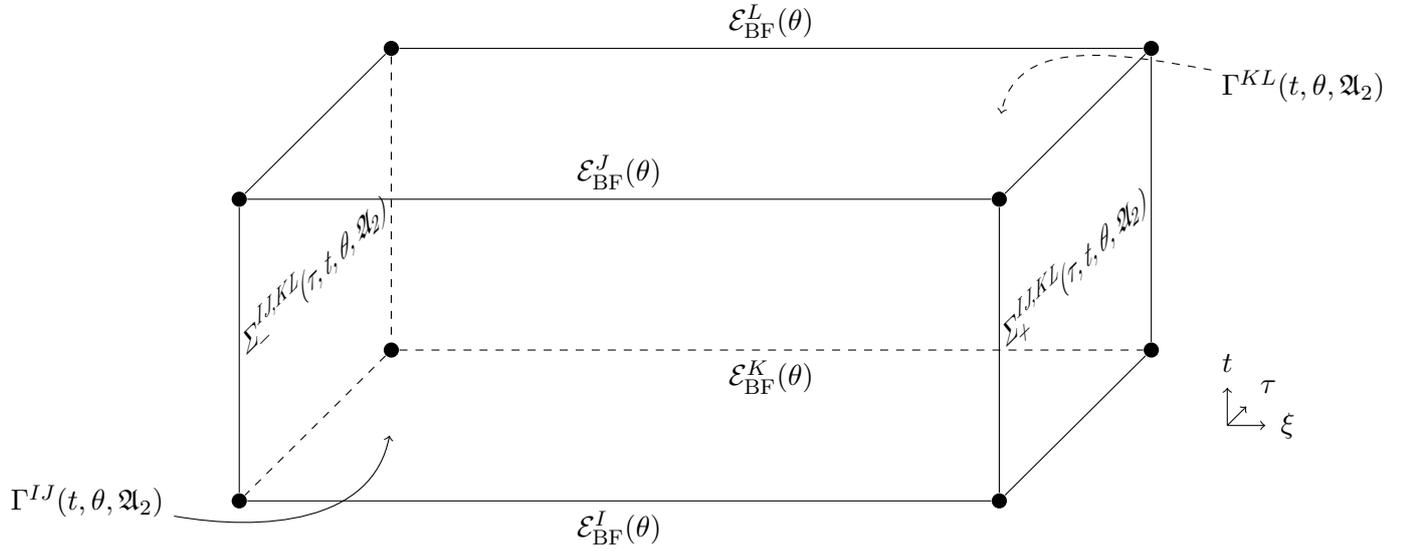

Since the $\mathcal{E}^{*}_{\text{BF}}(\theta)$'s and $\Gamma^{**}(\tau, \theta, \mathfrak{A}_2)$'s are $\xi$-invariant and therefore, have the same value for all $\xi$, we have BPS worldvolumes of the kind shown in \autoref{fig:m2 x r3:bps worldvolume}.

\subtitle{The 3d Model on $\R^3$ and an Open Membrane Theory in $\mathfrak{A}_2$}

Hence, one can understand the 3d gauged sigma model on $\R^3$ with target space $\mathfrak{A}_2$ to define an open membrane theory in $\mathfrak{A}_2$, with \emph{effective} worldvolume, boundaries, and edges shown in \autoref{fig:m2 x r3:bps worldvolume}, where $\xi$ is the temporal direction, and $\tau$ and $t$ are the spatial directions, respectively.

\subtitle{SQM Flow Lines, BPS Worldsheets, and BPS Worldvolumes}

In short, the classical trajectories of HW theory on $M_2 \times \R^3$ are the SQM flow lines in the equivalent 1d SQM, which will correspond to BPS worldsheets in the 2d gauged sigma model with target space $\mathcal{M}(\R_{\tau}, \mathfrak{A}_2)$ of the kind shown in~\autoref{fig:m2 x r3:bps worldsheet}, which, in turn, will correspond to BPS worldvolumes in the 3d gauged sigma model with target space $\mathfrak{A}_2$ of the kind shown in~\autoref{fig:m2 x r3:bps worldvolume}.

\subsection{Soliton String Theory, the HW Partition Function, and an FS type \texorpdfstring{$A_\infty$}{A-infinity}-category of \texorpdfstring{$\mathfrak{A}_2^{\theta}$}{A2-theta}-solitons}
\label{sec:m2 x r3:fs-cat}

\subtitle{The 2d Model as a 2d Gauged LG Model}

Notice that we can also express~\eqref{eq:m2 x r3:worldsheet:eqn} as
\begin{equation}
  \label{eq:m2 x r3:2d lg:worldsheet:eqn}
  \begin{aligned}
    \Dv{C^m}{\xi} - k \Dv{C^m}{t}
    - k \left( \Dv{(\tilde{A}_{\tau})^m}{\xi} - k \Dv{(\tilde{A}_{\tau})^m}{t} \right)
    - j F_{\xi t}
    &= - j (\tilde{D}_{\tau} C)^m
    - i e^{-i\theta} \mathscr{F}^m
      \, ,
    \\
    \Dv{\mathscr{A}^m}{\xi} - k_{\theta} \Dv{\mathscr{A}^m}{t}
    &= - j_{\theta} (\tilde{D}_{\tau} \mathscr{A})^m
    + i (\mathscr{D} C)^m
      \, .
  \end{aligned}
\end{equation}
In turn, this means that we can express the action of the 2d gauged sigma model with target space $\mathcal{M}(\R_{\tau}, \mathfrak{A}_2)$ in~\eqref{eq:m2 x r3:2d model action} as
\begin{equation}
  \label{eq:m2 x r3:2d lg:action}
  \begin{aligned}
    & S_{\text{2d-LG}, \mathcal{M}(\R_{\tau}, \mathfrak{A}_2)}
    \\
    & = \int dt d\xi \bigg(
      \bigg|
      \left( D_{\xi} - k D_t \right) \left( C^m - k (\tilde{A}_{\tau})^m \right)
      - j F_{\xi t}
      + j (\tilde{D}_{\tau} C)^m
      + i e^{-i\theta} \mathscr{F}^m
      \bigg|^2
    \\
    & \qquad \qquad \qquad
      + \left| ( D_{\xi} - k_{\theta} D_t) \mathscr{A}^m
      + j_{\theta} (\tilde{D}_{\tau} \mathscr{A})^m
      - i (\mathscr{D} C)^m
      \right|^2
      + \dots
      \bigg)
    \\
    & = \int dt d\xi \Bigg(
      \Bigg|
      \left( D_{\xi} - k D_t \right) \left( C^m - k (\tilde{A}_{\tau})^m \right)
      - j F_{\xi t}
      + g^{m\bar{n}}_{\mathcal{M}(\R_{\tau}, \mathfrak{A}_2)} \left( \frac{j \zeta}{2} \pdv{W_2}{C^n} \right)^*
      - k g^{m\bar{n}}_{\mathcal{M}(\R_{\tau}, \mathfrak{A}_2)} \left( \frac{j \zeta}{2} \pdv{W_2}{(\tilde{A}_{\tau})^n} \right)^*
      \Bigg|^2
    \\
    & \qquad \qquad \qquad
      + \left| (D_{\xi} - k_{\theta} D_t )\mathscr{A}^m
      + g^{m\bar{n}}_{\mathcal{M}(\R_{\tau}, \mathfrak{A}_2)} \left( \frac{j \zeta}{2} \pdv{W_2}{\mathscr{A}^n} \right)^*
      \right|^2
      + \dots
      \Bigg)
    \\
    &= \int dt d\xi \left(
      \left| D_{\rho} \mathscr{A}^m \right|^2
      + \left| D_{\rho} C^m \right|^2
      + \left| D_{\rho} (\tilde{A}_{\tau})^m \right|^2
      + \left| \pdv{W_2}{\mathscr{A}^{m}} \right|^2
      + \left| \pdv{W_2}{C^m} \right|^2
      + \left| \pdv{W_2}{(\tilde{A}_{\tau})^m} \right|^2
      + \left| F_{\xi t} \right|^2
      + \dots
      \right)
      \, ,
  \end{aligned}
\end{equation}
where $\rho$ is the index on the worldsheet, $g_{\mathcal{M}(\R_{\tau}, \mathfrak{A}_2)}$ is the metric on $\mathcal{M}(\R_{\tau}, \mathfrak{A}_2)$, and $\zeta \coloneq e^{i \theta}$.
In other words, the 2d gauged sigma model with target space $\mathcal{M}(\R_{\tau}, \mathfrak{A}_2)$ can also be interpreted as a 2d gauged LG model with target space $\mathcal{M}(\R_{\tau}, \mathfrak{A}_2)$ and a holomorphic superpotential $W_2(\mathscr{A}, C, \tilde{A}_{\tau})$.

By setting $d_\xi C^m = 0 = d_\xi \mathscr{A}^m$ and $A_\xi, A_t, (\tilde{A}_{\tau})^m \rightarrow 0$ in the expression within the squared terms in~\eqref{eq:m2 x r3:2d lg:action}, we can read off the LG $\mathcal{M}^{\theta}(\R_{\tau}, \mathfrak{A}_2)$-soliton equations corresponding to $\sigma^{IJ, KL}_{\pm}(t, \theta, \mathfrak{A}_2)$ (that re-expresses \eqref{eq:m2 x r3:m-soliton:eqns:no gauge}) as
\begin{equation}
  \label{eq:m2 x r3:2d lg:m-soliton:eqn}
  \begin{aligned}
    \dv{C^m}{t}
    &= - k g^{m\bar{n}}_{\mathcal{M}(\R_{\tau}, \mathfrak{A}_2)} \left( \frac{j \zeta}{2} \pdv{W_2}{C^n} \right)_{\tilde{A}_{\tau} = 0}^*
      - g^{m\bar{n}}_{\mathcal{M}(\R_{\tau}, \mathfrak{A}_2)} \left( \frac{j \zeta}{2} \pdv{W_2}{(\tilde{A}_{\tau})^n} \right)_{\tilde{A}_{\tau} = 0}^*
    \, ,
    \\
    \dv{\mathscr{A}^m}{t}
    &= - k_{\theta} g^{m\bar{n}}_{\mathcal{M}(\R_{\tau}, \mathfrak{A}_2)} \left( \frac{j \zeta}{2} \pdv{W_2}{\mathscr{A}^n} \right)_{\tilde{A}_{\tau} = 0}^*
    \, ,
  \end{aligned}
\end{equation}
where the subscript ``$\tilde{A}_{\tau} = 0$'' means to set $\tilde{A}_{\tau}$ to zero in those terms.

By setting $d_t C^m = 0 = d_t \mathscr{A}^m$ in \eqref{eq:m2 x r3:2d lg:m-soliton:eqn}, we get the LG $\mathcal{M}^{\theta}(\R_{\tau}, \mathfrak{A}_2)$-soliton endpoint equations corresponding to $\sigma^{IJ, KL}(\pm \infty, \theta, \mathfrak{A}_2)$ (that re-expresses \eqref{eq:m2 x r3:m-soliton:endpts:no gauge}) as
\begin{equation}
  \label{eq:m2 x r3:2d lg:m-soliton:endpts}
  \begin{aligned}
    k g^{m\bar{n}}_{\mathcal{M}(\R_{\tau}, \mathfrak{A}_2)} \left( \frac{j \zeta}{2} \pdv{W_2}{C^n} \right)_{\tilde{A}_{\tau} = 0}^*
    + g^{m\bar{n}}_{\mathcal{M}(\R_{\tau}, \mathfrak{A}_2)} \left( \frac{j \zeta}{2} \pdv{W_2}{(\tilde{A}_{\tau})^n} \right)_{\tilde{A}_{\tau} = 0}^*
    &= 0
    \, ,
    \\
    k_{\theta} g^{m\bar{n}}_{\mathcal{M}(\R_{\tau}, \mathfrak{A}_2)} \left( \frac{j \zeta}{2} \pdv{W_2}{\mathscr{A}^n} \right)_{\tilde{A}_{\tau} = 0}^*
    &= 0
    \, .
  \end{aligned}
\end{equation}

Recall from the end of \autoref{sec:m2 x r3:gc-bf} that we are only considering certain $M_2$ and $G$ such that the $\theta$-deformed $G_{\C}$-BF configurations are isolated and non-degenerate.
Next, recall also that such configurations will correspond to the endpoints of the $\mathfrak{A}_2^{\theta}$-solitons; therefore, just like their endpoints, these $\mathfrak{A}_2^{\theta}$-solitons would be isolated and non-degenerate.
As these $\mathfrak{A}_2^{\theta}$-solitons will correspond, in the 2d gauged sigma model, to the endpoints of the $\mathcal{M}^{\theta}(\R_{\tau}, \mathfrak{A}_2)$-solitons, i.e., $\sigma^{IJ, KL}(\pm \infty, \theta, \mathfrak{A}_2)$, this means the latter would also be isolated and non-degenerate.
Thus, from their definition in~\eqref{eq:m2 x r3:2d lg:m-soliton:endpts} which tells us that they are critical points of $W_2(\mathscr{A}, C, \tilde{A}_{\tau})$, we conclude that $W_2(\mathscr{A}, C, \tilde{A}_{\tau})$ can be regarded as a holomorphic Morse function in $\mathcal{M}(\R_{\tau}, \mathfrak{A}_2)$.

Since we are able to interpret the 2d gauged sigma model as a 2d gauged LG model with holomorphic superpotential $W_2(\mathscr{A}, C, \tilde{A}_{\tau})$, we can make use of the well-known fact that an LG $\mathcal{M}^{\theta}(\R_{\tau}, \mathfrak{A}_2)$-soliton  maps to a straight line segment in the complex $W_2$-plane.
Specifically, an LG $\mathcal{M}^{\theta}(\R_{\tau}, \mathfrak{A}_2)$-soliton defined in~\eqref{eq:m2 x r3:2d lg:m-soliton:eqn} will map to a straight line segment $[W_2^{IJ}(\theta), W_2^{KL}(\theta)]$ in the complex $W_2$-plane that starts and ends at the critical values $W_2^{IJ}(\theta) \equiv W_2(\sigma^{IJ}(- \infty, \theta, \mathfrak{A}_2))$ and $W_2^{KL}(\theta) \equiv W_2(\sigma^{KL}(+ \infty, \theta, \mathfrak{A}_2))$, respectively, where its slope depends on $\theta$ (via $\zeta$).
We shall also assume that $\Re(W_2^{IJ}(\theta)) < \Re(W_2^{KL}(\theta))$ and that $I \neq J$, $J \neq K$, and $K \neq L$.

\subtitle{The 2d Gauged LG Model as an LG SQM}

Last but not least, after suitable rescalings, we can recast~\eqref{eq:m2 x r3:2d lg:action} as a 1d LG SQM (that re-expresses \eqref{eq:m2 x r3:sqm action}), where its action will be given by\footnote{%
  In the following expression, we have integrated out $A_\xi$ and omitted the fields corresponding to the finite-energy gauge fields $A_{\{t, \tau\}}$ (as explained in~\autoref{ft:stokes theorem for m2 x r3:sqm}).
  \label{ft:stokes theorem for m2 x r3:2d-lg sqm}
}
\begin{equation}
  \label{eq:m2 x r3:2d lg:sqm:action}
  \begin{aligned}
     S_{\text{2d-LG SQM}, \mathcal{M}(\R_t, \mathcal{M}(\R_{\tau}, \mathfrak{A}_2))}
     = \int d\xi \Bigg(
    & \left| \left(
        \dv{C^u}{\xi}
        - k \dv{\breve{A}^u}{\xi}
      \right)
      + g_{\mathcal{M}(\R_t, \mathcal{M}(\R_{\tau}, \mathfrak{A}_2))}^{uv} \left(
        \pdv{H_2}{C^v}
        - k \pdv{H_2}{\breve{A}^v}
      \right)
      \right|^2
     \\
     & + \left| \dv{\mathscr{A}^u}{\xi}
      + g_{\mathcal{M}(\R_t, \mathcal{M}(\R_{\tau}, \mathfrak{A}_2))}^{uv} \pdv{H_2}{\mathscr{A}^v}
      \right|^2
      + \dots
      \Bigg)
      \, ,
  \end{aligned}
\end{equation}
where $H_2(\mathscr{A}, C, \breve{A})$ is the \emph{real-valued} potential in $\mathcal{M}(\R_t, \mathcal{M}(\R_{\tau}, \mathfrak{A}_2))$, and the subscript ``2d-LG SQM, $\mathcal{M}(\R_t, \mathcal{M}(\R_{\tau}, \mathfrak{A}_2))$'' is to specify that it is a 1d SQM in $\mathcal{M}(\R_t, \mathcal{M}(\R_{\tau}, \mathfrak{A}_2))$ obtained from the equivalent 2d LG model.
We will also refer to this \emph{1d} LG SQM as ``2d-LG SQM'' in the rest of this subsection, so as to distinguish it from something similar that will appear in the next subsection.

The 2d-LG SQM will localize onto configurations that \emph{simultaneously} set to zero the LHS and RHS of the expression within the squared terms in~\eqref{eq:m2 x r3:2d lg:sqm:action}.
In other words, it will localize onto $\xi$-invariant critical points of $H_2(\mathscr{A}, C, \breve{A})$ that will correspond, when $A_{\xi}, A_t, (\tilde{A}_{\tau})^m \rightarrow 0$, to the LG $\mathcal{M}^{\theta}(\R_{\tau}, \mathfrak{A}_2)$-solitons defined by~\eqref{eq:m2 x r3:2d lg:m-soliton:eqn}.
For our choice of $M_2$ and $G$, the LG $\mathcal{M}^{\theta}(\R_{\tau}, \mathfrak{A}_2)$-solitons, just like their endpoints, will be isolated and non-degenerate.
Thus, $H_2(\mathscr{A}, C, \breve{A})$ can be regarded as a real-valued Morse functional in $\mathcal{M}(\R_t, \mathcal{M}(\R_{\tau}, \mathfrak{A}_2))$.

\subtitle{Morphisms from $\gamma^{IJ}(\theta, \mathfrak{A}_2)$ to $\gamma^{KL}(\theta, \mathfrak{A}_2)$ as Floer Homology Classes of Intersecting Thimbles}

Note that we can also describe an LG $\mathcal{M}^{\theta}(\R_{\tau}, \mathfrak{A}_2)$-soliton in terms of the intersection of thimbles.
One can understand such thimbles as submanifolds of a certain fiber space over the complex $W_2$-plane.
In particular, solutions satisfying
\begin{equation}
  \label{eq:m2 x r3:left thimble}
  \lim_{t \rightarrow - \infty} \sigma_{\pm}(t, \theta, \mathfrak{A}_2) = \sigma^{IJ}(-\infty, \theta, \mathfrak{A}_2)
\end{equation}
are known as left thimbles with the collection of such solutions denoted as $S^{IJ}_{\text{BF}}$, whilst those satisfying
\begin{equation}
  \label{eq:m2 x r3:right thimble}
  \lim_{t \rightarrow + \infty} \sigma_{\pm}(t, \theta, \mathfrak{A}_2) = \sigma^{KL}(+\infty, \theta, \mathfrak{A}_2)
\end{equation}
are known as right thimbles with the collection of such solutions denoted as $S^{KL}_{\text{BF}}$.
Therefore, left and right thimbles would correspond, respectively, to the left and right endpoints of an LG $\mathcal{M}^{\theta}(\R_{\tau}, \mathfrak{A}_2)$-soliton solution $\sigma^{IJ, KL}_{\pm}(t, \theta, \mathfrak{A}_2)$.

Clearly, a $\sigma^{IJ, KL}_{\pm}(t, \theta, \mathfrak{A}_2)$-soliton must simultaneously be in a left and right thimble.
Thus, it can be represented as a transversal intersection of the left and right thimble in the fiber space over the line segment $[W_2^{IJ}(\theta), W_2^{KL}(\theta)]$.\footnote{%
  This intersection is guaranteed at some $\theta$, for which we can freely tune as the physical theory is symmetric under its variation.
  \label{ft:intersection of thimbles}
}
In other words, each $\sigma^{IJ, KL}_{\pm}(t, \theta, \mathfrak{A}_2)$-soliton pair, whose left and right endpoints correspond to $\gamma^{IJ}(\theta, \mathfrak{A}_2)$ and $\gamma^{KL}(\theta, \mathfrak{A}_2)$ on a BPS worldsheet as shown in \autoref{fig:m2 x r3:bps worldsheet}, will correspond to a pair of intersection points $p^{IJ, KL}_{\text{BF}, \pm}(\theta) \in S^{IJ}_{\text{BF}} \bigcap S^{KL}_{\text{BF}}$.

This therefore means that the 2d-LG SQM in $\mathcal{M}(\R_t, \mathcal{M}(\R_{\tau}, \mathfrak{A}_2))$ with action \eqref{eq:m2 x r3:2d lg:sqm:action}, will physically realize a Floer homology that we shall name an $\mathcal{M}(\R_{\tau}, \mathfrak{A}_2)$-LG Floer homology.
The chains of the $\mathcal{M}(\R_{\tau}, \mathfrak{A}_2)$-LG Floer complex will be generated by LG $\mathcal{M}^{\theta}(\R_{\tau}, \mathfrak{A}_2)$-solitons which we can identify with $p^{**, **}_{\text{BF}, \pm}(\theta)$, and the $\mathcal{M}(\R_{\tau}, \mathfrak{A}_2)$-LG Floer differential will be realized by the flow lines governed by the gradient flow equations satisfied by the $\xi$-varying configurations which set the expression within the squared terms in \eqref{eq:m2 x r3:2d lg:sqm:action} to zero.
The partition function of the 2d-LG SQM in $\mathcal{M}(\R_t, \mathcal{M}(\R_{\tau}, \mathfrak{A}_2))$ will then be given by\footnote{%
  The `$\theta$' label is omitted in the LHS of the following expression, as the physical theory is actually equivalent for all values of $\theta$.
  \label{ft:theta omission in m2-2d lg partition fn}
}
\begin{equation}
  \label{eq:m2 x r3:2d lg:partition function}
  \mathcal{Z}_{\text{2d-LG SQM}, \mathcal{M}(\R_t, \mathcal{M}(\R_{\tau}, \mathfrak{A}_2))}(G)
  = \sum^l_{I \neq J \neq K \neq L = 1}
  \,
  \sum_{%
    p^{IJ, KL}_{\text{BF}, \pm}
    \in S^{IJ}_{\text{BF}} \cap S^{KL}_{\text{BF}}
  }
  \text{HF}^G_{d_m} \left(
    p^{IJ, KL}_{\text{BF}, \pm}(\theta)
  \right)
  \, .
\end{equation}
Here, the contribution $\text{HF}^G_{d_m} (p^{IJ, KL}_{\text{BF}, \pm}(\theta))$ can be identified with a homology class in an $\mathcal{M}(\R_{\tau}, \mathfrak{A}_2)$-LG Floer homology generated by intersection points of thimbles.
These intersection points represent LG $\mathcal{M}^{\theta}(\R_{\tau}, \mathfrak{A}_2)$-solitons defined by \eqref{eq:m2 x r3:2d lg:m-soliton:eqn}, whose endpoints correspond to $\theta$-deformed, non-constant paths in $\mathcal{M}(\R_{\tau}, \mathfrak{A}_2)$ defined by \eqref{eq:m2 x r3:2d lg:m-soliton:endpts}.
The degree of each chain in the complex is $d_m$, and is counted by the number of outgoing flow lines from the fixed critical points of $H_2(\mathscr{A}, C, \breve{A})$ in $\mathcal{M}(\R_t, \mathcal{M}(\R_{\tau}, \mathfrak{A}_2))$ which can also be identified as $p^{IJ, KL}_{\text{BF}, \pm}(\theta)$.

Therefore, $\mathcal{Z}_{\text{2d-LG SQM}, \mathcal{M}(\R_t, \mathcal{M}(\R_{\tau}, \mathfrak{A}_2))}(G)$ in \eqref{eq:m2 x r3:2d lg:partition function} is a sum of LG $\mathcal{M}^{\theta}(\R_{\tau}, \mathfrak{A}_2)$-solitons defined by \eqref{eq:m2 x r3:2d lg:m-soliton:eqn} with endpoints \eqref{eq:m2 x r3:2d lg:m-soliton:endpts}, or equivalently, $\sigma^{IJ, KL}_{\pm}(t, \theta, \mathfrak{A}_2)$-solitons defined by \eqref{eq:m2 x r3:m-soliton:eqns:no gauge} with endpoints \eqref{eq:m2 x r3:m-soliton:endpts}, whose start and end correspond to the non-constant paths $\gamma^{IJ}(\theta, \mathfrak{A}_2)$ and $\gamma^{KL}(\theta, \mathfrak{A}_2)$, respectively.
In other words, we can write
\begin{equation}
  \label{eq:m2 x r3:2d lg:floer homology as vector}
  \text{CF}_{\mathcal{M}(\R_{\tau}, \mathfrak{A}_2)} \left(
    \gamma^{IJ}(\theta, \mathfrak{A}_2), \gamma^{KL}(\theta, \mathfrak{A}_2)
  \right)_{\pm}
  =
  \text{HF}^G_{d_m} \left(
    p^{IJ, KL}_{\text{BF}, \pm}(\theta)
  \right)
\end{equation}
where $\text{CF}_{\mathcal{M}(\R_{\tau}, \mathfrak{A}_2)} ( \gamma^{IJ}(\theta, \mathfrak{A}_2), \gamma^{KL}(\theta, \mathfrak{A}_2) )_{\pm}$ is a vector representing a $\sigma^{IJ, KL}_{\pm}(t, \theta, \mathfrak{A}_2)$-soliton, such that $\Re(W^{IJ}_2(\theta)) < \Re(W^{KL}_2(\theta))$.

Now, let us make use of the fact that a soliton can be regarded as a morphism between its endpoints.
Specifically, the pair of $\sigma^{IJ, KL}_{\pm}(t, \theta, \mathfrak{A}_2)$-solitons can be regarded as a pair of morphisms $\text{Hom}(\gamma^{IJ}(\theta, \mathfrak{A}_2), \gamma^{KL}(\theta, \mathfrak{A}_2))_{\pm}$ from $\gamma^{IJ}(\theta, \mathfrak{A}_2)$ to $\gamma^{KL}(\theta, \mathfrak{A}_2)$.
Thus, we have the following one-to-one identification\footnote{%
  The `$\theta$' label is omitted in the following expression, as the physical theory is actually equivalent for all values of $\theta$.
  \label{ft:omission of theta in m2 2d-lg}
}
\begin{equation}
  \label{eq:m2 x r3:2d lg:floer hom as morphism}
  \boxed{
    \text{Hom} \left(
      \gamma^{IJ}(\mathfrak{A}_2), \gamma^{KL}(\mathfrak{A}_2)
    \right)_{\pm}
    \Longleftrightarrow
    \text{HF}^G_{d_m} \left(
      p^{IJ, KL}_{\text{BF}, \pm}
    \right)
  }
\end{equation}
where the RHS is proportional to the identity class when $I = K$ and $J = L$, and zero when $I \leftrightarrow K$ and $J \leftrightarrow L$ (since the $\sigma^{IJ, KL}_{\pm}(t, \theta, \mathfrak{A}_2)$-soliton only moves in one direction from $\gamma^{IJ}(\theta, \mathfrak{A}_2)$ to $\gamma^{KL}(\theta, \mathfrak{A}_2)$ as depicted in \autoref{fig:m2 x r3:bps worldsheet}).

\subtitle{Soliton String Theory from the 2d LG Model}

Just like the 2d gauged sigma model, the equivalent 2d gauged LG model will define an open string theory in $\mathcal{M}(\R_{\tau}, \mathfrak{A}_2)$ with effective worldsheets and boundaries shown in~\autoref{fig:m2 x r3:bps worldsheet}, where $\xi$ and $t$ are the temporal and spatial directions, respectively.

The dynamics of this open string theory in $\mathcal{M}(\R_{\tau}, \mathfrak{A}_2)$ will be governed by the BPS worldsheet equations of \eqref{eq:m2 x r3:2d lg:worldsheet:eqn}, where $(\mathscr{A}^m, C^m, (\tilde{A}_{\tau})^m)$ are scalars on the worldsheet corresponding to the holomorphic coordinates of $\mathcal{M}(\R_{\tau}, \mathfrak{A}_2)$.
At an arbitrary instant in time whence $d_{\xi} \mathscr{A}^m = d_{\xi} C^m = 0 = d_{\xi} (\tilde{A}_{\tau})^m = d_{\xi} A_t$ in~\eqref{eq:m2 x r3:2d lg:worldsheet:eqn}, the dynamics of $(\mathscr{A}^m, \mathscr{C}^m, (\tilde{A}_{\tau})^m)$ and the 2d gauge fields $(A_t, A_{\xi})$ along $t$ will be governed by the soliton equations
\begin{equation}
  \label{eq:m2 x r3:2d lg:string m-soliton}
  \begin{aligned}
    \dv{(\tilde{A}_{\tau})^m}{t}
    + k \dv{C^m}{t}
    - j \dv{A_\xi}{t}
    =& - [A_t + k A_{\xi}, (\tilde{A}_{\tau})^m + k C^m]
      + j [A_t, A_\xi]
      \\
    & + g^{m\bar{n}}_{\mathcal{M}(\R_{\tau}, \mathfrak{A}_2)} \left( \frac{j \zeta}{2} \pdv{W_2}{C^{n}} \right)^*
      - k g^{m\bar{n}}_{\mathcal{M}(\R_{\tau}, \mathfrak{A}_2)} \left( \frac{j \zeta}{2} \pdv{W_2}{(\tilde{A}_{\tau})^{n}} \right)^*
      \, ,
    \\
    k_{\theta} \dv{\mathscr{A}^m}{t}
    =& [A_{\xi} - k_{\theta} A_{t}, \mathscr{A}^m]
      + g^{m\bar{n}}_{\mathcal{M}(\R_{\tau}, \mathfrak{A}_2)} \left( \frac{j \zeta}{2} \pdv{W_2}{\mathscr{A}^{n}} \right)^*
      \, .
  \end{aligned}
\end{equation}

Hence, just as a topological A-model with a worldsheet scalar $\mathscr{A}^m$ can be interpreted as an instanton string theory whose corresponding dynamics of the fields along the spatial $t$-direction will be governed by the instanton equation $d \mathscr{A}^m / d t = 0$, our LG model can be interpreted as a \emph{soliton} string theory.

\subtitle{The Normalized HW Partition Function, LG $\mathcal{M}^{\theta}(\R_{\tau}, \mathfrak{A}_2)$-soliton String Scattering, and Maps of an $A_{\infty}$-structure}

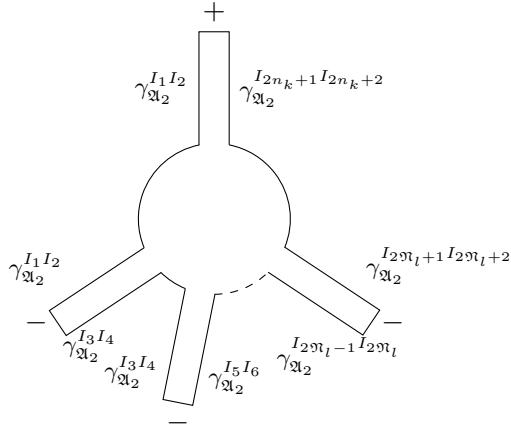
\begin{figure}
  \centering
  \begin{tikzpicture}[declare function={
      lenX(\legLength,\leftAngle,\rightAngle)
      = \legLength * cos((\leftAngle + \rightAngle)/2);
      lenY(\legLength,\leftAngle,\rightAngle)
      = \legLength * sin((\leftAngle + \rightAngle)/2);
      lenLX(\segAngle,\leftAngle,\rightAngle)
      = -2 * tan(\segAngle/2) * sin((\leftAngle + \rightAngle)/2);
      lenLY(\segAngle,\leftAngle,\rightAngle)
      = 2 * tan(\segAngle/2) * cos((\leftAngle + \rightAngle)/2);
    }]
    \def \NumSeg {8}                                
    \def \Rad {1}                                   
    \def \Leg {1.5}                                 
    \def \SegAngle {180/\NumSeg}                    
    \def \TpRtAngle {{(\NumSeg - 1)*\SegAngle/2}}   
    \def \TpLtAngle {{(\NumSeg + 1)*\SegAngle/2}}   
    \def \BaLtAngle {{(\NumSeg + 1)*\SegAngle}}     
    \def \BaRtAngle {{(\NumSeg + 2)*\SegAngle}}     
    \def \BbLtAngle {{(\NumSeg + 3)*\SegAngle}}     
    \def \BbRtAngle {{(\NumSeg + 4)*\SegAngle}}     
    \def \BcLtAngle {{(2 * \NumSeg - 2)*\SegAngle}} 
    \def \BcRtAngle {(2 * \NumSeg - 1)*\SegAngle}   
    \draw ([shift=({\BcRtAngle-360}:\Rad)]3,3) arc ({\BcRtAngle-360}:\TpRtAngle:\Rad);
    \draw ([shift=(\TpLtAngle:\Rad)]3,3) arc (\TpLtAngle:\BaLtAngle:\Rad);
    \draw ([shift=(\BaRtAngle:\Rad)]3,3) arc (\BaRtAngle:\BbLtAngle:\Rad);
    \draw[dashed] ([shift=(\BbRtAngle:\Rad)]3,3) arc (\BbRtAngle:\BcLtAngle:\Rad);
    \draw ([shift=(\TpRtAngle:\Rad)]3cm,3cm)
    -- node[right] {\footnotesize $\gamma^{I_{2n_l + 1} I_{2n_l + 2}}_{\mathfrak{A}_2}$}
    ++(
    {lenX(\Leg,\TpLtAngle,\TpRtAngle)},
    {lenY(\Leg,\TpLtAngle,\TpRtAngle)}
    )
    -- node[above] {$+$}
    ++(
    {lenLX(\SegAngle,\TpLtAngle,\TpRtAngle)},
    {lenLY(\SegAngle,\TpLtAngle,\TpRtAngle)}
    )
    -- node[left] {\footnotesize $\gamma^{I_1 I_2}_{\mathfrak{A}_2}$}
    ++(
    -{lenX(\Leg,\TpLtAngle,\TpRtAngle)},
    -{lenY(\Leg,\TpLtAngle,\TpRtAngle)}
    );
    \draw ([shift=(\BaLtAngle:\Rad)]3,3)
    -- node[near end, above left] {\footnotesize $\gamma^{I_1 I_2}_{\mathfrak{A}_2}$}
    ++(
    {lenX(\Leg,\BaLtAngle,\BaRtAngle)},
    {lenY(\Leg,\BaLtAngle,\BaRtAngle)}
    )
    -- node[left] {$-$}
    ++(
    {lenLX(\SegAngle,\BaLtAngle,\BaRtAngle)},
    {lenLY(\SegAngle,\BaLtAngle,\BaRtAngle)}
    )
    -- node[near start, below] {\footnotesize $\gamma^{I_3 I_4}_{\mathfrak{A}_2}$}
    ++(
    -{lenX(\Leg,\BaLtAngle,\BaRtAngle)},
    -{lenY(\Leg,\BaLtAngle,\BaRtAngle)}
    );
    \draw ([shift=(\BbLtAngle:\Rad)]3,3)
    -- node[near end, left] {\footnotesize $\gamma^{I_3 I_4}_{\mathfrak{A}_2}$}
    ++(
    {lenX(\Leg,\BbLtAngle,\BbRtAngle)},
    {lenY(\Leg,\BbLtAngle,\BbRtAngle)}
    )
    -- node[below] {$-$}
    ++(
    {lenLX(\SegAngle,\BbLtAngle,\BbRtAngle)},
    {lenLY(\SegAngle,\BbLtAngle,\BbRtAngle)}
    )
    -- node[near start, right] {\footnotesize $\gamma^{I_5 I_6}_{\mathfrak{A}_2}$}
    ++(
    -{lenX(\Leg,\BbLtAngle,\BbRtAngle)},
    -{lenY(\Leg,\BbLtAngle,\BbRtAngle)}
    );
    \draw ([shift=(\BcLtAngle:\Rad)]3,3)
    -- node[near end, below=3pt] {\footnotesize $\gamma^{I_{2\mathfrak{N}_l - 1} I_{2\mathfrak{N}_l}}_{\mathfrak{A}_2}$}
    ++(
    {lenX(\Leg,\BcLtAngle,\BcRtAngle)},
    {lenY(\Leg,\BcLtAngle,\BcRtAngle)}
    )
    -- node[right] {$-$}
    ++(
    {lenLX(\SegAngle,\BcLtAngle,\BcRtAngle)},
    {lenLY(\SegAngle,\BcLtAngle,\BcRtAngle)}
    )
    -- node[near start, above right] {\footnotesize $\gamma^{I_{2\mathfrak{N}_l + 1} I_{2\mathfrak{N}_l + 2}}_{\mathfrak{A}_2}$}
    ++(
    -{lenX(\Leg,\BcLtAngle,\BcRtAngle)},
    -{lenY(\Leg,\BcLtAngle,\BcRtAngle)}
    );
  \end{tikzpicture}
  \caption[]{Tree-level scattering BPS worldsheet of incoming ($-$) and outgoing ($+$) LG $\mathcal{M}^{\theta}(\R_{\tau}, \mathfrak{A}_2)$-soliton strings.}
  \label{fig:m2 x r3:mu-d maps}
\end{figure}

The spectrum of HW theory is given by the $\mathcal{Q}$-cohomology of operators.
In particular, its normalized 5d partition function will be a sum over the free-field correlation functions of these operators.\footnote{%
  HW theory is a balanced TQFT \cite{haydys-2015-fukay-seidel}, whence the 5d normalized HW partition function can be computed by bringing down interaction terms to absorb fermion pair zero-modes in the path integral measure.
  These interaction terms can be regarded as operators of the free-field theory that are necessarily in the $\mathcal{Q}$-cohomology (since the non-vanishing partition function ought to remain $\mathcal{Q}$-invariant), where their contribution to the normalized partition function can be understood as free-field correlation functions.
  \label{ft:normalized hw partition fn}
}
As our HW theory is semi-classical, these correlation functions will correspond to tree-level scattering only.
From the equivalent 2d-LG SQM and the 2d gauged LG model perspective, the $\mathcal{Q}$-cohomology will be spanned by the LG $\mathcal{M}^{\theta}(\R_{\tau}, \mathfrak{A}_2)$-soliton strings defined by \eqref{eq:m2 x r3:2d lg:m-soliton:eqn}.
In turn, this means that the normalized HW partition function can also be regarded as a sum over tree-level scattering amplitudes of these LG soliton strings.
The BPS worldsheet underlying such a tree-level scattering amplitude is shown in~\autoref{fig:m2 x r3:mu-d maps}.\footnote{%
  Here, we have exploited the topological and hence conformal invariance of the soliton string theory to replace the outgoing LG $\mathcal{M}^{\theta}(\R_{\tau}, \mathfrak{A}_{2})$-soliton strings with their vertex operators on the disc, then used their coordinate-independent operator products to reduce them to a single vertex operator, before finally translating it back as a single outgoing LG $\mathcal{M}^{\theta}(\R_{\tau}, \mathfrak{A}_{2})$-soliton string.
  \label{ft:reason for single outgoing lg soliton in mud map}
}

In other words, we can express the normalized HW partition function as
\begin{equation}
  \label{eq:m2 x r3:2d lg:normalized partition fn}
  \tilde{\mathcal{Z}}_{\text{HW}, M_2 \times \R^3}(G) = \sum_{\mathfrak{N}_l} \mu^{\mathfrak{N}_l}_{\mathfrak{A}_2}
  \, ,
  \qquad
  \mathfrak{N}_l = 1, 2, \dots, \left\lfloor \frac{l - 2}{2} \right\rfloor
\end{equation}
where each
\begin{equation}
  \label{eq:m2 x r3:2d lg:composition maps}
  \boxed{
    \mu^{\mathfrak{N}_l}_{\mathfrak{A}_2}: \bigotimes_{i = 1}^{\mathfrak{N}_l}
    \text{Hom} \left(
      \gamma^{I_{2i - 1} I_{2i}}(\mathfrak{A}_2), \gamma^{I_{2(i + 1) - 1} I_{2(i + 1)}}(\mathfrak{A}_2)
    \right)_-
    \longto
    \text{Hom} \left(
      \gamma^{I_1 I_2}(\mathfrak{A}_2), \gamma^{I_{2\mathfrak{N}_l + 1} I_{2\mathfrak{N}_l + 2}}(\mathfrak{A}_2)
    \right)_+
  }
\end{equation}
is a scattering amplitude of $\mathfrak{N}_l$ incoming LG $\mathcal{M}^{\theta}(\R_{\tau}, \mathfrak{A}_2)$-soliton strings $\text{Hom} (\gamma^{I_1 I_2}(\mathfrak{A}_2), \gamma^{I_3 I_4}(\mathfrak{A}_2) )_-$, $\dots$, $\text{Hom} (\gamma^{I_{2\mathfrak{N}_l - 1} I_{2\mathfrak{N}_l}}(\mathfrak{A}_2), \gamma^{I_{2\mathfrak{N}_l + 1} I_{2\mathfrak{N}_l + 2}}(\mathfrak{A}_2) )_-$, and a single outgoing LG $\mathcal{M}^{\theta}(\R_{\tau}, \mathfrak{A}_2)$-soliton string
$\text{Hom} (\gamma^{I_1 I_2}(\mathfrak{A}_2), \gamma^{I_{2\mathfrak{N}_l + 1} I_{2\mathfrak{N}_l + 2}}(\mathfrak{A}_2) )_+$, with left and right boundaries as labeled, whose underlying worldsheet shown in~\autoref{fig:m2 x r3:mu-d maps} can be regarded as a disc with $\mathfrak{N}_l + 1$ vertex operators at the boundary.
In short, $\mu^{\mathfrak{N}_l}_{\mathfrak{A}_2}$ counts pseudoholomorphic discs with $\mathfrak{N}_l + 1$ punctures at the boundary that are mapped to $\mathcal{M}(\R_{\tau}, \mathfrak{A}_2)$ according to the BPS worldsheet equations \eqref{eq:m2 x r3:2d lg:worldsheet:eqn}.

In turn, this means that $\mu^{\mathfrak{N}_l}_{\mathfrak{A}_2}$ counts the moduli of solutions to \eqref{eq:hw bps eqns:m2 x r3:rotated} (or equivalently \eqref{eq:hw bps eqns:m2 x r3}) with $\mathfrak{N}_l + 1$ boundary conditions that can be described as follows.
First, note that we can regard $\R^2$ as the effective worldsheet in \autoref{fig:m2 x r3:mu-d maps} that we shall denote as $\Omega$ such that $M_5$ can be interpreted as a trivial $\R_{\tau} \times M_2$ fibration over $\Omega$.
Then, at the $\mathfrak{N}_l + 1$ soliton strings on $\Omega$ where $\xi = \pm \infty$, \eqref{eq:hw bps eqns:m2 x r3:rotated} will become \eqref{eq:m2 x r3:hw configs} and \eqref{eq:m2 x r3:hw configs:aux conds} with $A_{\xi}, A_{\tau}, A_t \rightarrow 0$, and over the soliton string boundaries on $\Omega$ where $t = \pm \infty$, \eqref{eq:hw bps eqns:m2 x r3:rotated} will become \eqref{eq:m2 x r3:hw configs} and \eqref{eq:m2 x r3:hw configs:aux conds} with $A_\xi, A_{\tau}, A_t \rightarrow 0$ \emph{and} $\partial_t C = 0 = \partial_t \mathscr{A}_w$, which defines solitons along $\R_{\tau}$ whose endpoints are $\theta$-deformed $G_{\C}$-BF configurations on $M_2$.

Note at this point that the collection of $\mu^{\mathfrak{N}_l}_{\mathfrak{A}_2}$ maps in \eqref{eq:m2 x r3:2d lg:composition maps} can be regarded as composition maps defining an $A_{\infty}$-structure.

\subtitle{An FS type $A_{\infty}$-category of $\mathfrak{A}_2^{\theta}$-solitons}

Altogether, this means that the normalized partition function of HW theory on $M_2 \times \R^3$ as expressed in \eqref{eq:m2 x r3:2d lg:normalized partition fn}, manifests a \emph{novel} FS type $A_{\infty}$-category defined by the $\mu^{\mathfrak{N}_l}_{\mathfrak{A}_2}$ maps \eqref{eq:m2 x r3:2d lg:composition maps} and the one-to-one identification \eqref{eq:m2 x r3:2d lg:floer hom as morphism}, where the $\mathfrak{N}_l + 1$ number of objects $\big\{\gamma^{I_1 I_2}(\mathfrak{A}_2), \gamma^{I_3 I_4}(\mathfrak{A}_2), \dots,$ $\gamma^{I_{2\mathfrak{N}_l + 1} I_{2\mathfrak{N}_l + 2}}(\mathfrak{A}_2) \big\}$ correspond to $\mathfrak{A}_2^{\theta}$-solitons with endpoints themselves corresponding to ($\theta$-deformed) $G_{\C}$-BF configurations on $M_2$!

\subsection{Soliton Membrane Theory, the HW Partition Function, and a Fueter type \texorpdfstring{$A_\infty$}{A-infinity}-2-category 2-categorifying the 3d-HW Floer Homology of \texorpdfstring{$M_2$}{M2}}
\label{sec:m2 x r3:fueter-cat}

\subtitle{The 3d Model as a Gauged LG Model}

Notice that we can also express~\eqref{eq:m2 x r3:worldvolume:eqn} (corresponding to \eqref{eq:m2 x r3:worldsheet:eqn} in the 2d gauged sigma model) as
\begin{equation}
  \label{eq:m2 x r3:3d-lg:worldvolume:eqn}
  \begin{aligned}
    i \Dv{C^a}{\xi} + j \Dv{C^a}{t} + k \Dv{C^a}{\tau}
    - i F_{t \tau} - j F_{\tau \xi} - k F_{\xi t}
    &= e^{-i\theta} \mathscr{F}^a
      \, ,
    \\
    i e^{-i\theta} \Dv{\mathscr{A}^a}{\xi}
    + j \Dv{\mathscr{A}^a}{t}
    + k \Dv{\mathscr{A}^a}{\tau}
    &=
      - e^{-i\theta} (\mathscr{D}C)^a
      \, ,
  \end{aligned}
\end{equation}
which are non-constant, $\theta$-deformed, gauged Fueter equations for the $(\mathscr{A}^a, C^a)$ fields (that will correspond to \eqref{eq:m2 x r3:2d lg:worldsheet:eqn} in the 2d gauged LG model).

Doing so, the action of the 3d gauged sigma model with target space $\mathfrak{A}_2$ in~\eqref{eq:m2 x r3:3d model action} can be expressed as
\begin{equation}
  \label{eq:m2 x r3:3d lg:action}
  \begin{aligned}
    S_{\text{3d-LG}, \mathfrak{A}_2} = \frac{1}{e^2} \int dt d\tau d\xi \,
    \bigg(
    & \left| i D_{\xi} C^a + j D_t C^a + k D_{\tau} C^a
      - i F_{t \tau} - j F_{\tau \xi} - k F_{\xi t}
      - e^{-i\theta} \mathscr{F}^a
      \right|^2
    \\
    & + \left| i e^{-i\theta} D_{\xi} \mathscr{A}^a + j D_t \mathscr{A}^a + k D_{\tau} \mathscr{A}^a
      + e^{-i\theta} (\mathscr{D}C)^a
      \right|^2
      + \dots
      \bigg)
    \\
    = \frac{1}{e^2} \int dt d\tau d\xi \,
    \bigg(
    & \left| i D_{\xi} C^a + j D_t C^a + k D_{\tau} C^a
      - i F_{t \tau} - j F_{\tau \xi} - k F_{\xi t}
      - i g_{\mathfrak{A}_2}^{a\bar{b}} \left(
      \frac{i\zeta}{2} \pdv{\mathcal{W}_2}{C^b}
      \right)^{*}
      \right|^2
    \\
    & + \left| i e^{-i\theta} D_{\xi} \mathscr{A}^a + j D_t \mathscr{A}^a + k D_{\tau} \mathscr{A}^a
      + i g_{\mathfrak{A}_2}^{a\bar{b}} \left(
      \frac{i\zeta}{2} \pdv{\mathcal{W}_2}{\mathcal{A}^b}
      \right)^{*}
      \right|^2
      + \dots
      \bigg)
    \\
    = \frac{1}{e^2} \int dt d\tau d\xi \,
    \bigg(
    & \left| D_\varrho \mathcal{A}^a \right|^2
      + \left| D_\varrho C^a \right|^2
      + \left| \pdv{\mathcal{W}_2}{\mathcal{A}^a} \right|^2
      + \left| \pdv{\mathcal{W}_2}{C^a} \right|^2
      + \left| F_{t \tau} \right|^2
      + \left| F_{\tau \xi} \right|^2
      + \left| F_{\xi t} \right|^2
      + \dots
      \bigg)
      \, ,
  \end{aligned}
\end{equation}
where $\varrho$ is the index on the worldvolume, and $g_{\mathfrak{A}_2}$ is the metric on $\mathfrak{A}_2$.
In other words, our 3d gauged sigma model can also be interpreted as a 3d gauged LG model in $\mathfrak{A}_2$ with holomorphic superpotential $\mathcal{W}_2(\mathscr{A}, C) = \int_{M_2} C \wedge \mathscr{F}$.

By setting $d_{\xi} C^a = 0 = d_{\xi} \mathscr{A}^a$ and $A_\xi, A_\tau, A_t \rightarrow 0$ in the expression within the squared terms in~\eqref{eq:m2 x r3:3d lg:action}, we can read off the LG $\mathfrak{A}_2^{\theta}$-sheet equations corresponding to $\Sigma^{IJ, KL}_\pm(t, \tau, \theta, \mathfrak{A}_2)$ (that re-expresses \eqref{eq:m2 x r3:sheet:eqns:no gauge}) as
\begin{equation}
  \label{eq:m2 x r3:3d lg:sheet:eqns}
  \begin{aligned}
    j \dv{C^a}{t} + k \dv{C^a}{\tau}
    &= i g_{\mathfrak{A}_2}^{a\bar{b}} \left(
      \frac{i\zeta}{2} \pdv{\mathcal{W}_2}{C^b}
      \right)^{*}
      \, ,
    \\
    j \dv{\mathscr{A}^a}{t} + k \dv{\mathscr{A}^a}{\tau}
    &= - i g_{\mathfrak{A}_2}^{a\bar{b}} \left(
      \frac{i\zeta}{2} \pdv{\mathcal{W}_2}{\mathscr{A}^b}
      \right)^{*}
      \, .
  \end{aligned}
\end{equation}

By setting $d_t C^a = 0 = d_t \mathscr{A}^a$ in~\eqref{eq:m2 x r3:3d lg:sheet:eqns}, we can read off the LG $\mathfrak{A}_2^{\theta}$-soliton equations corresponding to $\Gamma^{IJ}(\tau, \theta, \mathfrak{A}_2)$ and $\Gamma^{KL}(\tau, \theta, \mathfrak{A}_2)$, or equivalently, the LG $\mathfrak{A}_2^{\theta}$-sheet edge equations corresponding to $\Sigma^{IJ, KL}(\pm \infty, \tau, \theta, \mathfrak{A}_2)$, (that re-expresses~\eqref{eq:m2 x r3:soliton}) as
\begin{equation}
  \label{eq:m2 x r3:3d lg:soliton:eqns}
  \begin{aligned}
    k \dv{C^a}{\tau}
    &= i g_{\mathfrak{A}_2}^{a\bar{b}} \left(
      \frac{i\zeta}{2} \pdv{\mathcal{W}_2}{C^b}
      \right)^{*}
      \, ,
    \\
    k \dv{\mathscr{A}^a}{\tau}
    &= - i g_{\mathfrak{A}_2}^{a\bar{b}} \left(
      \frac{i\zeta}{2} \pdv{\mathcal{W}_2}{\mathscr{A}^b}
      \right)^{*}
      \, .
  \end{aligned}
\end{equation}

By setting $d_{\tau} C^a = 0 = d_{\tau} \mathscr{A}^a$ in~\eqref{eq:m2 x r3:3d lg:soliton:eqns}, we can read off the LG $\mathfrak{A}_2^{\theta}$-soliton endpoint equations corresponding to $\Gamma^{**}(\pm \infty, \theta, \mathfrak{A}_2)$, or equivalently, the LG $\mathfrak{A}_2^{\theta}$-sheet vertex equations corresponding to $\Sigma^{IJ, KL}(\pm \infty, \pm \infty, \theta, \mathfrak{A}_2)$ and $\Sigma^{IJ, KL}(\pm \infty, \mp \infty, \theta, \mathfrak{A}_2)$, (that re-expresses~\eqref{eq:m2 x r3:sheet:endpts:no gauge}) as
\begin{equation}
  \label{eq:m2 x r3:3d lg:sheet:endpts}
  i g_{\mathfrak{A}_2}^{a\bar{b}} \left(
    \frac{i\zeta}{2} \pdv{\mathcal{W}_2}{C^b}
  \right)^{*}
  = 0
  \, ,
  \qquad
  i g_{\mathfrak{A}_2}^{a\bar{b}} \left(
    \frac{i\zeta}{2} \pdv{\mathcal{W}_2}{\mathscr{A}^b}
  \right)^{*}
  = 0
  \, .
\end{equation}

Recall from the end of \autoref{sec:m2 x r3:gc-bf} that we are only considering certain $M_2$ and $G$ such that (the endpoints $\Gamma^{**}(\pm \infty, \theta, \mathfrak{A}_2)$ and thus) the LG $\mathfrak{A}_2^{\theta}$-solitons, and effectively, (the vertices $\Sigma^{IJ, KL}(\pm \infty, \pm \infty, \theta, \mathfrak{A}_2)$, $\Sigma^{IJ, KL}(\pm \infty, \mp \infty, \theta, \mathfrak{A}_2)$ and thus) the LG $\mathfrak{A}_2^{\theta}$-sheets, are isolated and non-degenerate.
Therefore, from their definition in~\eqref{eq:m2 x r3:3d lg:sheet:endpts} which tells us that they correspond to critical points of $\mathcal{W}_2(\mathscr{A}, C)$, we conclude that $\mathcal{W}_2(\mathscr{A}, C)$ can be regarded as a holomorphic Morse function in $\mathfrak{A}_2$.

\subtitle{LG $\mathfrak{A}_2^{\theta}$-sheets Mapping to Quadrilaterals in the Complex $\mathcal{W}_2$-plane}

Just like in \autoref{sec:m2 x r3:fs-cat}, the LG $\mathfrak{A}_2^{\theta}$-solitons will map to straight line segments in the complex $\mathcal{W}_2$-plane.
Specifically, an LG $\mathfrak{A}_2^\theta$-soliton  defined in \eqref{eq:m2 x r3:3d lg:soliton:eqns} maps to a straight line segment $[\mathcal{W}_2^I(\theta), \mathcal{W}_2^J(\theta)]$ in the complex $\mathcal{W}_2$-plane that starts and ends at the critical values $\mathcal{W}_2^I(\theta) \equiv \mathcal{W}_2(\Gamma^I(-\infty, \theta, \mathfrak{A}_2))$ and $\mathcal{W}_2^J(\theta) \equiv \mathcal{W}_2(\Gamma^J(+\infty, \theta, \mathfrak{A}_2))$, respectively, where its slope depends on $\theta$ (via $\zeta$).

Therefore, the LG $\mathfrak{A}_2^{\theta}$-sheets, whose edges are LG $\mathfrak{A}_2^\theta$-solitons, can be understood to map to quadrilaterals in the complex $\mathcal{W}_2$-plane, whose edges are precisely the straight line segments that the LG $\mathfrak{A}_2^\theta$-solitons map to.
Specifically, an LG $\mathfrak{A}_2^{\theta}$-sheet defined in \eqref{eq:m2 x r3:3d lg:sheet:eqns} maps to a quadrilateral whose bottom-left, top-left, bottom-right, and top-right vertices $( \mathcal{W}_2^I(\theta), \mathcal{W}_2^J(\theta), \mathcal{W}_2^K(\theta), \mathcal{W}_2^L(\theta) )$ are the critical points $\mathcal{W}_2^I(\theta) \equiv \mathcal{W}_2( \Sigma^I(- \infty, -\infty, \theta, \mathfrak{A}_2))$, $\mathcal{W}_2^J(\theta) \equiv \mathcal{W}_2( \Sigma^J(- \infty, +\infty, \theta, \mathfrak{A}_2))$, $\mathcal{W}_2^K(\theta) \equiv \mathcal{W}_2( \Sigma^K(+ \infty, -\infty, \theta, \mathfrak{A}_2))$, and $\mathcal{W}_2^L(\theta) \equiv \mathcal{W}_2( \Sigma^L(+ \infty, +\infty, \theta, \mathfrak{A}_2))$, respectively, where the slope of the straight line segments between each left-right vertex pair depends on $\theta$ (via $\zeta$).

This fact will be useful shortly.
We shall also assume that $\text{Re}(\mathcal{W}_2^I(\theta)) < \{ \text{Re}(\mathcal{W}_2^J(\theta)), \text{Re}(\mathcal{W}_2^K(\theta)) \} < \text{Re}(\mathcal{W}_2^L(\theta))$.

\subtitle{The 3d Gauged LG Model as an LG SQM}

Last but not least, after suitable rescalings, we can recast~\eqref{eq:m2 x r3:3d lg:action} as a 1d LG SQM (that re-expresses~\eqref{eq:m2 x r3:sqm action}), where its action will be given by\footnote{%
  In the following expression, we have integrated out $A_\xi$ and omitted the fields corresponding to the finite-energy gauge fields $A_{\{t, \tau\}}$ (as explained in~\autoref{ft:stokes theorem for m2 x r3:sqm}).
  \label{ft:stokes theorem for m2 x r3:3d-lg sqm}
}
\begin{equation}
  \label{eq:m2 x r3:3d lg:sqm action}
  \begin{aligned}
    S_{\text{3d-LG SQM}, \mathcal{M}(\R^2, \mathfrak{A}_2)}
    = \frac{1}{e^2} \int d\xi \Bigg(
    &\left| i \left( \dv{C^u}{\xi}
      - k \dv{\breve{A}^u}{\xi}
      \right)
      + i g^{uv}_{\mathcal{M}(\R^2, \mathfrak{A}_2)} \left( \pdv{\mathfrak{F}_2}{C^v}
      - k \pdv{\mathfrak{F}_2}{\breve{A}^v}
      \right)
      \right|^2
    \\
    &+ \left| i e^{-i\theta} \dv{\mathscr{A}^u}{\xi}
      + i g^{uv}_{\mathcal{M}(\R^2, \mathfrak{A}_2)} \pdv{\mathfrak{F}_2}{\mathscr{A}^v}
      \right|^2
      + \dots
      \Bigg)
      \, .
  \end{aligned}
\end{equation}
Here, $\mathfrak{F}_2(\mathscr{A}, C, \breve{A})$ is the \emph{real-valued} potential in $\mathcal{M}(\R^2, \mathfrak{A}_2)$, and the subscript ``3d-LG SQM, $\mathcal{M}(\R^2, \mathfrak{A}_2)$'' is to specify that it is a 1d SQM with target space $\mathcal{M}(\R^2, \mathfrak{A}_2)$ obtained from the equivalent 3d LG model.
We will also refer to this \emph{1d} LG SQM as ``3d-LG SQM'' in the rest of this subsection, so as to distinguish it from the ``2d-LG SQM'' of \autoref{sec:m2 x r3:fs-cat}.

The 3d-LG SQM will localize onto configurations that \emph{simultaneously} set to zero the LHS and RHS of the expression within the squared terms in~\eqref{eq:m2 x r3:3d lg:sqm action}.
In other words, it will localize onto $\xi$-invariant critical points of $\mathfrak{F}_2(\mathscr{A}, C, \breve{A})$ that will correspond, when $A_{\xi}, A_{\tau}, A_t \rightarrow 0$, to the LG $\mathfrak{A}_2^{\theta}$-sheets defined by~\eqref{eq:m2 x r3:3d lg:sheet:eqns}.
For our choice of $M_2$ and $G$, the LG $\mathfrak{A}_2^{\theta}$-sheets, just like their vertices, will be isolated and non-degenerate.
Thus, $\mathfrak{F}_2(\mathscr{A}, C, \breve{A})$ can be regarded as a \emph{real-valued} Morse functional in $\mathcal{M}(\R^2, \mathfrak{A}_2)$.

\subtitle{Morphisms between  $\mathfrak{A}_2^{\theta}$-solitons as Intersection Floer Homology Classes}

At this point, note that an LG $\mathfrak{A}_2^{\theta}$-sheet in the 3d LG model defined by \eqref{eq:m2 x r3:3d lg:sheet:eqns} (whose edges are LG $\mathfrak{A}_2^\theta$-solitons) will correspond to an LG $\mathcal{M}^{\theta}(\R_{\tau}, \mathfrak{A}_2)$-soliton in the 2d LG model defined by \eqref{eq:m2 x r3:2d lg:m-soliton:eqn} (whose endpoints can be described by thimbles).
This means that similar to the latter, an LG $\mathfrak{A}_2^{\theta}$-sheet can also be described in terms of the intersection of thimble-intersections.
To elucidate this, first, let us refer to a solution satisfying
\begin{equation}
  \label{eq:m2 x r3:3d lg:left soliton}
  \lim_{t \rightarrow - \infty} \Sigma_{\pm}(t, \tau, \theta, \mathfrak{A}_2) = \Gamma^{IJ}(\tau, \theta, \mathfrak{A}_2)
\end{equation}
as a left thimble-intersection, and
\begin{equation}
  \label{eq:m2 x r3:3d lg:right soliton}
  \lim_{t \rightarrow + \infty} \Sigma_{\pm}(t, \tau, \theta, \mathfrak{A}_2) = \Gamma^{KL}(\tau, \theta, \mathfrak{A}_2)
\end{equation}
as a right thimble-intersection.
These correspond to LG $\mathfrak{A}_2^{\theta}$-solitons, and they are, respectively, the left and right edges of an LG $\mathfrak{A}_2^{\theta}$-sheet $\Sigma_{\pm}^{IJ,KL}(t, \tau, \theta, \mathfrak{A}_2)$.

Next, let us also refer to a solution satisfying (i)
\begin{equation}
  \label{eq:m2 x r3:3d lg:bottom-left thimble}
  \lim_{\substack{t \rightarrow - \infty \\ \tau \rightarrow - \infty}} \Sigma_{\pm}(t, \tau, \theta, \mathfrak{A}_2)
  = \Sigma^I( -\infty, -\infty, \theta, \mathfrak{A}_2)
\end{equation}
as a bottom-left thimble, (ii)
\begin{equation}
  \label{eq:m2 x r3:3d lg:top-left thimble}
  \lim_{\substack{t \rightarrow - \infty \\ \tau \rightarrow + \infty}} \Sigma_{\pm}(t, \tau, \theta, \mathfrak{A}_2)
  = \Sigma^J( -\infty, +\infty, \theta, \mathfrak{A}_2)
\end{equation}
as a top-left thimble, (iii)
\begin{equation}
  \label{eq:m2 x r3:3d lg:bottom-right thimble}
  \lim_{\substack{t \rightarrow + \infty \\ \tau \rightarrow - \infty}} \Sigma_{\pm}(t, \tau, \theta, \mathfrak{A}_2)
  = \Sigma^K( +\infty, -\infty, \theta, \mathfrak{A}_2)
\end{equation}
as a bottom-right thimble, and (iv)
\begin{equation}
  \label{eq:m2 x r3:3d lg:top-right thimble}
  \lim_{\substack{t \rightarrow + \infty \\ \tau \rightarrow + \infty}} \Sigma_{\pm}(t, \tau, \theta, \mathfrak{A}_2)
  = \Sigma^L( +\infty, +\infty, \theta, \mathfrak{A}_2)
\end{equation}
as a top-right thimble.
It is clear that~\eqref{eq:m2 x r3:3d lg:bottom-left thimble} and \eqref{eq:m2 x r3:3d lg:top-left thimble} correspond to the bottom and top endpoints of an LG $\mathfrak{A}_2^{\theta}$-soliton solution $\Gamma^{IJ}(\tau, \theta, \mathfrak{A}_2)$, i.e., the bottom and top vertices of the left edge of an LG $\mathfrak{A}_2^{\theta}$-sheet solution $\Sigma_{\pm}^{IJ, KL}(t, \tau, \theta, \mathfrak{A}_2)$. It is also clear that~\eqref{eq:m2 x r3:3d lg:bottom-right thimble} and \eqref{eq:m2 x r3:3d lg:top-right thimble} correspond to the bottom and top endpoints of an LG $\mathfrak{A}_2^{\theta}$-soliton solution $\Gamma^{KL}(\tau, \theta, \mathfrak{A}_2)$, i.e., the bottom and top vertices of the right edge of an LG $\mathfrak{A}_2^{\theta}$-sheet solution $\Sigma_{\pm}^{IJ, KL}(t, \tau, \theta, \mathfrak{A}_2)$.

Notice that a (i) $\Gamma^{IJ}(\tau, \theta, \mathfrak{A}_2)$-soliton or a (ii) $\Gamma^{KL}(\tau, \theta, \mathfrak{A}_2)$-soliton must simultaneously be in a (i) bottom-left and top-left thimble or (ii) bottom-right and top-right thimble.
They can thus be represented as a transversal intersection between their ``bottom'' and ``top'' thimbles in the fiber space over the line segment (i) $[\mathcal{W}_2^I(\theta), \mathcal{W}_2^J(\theta)]$ or (ii) $[\mathcal{W}_2^K(\theta), \mathcal{W}_2^L(\theta)]$.\footnote{%
  Just like in \autoref{ft:intersection of thimbles}, this intersection is guaranteed at some $\theta$, for which we can freely tune as the physical theory is symmetric under its variation.
  \label{fn:guaranteed intersections of thimbles}
}
Denoting such intersections as (i) $S^{IJ}_{\text{BF}}$ or (ii) $S^{KL}_{\text{BF}}$, each (i) $\Gamma^{IJ}(\tau, \theta, \mathfrak{A}_2)$-soliton or (ii) $\Gamma^{KL}(\tau, \theta, \mathfrak{A}_2)$-soliton, whose bottom and top endpoints correspond to (i) $\mathcal{E}^I_{\text{BF}}(\theta)$ and $\mathcal{E}^J_{\text{BF}}(\theta)$ or (ii) $\mathcal{E}^K_{\text{BF}}(\theta)$ and $\mathcal{E}^L_{\text{BF}}(\theta)$, will correspond to an intersection point (i) $q^{IJ}_{\text{BF}}(\theta) \in S^{IJ}_{\text{BF}}$ or (ii) $q^{KL}_{\text{BF}}(\theta) \in S^{KL}_{\text{BF}}$.

Moreover, from~\eqref{eq:m2 x r3:3d lg:left soliton} and~\eqref{eq:m2 x r3:3d lg:right soliton}, it is clear that a $\Sigma_{\pm}^{IJ, KL}(t, \tau, \theta, \mathfrak{A}_2)$-sheet must simultaneously be in a left thimble-intersection and right thimble-intersection.
It can therefore be represented as a transversal intersection between its left and right thimble-intersection in the fiber space over the quadrilateral with vertices $(\mathcal{W}_2^I(\theta), \mathcal{W}_2^J(\theta), \mathcal{W}_2^K(\theta), \mathcal{W}_2^L(\theta))$.
Each $\Sigma_{\pm}^{IJ, KL}(t, \tau, \theta, \mathfrak{A}_2)$-sheet pair, whose left and right edges correspond to a $\Gamma^{IJ}(\tau, \theta, \mathfrak{A}_2)$-soliton and a $\Gamma^{KL}(\tau, \theta, \mathfrak{A}_2)$-soliton, respectively, will correspond to a pair of intersection points $\{q_{\text{BF}}^{IJ}(\theta), q_{\text{BF}}^{KL}(\theta)\}_\pm \eqqcolon \mathfrak{P}_{\text{BF}, \pm}^{IJ, KL}(\theta) \in S^{IJ}_{\text{BF}}(\theta) \cap S^{KL}_{\text{BF}}(\theta)$.

At any rate, the 3d-LG SQM in $\mathcal{M}(\R^2, \mathfrak{A}_2)$ with action \eqref{eq:m2 x r3:3d lg:sqm action} will physically realize a Floer homology that we shall name an $\mathfrak{A}_2$-3d-LG Floer homology.
The chains of the $\mathfrak{A}_2$-3d-LG Floer complex are generated by LG $\mathfrak{A}_2^{\theta}$-sheets which we can thus identify with $\mathfrak{P}^{**, **}_{\text{BF}, \pm}(\theta)$, and the $\mathfrak{A}_2$-3d-LG Floer differential will be realized by the flow lines governed by the gradient flow equations satisfied by $\xi$-varying configurations which set the expression within the squared terms of \eqref{eq:m2 x r3:3d lg:sqm action} to zero.
The partition function of the 3d-LG SQM in $\mathcal{M}(\R^2, \mathfrak{A}_2)$ will be given by\footnote{%
  The `$\theta$' label is omitted in the LHS of the following expression (as explained in \autoref{ft:theta omission in m2-2d lg partition fn}).
  \label{ft:theta omission in 3d lg partition fn}
}
\begin{equation}
  \label{eq:m2 x r3:3d lg:sqm:partition fn}
  \mathcal{Z}_{\text{3d-LG SQM}, \mathcal{M}(\R^2, \mathfrak{A}_2)}(G)
  = \sum_{I \neq J \neq K \neq L = 1}^l \,
  \sum_{\substack{\mathfrak{P}^{IJ, KL}_{\text{BF}, \pm} \\ \in S^{IJ}_{\text{BF}} \cap S^{KL}_{\text{BF}}}}
  \text{HF}^G_{d_m} \left(
    \mathfrak{P}^{IJ, KL}_{\text{BF}}(\theta)
  \right)
  \, ,
\end{equation}
where the contribution $\text{HF}^G_{d_m}(\mathfrak{P}^{IJ, KL}_\pm(\theta))$ can be identified with a homology class in an $\mathfrak{A}_2$-3d-LG Floer homology generated by intersection points of thimble-intersections.
These intersection points represent LG $\mathfrak{A}_2^{\theta}$-sheets, whose edges correspond to LG $\mathfrak{A}_2^{\theta}$-solitons, and whose vertices will correspond to $\theta$-deformed $G_{\C}$-BF configurations on $M_2$.
The degree of each chain in the complex is $d_m$, and is counted by the number of outgoing flow lines from the fixed critical points of $\mathfrak{F}_2(\mathscr{A}, C, \breve{A})$ in $\mathcal{M}(\R^2, \mathfrak{A}_2)$ which can also be identified as $\mathfrak{P}^{IJ, KL}_{\text{BF},\pm}(\theta)$.

Therefore, $\mathcal{Z}_{\text{3d-LG SQM}, \mathcal{M}(\R^2, \mathfrak{A}_2)}(G)$ in \eqref{eq:m2 x r3:3d lg:sqm:partition fn} is a sum of LG $\mathfrak{A}_2^{\theta}$-sheets defined by (i) \eqref{eq:m2 x r3:3d lg:sheet:eqns} with (ii) edges \eqref{eq:m2 x r3:3d lg:soliton:eqns} and (iii) vertices \eqref{eq:m2 x r3:3d lg:sheet:endpts}, or equivalently, $\Sigma_{\pm}^{IJ, KL}(t, \tau, \theta, \mathfrak{A}_2)$-sheets defined by (i) \eqref{eq:m2 x r3:sheet:eqns:no gauge} and~\eqref{eq:m2 x r3:sheet eqns:aux conds} with (ii) edges \eqref{eq:m2 x r3:soliton} and~\eqref{eq:m2 x r3:sheet eqns:aux conds}, and (iii) vertices \eqref{eq:m2 x r3:soliton:endpts}, respectively.
In other words, we can write
\begin{equation}
  \label{eq:m2 x r3:3d lg:floer-hom as vector}
  \text{CF}_{\mathcal{M}(\R^2, \mathfrak{A}_2)} \left(
    \Gamma^{IJ}(\tau, \theta, \mathfrak{A}_2),
    \Gamma^{KL}(\tau, \theta, \mathfrak{A}_2)
  \right)_\pm
  =
  \text{HF}^G_{d_m} \left(
    \mathfrak{P}^{IJ, KL}_{\text{BF}}(\theta)
  \right)
  \, ,
\end{equation}
where $\text{CF}_{\mathcal{M}(\R^2, \mathfrak{A}_2)} (\Gamma^{IJ}(\tau, \theta, \mathfrak{A}_2), \Gamma^{KL}(\tau, \theta, \mathfrak{A}_2) )_\pm$ is a vector representing a $\Sigma^{IJ, KL}_{\pm}(t, \tau, \theta, \mathfrak{A}_2)$-sheet, whose left and right edges correspond to $\Gamma^{IJ}(\tau, \theta, \mathfrak{A}_2)$ and $\Gamma^{KL}(\tau, \theta, \mathfrak{A}_2)$, respectively, and whose bottom-left, top-left, bottom-right, and top-right vertices correspond to $\mathcal{E}^I_{\text{BF}}(\theta)$, $\mathcal{E}^J_{\text{BF}}(\theta)$, $\mathcal{E}^K_{\text{BF}}(\theta)$, and $\mathcal{E}^L_{\text{BF}}(\theta)$, respectively, such that $\text{Re}(\mathcal{W}_2^I(\theta)) < \{ \text{Re}(\mathcal{W}_2^J(\theta)), \text{Re}(\mathcal{W}_2^K(\theta)) \} < \text{Re}(\mathcal{W}_2^L(\theta))$.

Here, it is useful to note that a sheet can be regarded as a morphism between its edges, which in turn, can be regarded as a morphism between its endpoints.
In other words, the $\Sigma^{IJ, KL}_{\pm}(t, \tau, \theta, \mathfrak{A}_2)$-sheet can be regarded as a 1-morphism $\text{Hom} \big( \Gamma^{IJ}(\tau, \theta, \mathfrak{A}_2), \Gamma^{KL}(\tau, \theta, \mathfrak{A}_2) \big)_\pm$, from its left edge to its right edge.
In turn, the (i) $\Gamma^{IJ}(\tau, \theta, \mathfrak{A}_2)$-soliton and (ii) $\Gamma^{KL}(\tau, \theta, \mathfrak{A}_2)$-soliton can itself be regarded as a 1-morphism (i) $\text{Hom}( \mathcal{E}^I_{\text{BF}}(\theta), \mathcal{E}^J_{\text{BF}}(\theta) )$ and (ii) $\text{Hom}( \mathcal{E}^K_{\text{BF}}(\theta), \mathcal{E}^L_{\text{BF}}(\theta) )$, from its bottom to top endpoint.
Thus, we have the following one-to-one identifications\footnote{%
  The `$\theta$' label is omitted in the following expression (as explained in \autoref{ft:omission of theta in m2 2d-lg}).
  \label{ft:omission of theta in m2 3d-lg}
}
\begin{equation}
  \label{eq:m2 x r3:3d lg:2-morphism}
  \boxed{
    \text{Hom} \left(
      \Gamma^{IJ}(\tau, \mathfrak{A}_2),
      \Gamma^{KL}(\tau, \mathfrak{A}_2)
    \right)_\pm
    \Longleftrightarrow
    \text{Hom} \left(
      \text{Hom}(\mathcal{E}^I_{\text{BF}}, \mathcal{E}^J_{\text{BF}}),
      \text{Hom}(\mathcal{E}^K_{\text{BF}}, \mathcal{E}^L_{\text{BF}})
    \right)_\pm
    \Longleftrightarrow
    \text{HF}^G_{d_m} \left(
      \mathfrak{P}^{IJ, KL}_{\text{BF}, \pm}
    \right)
  }
\end{equation}
where the RHS is proportional to the identity class when $I = K$ and $J = L$, and zero when (i) $I \leftrightarrow K$ and $J \leftrightarrow L$ (since the $\Sigma^{IJ, KL}(t, \tau, \theta, \mathfrak{A}_2)$-sheet only moves in one direction from $\Gamma^{IJ}(\tau, \theta, \mathfrak{A}_2)$ to $\Gamma^{KL}(\tau, \theta, \mathfrak{A}_2)$ as depicted in~\autoref{fig:m2 x r3:frakA-sheet}), and (ii) $I \leftrightarrow J$ or $K \leftrightarrow L$ (since the $\Gamma^{**}(\tau, \theta, \mathfrak{A}_2)$-solitons only move in one direction from $\mathcal{E}^I_{\text{BF}}(\theta)$ to $\mathcal{E}^J_{\text{BF}}(\theta)$ or $\mathcal{E}^K_{\text{BF}}(\theta)$ to $\mathcal{E}^L_{\text{BF}}(\theta)$, as depicted in~\autoref{fig:m2 x r3:frakA-sheet}).

\subtitle{Soliton Membrane Theory from the 3d LG Model}

Just like the 3d gauged sigma model, the equivalent 3d gauged LG model will define an open membrane theory in $\mathfrak{A}_2$ with effective worldvolumes and boundaries shown in \autoref{fig:m2 x r3:bps worldvolume}, where $\xi$ is the temporal direction and $(\tau, t)$ are the spatial directions.

The dynamics of this open membrane theory in $\mathfrak{A}_2$ will be governed by the BPS worldvolume equations of \eqref{eq:m2 x r3:3d-lg:worldvolume:eqn}, where $(\mathscr{A}^a, C^a)$ are scalars on the worldvolume corresponding to the holomorphic coordinates of $\mathfrak{A}_2$.
At an arbitrary instant in time whence $d_{\xi} \mathscr{A}^a = d_{\xi} C^a = 0 = d_{\xi} A_{\{t, \tau\}}$ in \eqref{eq:m2 x r3:3d-lg:worldvolume:eqn}, the dynamics of $(\mathscr{A}^a, C^a)$ and the 3d gauge fields $(A_t, A_{\tau}, A_{\xi})$ along $(\tau, t)$ will be governed by the membrane equations
\begin{equation}
  \label{eq:m2 x r3:soliton sheet:eqn}
  \begin{aligned}
    j \dv{C^a}{t}
    + k \dv{C^a}{\tau}
    - i \dv{A_\tau}{t}
    + i \dv{A_t}{\tau}
    - j \dv{A_{\xi}}{\tau}
    + k \dv{A_{\xi}}{t}
    &= - [ i A_{\xi} + j A_{\tau} + k A_t, C^a]
      - i [A_t, A_\tau]
    \\
    & \qquad
      + [j A_t - k A_{\tau} , A_{\xi}]
      + i g^{a \bar{b}}_{\mathfrak{A}_2} \left(
      \frac{i \zeta}{2} \pdv{\mathcal{W}_2}{C^b}
      \right)^{*}
      \, ,
    \\
    j \dv{\mathscr{A}^a}{\tau} + k \dv{\mathscr{A}^a}{t}
    &= - [ ie^{-i\theta} A_{\xi} + j A_{\tau} + k A_t, \mathscr{A}^a]
      - i g^{a \bar{b}}_{\mathfrak{A}_2} \left(
      \frac{i \zeta}{2} \pdv{\mathcal{W}_2}{\mathscr{A}^b}
      \right)^{*}
      \, .
  \end{aligned}
\end{equation}
Notice that by introducing a complex worldvolume coordinate $z = \tau + i t$, \eqref{eq:m2 x r3:soliton sheet:eqn} can instead be interpreted as soliton equations in $z$.
However, since $z$ is two-dimensional, a more apt description of the equations would be that they are 2d soliton \emph{membrane} equations.

Hence, just as how our 2d gauged LG model in \autoref{sec:m2 x r3:fs-cat}, whose corresponding dynamics of the worldsheet scalars along the spatial $\tau$-direction are governed by 1d soliton equations, can be interpreted as a \emph{soliton string} theory, we can interpret our 3d gauged LG model as a \emph{soliton membrane} theory.

\subtitle{The Normalized HW Partition Function, Soliton Membrane Scattering, and Maps of an $A_{\infty}$-structure}

\begin{figure}
  \centering
  \begin{tikzpicture}[%
    auto,%
    every edge/.style={draw},%
    relation/.style={scale=1, sloped, anchor=center, align=center,%
      color=black},%
    vertRelation/.style={scale=1, anchor=center, align=center},%
    dot/.style={circle, fill, minimum size=2*\radius, node contents={},%
      inner sep=0pt},%
    ]
    \def \WsLength {0.8}  
    \def \WsDepth {0.2}   
    \draw (-2,0) arc (180:360:2 and 0.6);
    \draw[dashed] (2,0) arc (0:180:2 and 0.6);
    \draw[white, thick] (-2,0) arc (180:215:2 and 0.6);
    \draw[white, thick] (2,0) arc (360:325:2 and 0.6);
    \draw ({2*cos(10)}, {2*sin(10)}) arc (10:80:2);
    \coordinate (out-ws-bottom-front-left) at (-0.8, 1.5);
    \coordinate (out-ws-bottom-front-right) at (0, 1.5);  
    \coordinate (out-ws-bottom-back-left) at (-0.1, 1.7);
    \coordinate (out-ws-bottom-back-right) at (0.7, 1.9);
    \coordinate (out-ws-top-front-left) at (-0.8, 3.5);
    \coordinate (out-ws-top-front-right) at (0, 3.5);
    \coordinate (out-ws-top-back-left) at (-0.1, 3.7);
    \coordinate (out-ws-top-back-right) at (0.7, 3.7);
    \draw (out-ws-bottom-front-left)
    -- (out-ws-top-front-left)
    node[below left] {\footnotesize $\mathcal{E}^{I_1}_{\text{BF}}$}
    -- (out-ws-top-back-left)
    -- (out-ws-top-back-right)
    -- (out-ws-top-front-right)
    -- (out-ws-top-front-left)
    ;
    \draw (out-ws-bottom-front-right)
    -- (out-ws-top-front-right)
    ;
    \draw (out-ws-bottom-back-right)
    -- (out-ws-top-back-right)
    node[below right] {\footnotesize $\mathcal{E}^{I_{2 \mathfrak{N}_l + 2}}_{\text{BF}}$}
    ;
    \draw[dashed] (out-ws-bottom-back-left)
    -- (out-ws-top-back-left);
    \draw (-0.1, 3.8) node[above] {$+$}
    ;
    \draw[dashed] ({2*cos(170)}, {2*sin(170)}) arc (170:100:2);
    \draw ({2*cos(170)}, {2*sin(170)}) arc (170:115:2);
    \coordinate (in-ws1-bottom-front-left) at (-1.5, -0.8);
    \coordinate (in-ws1-bottom-front-right) at (-1.5, 0.0);
    \coordinate (in-ws1-bottom-back-left) at (-1.6, -0.1);
    \coordinate (in-ws1-bottom-back-right) at (-1.85, 0.7);
    \coordinate (in-ws1-top-front-left) at (-3.5, -0.8);
    \coordinate (in-ws1-top-front-right) at (-3.5, 0.0);
    \coordinate (in-ws1-top-back-left) at (-3.7, -0.1);
    \coordinate (in-ws1-top-back-right) at (-3.7, 0.7);
    \draw (in-ws1-bottom-front-right)
    -- (in-ws1-top-front-right)
    node[above right] {\footnotesize $\mathcal{E}^{I_1}_{\text{BF}}$}
    ;
    \draw (in-ws1-top-front-left)
    node[below right] {\footnotesize $\mathcal{E}^{I_3}_{\text{BF}}$}
    -- (in-ws1-bottom-front-left)
    ;
    \draw (in-ws1-bottom-back-right)
    -- (in-ws1-top-back-right)
    node[above right] {\footnotesize $\mathcal{E}^{I_2}_{\text{BF}}$}
    -- (in-ws1-top-front-right)
    -- (in-ws1-top-front-left)
    -- (in-ws1-top-back-left)
    -- (in-ws1-top-back-right)
    ;
    \draw[dashed] (in-ws1-bottom-back-left)
    -- (in-ws1-top-back-left);
    ;
    \draw (-3.9, 0) node[left] {$-$}
    ;
    \draw[dashed] ({2*cos(192)}, {2*sin(192)}) arc (192:228:2)
    ;
    \draw ({2*cos(203)}, {2*sin(203)}) arc (203:228:2)
    ;
    \coordinate (in-ws2-bottom-front-left) at (-1.52, -1.26);
    \coordinate (in-ws2-bottom-front-right) at (-0.7, -1.6);
    \coordinate (in-ws2-bottom-back-left) at (-0.7, -1.2);
    \coordinate (in-ws2-bottom-back-right) at (0, -1.6);
    \coordinate (in-ws2-top-front-left) at (-2.5, -2.9);
    \coordinate (in-ws2-top-front-right) at (-1.8, -3.3);
    \coordinate (in-ws2-top-back-left) at (-1.8, -2.9);
    \coordinate (in-ws2-top-back-right) at (-1.1, -3.3);
    \draw (in-ws2-bottom-front-left)
    -- (in-ws2-top-front-left)
    node[above right={0.7} and {-0.2} ] {\footnotesize $\mathcal{E}^{I_4}_{\text{BF}}$}
    ;
    \draw (in-ws2-bottom-back-right)
    -- (in-ws2-top-back-right)
    node[near end, right] {\footnotesize $\mathcal{E}^{I_5}_{\text{BF}}$}
    ;
    \draw (in-ws2-bottom-back-left)
    -- (in-ws2-top-back-left)
    node[above right={0.7} and {-0.15}] {\footnotesize $\mathcal{E}^{I_3}_{\text{BF}}$}
    -- (in-ws2-top-front-left)
    -- (in-ws2-top-front-right)
    -- (in-ws2-top-back-right)
    -- (in-ws2-top-back-left)
    ;
    \draw[dashed] (in-ws2-bottom-front-right)
    -- (in-ws2-top-front-right)
    ;
    \draw (-2, -3.5) node[left] {$-$}
    ;
    \draw[dashed] ({2*cos(255)}, {2*sin(255)}) arc (255:350:2)
    ;
    \coordinate (in-ws3-bottom-front-right) at (1.5, -0.8);
    \coordinate (in-ws3-bottom-front-left) at (1.5, 0.0);
    \coordinate (in-ws3-bottom-back-right) at (1.6, -0.1);
    \coordinate (in-ws3-bottom-back-left) at (1.85, 0.7);
    \coordinate (in-ws3-top-front-right) at (3.5, -0.8);
    \coordinate (in-ws3-top-front-left) at (3.5, 0.0);
    \coordinate (in-ws3-top-back-right) at (3.3, -0.1);
    \coordinate (in-ws3-top-back-left) at (3.3, 0.7);
    \draw (in-ws3-bottom-front-left)
    -- (in-ws3-top-front-left)
    node[above left=0.1cm] {\footnotesize $\mathcal{E}^{I_{2 \mathfrak{N}_l + 1}}_{\text{BF}}$}
    ;
    \draw (in-ws3-top-front-right)
    node[below left] {\footnotesize $\mathcal{E}^{I_{2 \mathfrak{N}_l - 1}}_{\text{BF}}$}
    -- (in-ws3-bottom-front-right)
    ;
    \draw (in-ws3-bottom-back-left)
    -- (in-ws3-top-back-left)
    node[above left] {\footnotesize $\mathcal{E}^{I_{2 \mathfrak{N}_l + 2}}_{\text{BF}}$}
    -- (in-ws3-top-front-left)
    -- (in-ws3-top-front-right)
    ;
    \draw[dashed] (in-ws3-top-back-left)
    -- (in-ws3-top-back-right)
    -- (in-ws3-top-front-right)
    ;
    \draw[dashed] (in-ws3-bottom-back-right)
    -- (in-ws3-top-back-right)
    ;
    \draw (4.3, 0) node[left] {$-$}
    ;
  \end{tikzpicture}
  \caption[]{%
    Tree-level scattering BPS worldvolume of incoming ($-$) and outgoing ($+$) LG $\mathfrak{A}_2^{\theta}$-soliton membranes.}
  \label{fig:m2 x r3:fueter composition maps}
\end{figure}
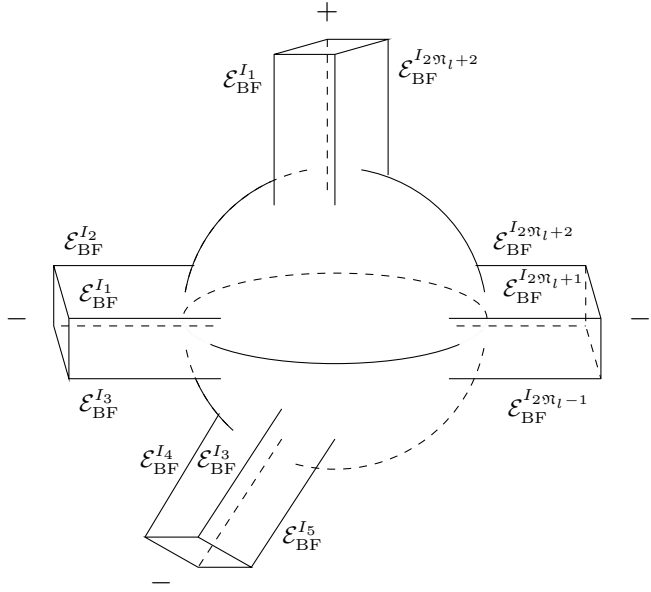

Just like in \autoref{sec:m2 x r3:fs-cat}, the normalized HW partition function will once again be a sum over the free-field correlation functions of operators that are in the $\mathcal{Q}$-cohomology.
From the equivalent 3d-LG SQM and the 3d gauged LG model perspective, the $\mathcal{Q}$-cohomology will be spanned by the LG $\mathfrak{A}_2^{\theta}$-sheets defined by \eqref{eq:m2 x r3:3d lg:sheet:eqns}.
In turn, this means that the normalized HW partition function can also be regarded as a sum over tree-level scattering amplitudes of these LG $\mathfrak{A}_2^{\theta}$-soliton membranes.
The BPS worldvolume underlying such a tree-level scattering amplitude is shown in \autoref{fig:m2 x r3:fueter composition maps}.

In other words, we can express the normalized HW partition function as
\begin{equation}
  \label{eq:m2 x r3:3d lg:partition fn}
  \mathcal{\tilde{Z}}_{\text{HW}, M_2 \times \R^3}(G)
  = \sum_{\mathfrak{N}_l} \varPi_{\mathfrak{A}_2}^{\mathfrak{N}_l}
  \, ,
  \qquad
  \mathfrak{N}_l = 1, 2, \dots, \left\lfloor  \frac{l - 2}{2} \right\rfloor
\end{equation}
where each
\begin{equation}
  \label{eq:m2 x r3:fueter composition maps}
  \boxed{
    \begin{aligned}
      \varPi^{\mathfrak{N}_l}_{\mathfrak{A}_2}: \bigotimes_{i = 1}^{\mathfrak{N}_l}
      & \text{Hom} \left(
        \text{Hom} \left( \mathcal{E}^{I_{2i - 1}}_{\text{BF}}, \mathcal{E}^{I_{2i}}_{\text{BF}} \right),
        \text{Hom} \left( \mathcal{E}^{I_{2(i + 1) - 1}}_{\text{BF}}, \mathcal{E}^{I_{2(i + 1)}}_{\text{BF}} \right)
        \right)_-
      \\
      &\longto
        \text{Hom} \left(
        \text{Hom} \left( \mathcal{E}^{I_1}_{\text{BF}}, \mathcal{E}^{I_2}_{\text{BF}} \right),
        \text{Hom} \left( \mathcal{E}^{I_{2 \mathfrak{N}_l + 1}}_{\text{BF}}, \mathcal{E}^{I_{2 \mathfrak{N}_l + 2}}_{\text{BF}} \right)
        \right)_+
    \end{aligned}
  }
\end{equation}
is a scattering amplitude of $\mathfrak{N}_l$ incoming LG $\mathfrak{A}_2^{\theta}$-soliton membranes $\text{Hom} \Big( \text{Hom} \big(\mathcal{E}^{I_1}_{\text{BF}}, \mathcal{E}^{I_2}_{\text{BF}} \big) , \text{Hom} \big(\mathcal{E}^{I_3}_{\text{BF}}, \mathcal{E}^{I_4}_{\text{BF}}\big) \Big)_-$, $\dots$, $\text{Hom} \Big( \text{Hom}\big(\mathcal{E}^{I_{2\mathfrak{N}_l - 1}}_{\text{BF}}, \mathcal{E}^{I_{2 \mathfrak{N}_l}}_{\text{BF}}\big) , \text{Hom}\big(\mathcal{E}^{I_{2\mathfrak{N}_l + 1}}_{\text{BF}}, \mathcal{E}^{I_{2\mathfrak{N}_l + 2}}_{\text{BF}}\big) \Big)_-$, and a single outgoing LG $\mathfrak{A}_2^{\theta}$-soliton membrane $\text{Hom} \Big( \text{Hom}(\mathcal{E}^{I_1}_{\text{BF}}, \mathcal{E}^{I_2}_{\text{BF}}) , \text{Hom}(\mathcal{E}^{I_{2\mathfrak{N}_l + 1}}_{\text{BF}}, \mathcal{E}^{I_{2\mathfrak{N}_l + 2}}_{\text{BF}}) \Big)_+$, with vertices as labeled, whose underlying worldvolume as shown in \autoref{fig:m2 x r3:fueter composition maps} will correspond to, in the 2d gauged LG model, the worldsheet as shown in \autoref{fig:m2 x r3:mu-d maps}.

Notice that the $\varPi^{\mathfrak{N}_l}_{\mathfrak{A}_2}$ maps in~\eqref{eq:m2 x r3:fueter composition maps} which involve 2-morphisms, like the $\mu^{\mathfrak{N}_l}_{\mathfrak{A}_2}$ maps in~\eqref{eq:m2 x r3:2d lg:composition maps} of the 2d gauged LG model, can also be regarded as composition maps defining an $A_{\infty}$-structure -- in particular, that of a 2-category whose $l$ objects $\{\mathcal{E}^1_{\text{BF}}, \mathcal{E}^2_{\text{BF}}, \dots, \mathcal{E}^l_{\text{BF}}\}$ correspond to ($\theta$-deformed) $G_\C$-BF configurations on $M_2$.

\subtitle{A Fueter type $A_\infty$-2-category 2-categorifying the 3d-HW Floer Homology of $M_2$}

As $G_\C$-BF configurations on $M_2$ are known to generate  the 3d-HW Floer homology of $M_2$ which is itself a 0-category, this 2-category is a 2-categorification of the said Floer homology.

Since this 2-category is determined by the gauged Fueter equation in~\eqref{eq:m2 x r3:3d-lg:worldvolume:eqn}, we shall name it a Fueter type 2-category.

Altogether, this means that the normalized partition function of HW theory on $M_5 = M_2 \times \R^3$, as expressed in~\eqref{eq:m2 x r3:3d lg:partition fn}, manifests a \emph{novel} Fueter type $A_\infty$-2-category, defined by the maps~\eqref{eq:m2 x r3:fueter composition maps} and the identifications~\eqref{eq:m2 x r3:3d lg:2-morphism}, which 2-categorifies the 3d-HW Floer homology of $M_2$!

\subtitle{An Equivalence Between a Fueter type $A_{\infty}$-2-category and an FS type $A_{\infty}$-category}

Recall from \autoref{sec:m2 x r3:fs-cat} that the normalized partition function of HW theory on $M_5 = M_2 \times \R^3$ also manifests the FS type $A_{\infty}$-category of $\mathfrak{A}_2^{\theta}$-solitons.
This means that we have a \emph{novel} equivalence between the Fueter type $A_\infty$-2-category 2-categorifying the 3d-HW Floer homology of $M_2$ and the FS type $A_\infty$-category of $\mathfrak{A}_2^\theta$-solitons!

\section{A Fueter type \texorpdfstring{$A_\infty$}{A-infinity}-2-category of Five-Manifolds}
\label{sec:cy2 x s x r3}

In this section, we will study Spin$(7)$ theory on $\text{Spin}(7) = M_5 \times \R^3$, where $M_5 = CY_2 \times S^1$ with $CY_2$ being a closed and compact Calabi-Yau twofold.
We will recast it as a 3d gauged LG model on $\R^3$, a 2d gauged LG model on $\R^2$, or a 1d LG SQM. Following the approach in \cite[$\S$9]{er-2024-topol-gauge-theor}, we will, via the 8d Spin$(7)$ partition function and its equivalent 2d gauged LG model, be able to physically realize a novel FS type $A_{\infty}$-category of solitons whose endpoints correspond to HW configurations on $CY_2 \times S^1$ that generate a holomorphic Donaldson-Thomas (DT) Floer homology.
Similarly, via the 8d Spin$(7)$ partition function and its equivalent 3d gauged LG model, we will be able to also physically realize a novel Fueter type $A_{\infty}$-2-category that 2-categorifies the holomorphic DT Floer homology of $CY_2 \times S^1$.

\subsection{\texorpdfstring{Spin$(7)$}{Spin(7)} Theory on \texorpdfstring{$CY_2 \times S^1 \times \R^3$}{CY2 x S1 x R3} as a 3d Model on \texorpdfstring{$\R^3$}{R3}, 2d Model on \texorpdfstring{$\R^2$}{R2}, or 1d SQM}
\label{sec:cy2 x s x r3:theory}

Recall from \autoref{sec:topo theories:spin7} that our Spin$(7)$ theory is defined on a Spin$(7)$-manifold.
Let us specialize to the case where it is actually a product manifold containing an $\R^3$ submanifold within.
One such possibility is $\text{Spin}(7) = CY_2 \times S^1 \times \R^3$ \cite{er-2024-topol-gauge-theor, esfahani-2022-monop-singul}.
We will consider this case, and study Spin$(7)$ theory on $\text{Spin}(7) = CY_2 \times S^1 \times \R^3$.

\subtitle{Spin$(7)$ Theory on $CY_2 \times S^1 \times \R^3$}

Relabeling $(x^0, x^1, x^2, x^3)$ as $(t, \tau, \xi, y)$ for future convenience, note that we can express \eqref{eq:spin7 action} as
\begin{equation}
  \label{eq:cy2 x s x r3:action:raw}
  \begin{aligned}
    S_{\text{Spin}(7), CY_2 \times S^1 \times \R^3}
    = \frac{1}{e^2} \int_{CY_2 \times S^1 \times \R^3} dt d\tau d\xi dy d^4 x \, \Tr
    \Bigg(
    & |F_{t \alpha}^+|^2
      + |F_{\tau \alpha}^+|^2
      + |F_{\xi \alpha}^+|^2
      + |F_{y \alpha}^+|^2
    \\
    & + |F_{\alpha\beta}^+|^2
      + |F_{t \tau}^+|^2
      + |F_{t \xi}^+|^2
      + |F_{t y}^+|^2
    \\
    & + |F_{\tau \xi}^+|^2
      + |F_{\tau y}^+|^2
      + |F_{\xi y}^+|^2
      + \dots
      \Bigg)
      \, ,
  \end{aligned}
\end{equation}
where $x^\alpha$ for $\alpha \in \{4, \dots, 7\}$ are the coordinates of $CY_2$, and the ``$\dots$'' contains the fermion terms in \eqref{eq:spin7 action}.
Self-duality of the $F^+$'s mean that we have several equalities amongst the terms in \eqref{eq:cy2 x s x r3:action:raw}, e.g., $|F_{t \tau}^+|^2 = |F_{\xi y}^+|^2 = |F_{45}^+|^2 = |F_{67}^+|^2$, and that it can be written more compactly as
\begin{equation}
  \label{eq:cy2 x s x r3:action}
  S_{\text{Spin}(7), CY_2 \times S^1 \times \R^3}
  = \frac{4}{e^2} \int_{CY_2 \times S^1 \times \R^3} dt d\tau d\xi dy d^4 x \, \Tr
  \Bigg(
  |F_{\xi \alpha}^+|^2
  + |F_{\xi t}^+|^2
  + |F_{\xi \tau}^+|^2
  + |F_{\xi y}^+|^2
  + \dots
  \Bigg)
  \, .
\end{equation}

The conditions the bosons that minimize the action \eqref{eq:cy2 x s x r3:action}, i.e., the BPS equations of Spin$(7)$ theory on $CY_2 \times S^1 \times \R^3$, are easily identified by setting to zero the expression within the squared terms therein.
Exploiting the fact that $CY_2$ is hyperkähler which thus has three complex structures satisfying the quaternionic relations, the BPS equations work out to be
\begin{equation}
  \label{eq:cy2 x s x r:bps eqns}
  \begin{aligned}
    D_{\xi} C + F_{t \tau} - \partial_y A_{\xi}
    &= i \omega^{\kappa \bar{\kappa}} \mathcal{F}_{\kappa \bar{\kappa}}
      \, ,
    \\
    D_{\tau} C - i D_t C - \partial_y A_{\tau} + i \partial_y A_t
    + i (F_{\xi \tau} - i F_{\xi t})
    &= - \frac{i}{2} \varepsilon^{\kappa\lambda} \mathcal{F}_{\kappa\lambda}
      \, ,
    \\
    D_{\xi} \mathcal{A}_\kappa - k D_{\tau} \mathcal{A}_\kappa + j D_t \mathcal{A}_\kappa - \partial_\kappa( A_{\xi} - k A_{\tau} + j A_t)
    &= i (\partial_y \mathcal{A}_\kappa - \mathcal{D}_\kappa C)
      \, ,
  \end{aligned}
\end{equation}
where (i) $(i, j, k)$ are the imaginary numbers with respect to the three complex structures $(I, J, K)$ of $CY_2$;
(ii) $(\kappa, \lambda)$ and $(\bar{\kappa}, \bar{\lambda})$ are indices of the holomorphic and anti-holomorphic coordinates of $CY_2$ (w.r.t.~$I$);
(iii) $\mathcal{A}_\kappa \in \Omega^{(1, 0)}(CY_2, \text{ad}(G)) \otimes \Omega^0(S^1 \times \R^3, \text{ad}(G))$ is a holomorphic gauge connection on $CY_2$ (w.r.t.~$I$), with $\mathcal{F}_{\kappa \bar{\kappa}} \in \Omega^{(1, 1)}(CY_2, \text{ad}(G)) \otimes \Omega^0(S^1 \times \R^3, \text{ad}(G))$ and $\mathcal{F}_{\kappa \lambda} \in \Omega^{(2, 0)}(CY_2, \text{ad}(G)) \otimes \Omega^0(S^1 \times \R^3, \text{ad}(G))$ its $(1, 1)$ and $(2, 0)$-form field strengths, and $\mathcal{D}_\kappa$ its covariant derivative;
(iv) $C = A_y \in \Omega^0(CY_2 \times \R^3, \text{ad}(G)) \otimes \Omega^1(S^1, \text{ad}(G))$ is the gauge connection on $S^1$;
and (v) $(\omega_{\kappa \bar{\kappa}}, \varepsilon_{\kappa\lambda})$ are components of the Kähler two-form and totally antisymmetric rank two tensor of $CY_2$ (w.r.t.~$I$).

Next, note that we are physically free to rotate the $(\tau, t)$-subplane of $\R^3$ about the origin by an angle $\theta$, whence \eqref{eq:cy2 x s x r:bps eqns} become
\begin{equation}
  \label{eq:cy2 x s x r3:bps eqns:rotated}
  \begin{aligned}
    D_{\xi} C + F_{t \tau} - \partial_y A_{\xi}
    &= i \omega^{\kappa \bar{\kappa}} \mathcal{F}_{\kappa \bar{\kappa}}
      \, ,
    \\
    D_{\tau} C - i D_t C - \partial_y A_{\tau} + i \partial_y A_t
    + i (F_{\xi \tau} - i F_{\xi t})
    &= - \frac{i e^{-i\theta}}{2} \varepsilon^{\kappa\lambda} \mathcal{F}_{\kappa\lambda}
      \, ,
    \\
    D_{\xi} \mathcal{A}_\kappa - k_{\theta} D_{\tau} \mathcal{A}_\kappa + j_{\theta} D_t \mathcal{A}_\kappa - \partial_\kappa( A_{\xi} - k_{\theta} A_{\tau} + j_{\theta} A_t)
    &= i (\partial_y \mathcal{A}_\kappa - \mathcal{D}_\kappa C)
      \, .
  \end{aligned}
\end{equation}
This allows us to write the action for Spin$(7)$ theory on $CY_2 \times S^1 \times \R^3$ as
\begin{equation}
  \label{eq:cy2 x s x r3:action:rotated}
  \begin{aligned}
    &S_{\text{Spin}(7), CY_2 \times S^1 \times \R^3}
    \\
    &= \frac{4}{e^2} \int_{\R^3} dt d\tau d\xi \int_{CY_2 \times S^1} dy |dz|^2 \,
      \Tr \Bigg(
      |D_{\xi} C + F_{t\tau} + p|^2
      + |D_{\tau} C - i D_t C + i(F_{\xi\tau} - i F_{\xi t}) + q|^2
    \\
    & \qquad \qquad \qquad \qquad \qquad \qquad \qquad \qquad
      + |D_{\xi} \mathcal{A}_\kappa - k_{\theta} D_{\tau} \mathcal{A}_\kappa + j_{\theta} D_t \mathcal{A}_\kappa + r_\kappa|^2
      + \dots
      \Bigg)
      \, ,
  \end{aligned}
\end{equation}
where
\begin{equation}
  \label{eq:cy2 x s x r3:action:rotated:components}
  \begin{aligned}
    p
    &= - \partial_y A_{\xi}
      - i \omega^{\kappa \bar{\kappa}} \mathcal{F}_{\kappa \bar{\kappa}}
      \, ,
    \\
    q
    &= - \partial_y A_{\tau} + i \partial_y A_t
      + \frac{ie^{-i\theta}}{2} \varepsilon^{\kappa\lambda} \mathcal{F}_{\kappa\lambda}
      \, ,
    \\
    r_\kappa
    &= - \partial_\kappa( A_{\xi} - k_{\theta} A_{\tau} + j_{\theta} A_t)
      - i (\partial_y \mathcal{A}_\kappa - \mathcal{D}_\kappa C)
      \, .
  \end{aligned}
\end{equation}

\subtitle{Spin$(7)$ Theory as a 3d Model}

After suitable rescalings, we can recast \eqref{eq:cy2 x s x r3:action:rotated} as a 3d model on $\R^3$, where its action now reads\footnote{%
  Just like in \autoref{ft:stokes theorem for m2 x r3:3d model}, to arrive at the following expression, we have (i) employed Stokes' theorem and the fact that $CY_2 \times S^1$ has no boundary to omit terms with $\partial_{\{\kappa, y\}} A_{\{t, \tau, \xi\}}$ as they will vanish when integrated over $CY_2 \times S^1$, and
  (ii) integrated out the scalar field $\mathfrak{h}_5(p) = i \omega^{\hat{a} \bar{\hat{a}}} \mathcal{F}_{\hat{a} \bar{\hat{a}}}$ corresponding to the scalar $p$, whose contribution to the action is $|\mathfrak{h}_5(p)|^2$.
  \label{ft:stokes theorem for cy2 x s x r3:3d model}
}
\begin{equation}
  \label{eq:cy2 x s x r3:3d model action}
  \begin{aligned}
    S_{\text{3d}, \mathfrak{A}_5}
    & = \frac{1}{e^2} \int_{\R^3} dt d\tau d\xi \Bigg(
    |D_{\xi} C^{\hat{a}} + F_{t\tau}|^2
    + |D_{\tau} C^{\hat{a}} - i D_t C^{\hat{a}} + i(F_{\xi\tau} - i F_{\xi t}) + q^{\hat{a}}|^2
    \\
    & \qquad \qquad \qquad \qquad + |D_{\xi} \mathcal{A}^{\hat{a}} - k_{\theta} D_{\tau} \mathcal{A}^{\hat{a}} + j_{\theta} D_t \mathcal{A}^{\hat{a}} + r^{\hat{a}}|^2
      + \dots
      \Bigg)
    \\
    & = \frac{1}{e^2} \int_{\R_{\tau} \times \R_{\xi}} d\tau d\xi \int_{\R_t} dt \Bigg(
    |D_{\xi} C^{\hat{a}} - D_{\tau} A_t + P + p^{\hat{a}}|^2
    + |D_{\tau} C^{\hat{a}} + iF_{\xi\tau} + D_{\xi} A_t + Q^{\hat{a}} + q^{\hat{a}}|^2
    \\
    & \qquad \qquad \qquad \qquad \qquad \quad + |D_{\xi} \mathcal{A}^{\hat{a}} - k_{\theta} D_{\tau} \mathcal{A}^{\hat{a}} + R^{\hat{a}} + r^{\hat{a}}|^2
      + \dots
      \Bigg)
      \, .
  \end{aligned}
\end{equation}
Here, $(\mathcal{A}^{\hat{a}}, C^{\hat{a}})$ and $\hat{a}$ are coordinates and indices on the space $\mathfrak{A}_5$ of irreducible $(\mathcal{A}_{\kappa}, C)$ fields on $CY_2 \times S^1$, and
\begin{equation}
  \label{eq:cy2 x s x r3:3d model action:components}
  P
  = \partial_t A_{\tau}
  \, ,
  \qquad
  Q^{\hat{a}}
  = - i D_t C^{\hat{a}} - \partial_t A_{\xi}
  \, ,
  \qquad
  R^{\hat{a}}
  = j_{\theta} D_t \mathcal{A}^{\hat{a}}
  \, ,
\end{equation}
with $(q^{\hat{a}}, r^{\hat{a}})$ corresponding to $(q, r_\kappa)$ in \eqref{eq:cy2 x s x r3:action:rotated:components}.

In other words, Spin$(7)$ theory on $CY_2 \times S^1 \times \R^3$ can be regarded as a 3d gauged sigma model along the $(t, \tau, \xi)$-directions with target space $\mathfrak{A}_5$ and action \eqref{eq:cy2 x s x r3:3d model action}.

\subtitle{Spin$(7)$ Theory as a 2d Model}

From \eqref{eq:cy2 x s x r3:3d model action}, one can see that we can, after suitable rescalings, also recast the 3d model action as an equivalent 2d model action\footnote{%
  Just like in \autoref{ft:stokes theorem for m2 x r3:2d model}, to arrive at the following expression, we have employed Stokes' theorem and the fact that the finite-energy gauge fields $A_{\{t, \tau, \xi\}}$ would vanish at $t \rightarrow \pm \infty$.
  \label{ft:stokes theorem for cy2 x s x r3:2d model}
}
\begin{equation}
  \label{eq:cy2 x s x r3:2d model action}
  \begin{aligned}
    S_{\text{2d}, \mathcal{M}(\R_t, \mathfrak{A}_5)} = \frac{1}{e^2} \int_{\R^2} d\tau d\xi \Bigg(
    & |D_{\xi} C^{\hat{m}} - D_{\tau} (\tilde{A}_t)^{\hat{m}}|^2
    + |D_{\tau} C^{\hat{m}} + iF_{\xi\tau} + D_{\xi} (\tilde{A}_t)^{\hat{m}} + Q^{\hat{m}} + q^{\hat{m}}|^2
    \\
    & + |D_{\xi} \mathcal{A}^{\hat{m}} - k_{\theta} D_{\tau} \mathcal{A}^{\hat{m}} + R^{\hat{m}} + r^{\hat{m}}|^2
      + \dots
      \Bigg)
      \, .
  \end{aligned}
\end{equation}
Here, $(\mathcal{A}^{\hat{m}}, C^{\hat{m}}, (\tilde{A}_t)^{\hat{m}})$ and $\hat{m}$ are coordinates and indices on the path space $\mathcal{M}(\R_t, \mathfrak{A}_5)$ of smooth paths from $\R_t$ to $\mathfrak{A}_5$, with
\begin{equation}
  \label{eq:cy2 x s x r3:2d model action:components}
  Q^{\hat{m}}
  = - i ( \tilde{D}_t C)^{\hat{m}}
  \, ,
  \qquad
  R^{\hat{m}}
  = j_{\theta} (\tilde{D}_t \mathcal{A})^{\hat{m}}
  \, ,
\end{equation}
corresponding to $(Q^{\hat{a}}, R^{\hat{a}})$ in \eqref{eq:cy2 x s x r3:3d model action:components}, $(q^{\hat{m}}, r^{\hat{m}})$ corresponding to $(q^{\hat{a}}, r^{\hat{a}})$, and $(\tilde{A}_t, \tilde{D}_t)$ corresponding to $(A_t, D_t)$, in the underlying 3d model.

In other words, Spin$(7)$ theory on $CY_2 \times S^1 \times \R^3$ can also be regarded as a 2d gauged sigma model along the $(\tau, \xi)$-directions with target space $\mathcal{M}(\R_t, \mathfrak{A}_5)$ and action \eqref{eq:cy2 x s x r3:2d model action}.

\subtitle{Spin$(7)$ Theory as a 1d SQM}

Singling out $\xi$ as the direction in ``time'', the equivalent SQM action can be obtained from~\eqref{eq:cy2 x s x r3:2d model action} after suitable rescalings as\footnote{%
  In the resulting SQM, we have integrated out $A_{\xi}$ and omitted the terms containing the fields corresponding to $A_{\{t, \tau\}}$ (as explained in \autoref{ft:stokes theorem for m2 x r3:sqm}).
  \label{ft:stokes theorem for cy2 x s x r3:sqm}
}
\begin{equation}
  \label{eq:cy2 x s x r3:sqm action}
  \begin{aligned}
    S_{\text{SQM}, \mathcal{M}(\R_{\tau}, \mathcal{M}(\R_t, \mathfrak{A}_5))} = \frac{1}{e^2} \int d\xi \Bigg(
    & \left| \partial_{\xi} C^{\hat{u}}
      + g^{\hat{u}\hat{v}}_{\mathcal{M}(\R_{\tau}, \mathcal{M}(\R_t, \mathfrak{A}_5))} \pdv{h_5}{C^{\hat{v}}}
      \right|^2
      + \left| \partial_{\xi} \breve{A}^{\hat{u}}
      + g^{\hat{u}\hat{v}}_{\mathcal{M}(\R_{\tau}, \mathcal{M}(\R_t, \mathfrak{A}_5))} \pdv{h_5}{\breve{A}^{\hat{v}}}
      \right|^2
    \\
    & + \left| \partial_{\xi} \mathcal{A}^{\hat{u}}
      + g^{\hat{u}\hat{v}}_{\mathcal{M}(\R_{\tau}, \mathcal{M}(\R_t, \mathfrak{A}_5))} \pdv{h_5}{\mathcal{A}^{\hat{v}}}
      \right|^2
      + \dots \Bigg)
      \, ,
  \end{aligned}
\end{equation}
where $(\mathcal{A}^{\hat{u}}, C^{\hat{u}}, \breve{A}^{\hat{u}})$ and $(\hat{u}, \hat{v})$ are coordinates on the path space $\mathcal{M}(\R_{\tau}, \mathcal{M}(\R_t, \mathfrak{A}_5))$ of smooth maps from $\R_{\tau}$ to $\mathcal{M}(\R_t, \mathfrak{A}_5)$ with $\breve{A}^{\hat{u}} \coloneqq (\breve{A}_t + i \breve{A}_{\tau})^{\hat{u}}$ in $\mathcal{M}(\R_{\tau}, \mathcal{M}(\R_t, \mathfrak{A}_5))$ corresponding to $(\tilde{A}_t)^{\hat{m}} + i A_{\tau}$ in the underlying 2d model, and to $A_t + i A_{\tau}$ in the underlying 3d model;
$g_{\mathcal{M}(\R_{\tau}, \mathcal{M}(\R_t, \mathfrak{A}_5))}$ is the metric of $\mathcal{M}(\R_{\tau}, \mathcal{M}(\R_t, \mathfrak{A}_5))$;
and $h_5(\mathcal{A}, C, \breve{A})$ is the SQM potential function.
Note that we can also interpret $\mathcal{M}(\R_{\tau}, \mathcal{M}(\R_t, \mathfrak{A}_5))$ as the double path space $\mathcal{M}(\R^2, \mathfrak{A}_5)$ of smooth maps from $\R^2$ to $\mathfrak{A}_5$.

In short, Spin$(7)$ theory on $CY_2 \times S^1 \times \R^3$ can also be regarded as a 1d SQM along the $\xi$-direction in $\mathcal{M}(\R^2, \mathfrak{A}_5)$ whose action is \eqref{eq:cy2 x s x r3:sqm action}.

\subsection{Non-constant Paths, Solitons, Sheets, and the Holomorphic DT Floer Homology of \texorpdfstring{$CY_2$}{CY2}}
\label{sec:cy2 x s x r3:hw}

By following the same analysis in \autoref{sec:m2 x r3:gc-bf}, we find that the equivalent 1d SQM of Spin$(7)$ theory on $CY_2 \times S^1 \times \R^3$ will localize onto \emph{$\xi$-invariant, $\theta$-deformed}, non-constant double paths in $\mathcal{M}(\R_{\tau}, \mathcal{M}(\R_t, \mathfrak{A}_5))$ which will correspond, in the 2d gauged sigma model with target space $\mathcal{M}(\R_t, \mathfrak{A}_5)$, to \emph{$\xi$-invariant, $\theta$-deformed} solitons along the $\tau$-direction that we shall refer to as $\mathcal{M}^{\theta}(\R_t, \mathfrak{A}_5)$-solitons.

\subtitle{$\mathcal{M}^{\theta}(\R_t, \mathfrak{A}_5)$-solitons in the 2d Gauged Model}

Specifically, such $\mathcal{M}^{\theta}(\R_t, \mathfrak{A}_5)$-solitons are defined by
\begin{equation}
  \label{eq:cy2 x s x r3:m-soliton eqns}
  \begin{aligned}
    \partial_{\tau} (\tilde{A}_t)^{\hat{m}}
    &= - [A_{\tau}, (\tilde{A}_t)^{\hat{m}}]
      + [A_{\xi}, C^{\hat{m}}]
      \, ,
    \\
    \partial_{\tau} C^{\hat{m}} - i \partial_{\tau} A_{\xi}
    &= - [A_{\tau}, C^{\hat{m}} - i A_{\xi}]
      + i (\tilde{D}_t C)^{\hat{m}}
      + [(\tilde{A}_t)^{\hat{m}}, A_{\xi}]
      - q^{\hat{m}}
      \, ,
    \\
    k_{\theta} \partial_{\tau} \mathcal{A}^{\hat{m}}
    &= [A_{\xi} - k_{\theta} A_{\tau}, \mathcal{A}^{\hat{m}}]
      + j_{\theta} (\tilde{D}_t \mathcal{A})^{\hat{m}}
      + r^{\hat{m}}
      \, .
  \end{aligned}
\end{equation}

\subtitle{$\mathfrak{A}^{\theta}_5$-sheets in the 3d Gauged Model}

In turn, they will correspond, in the 3d gauged sigma model with target space $\mathfrak{A}_5$, to $\xi$-invariant, $\theta$-deformed sheets along the $(\tau, t)$-directions that are defined by
\begin{equation}
  \label{eq:cy2 x s x r3:sheet eqns}
  \begin{aligned}
    \partial_{\tau} A_t - \partial_t A_{\tau}
    &= - [A_{\tau}, A_t]
      + [A_{\xi}, C^{\hat{a}}]
      \, ,
    \\
    \partial_{\tau} C^{\hat{a}} - i \partial_t C^{\hat{a}}
    - i \partial_{\tau} A_{\xi} - \partial_t A_{\xi}
    &= - [A_{\tau} - i A_t, C^{\hat{a}} - i A_{\xi}]
      - q^{\hat{a}}
      \, ,
    \\
    k_{\theta} \partial_{\tau} \mathcal{A}^{\hat{a}} - j_{\theta} \partial_t \mathcal{A}^{\hat{a}}
    &= [A_{\xi} - k_{\theta} A_{\tau} + j_{\theta} A_t, \mathcal{A}^{\hat{a}}]
      + r^{\hat{a}}
      \, ,
  \end{aligned}
\end{equation}
and the condition
\begin{equation}
  \label{eq:cy2 x s x r3:sheet eqns:aux cond}
  \mathfrak{h}_5(p) = 0
  \, ,
\end{equation}
where $\mathfrak{h}_5(p)$ is the auxiliary scalar field defined in \autoref{ft:stokes theorem for cy2 x s x r3:3d model}.
We shall refer to such sheets defined by~\eqref{eq:cy2 x s x r3:sheet eqns} and \eqref{eq:cy2 x s x r3:sheet eqns:aux cond} as $\mathfrak{A}^{\theta}_5$-sheets.

\subtitle{$\xi$-independent, $\theta$-deformed Spin$(7)$ Configurations in Spin$(7)$ Theory}

In turn, the 3d configurations defined by~\eqref{eq:cy2 x s x r3:sheet eqns} and \eqref{eq:cy2 x s x r3:sheet eqns:aux cond} will correspond, in Spin$(7)$ theory, to $\xi$-independent, $\theta$-deformed Spin$(7)$ configurations on $CY_2 \times S^1 \times \R^3$ that are defined, via~\eqref{eq:cy2 x s x r3:action:rotated:components}, by
\begin{equation}
  \label{eq:cy2 x s x r3:spin7 configs}
  \begin{aligned}
    \partial_{\tau} A_t - \partial_t A_{\tau}
    &= - [A_{\tau}, A_t]
      + [A_{\xi}, C]
      + \partial_y A_{\xi}
      \, ,
    \\
    \partial_{\tau} C - i \partial_t C
    - i \partial_{\tau} A_{\xi} - \partial_t A_{\xi}
    &= - [A_{\tau} - i A_t, C - i A_{\xi}]
      + \partial_y A_{\tau} - i \partial_y A_t
      - \frac{i e^{-i\theta}}{2} \varepsilon^{\kappa \lambda} \mathcal{F}_{\kappa \lambda}
      \, ,
    \\
    k_{\theta} \partial_{\tau} \mathcal{A}_\kappa - j_{\theta} \partial_t A_\kappa
    &= - \mathcal{D}_\kappa (A_{\xi} - k_{\theta} A_{\tau} + j_{\theta} A_t)
      - i (\partial_y \mathcal{A}_\kappa - \mathcal{D}_\kappa C)
      \, ,
    \end{aligned}
\end{equation}
and the condition
\begin{equation}
  \label{eq:cy2 x s x r3:spin7 configs:aux cond}
  i \omega^{\kappa \bar{\kappa}} \mathcal{F}_{\kappa \bar{\kappa}} = 0
  \, .
\end{equation}

\subtitle{Spin$(7)$ Configurations, $\mathfrak{A}_5^{\theta}$-sheets, $\mathcal{M}^{\theta}(\R_t, \mathfrak{A}_5)$-solitons, and Non-constant Double Paths}

In short, these \emph{$\xi$-independent, $\theta$-deformed} Spin$(7)$ configurations on $CY_2 \times S^1 \times \R^3$ that are defined by~\eqref{eq:cy2 x s x r3:spin7 configs} and \eqref{eq:cy2 x s x r3:spin7 configs:aux cond}, will correspond to the $\mathfrak{A}_5^{\theta}$-sheets defined by~\eqref{eq:cy2 x s x r3:sheet eqns} and \eqref{eq:cy2 x s x r3:sheet eqns:aux cond}, which, in turn, will correspond to the $\mathcal{M}^{\theta}(\R_t, \mathfrak{A}_5)$-solitons defined by~\eqref{eq:cy2 x s x r3:m-soliton eqns}, which, in turn, will correspond to the $\xi$-invariant, $\theta$-deformed, non-constant double paths in $\mathcal{M}(\R^2, \mathfrak{A}_5)$ defined by setting both the LHS and RHS of the expression within the squared terms of \eqref{eq:cy2 x s x r3:sqm action} \emph{simultaneously} to zero.

\subtitle{$\mathcal{M}^{\theta}(\R_t, \mathfrak{A}_5)$-soliton Endpoints Corresponding to Non-constant Paths}

Consider now the fixed endpoints of the $\mathcal{M}^{\theta}(\R_t, \mathfrak{A}_5)$-solitons at $t = \pm \infty$, where we also expect the fields in the 2d gauged sigma model corresponding to the finite-energy 3d gauge fields $A_\xi, A_{\tau}, A_t$ to decay to zero.
They are given by~\eqref{eq:cy2 x s x r3:m-soliton eqns} with $\partial_\tau C^{\hat{m}} = 0 = \partial_{\tau} \mathcal{A}^{\hat{m}}$ and $A_\xi, A_{\tau}, (\tilde{A}_t)^{\hat{m}} \rightarrow 0$, i.e,
\begin{equation}
  \label{eq:cy2 x s x r3:m-soliton:endpts}
  i (\tilde{\partial}_t C)^{\hat{m}}
  = q^{\hat{m}}
  \, ,
  \qquad
  j_{\theta} (\tilde{\partial}_t \mathcal{A})^{\hat{m}}
  = - r^{\hat{m}}
  \, .
\end{equation}
These are $(\xi, \tau)$-invariant, $\theta$-deformed, non-constant paths in $\mathcal{M}(\R_t, \mathfrak{A}_5)$.

\subtitle{$\mathfrak{A}_5^{\theta}$-sheet Edges Corresponding to $\mathfrak{A}_5^{\theta}$-solitons in the 3d Gauged Model}

In turn, \eqref{eq:cy2 x s x r3:m-soliton:endpts} will correspond, in the 3d gauged sigma model, to the fixed edges of the $\mathfrak{A}_5^{\theta}$-sheets at $\tau = \pm \infty$, i.e., $(\xi, \tau)$-invariant, $\theta$-deformed solitons along the $t$-direction that are defined by
\begin{equation}
  \label{eq:cy2 x s x r3:soliton}
  i \partial_t C^{\hat{a}}
  = q^{\hat{a}}
  \, ,
  \qquad
  j_{\theta} \partial_t \mathcal{A}^{\hat{a}}
  = - r^{\hat{a}}
  \, .
\end{equation}
Notice that these equations can also be obtained from \eqref{eq:cy2 x s x r3:sheet eqns} with $\partial_{\tau} C^{\hat{a}} = 0 = \partial_{\tau} \mathcal{A}^{\hat{a}}$ and $A_\xi, A_{\tau}, A_t \rightarrow 0$.
We shall henceforth refer to such solitons defined by~\eqref{eq:cy2 x s x r3:soliton} and \eqref{eq:cy2 x s x r3:sheet eqns:aux cond} as $\mathfrak{A}^{\theta}_5$-solitons.

\subtitle{$\mathfrak{A}^{\theta}_5$-soliton Endpoints or $\mathfrak{A}_5^{\theta}$-sheet Vertices Corresponding to $\theta$-deformed HW Configurations on $CY_2 \times S^1$}

Consider now (i) the fixed endpoints of the $\mathfrak{A}_5^{\theta}$-solitons at $t = \pm \infty$, or equivalently (ii) the vertices of the $\mathfrak{A}_5^{\theta}$-sheets at $\tau, t = \pm \infty$.
They are given by (i)~\eqref{eq:cy2 x s x r3:soliton} and \eqref{eq:cy2 x s x r3:sheet eqns:aux cond} with $\partial_t C^{\hat{a}} = 0 = \partial_t \mathcal{A}^{\hat{a}}$, or equivalently (ii)~\eqref{eq:cy2 x s x r3:sheet eqns} and \eqref{eq:cy2 x s x r3:sheet eqns:aux cond} with $\partial_{\{t, \tau\}} C^{\hat{a}} = 0 = \partial_{\{t, \tau\}} \mathcal{A}^{\hat{a}}$ and $A_{\xi}, A_{\tau}, A_t \rightarrow 0$.
In turn, they will correspond, in Spin$(7)$ theory, to $(\xi, \tau, t)$-independent, $\theta$-deformed configurations that obey~\eqref{eq:cy2 x s x r3:spin7 configs} and~\eqref{eq:cy2 x s x r3:spin7 configs:aux cond} with $\partial_{\{t, \tau\}} C = 0 = \partial_{\{t, \tau\}} \mathcal{A}_\kappa$ and $A_\xi, A_\tau, A_t \rightarrow 0$, i.e.,
\begin{equation}
  \label{eq:cy2 x s x r3:soliton:endpts:spin7}
  ie^{-i\theta} \varepsilon^{\kappa\lambda} \mathcal{F}_{\kappa\lambda}
  = 0
  \, ,
  \qquad
  i (\partial_y \mathcal{A}_\kappa - \mathcal{D}_\kappa C)
  = 0
  \, ,
  \qquad
  i \omega^{\kappa \bar{\kappa}} \mathcal{F}_{\kappa \bar{\kappa}}
  = 0
  \, .
\end{equation}

At $\theta = 0$ or $\pi$, \eqref{eq:cy2 x s x r3:soliton:endpts:spin7} can be written, in the real coordinates of $CY_2$, as
\begin{equation}
  \label{eq:cy2 x s x r3:vw eqns}
  F_{\alpha\beta}^+ = 0
  \, ,
  \qquad
  F_{y \alpha} = 0
  \, .
\end{equation}
These are the HW equations on $CY_2 \times S^1$ with the self-dual two-form field being zero (i.e., \eqref{eq:hw eqns} with $B = 0$ and $A_t = C$).\footnote{%
  Recall from our definition of $C$ below \eqref{eq:cy2 x s x r:bps eqns} that $C$ is not a scalar -- it is actually the gauge connection along the $y$-direction, i.e., $A_y$.
  \label{ft:reminder that C is scalar}
}
We shall, in the rest of this section, refer to configurations that span the space of solutions to these equations as HW configurations on $CY_2 \times S^1$.
One thing to note at this point is that HW configurations on $CY_2 \times S^1$ are known to generate the holomorphic DT Floer homology of $CY_2 \times S^1$ \cite[$\S$5]{er-2024-topol-gauge-theor}.

In other words, the $(\xi, \tau, t)$-independent, $\theta$-deformed Spin$(7)$ configurations corresponding to the endpoints of the $\mathfrak{A}^{\theta}_5$-solitons (or equivalently, the vertices of the $\mathfrak{A}^{\theta}_5$-sheets) are $\theta$-deformed HW configurations on $CY_2 \times S^1$.
We will also assume choices of $CY_2$ whereby such configurations are isolated and non-degenerate.\footnote{
  In \cite[footnote 17]{er-2024-topol-gauge-theor}, we (i) considered choices of $CY_2$ such that $CY_2 \times T^3$ satisfies appropriate transversality assumptions of $G_2$ instantons on $CY_2 \times T^3$ being acyclic (so that the actual dimension of its moduli space is zero~\cite[$\S$3]{walpuski-2013-g2-gener}), and (ii) added physically-inconsequential $\mathcal{Q}$-exact perturbations of the action, to ensure that HW configurations on $CY_2 \times S^1$ are (i) isolated and (ii) non-degenerate at $\theta = 0$.
  Therefore, at $\theta = 0$, the endpoints of the $\mathfrak{A}_5^{\theta}$-solitons will be isolated and non-degenerate.
  As the physical theory is symmetric under a variation of $\theta$, this observation of the endpoints of the $\mathfrak{A}_5^{\theta}$-solitons (or equivalently, the vertices of the $\mathfrak{A}_5^{\theta}$-sheets) will continue to hold true for any value of $\theta$.
  Hence, this presumption that the moduli space of $\theta$-deformed HW configurations on $CY_2 \times S^1$ will be made of isolated and non-degenerate points, is justified.
  We would like to thank D. Joyce for discussions on this point.
  \label{ft:isolation of hw}
}

\subtitle{Non-constant Double Paths, $\mathcal{M}^{\theta}(\R_t, \mathfrak{A}_5)$-solitons, $\mathfrak{A}_5^{\theta}$-solitons, and $\mathfrak{A}_5^{\theta}$-sheets}

In short, from the equivalent 1d SQM of Spin$(7)$ theory on $CY_2 \times S^1 \times \R^3$, the theory localizes onto $\xi$-invariant, $\theta$-deformed, non-constant double paths in $\mathcal{M}(\R^2, \mathfrak{A}_5)$, which, in turn, will correspond to $\mathcal{M}^{\theta}(\R_t, \mathfrak{A}_5)$-solitons in the 2d gauged sigma model whose endpoints are $(\xi, \tau)$-invariant, $\theta$-deformed, non-constant paths in $\mathcal{M}(\R_t, \mathfrak{A}_5)$.
In the 3d gauged sigma model, these $\mathcal{M}^{\theta}(\R_t, \mathfrak{A}_5)$-solitons will correspond to $\mathfrak{A}_5^{\theta}$-sheets, whose edges are $\mathfrak{A}_5^{\theta}$-solitons, and whose vertices will correspond to $\theta$-deformed HW configurations on $CY_2 \times S^1$ that generate the holomorphic DT Floer homology of $CY_2 \times S^1$.

\subsection{The 2d Model and Open Strings, the 3d Model and Open Membranes}
\label{sec:cy2 x s x r3:2d-3d model}

By following the same analysis in \autoref{sec:m2 x r3:2d-3d model}, we find that the 2d gauged sigma model with target space $\mathcal{M}(\R_t, \mathfrak{A}_5)$ whose action is \eqref{eq:cy2 x s x r3:2d model action}, will define an open string theory in $\mathcal{M}(\R_t, \mathfrak{A}_5)$.
Similarly, we find that the 3d gauged sigma model with target space $\mathfrak{A}_5$ whose action is \eqref{eq:cy2 x s x r3:3d model action}, will define an open membrane theory in $\mathfrak{A}_5$.
We will now work out the details pertaining to the BPS worldsheets and BPS worldvolumes (including their boundaries) that are necessary to define this open string and open membrane theory, respectively.

\subtitle{BPS Worldsheets of the 2d Model}

The BPS worldsheets of the 2d gauged sigma model with target space $\mathcal{M}(\R_t, \mathfrak{A}_5)$ correspond to its classical trajectories.
Specifically, these are defined by setting to zero the expression within the squared terms in \eqref{eq:cy2 x s x r3:2d model action}, i.e.,
\begin{equation}
  \label{eq:cy2 x s x r3:worldsheet:eqn}
  \begin{gathered}
    \Dv{C^{\hat{m}}}{\xi} - \Dv{(\tilde{A}_t)^{\hat{m}}}{\tau}
    = 0
    \, ,
    \qquad
    \Dv{(\tilde{A}_t)^{\hat{m}}}{\xi} + \Dv{C^{\hat{m}}}{\tau}
    + i F_{\xi\tau}
    = - Q^{\hat{m}} - q^{\hat{m}}
    \, ,
    \\
    \Dv{\mathcal{A}^{\hat{m}}}{\xi} - k_{\theta} \Dv{\mathcal{A}^{\hat{m}}}{\tau}
    = - R^{\hat{m}} - r^{\hat{m}}
    \, .
  \end{gathered}
\end{equation}

\subtitle{BPS Worldsheets with Boundaries Labeled by Non-constant Paths in $\mathcal{M}(\R_t, \mathfrak{A}_5)$}

The boundaries of the BPS worldsheets are traced out by the endpoints of the $\mathcal{M}^{\theta}(\R_t, \mathfrak{A}_5)$-solitons as they propagate in $\xi$.
As we have seen at the end of \autoref{sec:cy2 x s x r3:hw}, these endpoints correspond to $(\xi, \tau)$-invariant, $\theta$-deformed, non-constant paths in $\mathcal{M}(\R_t, \mathfrak{A}_5)$ that we shall, at $\xi = \pm \infty$, denote as $\gamma_{\pm}(\theta, \mathfrak{A}_5)$.
In turn, they will correspond, in the 3d gauged sigma model with target space $\mathfrak{A}_5$, to $\mathfrak{A}_5^{\theta}$-solitons that we shall, at $\xi = \pm \infty$, denote as $\Gamma_{\pm}(t, \theta, \mathfrak{A}_5)$, whose endpoints will correspond to $\theta$-deformed HW configurations on $CY_2 \times S^1$.

If there are $m \geq 4$ such configurations $\{\mathcal{E}^1_{\text{HW}}(\theta), \mathcal{E}^2_{\text{HW}}(\theta), \dots, \mathcal{E}^m_{\text{HW}}(\theta)\}$, just as in \autoref{sec:m2 x r3:2d-3d model}, we can further specify any $\Gamma_{\pm}(t, \theta, \mathfrak{A}_5)$ as $\Gamma^{IJ}_{\pm}(t, \theta, \mathfrak{A}_5)$, where its endpoints would correspond to $\mathcal{E}^I_{\text{HW}}(\theta)$ and $\mathcal{E}^J_{\text{HW}}(\theta)$.
Consequently, in the 2d gauged sigma model, we can further specify any $\gamma_{\pm}(\theta, \mathfrak{A}_5)$ as $\gamma^{IJ}_{\pm}(\theta, \mathfrak{A}_5)$, where the latter will correspond to a $\Gamma^{IJ}_{\pm}(t, \theta, \mathfrak{A}_5)$ $\mathfrak{A}_5^{\theta}$-soliton in the equivalent 3d gauged sigma model.

Since the endpoints of an $\mathcal{M}^{\theta}(\R_t, \mathfrak{A}_5)$-soliton are now denoted as $\gamma^{**}_{\pm}(\theta, \mathfrak{A}_5)$, we can also denote and specify the former at $\xi = \pm \infty$ as $\sigma^{IJ, KL}_{\pm}(\tau, \theta, \mathfrak{A}_5)$,\footnote{%
  Just like in \autoref{ft:fixing m-A2-soliton centre of mass dof}, the $\xi$-invariant $\mathcal{M}^{\theta}(\R_t, \mathfrak{A}_5)$-solitons can be fixed at $\xi = \pm \infty$ by adding physically-inconsequential $\mathcal{Q}$-exact terms to the SQM action.
  \label{ft:fixing m-A5-soliton centre of mass dof}
} where its left and right endpoints would be $\gamma^{IJ}_{\pm}(\theta, \mathfrak{A}_5)$ and $\gamma^{KL}_{\pm}(\theta, \mathfrak{A}_5)$, respectively.

As the $\gamma^{**}(\theta, \mathfrak{A}_5)$'s are $\xi$-invariant and therefore, have the same value for all $\xi$, we have BPS worldsheets of the kind similar to \autoref{fig:m2 x r3:bps worldsheet}.
This time, however, instead of the boundaries being labeled $\gamma^{**}(\theta, \mathfrak{A}_2)$, they will be labeled $\gamma^{**}(\theta, \mathfrak{A}_5)$.
And, at $\xi = \pm \infty$, instead of the $\mathcal{M}^{\theta}(\R_{\tau}, \mathfrak{A}_2)$-solitons $\sigma^{IJ, KL}_{\pm}(t, \theta, \mathfrak{A}_2)$, we will have $\mathcal{M}^{\theta}(\R_t, \mathfrak{A}_5)$-solitons $\sigma^{IJ, KL}_{\pm}(\tau, \theta, \mathfrak{A}_5)$.

\subtitle{The 2d Model on $\R^2$ and an Open String Theory in $\mathcal{M}(\R_t, \mathfrak{A}_5)$}

Thus, like in \autoref{sec:m2 x r3:2d-3d model}, one can understand the 2d gauged sigma model with target space $\mathcal{M}(\R_t, \mathfrak{A}_5)$ to define an open string theory in $\mathcal{M}(\R_t, \mathfrak{A}_5)$, whose \emph{effective} worldsheet and boundaries are similar to \autoref{fig:m2 x r3:bps worldsheet}, where $\xi$ and $\tau$ are the temporal and spatial directions, respectively.

\subtitle{BPS Worldvolumes of the 3d Model}

The BPS worldvolumes of the 3d gauged sigma model with target space $\mathfrak{A}_5$ correspond to its classical trajectories.
Specifically, these are defined by setting to zero the expression within the squared terms in \eqref{eq:cy2 x s x r3:3d model action}, i.e.,
\begin{equation}
  \label{eq:cy2 x s x r3:worldvolume:eqn}
  \begin{aligned}
    \Dv{C^{\hat{a}}}{\xi} + F_{t\tau}
    &= 0
      \, ,
    & \qquad
    \Dv{C^{\hat{a}}}{\tau} - i \Dv{C^{\hat{a}}}{t}
    + i(F_{\xi\tau} - i F_{\xi t})
    &= - q^{\hat{a}}
      \, ,
    \\
    0
    &= \mathfrak{h}_5(p)
      \, ,
    & \qquad
    \Dv{\mathcal{A}^{\hat{a}}}{\xi} - k_{\theta} \Dv{\mathcal{A}^{\hat{a}}}{\tau} + j_{\theta} \Dv{\mathcal{A}^{\hat{a}}}{t}
    &= - r^{\hat{a}}
      \, .
  \end{aligned}
\end{equation}

\subtitle{BPS Worldvolumes with Boundaries Labeled by $\mathfrak{A}_5^{\theta}$-solitons, and Edges Labeled by HW Configurations on $CY_2 \times S^1$}

The boundaries and edges of the BPS worldvolumes are traced out by the edges and vertices of the $\mathfrak{A}_5^{\theta}$-sheets, respectively, as they propagate in $\xi$.
As we have seen at the end of \autoref{sec:cy2 x s x r3:hw}, these edges and vertices would correspond to $\mathfrak{A}_5^{\theta}$-solitons and $\theta$-deformed HW configurations on $CY_2 \times S^1$, respectively.

This means that we can denote and specify any $\mathfrak{A}_5^{\theta}$-sheet at $\xi = \pm \infty$ as $\Sigma^{IJ,KL}_{\pm}(\tau, t, \theta, \mathfrak{A}_5)$,\footnote{%
  Just like in \autoref{ft:fixing A2-sheet centre of mass dof}, the $\xi$-invariant $\mathfrak{A}_5^{\theta}$-sheets can be fixed at $\xi = \pm \infty$ by adding physically-inconsequential $\mathcal{Q}$-exact terms to the SQM action.
  \label{ft:fixing A5-sheet centre of mass dof}
} where
(i) its left and right edges correspond to the $\mathfrak{A}_5^{\theta}$-solitons $\Gamma^{IJ}(t, \theta, \mathfrak{A}_5)$ and $\Gamma^{KL}(t, \theta, \mathfrak{A}_5)$, respectively, and
(ii) its four vertices would correspond to $\mathcal{E}^I_{\text{HW}}(\theta)$, $\mathcal{E}^J_{\text{HW}}(\theta)$, $\mathcal{E}^K_{\text{HW}}(\theta)$, and $\mathcal{E}^L_{\text{HW}}(\theta)$, similar to the kind shown in \autoref{fig:m2 x r3:frakA-sheet}.
However, instead of the edges being labeled $\Gamma^{**}(\tau, \theta, \mathfrak{A}_2)$, they will be labeled $\Gamma^{**}(t, \theta, \mathfrak{A}_5)$.
And, instead of the vertices being labeled $\mathcal{E}^*_{\text{BF}}(\theta)$, they will be labeled $\mathcal{E}^*_{\text{HW}}(\theta)$.

Since the $\mathcal{E}^{*}_{\text{HW}}(\theta)$'s and $\Gamma^{**}(t, \theta, \mathfrak{A}_5)$'s are $\xi$-invariant and therefore, have the same value for all $\xi$, we have BPS worldvolumes of the kind similar to \autoref{fig:m2 x r3:bps worldvolume}.
This time, however, instead of the faces being labeled $\Sigma^{**,**}_{\pm}(t, \tau, \theta, \mathfrak{A}_2)$ and $\Gamma^{**}(\tau, \theta, \mathfrak{A}_2)$, they will be labeled $\Sigma^{**,**}_{\pm}(\tau, t, \theta, \mathfrak{A}_5)$ and $\Gamma^{**}(t, \theta, \mathfrak{A}_5)$.
And, instead of the edges being labeled $\mathcal{E}^*_{\text{BF}}(\theta)$, they will be labeled $\mathcal{E}^*_{\text{HW}}(\theta)$.

\subtitle{The 3d Model on $\R^3$ and an Open Membrane Theory in $\mathfrak{A}_5$}

Thus, like in \autoref{sec:m2 x r3:2d-3d model}, one can understand the 3d gauged sigma model with target space $\mathfrak{A}_5$ to define an open membrane theory in $\mathfrak{A}_5$, whose \emph{effective} worldvolume and boundaries are similar to \autoref{fig:m2 x r3:bps worldvolume}, where $\xi$ is the temporal direction, and $\tau$ and $t$ are the spatial directions.

\subsection{Soliton String Theory, the \texorpdfstring{Spin$(7)$}{Spin(7)} Partition Function, and an FS type \texorpdfstring{$A_{\infty}$}{A-infinity}-category of \texorpdfstring{$\mathfrak{A}_5^{\theta}$}{A5-theta}-solitons}
\label{sec:cy2 x s x r3:fs-cat}

\subtitle{The 2d Model as a 2d Gauged LG Model}

Notice that we can also express \eqref{eq:cy2 x s x r3:worldsheet:eqn} as
\begin{equation}
  \label{eq:cy2 x s x r3:2d lg:worldsheet:eqn}
  \begin{aligned}
    \Dv{C^{\hat{m}}}{\xi} - k \Dv{C^{\hat{m}}}{\tau}
    - k \left( \Dv{(\tilde{A}_t)^{\hat{m}}}{\xi} - k \Dv{(\tilde{A}_t)^{\hat{m}}}{\tau} \right)
    - j F_{\xi \tau}
    &= k Q^{\hat{m}} + k q^{\hat{m}}
      \, ,
    \\
    \Dv{\mathcal{A}^{\hat{m}}}{\xi} - k_{\theta} \Dv{\mathcal{A}^{\hat{m}}}{\tau}
    &= - R^{\hat{m}} - r^{\hat{m}}
      \, .
  \end{aligned}
\end{equation}
In turn, this means that we can express the action of the 2d gauged sigma model with target space $\mathcal{M}(\R_t, \mathfrak{A}_5)$ in~\eqref{eq:cy2 x s x r3:2d model action} as
\begin{equation}
  \label{eq:cy2 x s x r3:2d lg:action}
  \begin{aligned}
    & S_{\text{2d-LG}, \mathcal{M}(\R_t, \mathfrak{A}_5)}
    \\
    & = \int d\tau d\xi \bigg(
      \bigg|
      \left( D_{\xi} - k D_\tau \right) \left( C^{\hat{m}} - k (\tilde{A}_t)^{\hat{m}} \right)
      - j F_{\xi \tau}
      - k Q^{\hat{m}}
      - k q^{\hat{m}}
      \bigg|^2
    \\
    & \qquad \qquad \qquad
      + \left| ( D_{\xi} - k_{\theta} D_{\tau}) \mathcal{A}^{\hat{m}}
      + R^{\hat{m}}
      + r^{\hat{m}}
      \right|^2
      + \dots
      \bigg)
    \\
    & = \int d\tau d\xi \Bigg(
      \Bigg|
      \left( D_{\xi} - k D_\tau \right) \left( C^{\hat{m}} - k (\tilde{A}_t)^{\hat{m}} \right)
      - j F_{\xi \tau}
      + g^{\hat{m}\bar{\hat{n}}}_{\mathcal{M}(\R_{t}, \mathfrak{A}_5)} \left( \frac{j \zeta}{2} \pdv{W_5}{C^{\hat{n}}} \right)^*
      - k g^{\hat{m}\bar{\hat{n}}}_{\mathcal{M}(\R_{t}, \mathfrak{A}_5)} \left( \frac{j \zeta}{2} \pdv{W_5}{(\tilde{A}_t)^{\hat{n}}} \right)^*
      \Bigg|^2
    \\
    & \qquad \qquad \qquad
      + \left| (D_{\xi} - k_{\theta} D_{\tau} )\mathcal{A}^{\hat{m}}
      + g^{\hat{m}\bar{\hat{n}}}_{\mathcal{M}(\R_{t}, \mathfrak{A}_5)} \left( \frac{j \zeta}{2} \pdv{W_5}{\mathcal{A}^{\hat{n}}} \right)^*
      \right|^2
      + \dots
      \Bigg)
    \\
    &= \int d\tau d\xi \left(
      \left| D_{\rho} \mathcal{A}^{\hat{m}} \right|^2
      + \left| D_{\rho} C^{\hat{m}} \right|^2
      + \left| D_{\rho} (\tilde{A}_t)^{\hat{m}} \right|^2
      + \left| \pdv{W_5}{\mathcal{A}^{\hat{m}}} \right|^2
      + \left| \pdv{W_5}{C^{\hat{m}}} \right|^2
      + \left| \pdv{W_5}{(\tilde{A}_t)^{\hat{m}}} \right|^2
      + \left| F_{\xi \tau} \right|^2
      + \dots
      \right)
      \, ,
  \end{aligned}
\end{equation}
where $g_{\mathcal{M}(\R_{t}, \mathfrak{A}_5)}$ is the metric on $\mathcal{M}(\R_t, \mathfrak{A}_5)$.
In other words, the 2d gauged sigma model with target space $\mathcal{M}(\R_t, \mathfrak{A}_5)$ can also be interpreted as a 2d gauged LG model with target space $\mathcal{M}(\R_t, \mathfrak{A}_5)$ and a holomorphic superpotential $W_5(\mathcal{A}, C, \tilde{A}_t)$.

By setting $d_\xi C^{\hat{m}} = 0 = d_\xi \mathcal{A}^{\hat{m}}$ and $A_\xi, A_\tau, (\tilde{A}_t)^{\hat{m}} \rightarrow 0$ in the expression within the squared terms in~\eqref{eq:cy2 x s x r3:2d lg:action}, we can read off the LG $\mathcal{M}^{\theta}(\R_t, \mathfrak{A}_5)$-soliton equations corresponding to $\sigma^{IJ, KL}_{\pm}(\tau, \theta, \mathfrak{A}_5)$ (that re-expresses \eqref{eq:cy2 x s x r3:m-soliton eqns} with $A_{\xi}, A_{\tau}, (\tilde{A}_t)^{\hat{m}} \rightarrow 0$) as
\begin{equation}
  \label{eq:cy2 x s x r3:2d lg:m-soliton:eqn}
  \begin{aligned}
    \dv{C^{\hat{m}}}{\tau}
    &= - k g^{\hat{m}\bar{\hat{n}}}_{\mathcal{M}(\R_{t}, \mathfrak{A}_5)} \left( \frac{j \zeta}{2} \pdv{W_5}{C^{\hat{n}}} \right)^*_{\tilde{A}_t = 0}
      - g^{\hat{m}\bar{\hat{n}}}_{\mathcal{M}(\R_{t}, \mathfrak{A}_5)} \left( \frac{j \zeta}{2} \pdv{W_5}{(\tilde{A}_t)^{\hat{n}}} \right)^*_{\tilde{A}_t = 0}
    \, ,
    \\
    \dv{\mathcal{A}^{\hat{m}}}{\tau}
    &= - k_{\theta} g^{\hat{m}\bar{\hat{n}}}_{\mathcal{M}(\R_{t}, \mathfrak{A}_5)} \left( \frac{j \zeta}{2} \pdv{W_5}{\mathcal{A}^{\hat{n}}} \right)^*_{\tilde{A}_t = 0}
    \, .
  \end{aligned}
\end{equation}

By setting $d_{\tau} C^{\hat{m}} = 0 = d_{\tau} \mathcal{A}^{\hat{m}}$ in \eqref{eq:cy2 x s x r3:2d lg:m-soliton:eqn}, we get the LG $\mathcal{M}^{\theta}(\R_t, \mathfrak{A}_5)$-soliton endpoint equations corresponding to $\sigma^{IJ, KL}(\pm \infty, \theta, \mathfrak{A}_5)$ (that re-expresses \eqref{eq:cy2 x s x r3:m-soliton:endpts}) as
\begin{equation}
  \label{eq:cy2 x s x r3:2d lg:m-soliton:endpts}
  \begin{aligned}
    k g^{\hat{m}\bar{\hat{n}}}_{\mathcal{M}(\R_{t}, \mathfrak{A}_5)} \left( \frac{j \zeta}{2} \pdv{W_5}{C^{\hat{n}}} \right)^*_{\tilde{A}_t = 0}
    + g^{\hat{m}\bar{\hat{n}}}_{\mathcal{M}(\R_{t}, \mathfrak{A}_5)} \left( \frac{j \zeta}{2} \pdv{W_5}{(\tilde{A}_t)^{\hat{n}}} \right)^*_{\tilde{A}_t = 0}
    &= 0
    \, ,
    \\
    k_{\theta} g^{\hat{m}\bar{\hat{n}}}_{\mathcal{M}(\R_{t}, \mathfrak{A}_5)} \left( \frac{j \zeta}{2} \pdv{W_5}{\mathcal{A}^{\hat{n}}} \right)^*_{\tilde{A}_t = 0}
    &= 0
    \, .
  \end{aligned}
\end{equation}

Recall from the end of \autoref{sec:cy2 x s x r3:hw} that we are only considering certain $CY_2$ such that the $\theta$-deformed HW configurations are isolated and non-degenerate.
Next, recall also that such configurations will correspond to the endpoints of the $\mathfrak{A}_5^{\theta}$-solitons; therefore, just like their endpoints, these $\mathfrak{A}_5^{\theta}$-solitons would be isolated and non-degenerate.
As these $\mathfrak{A}_5^{\theta}$-solitons will correspond, in the 2d gauged sigma model, to the endpoints of the $\mathcal{M}^{\theta}(\R_t, \mathfrak{A}_5)$-solitons, i.e., $\sigma^{IJ, KL}(\pm \infty, \theta, \mathfrak{A}_5)$, this means the latter would also be isolated and non-degenerate.
Thus, from their definition in~\eqref{eq:cy2 x s x r3:2d lg:m-soliton:endpts} which tells us that they are critical points of $W_5(\mathcal{A}, C, \tilde{A}_t)$, we conclude that $W_5(\mathcal{A}, C, \tilde{A}_t)$ can be regarded as a holomorphic Morse function in $\mathcal{M}(\R_t, \mathfrak{A}_5)$.

Just like in \autoref{sec:m2 x r3:fs-cat}, an LG $\mathcal{M}^{\theta}(\R_t, \mathfrak{A}_5)$-soliton defined in \eqref{eq:cy2 x s x r3:2d lg:m-soliton:eqn} maps to a straight line in the complex $W_5$-plane that starts and ends at the critical values $W_5^{IJ}(\theta) \equiv W_5(\sigma^{IJ}(-\infty, \theta, \mathfrak{A}_5))$ and $W_5^{KL}(\theta) \equiv W_5(\sigma^{KL}(+\infty, \theta, \mathfrak{A}_5))$, respectively, where its slope depends on $\theta$ (via $\zeta$).
We shall also assume that $\Re(W_5^{IJ}(\theta)) < \Re(W_5^{KL}(\theta))$ and that $I \neq J$, $J \neq K$, and $K \neq L$.

\subtitle{The 2d Gauged LG Model as an LG SQM}

With suitable rescalings, we can recast \eqref{eq:cy2 x s x r3:2d lg:action} as a 1d LG SQM (that re-expresses \eqref{eq:cy2 x s x r3:sqm action}), where its action will be given by\footnote{%
  In the following expression, we have integrated out $A_\xi$ and omitted the fields corresponding to the finite-energy gauge fields $A_{\{t, \tau\}}$ (as explained in~\autoref{ft:stokes theorem for m2 x r3:sqm}).
  \label{ft:stokes theorem for cy2 x s x r3:2d-lg sqm}
}
\begin{equation}
  \label{eq:cy2 x s x r3:2d lg:sqm:action}
  \begin{aligned}
    S_{\text{2d-LG SQM}, \mathcal{M}(\R_{\tau}, \mathcal{M}(\R_t, \mathfrak{A}_5))}
    = \int d\xi \Bigg(
    & \left| \left(
      \dv{C^{\hat{u}}}{\xi}
      - k \dv{\breve{A}^{\hat{u}}}{\xi}
      \right)
      + g_{\mathcal{M}(\R_{\tau}, \mathcal{M}(\R_{t}, \mathfrak{A}_5))}^{\hat{u}\hat{v}} \left(
      \pdv{H_5}{C^{\hat{v}}}
      - k \pdv{H_5}{\breve{A}^{\hat{v}}}
      \right)
      \right|^2
    \\
    & + \left| \dv{\mathcal{A}^{\hat{u}}}{\xi}
      + g_{\mathcal{M}(\R_{\tau}, \mathcal{M}(\R_t, \mathfrak{A}_5))}^{\hat{u}\hat{v}} \pdv{H_5}{\mathcal{A}^{\hat{v}}}
      \right|^2
      + \dots
      \Bigg)
      \, ,
  \end{aligned}
\end{equation}
where $H_5(\mathcal{A}, C, \breve{A})$ is the \emph{real-valued} potential in $\mathcal{M}(\R_{\tau}, \mathcal{M}(\R_t, \mathfrak{A}_5))$, and the subscript ``2d-LG SQM, $\mathcal{M}(\R_{\tau}, \mathcal{M}(\R_t, \mathfrak{A}_5))$'' is to specify that it is a 1d SQM in $\mathcal{M}(\R_{\tau}, \mathcal{M}(\R_t, \mathfrak{A}_5))$ obtained from the equivalent 2d LG model.
We will also refer to this \emph{1d} LG SQM as ``2d-LG SQM'' in the rest of this subsection.

The 2d-LG SQM will localize onto configurations that \emph{simultaneously} set to zero the LHS and RHS of the expression within the squared terms in~\eqref{eq:cy2 x s x r3:2d lg:sqm:action}.
In other words, it will localize onto $\xi$-invariant critical points of $H_5(\mathcal{A}, C, \breve{A})$ that will correspond, when $A_{\xi}, A_{\tau}, (\tilde{A}_t)^{\hat{m}} \rightarrow 0$, to the LG $\mathcal{M}^{\theta}(\R_t, \mathfrak{A}_5)$-solitons defined by~\eqref{eq:cy2 x s x r3:2d lg:m-soliton:eqn}.
For our choice of $CY_2$, the LG $\mathcal{M}^{\theta}(\R_t, \mathfrak{A}_5)$-solitons, just like their endpoints, will be isolated and non-degenerate.
Thus, $H_5(\mathcal{A}, C, \breve{A})$ can be regarded as a real-valued Morse functional in $\mathcal{M}(\R_{\tau}, \mathcal{M}(\R_t, \mathfrak{A}_5))$.

\subtitle{Morphisms from $\gamma^{IJ}(\theta, \mathfrak{A}_5)$ to $\gamma^{KL}(\theta, \mathfrak{A}_5)$ as Floer Homology Classes of Intersecting Thimbles}

Repeating here the analysis in \autoref{sec:m2 x r3:fs-cat} with \eqref{eq:cy2 x s x r3:2d lg:sqm:action} as the action of the 2d-LG SQM, we find that we can interpret the LG $\mathcal{M}^{\theta}(\R_t, \mathfrak{A}_5)$-soliton solution $\sigma^{IJ, KL}_{\pm}(\tau, \theta, \mathfrak{A}_5)$ as intersections of thimbles.
Specifically, a $\sigma^{IJ, KL}_{\pm}(\tau, \theta, \mathfrak{A}_5)$-soliton pair, whose left and right endpoints correspond to $\gamma^{IJ}(\theta, \mathfrak{A}_5)$ and $\gamma^{KL}(\theta, \mathfrak{A}_5)$, respectively, can be identified as a pair of intersection points $p^{IJ, KL}_{\text{HW}, \pm}(\theta) \in S^{IJ}_{\text{HW}} \bigcap S^{KL}_{\text{HW}}$ of a left and right thimble in the fiber space over the line segment $[W^{IJ}_5(\theta), W^{KL}_5(\theta)]$.

This means that the 2d-LG SQM in $\mathcal{M}(\R_{\tau}, \mathcal{M}(\R_t, \mathfrak{A}_5))$ with action \eqref{eq:cy2 x s x r3:2d lg:sqm:action}, will physically realize a Floer homology that we shall name an $\mathcal{M}(\R_t, \mathfrak{A}_5)$-LG Floer homology.
The chains of the $\mathcal{M}(\R_t, \mathfrak{A}_5)$-LG Floer complex will be generated by LG $\mathcal{M}^{\theta}(\R_t, \mathfrak{A}_5)$-solitons which we can identify with $p^{**, **}_{\text{HW}, \pm}(\theta)$, and the $\mathcal{M}(\R_t, \mathfrak{A}_5)$-LG Floer differential will be realized by the flow lines governed by the gradient flow equations satisfied by the $\xi$-varying configurations which set the expression within the squared terms in \eqref{eq:cy2 x s x r3:2d lg:sqm:action} to zero.
The partition function of the 2d-LG SQM in $\mathcal{M}(\R_\tau, \mathcal{M}(\R_t, \mathfrak{A}_5))$ will then be given by\footnote{%
  The `$\theta$' label is omitted in the LHS of the following expression (as explained in \autoref{ft:theta omission in m2-2d lg partition fn}).
  \label{ft:theta omission in cy2 x s-2d lg partition fn}
}
\begin{equation}
  \label{eq:cy2 x s x r3:2d lg:partition function}
  \mathcal{Z}_{\text{2d-LG SQM}, \mathcal{M}(\R_\tau, \mathcal{M}(\R_t, \mathfrak{A}_5))}(G)
  = \sum^m_{I \neq J \neq K \neq L = 1}
  \,
  \sum_{%
    p^{IJ, KL}_{\text{HW}, \pm}
    \in S^{IJ}_{\text{HW}} \cap S^{KL}_{\text{HW}}
  }
  \text{HF}^G_{d_u} \left(
    p^{IJ, KL}_{\text{HW}, \pm}(\theta)
  \right)
  \, .
\end{equation}
Here, the contribution $\text{HF}^G_{d_u} (p^{IJ, KL}_{\text{HW}, \pm}(\theta))$ can be identified with a homology class in an $\mathcal{M}(\R_t, \mathfrak{A}_5)$-LG Floer homology generated by intersection points of thimbles.
These intersection points represent LG $\mathcal{M}^{\theta}(\R_t, \mathfrak{A}_5)$-solitons defined by \eqref{eq:cy2 x s x r3:2d lg:m-soliton:eqn}, whose endpoints correspond to $\theta$-deformed, non-constant paths in $\mathcal{M}(\R_t, \mathfrak{A}_5)$ defined by \eqref{eq:cy2 x s x r3:2d lg:m-soliton:endpts}.
The degree of each chain in the complex is $d_u$, and is counted by the number of outgoing flow lines from the fixed critical points of $H_5(\mathcal{A}, C, \breve{A})$ in $\mathcal{M}(\R_{\tau}, \mathcal{M}(\R_t, \mathfrak{A}_5))$ which can also be identified as $p^{IJ, KL}_{\text{HW}, \pm}(\theta)$.

Therefore, $\mathcal{Z}_{\text{2d-LG SQM}, \mathcal{M}(\R_\tau, \mathcal{M}(\R_t, \mathfrak{A}_5))}(G)$ in \eqref{eq:cy2 x s x r3:2d lg:partition function} is a sum of LG $\mathcal{M}^{\theta}(\R_t, \mathfrak{A}_5)$-solitons defined by \eqref{eq:cy2 x s x r3:2d lg:m-soliton:eqn} with endpoints \eqref{eq:cy2 x s x r3:2d lg:m-soliton:endpts}, or equivalently, $\sigma^{IJ, KL}_{\pm}(\tau, \theta, \mathfrak{A}_5)$-solitons defined by \eqref{eq:cy2 x s x r3:m-soliton eqns} (with $A_{\xi}, A_{\tau}, (\tilde{A}_t)^{\hat{m}} \rightarrow 0$) with endpoints \eqref{eq:cy2 x s x r3:m-soliton:endpts}, whose start and end correspond to the non-constant paths $\gamma^{IJ}(\theta, \mathfrak{A}_5)$ and $\gamma^{KL}(\theta, \mathfrak{A}_5)$, respectively.
In other words, we can write
\begin{equation}
  \label{eq:cy2 x s x r3:2d lg:floer homology as vector}
  \text{CF}_{\mathcal{M}(\R_t, \mathfrak{A}_5)} \left(
    \gamma^{IJ}(\theta, \mathfrak{A}_5), \gamma^{KL}(\theta, \mathfrak{A}_5)
  \right)_{\pm}
  =
  \text{HF}^G_{d_u} \left(
    p^{IJ, KL}_{\text{HW}, \pm}(\theta)
  \right)
\end{equation}
where $\text{CF}_{\mathcal{M}(\R_t, \mathfrak{A}_5)} ( \gamma^{IJ}(\theta, \mathfrak{A}_5), \gamma^{KL}(\theta, \mathfrak{A}_5) )_{\pm}$ is a vector representing a $\sigma^{IJ, KL}_{\pm}(\tau, \theta, \mathfrak{A}_5)$-soliton, such that $\Re(W^{IJ}_5(\theta)) < \Re(W^{KL}_5(\theta))$.
This will lead us to the following one-to-one identification\footnote{%
  The `$\theta$' label is omitted in the following expression (as explained in \autoref{ft:omission of theta in m2 2d-lg}).
  \label{ft:omission of theta in cy2 x s 2d-lg}
}
\begin{equation}
  \label{eq:cy2 x s x r3:2d lg:floer hom as morphism}
  \boxed{
    \text{Hom} \left(
      \gamma^{IJ}(\mathfrak{A}_5), \gamma^{KL}(\mathfrak{A}_5)
    \right)_{\pm}
    \Longleftrightarrow
    \text{HF}^G_{d_u} \left(
      p^{IJ, KL}_{\text{HW}, \pm}
    \right)
  }
\end{equation}
where the RHS is proportional to the identity class when $I = K$ and $J = L$, and zero when $I \leftrightarrow K$ and $J \leftrightarrow L$ (since the $\sigma^{IJ, KL}_{\pm}(\tau, \theta, \mathfrak{A}_5)$-soliton only moves in one direction from $\gamma^{IJ}(\theta, \mathfrak{A}_5)$ to $\gamma^{KL}(\theta, \mathfrak{A}_5)$).

\subtitle{Soliton String Theory from the 2d LG Model}

Just like in \autoref{sec:m2 x r3:fs-cat}, the 2d gauged LG model in $\mathcal{M}(\R_t, \mathfrak{A}_5)$ with action \eqref{eq:cy2 x s x r3:2d lg:action} can be interpreted as a soliton string theory in $\mathcal{M}(\R_t, \mathfrak{A}_5)$.
The dynamics of this soliton string theory in $\mathcal{M}(\R_t, \mathfrak{A}_5)$ will be governed by the BPS worldsheet equations of \eqref{eq:cy2 x s x r3:2d lg:worldsheet:eqn}, where $(\mathcal{A}^{\hat{m}}, C^{\hat{m}}, (\tilde{A}_t)^{\hat{m}})$ are scalars on the worldsheet corresponding to the holomorphic coordinates of $\mathcal{M}(\R_t, \mathfrak{A}_5)$.
At an arbitrary instant in time whence $d_{\xi} \mathcal{A}^{\hat{m}} = d_{\xi} C^{\hat{m}} = 0 = d_{\xi} (\tilde{A}_t)^{\hat{m}} = d_{\xi} A_\tau$ in~\eqref{eq:cy2 x s x r3:2d lg:worldsheet:eqn}, the dynamics of $(\mathcal{A}^{\hat{m}}, C^{\hat{m}}, (\tilde{A}_t)^{\hat{m}})$ and the 2d gauge fields $(A_{\tau}, A_{\xi})$ along $\tau$ will be governed by the soliton equations
\begin{equation}
  \label{eq:cy2 x s x r3:2d lg:string m-soliton}
  \begin{aligned}
    \dv{(\tilde{A}_t)^{\hat{m}}}{\tau}
    + k \dv{C^{\hat{m}}}{\tau}
    - j \dv{A_\xi}{\tau}
    =& [A_{\xi} - k A_{\tau}, C^{\hat{m}} - k (\tilde{A}_t)^{\hat{m}}]
      + j [A_\tau, A_\xi]
      \\
    & + g^{\hat{m}\bar{\hat{n}}}_{\mathcal{M}(\R_{t}, \mathfrak{A}_5)} \left( \frac{j \zeta}{2} \pdv{W_5}{C^{\hat{n}}} \right)^*
      - k g^{\hat{m}\bar{\hat{n}}}_{\mathcal{M}(\R_{t}, \mathfrak{A}_5)} \left( \frac{j \zeta}{2} \pdv{W_5}{(\tilde{A}_t)^{\hat{n}}} \right)^*
      \, ,
    \\
    k_{\theta} \dv{\mathcal{A}^{\hat{m}}}{\tau}
    =& [A_{\xi} - k_{\theta} A_{\tau}, \mathcal{A}^{\hat{m}}]
      + g^{\hat{m}\bar{\hat{n}}}_{\mathcal{M}(\R_{t}, \mathfrak{A}_5)} \left( \frac{j \zeta}{2} \pdv{W_5}{\mathcal{A}^{\hat{n}}} \right)^*
      \, .
  \end{aligned}
\end{equation}

\subtitle{The Normalized Spin$(7)$ Partition Function, LG $\mathcal{M}^{\theta}(\R_t, \mathfrak{A}_5)$-soliton String Scattering, and Maps of an $A_{\infty}$-structure}

Since our Spin$(7)$ theory is semi-classical, its normalized 8d partition function will be a sum over tree-level scattering amplitudes of the LG $\mathcal{M}^{\theta}(\R_t, \mathfrak{A}_5)$-soliton strings defined by \eqref{eq:cy2 x s x r3:2d lg:m-soliton:eqn}.\footnote{%
  With an appropriate choice of a Spin$(7)$-manifold and $G$, Spin$(7)$ theory can be made to be a balanced TQFT just like HW theory \cite[footnote 5]{er-2024-topol-gauge-theor}; we shall henceforth assume such a choice.
  Therefore, the contributions to the normalized partition function can be understood as free-field correlation functions of operators that are in the $\mathcal{Q}$-cohomology (as explained in \autoref{ft:normalized hw partition fn}).
  Only tree-level scattering contributions need to be considered, as the theory is semi-classical.
  \label{ft:normalized spin7 partition fn}
}
The BPS worldsheet underlying such a tree-level scattering is similar to \autoref{fig:m2 x r3:mu-d maps}, where instead of the endpoints of each string being labeled $\gamma^{**}(\mathfrak{A}_2)$, they will now be labeled $\gamma^{**}(\mathfrak{A}_5)$.

In other words, we can, like in~\eqref{eq:m2 x r3:2d lg:normalized partition fn}, express the normalized Spin$(7)$ partition function as
\begin{equation}
  \label{eq:cy2 x s x r3:2d lg:normalized partition fn}
  \tilde{\mathcal{Z}}_{\text{Spin}(7), CY_2 \times S^1 \times \R^3}(G) = \sum_{\mathfrak{N}_m} \mu^{\mathfrak{N}_m}_{\mathfrak{A}_5}
  \, ,
  \qquad
  \mathfrak{N}_m = 1, 2, \dots, \left\lfloor \frac{m - 2}{2} \right\rfloor
\end{equation}
where each
\begin{equation}
  \label{eq:cy2 x s x r3:2d lg:composition maps}
  \boxed{
    \mu^{\mathfrak{N}_m}_{\mathfrak{A}_5}: \bigotimes_{i = 1}^{\mathfrak{N}_m}
    \text{Hom} \left(
      \gamma^{I_{2i - 1} I_{2i}}(\mathfrak{A}_5), \gamma^{I_{2(i + 1) - 1} I_{2(i + 1)}}(\mathfrak{A}_5)
    \right)_-
    \longto
    \text{Hom} \left(
      \gamma^{I_1 I_2}(\mathfrak{A}_5), \gamma^{I_{2\mathfrak{N}_m + 1} I_{2\mathfrak{N}_m + 2}}(\mathfrak{A}_5)
    \right)_+
  }
\end{equation}
is a scattering amplitude of $\mathfrak{N}_m$ incoming LG $\mathcal{M}^{\theta}(\R_t, \mathfrak{A}_5)$-soliton strings $\text{Hom} (\gamma^{I_1 I_2}(\mathfrak{A}_5), \gamma^{I_3 I_4}(\mathfrak{A}_5) )_-$, $\dots$, $\text{Hom} (\gamma^{I_{2\mathfrak{N}_m - 1} I_{2\mathfrak{N}_m}}(\mathfrak{A}_5), \gamma^{I_{2\mathfrak{N}_m + 1} I_{2\mathfrak{N}_m + 2}}(\mathfrak{A}_5) )_-$, and a single outgoing LG $\mathcal{M}^{\theta}(\R_t, \mathfrak{A}_5)$-soliton string $\text{Hom} (\gamma^{I_1 I_2}(\mathfrak{A}_5), \gamma^{I_{2\mathfrak{N}_m + 1} I_{2\mathfrak{N}_m + 2}}(\mathfrak{A}_5) )_+$, with left and right boundaries as labeled, whose underlying worldsheet can be regarded as a disc with $\mathfrak{N}_m + 1$ vertex operators at the boundary.
In short, $\mu^{\mathfrak{N}_m}_{\mathfrak{A}_5}$ counts pseudoholomorphic discs with $\mathfrak{N}_m + 1$ punctures at the boundary that are mapped to $\mathcal{M}(\R_t, \mathfrak{A}_5)$ according to the BPS worldsheet equations \eqref{eq:cy2 x s x r3:2d lg:worldsheet:eqn}.

Just as in \autoref{sec:m2 x r3:fs-cat}, the collection of $\mu^{\mathfrak{N}_m}_{\mathfrak{A}_5}$ maps in \eqref{eq:cy2 x s x r3:2d lg:composition maps} can be regarded as composition maps defining an $A_{\infty}$-structure.

\subtitle{An FS type $A_{\infty}$-category of $\mathfrak{A}_5^{\theta}$-solitons}

Altogether, this means that the normalized partition function of Spin$(7)$ theory on $CY_2 \times S^1 \times \R^3$ as expressed in \eqref{eq:cy2 x s x r3:2d lg:normalized partition fn}, manifests a \emph{novel} FS type $A_{\infty}$-category defined by the $\mu^{\mathfrak{N}_m}_{\mathfrak{A}_5}$ maps \eqref{eq:cy2 x s x r3:2d lg:composition maps} and the one-to-one identification \eqref{eq:cy2 x s x r3:2d lg:floer hom as morphism}, where the $\mathfrak{N}_m + 1$ number of objects $\big\{\gamma^{I_1 I_2}(\mathfrak{A}_5), \gamma^{I_3 I_4}(\mathfrak{A}_5), \dots,$ $\gamma^{I_{2\mathfrak{N}_m + 1} I_{2\mathfrak{N}_m + 2}}(\mathfrak{A}_5) \big\}$ correspond to $\mathfrak{A}_5^{\theta}$-solitons with endpoints themselves corresponding to ($\theta$-deformed) HW configurations on $CY_2 \times S^1$!

\subsection{Soliton Membrane Theory, the \texorpdfstring{Spin$(7)$}{Spin(7)} Partition Function, and a Fueter type \texorpdfstring{$A_{\infty}$}{A-infinity}-2-category 2-categorifying the Holomorphic DT Floer Homology of \texorpdfstring{$CY_2 \times S^1$}{CY2 x S1}}
\label{sec:cy2 x s x r3:fueter-cat}

Note that we can also express \eqref{eq:cy2 x s x r3:worldvolume:eqn} (corresponding to \eqref{eq:cy2 x s x r3:worldsheet:eqn} in the 2d gauged sigma model) as
\begin{equation}
  \label{eq:cy2 x s x r3:3d-lg:worldvolume:eqn}
  \begin{aligned}
    i \Dv{C^{\hat{a}}}{\xi} + j \Dv{C^{\hat{a}}}{\tau} + k \Dv{C^{\hat{a}}}{t}
    + i F_{t \tau} + j F_{\xi t} + k F_{\tau \xi}
    &= - j q^{\hat{a}}
      \, ,
    \\
    i e^{-i\theta} \Dv{\mathcal{A}^{\hat{a}}}{\xi}
    + j \Dv{\mathcal{A}^{\hat{a}}}{\tau}
    + k \Dv{\mathcal{A}^{\hat{a}}}{t}
    &=
      - i e^{-i\theta} r^{\hat{a}}
      \, ,
  \end{aligned}
\end{equation}
which are non-constant, $\theta$-deformed, gauged Fueter equations for the $(\mathcal{A}^{\hat{a}}, C^{\hat{a}})$ fields (that will correspond to \eqref{eq:cy2 x s x r3:2d lg:worldsheet:eqn} in the 2d gauged LG model).
In turn, this means that we can express the action of the 3d gauged sigma model with target space $\mathfrak{A}_5$ in \eqref{eq:cy2 x s x r3:3d model action} as
\begin{equation}
  \label{eq:cy2 x s x r3:3d lg:action}
  \begin{aligned}
    S_{\text{3d-LG}, \mathfrak{A}_5} = \frac{1}{e^2} \int dt d\tau d\xi \,
    \bigg(
    & \left| i D_{\xi} C^{\hat{a}} + j D_{\tau} C^{\hat{a}} + k D_{t} C^{\hat{a}}
      + i F_{t \tau} + j F_{\xi t} + k F_{\tau \xi}
      + j q^{\hat{a}}
      \right|^2
    \\
    & + \left| i e^{-i\theta} D_{\xi} \mathcal{A}^{\hat{a}} + j D_{\tau} \mathcal{A}^{\hat{a}} + k D_t \mathcal{A}^{\hat{a}}
      + i e^{-i\theta} r^{\hat{a}}
      \right|^2
      + \dots
      \bigg)
    \\
    = \frac{1}{e^2} \int dt d\tau d\xi \,
    \bigg(
    & \left| i D_{\xi} C^{\hat{a}} + j D_{\tau} C^{\hat{a}} + k D_{t} C^{\hat{a}}
      + i F_{t \tau} + j F_{\xi t} + k F_{\tau \xi}
      - g_{\mathfrak{A}_5}^{\hat{a}\bar{\hat{b}}} \left(
      \frac{j\zeta}{2} \pdv{\mathcal{W}_5}{C^{\hat{b}}}
      \right)^{*}
      \right|^2
    \\
    & + \left| i e^{-i\theta} D_{\xi} \mathcal{A}^{\hat{a}} + j D_{\tau} \mathcal{A}^{\hat{a}} + k D_t \mathcal{A}^{\hat{a}}
      - g_{\mathfrak{A}_5}^{\hat{a}\bar{\hat{b}}} \left(
      \frac{i\zeta}{2} \pdv{\mathcal{W}_5}{\mathcal{A}^{\hat{b}}}
      \right)^{*}
      \right|^2
      + \dots
      \bigg)
    \\
    = \frac{1}{e^2} \int dt d\tau d\xi \,
    \bigg(
    & \left| D_\varrho \mathcal{A}^{\hat{a}} \right|^2
      + \left| D_\varrho C^{\hat{a}} \right|^2
      + \left| \pdv{\mathcal{W}_5}{\mathcal{A}^{\hat{a}}} \right|^2
      + \left| \pdv{\mathcal{W}_5}{C^{\hat{a}}} \right|^2
      + \left| F_{t \tau} \right|^2
      + \left| F_{\xi t} \right|^2
      + \left| F_{\tau \xi} \right|^2
      + \dots
      \bigg)
      \, ,
  \end{aligned}
\end{equation}
where $g_{\mathfrak{A}_5}$ is the metric on $\mathfrak{A}_5$.
In other words, our 3d gauged sigma model can also be interpreted as a 3d gauged LG model in $\mathfrak{A}_5$ with holomorphic superpotential $\mathcal{W}_5(\mathcal{A}, C)$.

By setting $d_{\xi} C^{\hat{a}} = 0 = d_{\xi} \mathcal{A}^{\hat{a}}$ and $A_\xi, A_\tau, A_t \rightarrow 0$ in the expression within the squared terms in~\eqref{eq:cy2 x s x r3:3d lg:action}, we can read off the LG $\mathfrak{A}_5^{\theta}$-sheet equations corresponding to $\Sigma^{IJ, KL}_\pm(\tau, t, \theta, \mathfrak{A}_5)$ (that re-expresses \eqref{eq:cy2 x s x r3:sheet eqns} with $A_\xi, A_\tau, A_t \rightarrow 0$) as
\begin{equation}
  \label{eq:cy2 x s x r3:3d lg:sheet:eqns}
  \begin{aligned}
    j \dv{C^{\hat{a}}}{\tau} + k \dv{C^{\hat{a}}}{t}
    &= g_{\mathfrak{A}_5}^{\hat{a}\bar{\hat{b}}} \left(
      \frac{j\zeta}{2} \pdv{\mathcal{W}_5}{C^{\hat{a}}}
      \right)^{*}
      \, ,
    \\
    j \dv{\mathcal{A}^{\hat{a}}}{\tau} + k \dv{\mathcal{A}^{\hat{a}}}{t}
    &= g_{\mathfrak{A}_5}^{\hat{a}\bar{\hat{b}}} \left(
      \frac{i\zeta}{2} \pdv{\mathcal{W}_5}{\mathcal{A}^{\hat{a}}}
      \right)^{*}
      \, .
  \end{aligned}
\end{equation}

By setting $d_{\tau} C^{\hat{a}} = 0 = d_{\tau} \mathcal{A}^{\hat{a}}$ in~\eqref{eq:cy2 x s x r3:3d lg:sheet:eqns}, we can read off the LG $\mathfrak{A}_5^{\theta}$-soliton equations corresponding to $\Gamma^{IJ}(t, \theta, \mathfrak{A}_5)$ and $\Gamma^{KL}(t, \theta, \mathfrak{A}_5)$, or equivalently, the LG $\mathfrak{A}_5^{\theta}$-sheet edge equations corresponding to $\Sigma^{IJ, KL}(\pm \infty, t, \theta, \mathfrak{A}_5)$, (that re-expresses~\eqref{eq:cy2 x s x r3:soliton}) as
\begin{equation}
  \label{eq:cy2 x s x r3:3d lg:soliton:eqns}
  \begin{aligned}
    k \dv{C^{\hat{a}}}{t}
    &= g_{\mathfrak{A}_5}^{\hat{a}\bar{\hat{b}}} \left(
      \frac{j\zeta}{2} \pdv{\mathcal{W}_5}{C^{\hat{a}}}
      \right)^{*}
      \, ,
    \\
    k \dv{\mathcal{A}^{\hat{a}}}{t}
    &= g_{\mathfrak{A}_5}^{\hat{a}\bar{\hat{b}}} \left(
      \frac{i\zeta}{2} \pdv{\mathcal{W}_5}{\mathcal{A}^{\hat{a}}}
      \right)^{*}
      \, .
  \end{aligned}
\end{equation}

By setting $d_t C^{\hat{a}} = 0 = d_t \mathcal{A}^{\hat{a}}$ in~\eqref{eq:cy2 x s x r3:3d lg:soliton:eqns}, we can read off the LG $\mathfrak{A}_5^{\theta}$-soliton endpoint equations corresponding to $\Gamma^{**}(\pm \infty, \theta, \mathfrak{A}_5)$, or equivalently, the LG $\mathfrak{A}_5^{\theta}$-sheet vertex equations corresponding to $\Sigma^{IJ, KL}(\pm \infty, \pm \infty, \theta, \mathfrak{A}_5)$ and $\Sigma^{IJ, KL}(\pm \infty, \mp \infty, \theta, \mathfrak{A}_5)$, (that re-expresses~\eqref{eq:cy2 x s x r3:soliton} with $d_t C^{\hat{a}} = 0 = d_t \mathcal{A}^{\hat{a}}$) as
\begin{equation}
  \label{eq:cy2 x s x r3:3d lg:sheet:endpts}
  g_{\mathfrak{A}_5}^{\hat{a}\bar{\hat{b}}} \left(
    \frac{j\zeta}{2} \pdv{\mathcal{W}_5}{C^{\hat{a}}}
  \right)^{*}
  = 0
  \, ,
  \qquad
  g_{\mathfrak{A}_5}^{\hat{a}\bar{\hat{b}}} \left(
    \frac{i\zeta}{2} \pdv{\mathcal{W}_5}{\mathcal{A}^{\hat{a}}}
  \right)^{*}
  = 0
  \, .
\end{equation}

Recall from the end of \autoref{sec:cy2 x s x r3:hw} that we are only considering certain $CY_2$ such that (the endpoints $\Gamma^{**}(\pm \infty, \theta, \mathfrak{A}_5)$ and thus) the LG $\mathfrak{A}_5^{\theta}$-solitons, and effectively, (the vertices $\Sigma^{IJ, KL}(\pm \infty, \pm \infty, \theta, \mathfrak{A}_5)$, $\Sigma^{IJ, KL}(\pm \infty, \mp \infty, \theta, \mathfrak{A}_5)$ and thus) the LG $\mathfrak{A}_5^{\theta}$-sheets, are isolated and non-degenerate.
Therefore, from their definition in~\eqref{eq:cy2 x s x r3:3d lg:sheet:endpts} which tells us that they correspond to critical points of $\mathcal{W}_5(\mathcal{A}, C)$, we conclude that $\mathcal{W}_5(\mathcal{A}, C)$ can be regarded as a holomorphic Morse function in $\mathfrak{A}_5$.

Just like in~\autoref{sec:m2 x r3:fueter-cat}, this means that an LG $\mathfrak{A}_5^{\theta}$-soliton $\Gamma^{IJ}(t, \theta, \mathfrak{A}_5)$ defined in \eqref{eq:cy2 x s x r3:3d lg:soliton:eqns} maps to a straight line segment $[\mathcal{W}_5^I(\theta), \mathcal{W}_5^J(\theta)]$ in the complex $\mathcal{W}_5$-plane that starts and ends at critical values $\mathcal{W}^I_5(\theta) \equiv \mathcal{W}_5(\Gamma^I(- \infty, \theta, \mathfrak{A}_5))$ and $\mathcal{W}^J_5(\theta) \equiv \mathcal{W}_5(\Gamma^J(+ \infty, \theta, \mathfrak{A}_5))$, respectively, where its slope depends on $\theta$ (via $\zeta$).
Therefore, an LG $\mathfrak{A}_5^{\theta}$-sheet defined in \eqref{eq:cy2 x s x r3:3d lg:sheet:eqns} maps to a quadrilateral in the complex $\mathcal{W}_5$-plane, whose edges are the straight line segments that the LG $\mathfrak{A}_5^{\theta}$-solitons map to, and whose bottom-left, top-left, bottom-right, and top-right vertices are the critical points $\mathcal{W}^I_5(\theta) \equiv \mathcal{W}_5(\Sigma^I(- \infty, - \infty, \theta, \mathfrak{A}_5))$, $\mathcal{W}^J_5(\theta) \equiv \mathcal{W}_5(\Sigma^J(- \infty, + \infty, \theta, \mathfrak{A}_5))$, $\mathcal{W}^K_5(\theta) \equiv \mathcal{W}_5(\Sigma^K(+ \infty, - \infty, \theta, \mathfrak{A}_5))$, and $\mathcal{W}^L_5(\theta) \equiv $ $\mathcal{W}_5(\Sigma^L(+ \infty, + \infty, \theta, \mathfrak{A}_5))$, respectively, where the slope of the straight line segments between each left-right vertex pair depends on $\theta$ (via $\zeta$).

We shall also assume that $\text{Re}(\mathcal{W}_5^I(\theta)) < \{ \text{Re}(\mathcal{W}_5^J(\theta)), \text{Re}(\mathcal{W}_5^K(\theta)) \} < \text{Re}(\mathcal{W}_5^L(\theta))$.

\subtitle{The 3d Gauged LG Model as an LG SQM}

Last but not least, after suitable rescalings, we can recast~\eqref{eq:cy2 x s x r3:3d lg:action} as a 1d LG SQM (that re-expresses~\eqref{eq:cy2 x s x r3:sqm action}), where its action will be given by\footnote{%
  In the following expression, we have integrated out $A_\xi$ and omitted the fields corresponding to the finite-energy gauge fields $A_{\{t, \tau\}}$ (as explained in~\autoref{ft:stokes theorem for m2 x r3:sqm}).
  \label{ft:stokes theorem for cy2 x s x r3:3d-lg sqm}
}
\begin{equation}
  \label{eq:cy2 x s x r3:3d lg:sqm action}
  \begin{aligned}
    S_{\text{3d-LG SQM}, \mathcal{M}(\R^2, \mathfrak{A}_5)}
    = \frac{1}{e^2} \int d\xi \Bigg(
    &\left| i \left(
      \dv{C^{\hat{u}}}{\xi}
      - k \dv{\breve{A}^{\hat{u}}}{\xi}
      \right)
      + g^{\hat{u}\hat{v}}_{\mathcal{M}(\R^2, \mathfrak{A}_5)} \left(
      \pdv{\mathfrak{F}_5}{C^{\hat{v}}}
      - k \pdv{\mathfrak{F}_5}{\breve{A}^{\hat{v}}}
      \right)
      \right|^2
    \\
    & + \left| ie^{-i\theta} \dv{\mathcal{A}^{\hat{u}}}{\xi}
      + g^{\hat{u}\hat{v}}_{\mathcal{M}(\R^2, \mathfrak{A}_5)} \pdv{\mathfrak{F}_5}{\mathcal{A}^{\hat{v}}}
      \right|^2
      + \dots
      \Bigg)
      \, .
  \end{aligned}
\end{equation}
Here, $\mathfrak{F}_5(\mathcal{A}, C, \breve{A})$ is the \emph{real-valued} potential in $\mathcal{M}(\R^2, \mathfrak{A}_5)$, and the subscript ``3d-LG SQM, $\mathcal{M}(\R^2, \mathfrak{A}_5)$'' is to specify that it is a 1d SQM with target space $\mathcal{M}(\R^2, \mathfrak{A}_5)$ obtained from the equivalent 3d LG model.
We will also refer to this \emph{1d} LG SQM as ``3d-LG SQM'' in the rest of this subsection.

The 3d-LG SQM will localize onto configurations that \emph{simultaneously} set to zero the LHS and RHS of the expression within the squared terms in~\eqref{eq:cy2 x s x r3:3d lg:sqm action}.
In other words, it will localize onto $\xi$-invariant critical points of $\mathfrak{F}_5(\mathcal{A}, C, \breve{A})$ that will correspond, when $A_{\xi}, A_{\tau}, A_t \rightarrow 0$, to the LG $\mathfrak{A}_5^{\theta}$-sheets defined by~\eqref{eq:cy2 x s x r3:3d lg:sheet:eqns}.
For our choice of $CY_2$, the LG $\mathfrak{A}_5^{\theta}$-sheets, just like their vertices, will be isolated and non-degenerate.
Thus, $\mathfrak{F}_5(\mathcal{A}, C, \breve{A})$ can be regarded as a \emph{real-valued} Morse functional in $\mathcal{M}(\R^2, \mathfrak{A}_5)$.

\subtitle{Morphisms between $\mathfrak{A}_5^{\theta}$-solitons as Intersection Floer Homology Classes}

Repeating here the analysis in \autoref{sec:m2 x r3:fueter-cat} with \eqref{eq:cy2 x s x r3:3d lg:sqm action} as the action of the 3d-LG SQM, we find that we can interpret the LG $\mathfrak{A}_5^{\theta}$-soliton solution $\Gamma^{IJ}(t, \theta, \mathfrak{A}_5)$ as a thimble-intersection, and the LG $\mathfrak{A}_5^{\theta}$-sheet solution $\Sigma^{IJ,KL}_{\pm}(\tau, t, \theta, \mathfrak{A}_5)$ as an intersection of thimble-intersections.

Specifically, a $\Gamma^{IJ}(t, \theta, \mathfrak{A}_5)$-soliton, whose bottom and top endpoints correspond to $\mathcal{E}^I_{\text{HW}}(\theta)$ and $\mathcal{E}^J_{\text{HW}}(\theta)$, respectively, can be identified as an intersection point $q^{IJ}_{\text{HW}, \pm}(\theta) \in S^{IJ}_{\text{HW}}(\theta)$ of a bottom and top thimble in the fiber space over the line segment $[\mathcal{W}_5^I(\theta), \mathcal{W}_5^J(\theta)]$.
As a result, a $\Sigma^{IJ,KL}_{\pm}(\tau, t, \theta, \mathfrak{A}_5)$-sheet pair, whose left and right edges correspond to $\Gamma^{IJ}(t, \theta, \mathfrak{A}_5)$ and $\Gamma^{KL}(t, \theta, \mathfrak{A}_5)$, respectively, can be identified as a pair of intersection points $\{q^{IJ}_{\text{HW}, \pm}(\theta), q^{KL}_{\text{HW}, \pm}(\theta) \} \eqqcolon \mathfrak{P}^{IJ, KL}_{\text{HW}, \pm}(\theta) \in S^{IJ}_{\text{HW}}(\theta) \bigcap S^{KL}_{\text{HW}}(\theta)$ of a left and right thimble-intersection in the fiber space over the quadrilateral with vertices $(\mathcal{W}_5^I(\theta), \mathcal{W}_5^J(\theta), \mathcal{W}_5^K(\theta), \mathcal{W}_5^L(\theta))$.

At any rate, the 3d-LG SQM in $\mathcal{M}(\R^2, \mathfrak{A}_5)$ with action \eqref{eq:cy2 x s x r3:3d lg:sqm action} will physically realize a Floer homology that we shall name an $\mathfrak{A}_5$-3d-LG Floer homology.
The chains of the $\mathfrak{A}_5$-3d-LG Floer complex are generated by LG $\mathfrak{A}_5^{\theta}$-sheets which we can thus identify with $\mathfrak{P}^{**, **}_{\text{HW}, \pm}(\theta)$, and the $\mathfrak{A}_5$-3d-LG Floer differential will be realized by the flow lines governed by the gradient flow equations satisfied by $\xi$-varying configurations which set the expression within the squared terms of \eqref{eq:cy2 x s x r3:3d lg:sqm action} to zero.
The partition function of the 3d-LG SQM in $\mathcal{M}(\R^2, \mathfrak{A}_5)$ will be given by\footnote{%
  The `$\theta$' label is omitted in the LHS of the following expression (as explained in \autoref{ft:theta omission in m2-2d lg partition fn}).
  \label{ft:theta omission in 3d lg partition fn:cy2 x s x r3}
}
\begin{equation}
  \label{eq:cy2 x s x r3:3d lg:sqm:partition fn}
  \mathcal{Z}_{\text{3d-LG SQM}, \mathcal{M}(\R^2, \mathfrak{A}_5)}(G)
  = \sum_{I \neq J \neq K \neq L = 1}^m \,
  \sum_{\substack{\mathfrak{P}^{IJ, KL}_{\text{HW}, \pm} \\ \in S^{IJ}_{\text{HW}} \cap S^{KL}_{\text{HW}}}}
  \text{HF}^G_{d_u} \left(
    \mathfrak{P}^{IJ, KL}_{\text{HW}, \pm}(\theta)
  \right)
  \, ,
\end{equation}
where the contribution $\text{HF}^G_{d_u}(\mathfrak{P}^{IJ, KL}_{\text{HW}, \pm}(\theta))$ can be identified with a homology class in an $\mathfrak{A}_5$-3d-LG Floer homology generated by intersection points of thimble-intersections.
These intersection points represent LG $\mathfrak{A}_5^{\theta}$-sheets defined by \eqref{eq:cy2 x s x r3:3d lg:sheet:eqns}, whose edges correspond to LG $\mathfrak{A}_5^{\theta}$-solitons defined by \eqref{eq:cy2 x s x r3:3d lg:soliton:eqns}, and whose vertices defined by \eqref{eq:cy2 x s x r3:3d lg:sheet:endpts} will correspond to $\theta$-deformed HW configurations on $CY_2 \times S^1$.
The degree of each chain in the complex is $d_u$, and is counted by the number of outgoing flow lines from the fixed critical points of $\mathfrak{F}_5(\mathcal{A}, C, \breve{A})$ in $\mathcal{M}(\R^2, \mathfrak{A}_5)$ which can also be identified as $\mathfrak{P}^{IJ, KL}_{\text{HW},\pm}(\theta)$.

Therefore, $\mathcal{Z}_{\text{3d-LG SQM}, \mathcal{M}(\R^2, \mathfrak{A}_5)}(G)$ in \eqref{eq:cy2 x s x r3:3d lg:sqm:partition fn} is a sum of LG $\mathfrak{A}_5^{\theta}$-sheets defined by (i) \eqref{eq:cy2 x s x r3:3d lg:sheet:eqns} with (ii) edges \eqref{eq:cy2 x s x r3:3d lg:soliton:eqns} and (iii) vertices \eqref{eq:cy2 x s x r3:3d lg:sheet:endpts}, or equivalently, $\Sigma_{\pm}^{IJ, KL}(\tau, t, \theta, \mathfrak{A}_5)$-sheets defined by (i) \eqref{eq:cy2 x s x r3:sheet eqns} and~\eqref{eq:cy2 x s x r3:sheet eqns:aux cond} (with $A_{\xi}, A_{\tau}, A_t \rightarrow 0$) with (ii) edges \eqref{eq:cy2 x s x r3:soliton} and~\eqref{eq:cy2 x s x r3:sheet eqns:aux cond}, and (iii) vertices \eqref{eq:cy2 x s x r3:soliton} and~\eqref{eq:cy2 x s x r3:sheet eqns:aux cond} (with $d_t C^{\hat{a}} = 0 = d_t \mathcal{A}^{\hat{a}}$), respectively.
In other words, we can write
\begin{equation}
  \label{eq:cy2 x s x r3:3d lg:floer-hom as vector}
  \text{CF}_{\mathcal{M}(\R^2, \mathfrak{A}_5)} \left(
    \Gamma^{IJ}(t, \theta, \mathfrak{A}_5),
    \Gamma^{KL}(t, \theta, \mathfrak{A}_5)
  \right)_\pm
  =
  \text{HF}^G_{d_u} \left(
    \mathfrak{P}^{IJ, KL}_{\text{HW}, \pm}(\theta)
  \right)
  \, ,
\end{equation}
where $\text{CF}_{\mathcal{M}(\R^2, \mathfrak{A}_5)} (\Gamma^{IJ}(t, \theta, \mathfrak{A}_5), \Gamma^{KL}(t, \theta, \mathfrak{A}_5) )_\pm$ is a vector representing a $\Sigma^{IJ, KL}_{\pm}(\tau, t, \theta, \mathfrak{A}_5)$-sheet, whose left and right edges correspond to $\Gamma^{IJ}(t, \theta, \mathfrak{A}_5)$ and $\Gamma^{KL}(t, \theta, \mathfrak{A}_5)$, respectively, and whose bottom-left, top-left, bottom-right, and top-right vertices correspond to $\mathcal{E}^I_{\text{HW}}(\theta)$, $\mathcal{E}^J_{\text{HW}}(\theta)$, $\mathcal{E}^K_{\text{HW}}(\theta)$, and $\mathcal{E}^L_{\text{HW}}(\theta)$, respectively, such that $\text{Re}(\mathcal{W}_5^I(\theta)) < \{ \text{Re}(\mathcal{W}_5^J(\theta)), \text{Re}(\mathcal{W}_5^K(\theta)) \} < \text{Re}(\mathcal{W}_5^L(\theta))$.
This will lead us to the following one-to-one identifications\footnote{%
  The `$\theta$' label is omitted in the following expression (as explained in \autoref{ft:omission of theta in m2 2d-lg}).
  \label{ft:omission of theta in m2 3d-lg:cy2 x s x r3}
}
\begin{equation}
  \label{eq:cy2 x s x r3:3d lg:2-morphism}
  \boxed{
    \text{Hom} \left(
      \Gamma^{IJ}(t, \mathfrak{A}_5),
      \Gamma^{KL}(t, \mathfrak{A}_5)
    \right)_\pm
    \Longleftrightarrow
    \text{Hom} \left(
      \text{Hom}(\mathcal{E}^I_{\text{HW}}, \mathcal{E}^J_{\text{HW}}),
      \text{Hom}(\mathcal{E}^K_{\text{HW}}, \mathcal{E}^L_{\text{HW}})
    \right)_\pm
    \Longleftrightarrow
    \text{HF}^G_{d_u} \left(
      \mathfrak{P}^{IJ, KL}_{\text{HW}, \pm}
    \right)
  }
\end{equation}
where the RHS is proportional to the identity class when $I = K$ and $J = L$, and zero when (i) $I \leftrightarrow K$ and $J \leftrightarrow L$ (since the $\Sigma^{IJ, KL}(\tau, t, \theta, \mathfrak{A}_5)$-sheet only moves in one direction from $\Gamma^{IJ}(t, \theta, \mathfrak{A}_5)$ to $\Gamma^{KL}(t, \theta, \mathfrak{A}_5)$), and (ii) $I \leftrightarrow J$ or $K \leftrightarrow L$ (since the $\Gamma^{**}(t, \theta, \mathfrak{A}_5)$-solitons only move in one direction from $\mathcal{E}^I_{\text{HW}}(\theta)$ to $\mathcal{E}^J_{\text{HW}}(\theta)$ or $\mathcal{E}^K_{\text{HW}}(\theta)$ to $\mathcal{E}^L_{\text{HW}}(\theta)$).

\subtitle{Soliton Membrane Theory from the 3d LG Model}

Just like in \autoref{sec:m2 x r3:fueter-cat}, the 3d gauged LG model in $\mathfrak{A}_5$ with action \eqref{eq:cy2 x s x r3:3d lg:action} can be interpreted as a soliton membrane theory in $\mathfrak{A}_5$.
The dynamics of this soliton membrane theory in $\mathfrak{A}_5$ will be governed by the BPS worldvolume equations of \eqref{eq:cy2 x s x r3:3d-lg:worldvolume:eqn}, where $(\mathcal{A}^{\hat{a}}, C^{\hat{a}})$ are scalars on the worldvolume corresponding to the holomorphic coordinates of $\mathfrak{A}_5$.
At an arbitrary instant in time whence $d_{\xi} \mathcal{A}^{\hat{a}} = d_{\xi} C^{\hat{a}} = 0 = d_{\xi} A_{\{t, \tau\}}$ in \eqref{eq:cy2 x s x r3:3d-lg:worldvolume:eqn}, the dynamics of $(\mathcal{A}^{\hat{a}}, C^{\hat{a}})$ and the 3d gauge fields $(A_t, A_{\tau}, A_{\xi})$ along $(\tau, t)$ will be governed by the membrane equations
\begin{equation}
  \label{eq:cy2 x s x r3:soliton sheet:eqn}
  \begin{aligned}
    j \dv{C^{\hat{a}}}{\tau} + k \dv{C^{\hat{a}}}{t}
    + i \dv{A_\tau}{t} - i \dv{A_t}{\tau}
    - j \dv{A_{\xi}}{t} + k \dv{A_{\xi}}{\tau}
    &= - [ i A_{\xi} + j A_{\tau} + k A_t, C^{\hat{a}}]
      - i [A_t, A_\tau]
    \\
    & \qquad
      + [j A_t - k A_{\tau} , A_{\xi}]
      + g^{\hat{a} \bar{\hat{b}}}_{\mathfrak{A}_5} \left(
      \frac{j \zeta}{2} \pdv{\mathcal{W}_5}{C^{\hat{b}}}
      \right)^{*}
      \, ,
    \\
    j \dv{\mathcal{A}^{\hat{a}}}{\tau} + k \dv{\mathcal{A}^{\hat{a}}}{t}
    &= - [ i e^{-i\theta} A_{\xi} + j A_{\tau} + k A_t, \mathcal{A}^{\hat{a}}]
      + g^{\hat{a} \bar{\hat{b}}}_{\mathfrak{A}_5} \left(
      \frac{i \zeta}{2} \pdv{\mathcal{W}_5}{\mathcal{A}^{\hat{a}}}
      \right)^{*}
      \, .
  \end{aligned}
\end{equation}

\subtitle{The Normalized Spin$(7)$ Partition Function, Soliton Membrane Scattering, and Maps of an $A_{\infty}$-structure}

According to \autoref{ft:normalized spin7 partition fn}, the normalized Spin$(7)$ partition function can be regarded as a sum over tree-level scattering amplitudes of the LG $\mathfrak{A}_5^{\theta}$-sheets defined by \eqref{eq:cy2 x s x r3:3d lg:sheet:eqns}.
The BPS worldvolume underlying such a tree-level scattering amplitude is similar to \autoref{fig:m2 x r3:fueter composition maps}, where instead of the vertices of each sheet being labeled $\mathcal{E}^{*}_{\text{BF}}$, they are now labeled $\mathcal{E}^{*}_{\text{HW}}$.

In other words, we can, like in \eqref{eq:m2 x r3:3d lg:partition fn}, express the normalized Spin$(7)$ partition function as
\begin{equation}
  \label{eq:cy2 x s x r3:3d lg:partition fn}
  \mathcal{\tilde{Z}}_{\text{Spin}(7), CY_2 \times S^1 \times \R^3}(G)
  = \sum_{\mathfrak{N}_m} \varPi_{\mathfrak{A}_5}^{\mathfrak{N}_m}
  \, ,
  \qquad
  \mathfrak{N}_m = 1, 2, \dots, \left\lfloor \frac{m - 2}{2} \right\rfloor
\end{equation}
where each
\begin{equation}
  \label{eq:cy2 x s x r3:fueter composition maps}
  \boxed{
    \begin{aligned}
      \varPi^{\mathfrak{N}_m}_{\mathfrak{A}_5}: \bigotimes_{i = 1}^{\mathfrak{N}_m}
      & \text{Hom} \left(
        \text{Hom} \left( \mathcal{E}^{I_{2i - 1}}_{\text{HW}}, \mathcal{E}^{I_{2i}}_{\text{HW}} \right),
        \text{Hom} \left( \mathcal{E}^{I_{2(i + 1) - 1}}_{\text{HW}}, \mathcal{E}^{I_{2(i + 1)}}_{\text{HW}} \right)
        \right)_-
      \\
      &\longto
        \text{Hom} \left(
        \text{Hom} \left( \mathcal{E}^{I_1}_{\text{HW}}, \mathcal{E}^{I_2}_{\text{HW}} \right),
        \text{Hom} \left( \mathcal{E}^{I_{2 \mathfrak{N}_m + 1}}_{\text{HW}}, \mathcal{E}^{I_{2 \mathfrak{N}_m + 2}}_{\text{HW}} \right)
        \right)_+
    \end{aligned}
  }
\end{equation}
is a scattering amplitude of $\mathfrak{N}_m$ incoming LG $\mathfrak{A}_5^{\theta}$-soliton membranes $\text{Hom} \Big( \text{Hom} \big(\mathcal{E}^{I_1}_{\text{HW}}, \mathcal{E}^{I_2}_{\text{HW}} \big) , \text{Hom} \big(\mathcal{E}^{I_3}_{\text{HW}}, \mathcal{E}^{I_4}_{\text{HW}}\big) \Big)_-$, $\dots$, $\text{Hom} \Big( \text{Hom}\big(\mathcal{E}^{I_{2\mathfrak{N}_m - 1}}_{\text{HW}}, \mathcal{E}^{I_{2 \mathfrak{N}_m}}_{\text{HW}}\big) , \text{Hom}\big(\mathcal{E}^{I_{2\mathfrak{N}_m + 1}}_{\text{HW}}, \mathcal{E}^{I_{2\mathfrak{N}_m + 2}}_{\text{HW}}\big) \Big)_-$, and a single outgoing LG $\mathfrak{A}_5^{\theta}$-soliton membrane $\text{Hom} \Big( \text{Hom}(\mathcal{E}^{I_1}_{\text{HW}}, \mathcal{E}^{I_2}_{\text{HW}}) , \text{Hom}(\mathcal{E}^{I_{2\mathfrak{N}_m + 1}}_{\text{HW}}, \mathcal{E}^{I_{2\mathfrak{N}_m + 2}}_{\text{HW}}) \Big)_+$, with vertices as labeled.

Just as in \autoref{sec:m2 x r3:fueter-cat}, the collection of $\varPi^{\mathfrak{N}_m}_{\mathfrak{A}_5}$ maps in~\eqref{eq:cy2 x s x r3:fueter composition maps} which involve 2-morphisms, can also be regarded as composition maps defining an $A_{\infty}$-structure of a 2-category whose $m$ objects $\{\mathcal{E}^1_{\text{HW}}, \mathcal{E}^2_{\text{HW}}, \dots, \mathcal{E}^m_{\text{HW}}\}$ correspond to ($\theta$-deformed) HW configurations on $CY_2 \times S^1$.

\subtitle{A Fueter type $A_\infty$-2-category 2-categorifying the Holomorphic DT Floer Homology of $CY_2 \times S^1$}

As HW configurations on $CY_2 \times S^1$ are known to generate  the holomorphic DT Floer homology of $CY_2 \times S^1$ which is itself a 0-category, this 2-category is a 2-categorification of the said Floer homology.

Altogether, this means that the normalized partition function of Spin$(7)$ theory on $CY_2 \times S^1 \times \R^3$, as expressed in~\eqref{eq:cy2 x s x r3:3d lg:partition fn}, manifests a \emph{novel} Fueter type $A_\infty$-2-category, defined by the maps~\eqref{eq:cy2 x s x r3:fueter composition maps} and the identifications~\eqref{eq:cy2 x s x r3:3d lg:2-morphism}, which 2-categorifies the holomorphic DT Floer homology of $CY_2 \times S^1$!

\subtitle{An Equivalence Between a Fueter type $A_{\infty}$-2-category and an FS type $A_{\infty}$-category}

Recall from \autoref{sec:cy2 x s x r3:fs-cat} that the normalized partition function of Spin$(7)$ theory on $CY_2 \times S^1 \times \R^3$ also manifests the FS type $A_{\infty}$-category of $\mathfrak{A}_5^{\theta}$-solitons.
This means that we have a \emph{novel} equivalence between the Fueter type $A_\infty$-2-category 2-categorifying the holomorphic DT Floer homology of $CY_2 \times S^1$ and the FS type $A_\infty$-category of $\mathfrak{A}_5^\theta$-solitons!

\section{A Fueter type \texorpdfstring{$A_\infty$}{A-infinity}-2-category of Four-Manifolds}
\label{sec:cy2 x r3}

In this section, we will perform a Kaluza-Klein (KK) dimensional reduction of Spin$(7)$ theory on $\text{Spin}(7) = CY_2 \times S^1 \times \R^3$ along the $S^1$ circle by shrinking it to be infinitesimally small.
We will recast the resulting 7d-Spin$(7)$ theory as a 3d gauged LG model on $\R^3$, a 2d gauged LG model on $\R^2$, or a 1d LG SQM.
Following the approach in \cite[$\S$10]{er-2024-topol-gauge-theor}, we will, via the 7d-Spin$(7)$ partition function and its equivalent 2d gauged LG model, be able to physically realize a novel FS type $A_{\infty}$-category of solitons whose endpoints correspond to Vafa-Witten (VW) configurations on $CY_2$ that generate a HW Floer homology.
Similarly, via the 7d-Spin$(7)$ partition function and its equivalent 3d gauged LG model, we will be able to also physically realize a novel Fueter type $A_{\infty}$-2-category that 2-categorifies the HW Floer homology of $CY_2$.

\subsection{7d-\texorpdfstring{$\text{Spin}(7)$}{Spin(7)} Theory on \texorpdfstring{$CY_2 \times \R^3$}{CY2 x R3} as a 3d Model on \texorpdfstring{$\R^3$}{R3}, 2d Model on \texorpdfstring{$\R^2$}{R2}, or 1d SQM}
\label{sec:cy2 x r3:theory}

Let us perform a KK reduction of Spin$(7)$ theory on $CY_2 \times S^1 \times \R^3$ along the $S^1$ circle in the $y$-direction. The action of the resulting 7d-Spin$(7)$ theory on $CY_2 \times \R^3$ can be obtained from~\eqref{eq:cy2 x s x r3:action} by setting $\partial_y \rightarrow 0$ therein.

\subtitle{7d-Spin$(7)$ Theory on $CY_2 \times \R^3$}

The conditions on (the zero modes of) the bosons that minimize the resulting action are also obtained by performing a KK reduction along the circle in the $y$-direction of the conditions on (the zero modes of) the bosons that minimize the action of Spin$(7)$ theory, i.e., by the KK reduction of \eqref{eq:cy2 x s x r:bps eqns}.
They are given by
\begin{equation}
  \label{eq:cy2 x r:bps eqns}
  \begin{aligned}
    D_{\xi} C + F_{t \tau}
    &= i \omega^{\kappa \bar{\kappa}} \mathcal{F}_{\kappa \bar{\kappa}}
      \, ,
    \\
    D_{\tau} C - i D_t C
    + i (F_{\xi \tau} - i F_{\xi t})
    &= - \frac{i}{2} \varepsilon^{\kappa\lambda} \mathcal{F}_{\kappa\lambda}
      \, ,
    \\
    D_{\xi} \mathcal{A}_\kappa - k D_{\tau} \mathcal{A}_\kappa + j D_t \mathcal{A}_\kappa - \partial_\kappa( A_{\xi} - k A_{\tau} + j A_t)
    &= - i \mathcal{D}_\kappa C
      \, .
  \end{aligned}
\end{equation}

Next, note that we are physically free to rotate the $(\tau, t)$-subplane of $\R^3$ about the origin by an angle $\theta$, whence \eqref{eq:cy2 x r:bps eqns} become
\begin{equation}
  \label{eq:cy2 x r3:bps eqns:rotated}
  \begin{aligned}
    D_{\xi} C + F_{t \tau}
    &= i \omega^{\kappa \bar{\kappa}} \mathcal{F}_{\kappa \bar{\kappa}}
      \, ,
    \\
    D_{\tau} C - i D_t C
    + i (F_{\xi \tau} - i F_{\xi t})
    &= - \frac{i e^{-i\theta}}{2} \varepsilon^{\kappa\lambda} \mathcal{F}_{\kappa\lambda}
      \, ,
    \\
    D_{\xi} \mathcal{A}_\kappa - k_{\theta} D_{\tau} \mathcal{A}_\kappa + j_{\theta} D_t \mathcal{A}_\kappa - \partial_\kappa( A_{\xi} - k_{\theta} A_{\tau} + j_{\theta} A_t)
    &= - i \mathcal{D}_\kappa C
      \, .
  \end{aligned}
\end{equation}
This allows us to write the action for 7d-Spin$(7)$ theory on $CY_2 \times \R^3$ as
\begin{equation}
  \label{eq:cy2 x r3:action:rotated}
  \begin{aligned}
    S_{\text{7d-Spin}(7), CY_2 \times \R^3}
    = \frac{4}{e^2} \int_{\R^3} dt d\tau d\xi \int_{CY_2} d^4 x \, \Tr
    \Bigg(
    & |D_{\xi} C + F_{t\tau} + p|^2
      + |D_{\tau} C - i D_t C + i(F_{\xi\tau} - i F_{\xi t}) + q|^2
    \\
    & + |D_{\xi} \mathcal{A}_\kappa - k_{\theta} D_{\tau} \mathcal{A}_\kappa + j_{\theta} D_t \mathcal{A}_\kappa + r_\kappa|^2
      + \dots
      \Bigg)
      \, ,
  \end{aligned}
\end{equation}
where
\begin{equation}
  \label{eq:cy2 x r3:action:rotated:components}
  \begin{aligned}
    p
    &= - i \omega^{\kappa \bar{\kappa}} \mathcal{F}_{\kappa \bar{\kappa}}
      \, ,
    \\
    q
    &= \frac{i e^{-i\theta}}{2} \varepsilon^{\kappa\lambda} \mathcal{F}_{\kappa\lambda}
      \, ,
    \\
    r_\kappa
    &= - \partial_\kappa( A_{\xi} - k_{\theta} A_{\tau} + j_{\theta} A_t)
      + i \mathcal{D}_\kappa C
      \, .
  \end{aligned}
\end{equation}

\subtitle{7d-Spin$(7)$ Theory as a 3d Model}

After suitable rescalings, we can recast \eqref{eq:cy2 x r3:action:rotated} as a 3d model on $\R^3$, where its action now reads\footnote{%
  Just like in \autoref{ft:stokes theorem for m2 x r3:3d model}, to arrive at the following expression, we have (i) employed Stokes' theorem and the fact that $CY_2$ has no boundary to omit terms with $\partial_\kappa A_{\{t, \tau, \xi\}}$ as they will vanish when integrated over $CY_2$, and
  (ii) integrated out the scalar field $\mathfrak{h}_4(p) = i \omega^{\mathring{a} \bar{\mathring{a}}} \mathcal{F}_{\mathring{a} \bar{\mathring{a}}}$ corresponding to the scalar $p$, whose contribution to the action is $|\mathfrak{h}_4(p)|^2$.
  \label{ft:stokes theorem for cy2 x r3:3d model}
}
\begin{equation}
  \label{eq:cy2 x r3:3d model action}
  \begin{aligned}
    S_{\text{3d}, \mathfrak{A}_4}
    & = \frac{1}{e^2} \int_{\R^3} dt d\tau d\xi \Bigg(
    |D_{\xi} C^{\mathring{a}} + F_{t\tau}|^2
    + |D_{\tau} C^{\mathring{a}} - i D_t C^{\mathring{a}} + i(F_{\xi\tau} - i F_{\xi t}) + q^{\mathring{a}}|^2
    \\
    & \qquad \qquad \qquad \qquad + |D_{\xi} \mathcal{A}^{\mathring{a}} - k_{\theta} D_{\tau} \mathcal{A}^{\mathring{a}} + j_{\theta} D_t \mathcal{A}^{\mathring{a}} + r^{\mathring{a}}|^2
      + \dots
      \Bigg)
    \\
    & = \frac{1}{e^2} \int_{\R_{\tau} \times \R_{\xi}} d\tau d\xi \int_{\R_t} dt \Bigg(
    |D_{\xi} C^{\mathring{a}} - D_{\tau} A_t + P|^2
    + |D_{\tau} C^{\mathring{a}} + iF_{\xi\tau} + D_{\xi} A_t + Q^{\mathring{a}} + q^{\mathring{a}}|^2
    \\
    & \qquad \qquad \qquad \qquad \qquad \quad + |D_{\xi} \mathcal{A}^{\mathring{a}} - k_{\theta} D_{\tau} \mathcal{A}^{\mathring{a}} + R^{\mathring{a}} + r^{\mathring{a}}|^2
      + \dots
      \Bigg)
      \, .
  \end{aligned}
\end{equation}
Here, $(\mathcal{A}^{\mathring{a}}, C^{\mathring{a}})$ and $\mathring{a}$ are coordinates and indices on the space $\mathfrak{A}_4$ of irreducible $(\mathcal{A}_{\kappa}, C)$ fields on $CY_2$, and
\begin{equation}
  \label{eq:cy2 x r3:3d model action:components}
  P
  = \partial_t A_{\tau}
  \, ,
  \qquad
  Q^{\mathring{a}}
  = - i D_t C^{\mathring{a}} - \partial_t A_{\xi}
  \, ,
  \qquad
  R^{\mathring{a}}
  = j_{\theta} D_t \mathcal{A}^{\mathring{a}}
  \, ,
\end{equation}
with $(q^{\mathring{a}}, r^{\mathring{a}})$ corresponding to $(q, r_{\kappa})$ in \eqref{eq:cy2 x r3:action:rotated:components}.

In other words, 7d-Spin$(7)$ theory on $CY_2 \times \R^3$ can be regarded as a 3d gauged sigma model along the $(t, \tau, \xi)$-directions with target space $\mathfrak{A}_4$ and action \eqref{eq:cy2 x r3:3d model action}.

\subtitle{7d-Spin$(7)$ Theory as a 2d Model}

From \eqref{eq:cy2 x r3:3d model action}, one can see that we can, after suitable rescalings, also recast the 3d model action as an equivalent 2d model action\footnote{%
  Just like in \autoref{ft:stokes theorem for m2 x r3:2d model}, to arrive at the following expression, we have employed Stokes' theorem and the fact that the finite-energy gauge fields $A_{\{t, \tau, \xi\}}$ would vanish at $t \rightarrow \pm \infty$.
  \label{ft:stokes theorem for cy2 x r3:2d model}
}
\begin{equation}
  \label{eq:cy2 x r3:2d model action}
  \begin{aligned}
    S_{\text{2d}, \mathcal{M}(\R_t, \mathfrak{A}_4)} = \frac{1}{e^2} \int_{\R^2} d\tau d\xi \Bigg(
    & |D_{\xi} C^{\mathring{m}} - D_{\tau} (\tilde{A}_t)^{\mathring{m}}|^2
    + |D_{\tau} C^{\mathring{m}} + iF_{\xi\tau} + D_{\xi} (\tilde{A}_t)^{\mathring{m}} + Q^{\mathring{m}} + q^{\mathring{m}}|^2
    \\
    & + |D_{\xi} \mathcal{A}^{\mathring{m}} - k_{\theta} D_{\tau} \mathcal{A}^{\mathring{m}} + R^{\mathring{m}} + r^{\mathring{m}}|^2
      + \dots
      \Bigg)
      \, .
  \end{aligned}
\end{equation}
Here, $(\mathcal{A}^{\mathring{m}}, C^{\mathring{m}}, (\tilde{A}_t)^{\mathring{m}})$ and $\mathring{m}$ are coordinates and indices on the path space $\mathcal{M}(\R_t, \mathfrak{A}_4)$ of smooth paths from $\R_t$ to $\mathfrak{A}_4$, with
\begin{equation}
  \label{eq:cy2 x r3:2d model action:components}
  Q^{\mathring{m}}
  = - i ( \tilde{D}_t C)^{\mathring{m}}
  \, ,
  \qquad
  R^{\mathring{m}}
  = j_{\theta} (\tilde{D}_t \mathcal{A})^{\mathring{m}}
  \, ,
\end{equation}
corresponding to $(Q^{\mathring{a}}, R^{\mathring{a}})$ in \eqref{eq:cy2 x r3:3d model action:components}, $(q^{\mathring{m}}, r^{\mathring{m}})$ corresponding to $(q^{\mathring{a}}, r^{\mathring{a}})$, and $(\tilde{A}_t, \tilde{D}_t)$ corresponding to $(A_t, D_t)$, in the underlying 3d model.

In other words, 7d-Spin$(7)$ theory on $CY_2 \times \R^3$ can also be regarded as a 2d gauged sigma model along the $(\tau, \xi)$-directions with target space $\mathcal{M}(\R_t, \mathfrak{A}_4)$ and action \eqref{eq:cy2 x r3:2d model action}.

\subtitle{7d-Spin$(7)$ Theory as a 1d SQM}

Singling out $\xi$ as the direction in ``time'', the equivalent SQM action can be obtained from~\eqref{eq:cy2 x r3:2d model action} after suitable rescalings as\footnote{%
  In the resulting SQM, we have integrated out $A_{\xi}$ and omitted the terms containing the fields corresponding to $A_{\{t, \tau\}}$ (as explained in \autoref{ft:stokes theorem for m2 x r3:sqm}).
  \label{ft:stokes theorem for cy2 x r3:sqm}
}
\begin{equation}
  \label{eq:cy2 x r3:sqm action}
  \begin{aligned}
    S_{\text{SQM}, \mathcal{M}(\R_{\tau}, \mathcal{M}(\R_t, \mathfrak{A}_4))} = \frac{1}{e^2} \int d\xi \Bigg(
    & \left| \partial_{\xi} C^{\mathring{u}}
      + g^{\mathring{u}\mathring{v}}_{\mathcal{M}(\R_{\tau}, \mathcal{M}(\R_t, \mathfrak{A}_4))} \pdv{h_4}{C^{\mathring{v}}}
      \right|^2
      + \left| \partial_{\xi} \breve{A}^{\mathring{u}}
      + g^{\mathring{u}\mathring{v}}_{\mathcal{M}(\R_{\tau}, \mathcal{M}(\R_t, \mathfrak{A}_4))} \pdv{h_4}{\breve{A}^{\mathring{v}}}
      \right|^2
    \\
    & + \left| \partial_{\xi} \mathcal{A}^{\mathring{u}}
      + g^{\mathring{u}\mathring{v}}_{\mathcal{M}(\R_{\tau}, \mathcal{M}(\R_t, \mathfrak{A}_4))} \pdv{h_4}{\mathcal{A}^{\mathring{v}}}
      \right|^2
      + \dots \Bigg)
      \, ,
  \end{aligned}
\end{equation}
where $(\mathcal{A}^{\mathring{u}}, C^{\mathring{u}}, \breve{A}^{\mathring{u}})$ and $(\mathring{u}, \mathring{v})$ are coordinates on the path space $\mathcal{M}(\R_{\tau}, \mathcal{M}(\R_t, \mathfrak{A}_4))$ of smooth maps from $\R_{\tau}$ to $\mathcal{M}(\R_t, \mathfrak{A}_4)$ with $\breve{A}^{\mathring{u}} \coloneqq (\breve{A}_t + i \breve{A}_{\tau})^{\mathring{u}}$ in $\mathcal{M}(\R_{\tau}, \mathcal{M}(\R_t, \mathfrak{A}_4))$ corresponding to $(\tilde{A}_t)^{\mathring{m}} + i A_{\tau}$ in the underlying 2d model, and to $A_t + i A_{\tau}$ in the underlying 3d model;
$g_{\mathcal{M}(\R_{\tau}, \mathcal{M}(\R_t, \mathfrak{A}_4))}$ is the metric of $\mathcal{M}(\R_{\tau}, \mathcal{M}(\R_t, \mathfrak{A}_4))$;
and $h_4(\mathcal{A}, C, \breve{A})$ is the SQM potential function.
Note that we can also interpret $\mathcal{M}(\R_{\tau}, \mathcal{M}(\R_t, \mathfrak{A}_4))$ as the double path space $\mathcal{M}(\R^2, \mathfrak{A}_4)$ of smooth maps from $\R^2$ to $\mathfrak{A}_4$.

In short, 7d-Spin$(7)$ theory on $CY_2 \times \R^3$ can also be regarded as a 1d SQM along the $\xi$-direction in $\mathcal{M}(\R^2, \mathfrak{A}_4)$ whose action is \eqref{eq:cy2 x r3:sqm action}.

\subsection{Non-constant Paths, Solitons, Sheets, and the HW Floer Homology of \texorpdfstring{$CY_2$}{CY2}}
\label{sec:cy2 x r3:vw}

By following the analysis in \autoref{sec:m2 x r3:gc-bf}, we find that the equivalent 1d SQM of 7d-Spin$(7)$ theory on $CY_2 \times \R^3$ will localize onto \emph{$\xi$-invariant, $\theta$-deformed}, non-constant double paths in $\mathcal{M}(\R_{\tau}, \mathcal{M}(\R_t, \mathfrak{A}_4))$ which will correspond, in the 2d gauged sigma model with target space $\mathcal{M}(\R_t, \mathfrak{A}_4)$, to \emph{$\xi$-invariant, $\theta$-deformed} solitons along the $\tau$-direction that we shall refer to as $\mathcal{M}^{\theta}(\R_t, \mathfrak{A}_4)$-solitons.

\subtitle{$\mathcal{M}^{\theta}(\R_t, \mathfrak{A}_4)$-solitons in the 2d Gauged Model}

Specifically, such $\mathcal{M}^{\theta}(\R_t, \mathfrak{A}_4)$-solitons are defined by
\begin{equation}
  \label{eq:cy2 x r3:m-soliton eqns}
  \begin{aligned}
    \partial_{\tau} (\tilde{A}_t)^{\mathring{m}}
    &= - [A_{\tau}, (\tilde{A}_t)^{\mathring{m}}]
      + [A_{\xi}, C^{\mathring{m}}]
      \, ,
    \\
    \partial_{\tau} C^{\mathring{m}} - i \partial_{\tau} A_{\xi}
    &= - [A_{\tau}, C^{\mathring{m}} - i A_{\xi}]
      + i (\tilde{D}_t C)^{\mathring{m}}
      + [(\tilde{A}_t)^{\mathring{m}}, A_{\xi}]
      - q^{\mathring{m}}
      \, ,
    \\
    k_{\theta} \partial_{\tau} \mathcal{A}^{\mathring{m}}
    &= [A_{\xi} - k_{\theta} A_{\tau}, \mathcal{A}^{\mathring{m}}]
      + j_{\theta} (\tilde{D}_t \mathcal{A})^{\mathring{m}}
      + r^{\mathring{m}}
      \, .
  \end{aligned}
\end{equation}

\subtitle{$\mathfrak{A}^{\theta}_4$-sheets in the 3d Gauged Model}

In turn, they will correspond, in the 3d gauged sigma model with target space $\mathfrak{A}_4$, to $\xi$-invariant, $\theta$-deformed sheets along the $(\tau, t)$-directions that are defined by
\begin{equation}
  \label{eq:cy2 x r3:sheet eqns}
  \begin{aligned}
    \partial_{\tau} A_t - \partial_t A_{\tau}
    &= - [A_{\tau}, A_t]
      + [A_{\xi}, C^{\mathring{a}}]
      \, ,
    \\
    \partial_{\tau} C^{\mathring{a}} - i \partial_t C^{\mathring{a}}
    - i \partial_{\tau} A_{\xi} - \partial_t A_{\xi}
    &= - [A_{\tau} - i A_t, C^{\mathring{a}} - i A_{\xi}]
      - q^{\mathring{a}}
      \, ,
    \\
    k_{\theta} \partial_{\tau} \mathcal{A}^{\mathring{a}} - j_{\theta} \partial_t \mathcal{A}^{\mathring{a}}
    &= [A_{\xi} - k_{\theta} A_{\tau} + j_{\theta} A_t, \mathcal{A}^{\mathring{a}}]
      + r^{\mathring{a}}
      \, ,
  \end{aligned}
\end{equation}
and the condition
\begin{equation}
  \label{eq:cy2 x r3:sheet eqns:aux cond}
  \mathfrak{h}_4(p) = 0
  \, ,
\end{equation}
where $\mathfrak{h}_4(p)$ is the auxiliary scalar field defined in \autoref{ft:stokes theorem for cy2 x r3:3d model}.
We shall refer to such sheets defined by~\eqref{eq:cy2 x r3:sheet eqns} and \eqref{eq:cy2 x r3:sheet eqns:aux cond} as $\mathfrak{A}^{\theta}_4$-sheets.

\subtitle{$\xi$-independent, $\theta$-deformed 7d-Spin$(7)$ Configurations in 7d-Spin$(7)$ Theory}

In turn, the 3d configurations defined by~\eqref{eq:cy2 x r3:sheet eqns} and \eqref{eq:cy2 x r3:sheet eqns:aux cond} will correspond, in 7d-Spin$(7)$ theory, to $\xi$-independent, $\theta$-deformed 7d-Spin$(7)$ configurations on $CY_2 \times R^3$ that are defined, via \eqref{eq:cy2 x r3:action:rotated:components}, by
\begin{equation}
  \label{eq:cy2 x r3:7d-spin7 configs}
  \begin{aligned}
    \partial_{\tau} A_t - \partial_t A_{\tau}
    &= - [A_{\tau}, A_t]
      + [A_{\xi}, C]
      \, ,
    \\
    \partial_{\tau} C - i \partial_t C
    - i \partial_{\tau} A_{\xi} - \partial_t A_{\xi}
    &= - [A_{\tau} - i A_t, C - i A_{\xi}]
      - \frac{ie^{-i\theta}}{2} \varepsilon^{\kappa\lambda} \mathcal{F}_{\kappa\lambda}
      \, ,
    \\
    k_{\theta} \partial_{\tau} \mathcal{A}_{\kappa} - j_{\theta} \partial_t \mathcal{A}_{\kappa}
    &= - \mathcal{D}_\kappa (A_{\xi} - k_{\theta} A_{\tau} + j_{\theta} A_t)
      + i \mathcal{D}_\kappa C
      \, ,
    \end{aligned}
\end{equation}
and the condition
\begin{equation}
  \label{eq:cy2 x r3:7d-spin7 configs:aux cond}
  i \omega^{\kappa \bar{\kappa}} \mathcal{F}_{\kappa \bar{\kappa}} = 0
  \, .
\end{equation}

\subtitle{7d-Spin$(7)$ Configurations, $\mathfrak{A}_4^{\theta}$-sheets, $\mathcal{M}^{\theta}(\R_t, \mathfrak{A}_4)$-solitons, and Non-constant Double Paths}

In short, these \emph{$\xi$-independent, $\theta$-deformed} 7d-Spin$(7)$ configurations on $CY_2 \times \R^3$ that are defined by~\eqref{eq:cy2 x r3:7d-spin7 configs} and \eqref{eq:cy2 x r3:7d-spin7 configs:aux cond}, will correspond to the $\mathfrak{A}_4^{\theta}$-sheets defined by~\eqref{eq:cy2 x r3:sheet eqns} and \eqref{eq:cy2 x r3:sheet eqns:aux cond}, which, in turn, will correspond to the $\mathcal{M}^{\theta}(\R_t, \mathfrak{A}_4)$-solitons defined by~\eqref{eq:cy2 x r3:m-soliton eqns}, which, in turn, will correspond to the $\xi$-invariant, $\theta$-deformed, non-constant double paths in $\mathcal{M}(\R^2, \mathfrak{A}_4)$ defined by setting both the LHS and RHS of the expression within the squared terms of \eqref{eq:cy2 x r3:sqm action} \emph{simultaneously} to zero.

\subtitle{$\mathcal{M}^{\theta}(\R_t, \mathfrak{A}_4)$-soliton Endpoints Corresponding to Non-constant Paths}

Consider now the fixed endpoints of the $\mathcal{M}^{\theta}(\R_t, \mathfrak{A}_4)$-solitons at $\tau = \pm \infty$, where we also expect the fields in the 2d gauged sigma model corresponding to the finite-energy 3d gauge fields $A_\xi, A_{\tau}, A_t$ to decay to zero.
They are given by~\eqref{eq:cy2 x r3:m-soliton eqns} with $\partial_\tau C^{\mathring{m}} = 0 = \partial_{\tau} \mathcal{A}^{\mathring{m}}$ and $A_\xi, A_{\tau}, (\tilde{A}_t)^{\mathring{m}} \rightarrow 0$, i.e,
\begin{equation}
  \label{eq:cy2 x r3:m-soliton:endpts}
  i (\tilde{\partial}_t C)^{\mathring{m}}
  = q^{\mathring{m}}
  \, ,
  \qquad
  j_{\theta} (\tilde{\partial}_t \mathcal{A})^{\mathring{m}}
  = - r^{\mathring{m}}
  \, .
\end{equation}
These are $(\xi, \tau)$-invariant, $\theta$-deformed, non-constant paths in $\mathcal{M}(\R_t, \mathfrak{A}_4)$.

\subtitle{$\mathfrak{A}_4^{\theta}$-sheet Edges Corresponding to $\mathfrak{A}_4^{\theta}$-solitons in the 3d Gauged Model}

In turn, \eqref{eq:cy2 x r3:m-soliton:endpts} will correspond, in the 3d gauged sigma model, to the fixed edges of the $\mathfrak{A}_4^{\theta}$-sheets at $\tau = \pm \infty$, i.e., $(\xi, \tau)$-invariant, $\theta$-deformed solitons along the $t$-direction that are defined by
\begin{equation}
  \label{eq:cy2 x r3:soliton}
  i \partial_t C^{\mathring{a}}
  = q^{\mathring{a}}
  \, ,
  \qquad
  j_{\theta} \partial_t \mathcal{A}^{\mathring{a}}
  = - r^{\mathring{a}}
  \, .
\end{equation}
Notice that these equations can also be obtained from \eqref{eq:cy2 x r3:sheet eqns} with $\partial_{\tau} C^{\mathring{a}} = 0 = \partial_{\tau} \mathcal{A}^{\mathring{a}}$ and $A_\xi, A_{\tau}, A_t \rightarrow 0$.
We shall henceforth refer to such solitons defined by~\eqref{eq:cy2 x r3:soliton} and \eqref{eq:cy2 x r3:sheet eqns:aux cond} as $\mathfrak{A}^{\theta}_4$-solitons.

\subtitle{$\mathfrak{A}^{\theta}_4$-soliton Endpoints or $\mathfrak{A}_4^{\theta}$-sheet Vertices Corresponding to $\theta$-deformed VW Configurations on $CY_2$}

Consider now (i) the fixed endpoints of the $\mathfrak{A}_4^{\theta}$-solitons at $t = \pm \infty$, or equivalently (ii) the vertices of the $\mathfrak{A}_4^{\theta}$-sheets at $\tau, t = \pm \infty$.
They are given by (i)~\eqref{eq:cy2 x r3:soliton} and \eqref{eq:cy2 x r3:sheet eqns:aux cond} with $\partial_t C^{\mathring{a}} = 0 = \partial_t \mathcal{A}^{\mathring{a}}$, or equivalently (ii)~\eqref{eq:cy2 x r3:sheet eqns} and \eqref{eq:cy2 x r3:sheet eqns:aux cond} with $\partial_{\{t, \tau\}} C^{\mathring{a}} = 0 = \partial_{\{t, \tau\}} \mathcal{A}^{\mathring{a}}$ and $A_{\xi}, A_{\tau}, A_t \rightarrow 0$.
In turn, they will correspond, in 7d-Spin$(7)$ theory, to $(\xi, \tau, t)$-independent, $\theta$-deformed configurations that obey~\eqref{eq:cy2 x r3:7d-spin7 configs} and \eqref{eq:cy2 x r3:7d-spin7 configs:aux cond} with $\partial_{\{t, \tau\}} C = 0 = \partial_{\{t, \tau\}} \mathcal{A}_{\kappa}$ and $A_\xi, A_\tau, A_t \rightarrow 0$, i.e.,
\begin{equation}
  \label{eq:cy2 x r3:soliton:endpts:spin7}
  ie^{-i\theta} \varepsilon^{\kappa\lambda} \mathcal{F}_{\kappa\lambda}
  = 0
  \, ,
  \qquad
  i \mathcal{D}_{\kappa} C
  = 0
  \, ,
  \qquad
  \omega^{\kappa \bar{\kappa}} \mathcal{F}_{\kappa \bar{\kappa}}
  = 0
  \, .
\end{equation}

At $\theta = 0$ or $\pi$, \eqref{eq:cy2 x r3:soliton:endpts:spin7} can be written, in the real coordinates of $CY_2$, as
\begin{equation}
  \label{eq:cy2 x r3:vw eqns}
  F_{\alpha\beta}^+ = 0
  \, ,
  \qquad
  D_{\alpha} B^{\alpha\beta} = 0
  \, ,
\end{equation}
where $B_{\alpha\beta} \in \Omega^{2, +}(CY_2, \text{ad}(G)) \otimes \Omega^0(\R^3, \text{ad}(G))$ is a self-dual two-form on $CY_2$ with one of its three linearly-independent components being $C$, and the other two being zero.
These are the Vafa-Witten (VW) equations on $CY_2$ with the above-described self-dual two-form, and scalar being zero.
We shall, in the rest of this section, refer to configurations that span the space of solutions to these equations as VW configurations on $CY_2$.
One thing to note at this point is that VW configurations on $CY_2$ are known to generate the HW Floer homology of $CY_2$~\cite[$\S$3]{er-2023-topol-n}.

In other words, the $(\xi, \tau, t)$-independent, $\theta$-deformed 7d-Spin$(7)$ configurations corresponding to the endpoints of the $\mathfrak{A}^{\theta}_4$-solitons (or equivalently, the vertices of the $\mathfrak{A}^{\theta}_4$-sheets) are $\theta$-deformed VW configurations on $CY_2$.
We will also assume choices of $CY_2$ whereby such configurations are isolated and non-degenerate.\footnote{
  At $\theta = 0$, these configurations are undeformed VW configurations on $CY_2$.
  Note that such configurations are obtained by a KK reduction along $S^1$ of the undeformed HW configurations on $CY_2 \times S^1$ from \autoref{sec:cy2 x s x r3}.
  Since our choice of $CY_2$ is one such that these undeformed HW configurations on $CY_2 \times S^1$ are isolated, it would mean that the undeformed VW configurations on $CY_2$ must also be isolated.
  We can then apply the same reasoning in \autoref{ft:isolation of hw} to see that the endpoints of the $\mathfrak{A}_4^{\theta}$-solitons (or equivalently, the vertices of the $\mathfrak{A}_4^{\theta}$-sheets) will be isolated and non-degenerate.
  \label{ft:isolation of vw}
}

\subtitle{Non-constant Double Paths, $\mathcal{M}^{\theta}(\R_t, \mathfrak{A}_4)$-solitons, $\mathfrak{A}_4^{\theta}$-solitons, and $\mathfrak{A}_4^{\theta}$-sheets}

In short, from the equivalent 1d SQM of 7d-Spin$(7)$ theory on $CY_2 \times \R^3$, the theory localizes onto $\xi$-invariant, $\theta$-deformed, non-constant double paths in $\mathcal{M}(\R^2, \mathfrak{A}_4)$, which, in turn, will correspond to $\mathcal{M}^{\theta}(\R_t, \mathfrak{A}_4)$-solitons in the 2d gauged sigma model whose endpoints are $(\xi, \tau)$-invariant, $\theta$-deformed, non-constant paths in $\mathcal{M}(\R_t, \mathfrak{A}_4)$.
In the 3d gauged sigma model, these $\mathcal{M}^{\theta}(\R_t, \mathfrak{A}_4)$-solitons will correspond to $\mathfrak{A}_4^{\theta}$-sheets, whose edges are $\mathfrak{A}_4^{\theta}$-solitons, and whose vertices will correspond to $\theta$-deformed VW configurations on $CY_2$ that generate the HW Floer homology of $CY_2$.

\subsection{The 2d Model and Open Strings, the 3d Model and Open Membranes}
\label{sec:cy2 x r3:2d-3d model}

By following the same analysis in \autoref{sec:m2 x r3:2d-3d model}, we find that the 2d gauged sigma model with target space $\mathcal{M}(\R_t, \mathfrak{A}_4)$ whose action is \eqref{eq:cy2 x r3:2d model action}, will define an open string theory in $\mathcal{M}(\R_t, \mathfrak{A}_4)$.
Similarly, we find that the 3d gauged sigma model with target space $\mathfrak{A}_4$ whose action is \eqref{eq:cy2 x r3:3d model action}, will define an open membrane theory in $\mathfrak{A}_4$.
We will now work out the details pertaining to the BPS worldsheets and BPS worldvolumes (including their boundaries) that are necessary to define this open string and open membrane theory, respectively.

\subtitle{BPS Worldsheets of the 2d Model}

The BPS worldsheets of the 2d gauged sigma model with target space $\mathcal{M}(\R_t, \mathfrak{A}_4)$ correspond to its classical trajectories.
Specifically, these are defined by setting to zero the expression within the squared terms in \eqref{eq:cy2 x r3:2d model action}, i.e.,
\begin{equation}
  \label{eq:cy2 x r3:worldsheet:eqn}
  \begin{gathered}
    \Dv{C^{\mathring{m}}}{\xi} - \Dv{(\tilde{A}_t)^{\mathring{m}}}{\tau}
    = 0
    \, ,
    \qquad
    \Dv{(\tilde{A}_t)^{\mathring{m}}}{\xi} + \Dv{C^{\mathring{m}}}{\tau}
    + i F_{\xi\tau}
    = - Q^{\mathring{m}} - q^{\mathring{m}}
    \, ,
    \\
    \Dv{\mathcal{A}^{\mathring{m}}}{\xi} - k_{\theta} \Dv{\mathcal{A}^{\mathring{m}}}{\tau}
    = - R^{\mathring{m}} - r^{\mathring{m}}
    \, .
  \end{gathered}
\end{equation}

\subtitle{BPS Worldsheets with Boundaries Labeled by Non-constant Paths in $\mathcal{M}(\R_t, \mathfrak{A}_4)$}

The boundaries of the BPS worldsheets are traced out by the endpoints of the $\mathcal{M}^{\theta}(\R_t, \mathfrak{A}_4)$-solitons as they propagate in $\xi$.
As we have seen at the end of \autoref{sec:cy2 x r3:vw}, these endpoints correspond to $(\xi, \tau)$-invariant, $\theta$-deformed, non-constant paths in $\mathcal{M}(\R_t, \mathfrak{A}_4)$ that we shall, at $\xi = \pm \infty$, denote as $\gamma_{\pm}(\theta, \mathfrak{A}_4)$.
In turn, they will correspond, in the 3d gauged sigma model with target space $\mathfrak{A}_4$, to $\mathfrak{A}_4^{\theta}$-solitons that we shall, at $\xi = \pm \infty$, denote as $\Gamma_{\pm}(t, \theta, \mathfrak{A}_4)$, whose endpoints will correspond to $\theta$-deformed VW configurations on $CY_2$.

If there are $n \geq 4$ such configurations $\{\mathcal{E}^1_{\text{VW}}(\theta), \mathcal{E}^2_{\text{VW}}(\theta), \dots, \mathcal{E}^n_{\text{VW}}(\theta)\}$, just as in \autoref{sec:m2 x r3:2d-3d model}, we can further specify any $\Gamma_{\pm}(t, \theta, \mathfrak{A}_4)$ as $\Gamma^{IJ}_{\pm}(t, \theta, \mathfrak{A}_4)$, where its endpoints would correspond to $\mathcal{E}^I_{\text{VW}}(\theta)$ and $\mathcal{E}^J_{\text{VW}}(\theta)$.
Consequently, in the 2d gauged sigma model, we can further specify any $\gamma_{\pm}(\theta, \mathfrak{A}_4)$ as $\gamma^{IJ}_{\pm}(\theta, \mathfrak{A}_4)$, where the latter will correspond to a $\Gamma^{IJ}_{\pm}(t, \theta, \mathfrak{A}_4)$ $\mathfrak{A}_4^{\theta}$-soliton in the equivalent 3d gauged sigma model.

Since the endpoints of an $\mathcal{M}^{\theta}(\R_t, \mathfrak{A}_4)$-soliton are now denoted as $\gamma^{**}_{\pm}(\theta, \mathfrak{A}_4)$, we can also denote and specify the former at $\xi = \pm \infty$ as $\sigma^{IJ, KL}_{\pm}(\tau, \theta, \mathfrak{A}_4)$,\footnote{%
  Just like in \autoref{ft:fixing m-A2-soliton centre of mass dof}, the $\xi$-invariant $\mathcal{M}^{\theta}(\R_t, \mathfrak{A}_4)$-solitons can be fixed at $\xi = \pm \infty$ by adding physically-inconsequential $\mathcal{Q}$-exact terms to the SQM action.
  \label{ft:fixing m-A4-soliton centre of mass dof}
} where its left and right endpoints would be $\gamma^{IJ}_{\pm}(\theta, \mathfrak{A}_4)$ and $\gamma^{KL}_{\pm}(\theta, \mathfrak{A}_4)$, respectively.

As the $\gamma^{**}(\theta, \mathfrak{A}_4)$'s are $\xi$-invariant and therefore, have the same value for all $\xi$, we have BPS worldsheets of the kind similar to \autoref{fig:m2 x r3:bps worldsheet}.
This time, however, instead of the boundaries being labeled $\gamma^{**}(\theta, \mathfrak{A}_2)$, they will be labeled $\gamma^{**}(\theta, \mathfrak{A}_4)$.
And, at $\xi = \pm \infty$, instead of the $\mathcal{M}^{\theta}(\R_{\tau}, \mathfrak{A}_2)$-solitons $\sigma^{IJ, KL}_{\pm}(t, \theta, \mathfrak{A}_2)$, we will have $\mathcal{M}^{\theta}(\R_t, \mathfrak{A}_4)$-solitons $\sigma^{IJ, KL}_{\pm}(\tau, \theta, \mathfrak{A}_4)$.

\subtitle{The 2d Model on $\R^2$ and an Open String Theory in $\mathcal{M}(\R_t, \mathfrak{A}_4)$}

Thus, like in \autoref{sec:m2 x r3:2d-3d model}, one can understand the 2d gauged sigma model with target space $\mathcal{M}(\R_t, \mathfrak{A}_4)$ to define an open string theory in $\mathcal{M}(\R_t, \mathfrak{A}_4)$, whose \emph{effective} worldsheet and boundaries are similar to \autoref{fig:m2 x r3:bps worldsheet}, where $\xi$ and $\tau$ are the temporal and spatial directions, respectively.

\subtitle{BPS Worldvolumes of the 3d Model}

The BPS worldvolumes of the 3d gauged sigma model with target space $\mathfrak{A}_4$ correspond to its classical trajectories.
Specifically, these are defined by setting to zero the expression within the squared terms in \eqref{eq:cy2 x r3:3d model action}, i.e.,
\begin{equation}
  \label{eq:cy2 x r3:worldvolume:eqn}
  \begin{aligned}
    \Dv{C^{\mathring{a}}}{\xi} + F_{t\tau}
    &= 0
      \, ,
    & \qquad
    \Dv{C^{\mathring{a}}}{\tau} - i \Dv{C^{\mathring{a}}}{t}
    + i(F_{\xi\tau} - i F_{\xi t})
    &= - q^{\mathring{a}}
      \, ,
    \\
    0
    &= \mathfrak{h}_5(p)
      \, ,
    & \qquad
    \Dv{\mathcal{A}^{\mathring{a}}}{\xi} - k_{\theta} \Dv{\mathcal{A}^{\mathring{a}}}{\tau} + j_{\theta} \Dv{\mathcal{A}^{\mathring{a}}}{t}
    &= - r^{\mathring{a}}
      \, .
  \end{aligned}
\end{equation}

\subtitle{BPS Worldvolumes with Boundaries Labeled by $\mathfrak{A}_4^{\theta}$-solitons, and Edges Labeled by VW Configurations on $CY_2$}

The boundaries and edges of the BPS worldvolumes are traced out by the edges and vertices of the $\mathfrak{A}_4^{\theta}$-sheets, respectively, as they propagate in $\xi$.
As we have seen at the end of \autoref{sec:cy2 x r3:vw}, these edges and vertices would correspond to $\mathfrak{A}_4^{\theta}$-solitons and $\theta$-deformed VW configurations on $CY_2$, respectively.

This means that we can denote and specify any $\mathfrak{A}_4^{\theta}$-sheet at $\xi = \pm \infty$ as $\Sigma^{IJ,KL}_{\pm}(\tau, t, \theta, \mathfrak{A}_4)$,\footnote{%
  Just like in \autoref{ft:fixing A2-sheet centre of mass dof}, the $\xi$-invariant $\mathfrak{A}_4^{\theta}$-sheets can be fixed at $\xi = \pm \infty$ by adding physically-inconsequential $\mathcal{Q}$-exact terms to the SQM action.
  \label{ft:fixing A4-sheet centre of mass dof}
} where
(i) its left and right edges correspond to the $\mathfrak{A}_4^{\theta}$-solitons $\Gamma^{IJ}(t, \theta, \mathfrak{A}_4)$ and $\Gamma^{KL}(t, \theta, \mathfrak{A}_4)$, respectively, and
(ii) its four vertices would correspond to $\mathcal{E}^I_{\text{VW}}(\theta)$, $\mathcal{E}^J_{\text{VW}}(\theta)$, $\mathcal{E}^K_{\text{VW}}(\theta)$, and $\mathcal{E}^L_{\text{VW}}(\theta)$, similar to the kind shown in \autoref{fig:m2 x r3:frakA-sheet}.
However, instead of the edges being labeled $\Gamma^{**}(\tau, \theta, \mathfrak{A}_2)$, they will be labeled $\Gamma^{**}(t, \theta, \mathfrak{A}_4)$.
And, instead of the vertices being labeled $\mathcal{E}^*_{\text{BF}}(\theta)$, they will be labeled $\mathcal{E}^*_{\text{VW}}(\theta)$.

Since the $\mathcal{E}^{*}_{\text{VW}}(\theta)$'s and $\Gamma^{**}(t, \theta, \mathfrak{A}_4)$'s are $\xi$-invariant and therefore, have the same value for all $\xi$, we have BPS worldvolumes of the kind similar to \autoref{fig:m2 x r3:bps worldvolume}.
This time, however, instead of the faces being labeled $\Sigma^{**,**}_{\pm}(t, \tau, \theta, \mathfrak{A}_2)$ and $\Gamma^{**}(\tau, \theta, \mathfrak{A}_2)$, they will be labeled $\Sigma^{**,**}_{\pm}(\tau, t, \theta, \mathfrak{A}_4)$ and $\Gamma^{**}(t, \theta, \mathfrak{A}_4)$.
And, instead of the edges being labeled $\mathcal{E}^*_{\text{BF}}(\theta)$, they will be labeled $\mathcal{E}^*_{\text{VW}}(\theta)$.

\subtitle{The 3d Model on $\R^3$ and an Open Membrane Theory in $\mathfrak{A}_4$}

Thus, like in \autoref{sec:m2 x r3:2d-3d model}, one can understand the 3d gauged sigma model with target space $\mathfrak{A}_4$ to define an open membrane theory in $\mathfrak{A}_4$, whose \emph{effective} worldvolume and boundaries are similar to \autoref{fig:m2 x r3:bps worldvolume}, where $\xi$ is the temporal direction, and $\tau$ and $t$ are the spatial directions.

\subsection{Soliton String Theory, the 7d-\texorpdfstring{Spin$(7)$}{Spin(7)} Partition Function, and an FS type \texorpdfstring{$A_{\infty}$}{A-infinity}-category of \texorpdfstring{$\mathfrak{A}_4^{\theta}$}{A4-theta}-solitons}
\label{sec:cy2 x r3:fs-cat}

\subtitle{The 2d Model as a 2d Gauged LG Model}

Notice that we can also express \eqref{eq:cy2 x r3:worldsheet:eqn} as
\begin{equation}
  \label{eq:cy2 x r3:2d lg:worldsheet:eqn}
  \begin{aligned}
    \Dv{C^{\mathring{m}}}{\xi} - k \Dv{C^{\mathring{m}}}{\tau}
    - k \left( \Dv{(\tilde{A}_t)^{\mathring{m}}}{\xi} - k \Dv{(\tilde{A}_{\tau})^{\mathring{m}}}{\tau} \right)
    - j F_{\xi \tau}
    &= k Q^{\mathring{m}} + k q^{\mathring{m}}
      \, ,
    \\
    \Dv{\mathcal{A}^{\mathring{m}}}{\xi} - k_{\theta} \Dv{\mathcal{A}^{\mathring{m}}}{\tau}
    &= - R^{\mathring{m}} - r^{\mathring{m}}
      \, .
  \end{aligned}
\end{equation}
In turn, this means that we can express the action of the 2d gauged sigma model with target space $\mathcal{M}(\R_t, \mathfrak{A}_4)$ in~\eqref{eq:cy2 x r3:2d model action} as
\begin{equation}
  \label{eq:cy2 x r3:2d lg:action}
  \begin{aligned}
    & S_{\text{2d-LG}, \mathcal{M}(\R_t, \mathfrak{A}_4)}
    \\
    & = \int d\tau d\xi \bigg(
      \bigg|
      \left( D_{\xi} - k D_\tau \right) \left( C^{\mathring{m}} - k (\tilde{A}_t)^{\mathring{m}} \right)
      - j F_{\xi \tau}
      - k Q^{\mathring{m}}
      - k q^{\mathring{m}}
      \bigg|^2
    \\
    & \qquad \qquad \qquad
      + \left| ( D_{\xi} - k_{\theta} D_{\tau}) \mathcal{A}^{\mathring{m}}
      + R^{\mathring{m}}
      + r^{\mathring{m}}
      \right|^2
      + \dots
      \bigg)
    \\
    & = \int d\tau d\xi \Bigg(
      \Bigg|
      \left( D_{\xi} - k D_\tau \right) \left( C^{\mathring{m}} - k (\tilde{A}_t)^{\mathring{m}} \right)
      - j F_{\xi \tau}
      + g^{\mathring{m}\bar{\mathring{n}}}_{\mathcal{M}(\R_{t}, \mathfrak{A}_4)} \left( \frac{j \zeta}{2} \pdv{W_4}{C^{\mathring{n}}} \right)^*
      - k g^{\mathring{m}\bar{\mathring{n}}}_{\mathcal{M}(\R_{t}, \mathfrak{A}_4)} \left( \frac{j \zeta}{2} \pdv{W_4}{(\tilde{A}_t)^{\mathring{n}}} \right)^*
      \Bigg|^2
    \\
    & \qquad \qquad \qquad
      + \left| (D_{\xi} - k_{\theta} D_{\tau} )\mathcal{A}^{\mathring{m}}
      + g^{\mathring{m}\bar{\mathring{n}}}_{\mathcal{M}(\R_{t}, \mathfrak{A}_4)} \left( \frac{j \zeta}{2} \pdv{W_4}{\mathcal{A}^{\mathring{n}}} \right)^*
      \right|^2
      + \dots
      \Bigg)
    \\
    &= \int d\tau d\xi \left(
      \left| D_{\rho} \mathcal{A}^{\mathring{m}} \right|^2
      + \left| D_{\rho} C^{\mathring{m}} \right|^2
      + \left| D_{\rho} (\tilde{A}_t)^{\mathring{m}} \right|^2
      + \left| \pdv{W_4}{\mathcal{A}^{\mathring{m}}} \right|^2
      + \left| \pdv{W_4}{C^{\mathring{m}}} \right|^2
      + \left| \pdv{W_4}{(\tilde{A}_t)^{\mathring{m}}} \right|^2
      + \left| F_{\xi \tau} \right|^2
      + \dots
      \right)
      \, ,
  \end{aligned}
\end{equation}
where $g_{\mathcal{M}(\R_{t}, \mathfrak{A}_4)}$ is the metric on $\mathcal{M}(\R_t, \mathfrak{A}_4)$.
In other words, the 2d gauged sigma model with target space $\mathcal{M}(\R_t, \mathfrak{A}_4)$ can also be interpreted as a 2d gauged LG model with target space $\mathcal{M}(\R_t, \mathfrak{A}_4)$ and a holomorphic superpotential $W_4(\mathcal{A}, C, \tilde{A}_t)$.

By setting $d_\xi C^{\mathring{m}} = 0 = d_\xi \mathcal{A}^{\mathring{m}}$ and $A_\xi, A_\tau, (\tilde{A}_t)^{\mathring{m}} \rightarrow 0$ in the expression within the squared terms in~\eqref{eq:cy2 x r3:2d lg:action}, we can read off the LG $\mathcal{M}^{\theta}(\R_t, \mathfrak{A}_4)$-soliton equations corresponding to $\sigma^{IJ, KL}_{\pm}(\tau, \theta, \mathfrak{A}_4)$ (that re-expresses \eqref{eq:cy2 x r3:m-soliton eqns} with $A_{\xi}, A_{\tau}, (\tilde{A}_t)^{\mathring{m}} \rightarrow 0$) as
\begin{equation}
  \label{eq:cy2 x r3:2d lg:m-soliton:eqn}
  \begin{aligned}
    \dv{C^{\mathring{m}}}{\tau}
    &= - k g^{\mathring{m}\bar{\mathring{n}}}_{\mathcal{M}(\R_{t}, \mathfrak{A}_4)} \left( \frac{j \zeta}{2} \pdv{W_4}{C^{\mathring{n}}} \right)^*_{\tilde{A}_t = 0}
      - g^{\mathring{m}\bar{\mathring{n}}}_{\mathcal{M}(\R_{t}, \mathfrak{A}_4)} \left( \frac{j \zeta}{2} \pdv{W_4}{(\tilde{A}_t)^{\mathring{n}}} \right)^*_{\tilde{A}_t = 0}
    \, ,
    \\
    \dv{\mathcal{A}^{\mathring{m}}}{\tau}
    &= - k_{\theta} g^{\mathring{m}\bar{\mathring{n}}}_{\mathcal{M}(\R_{t}, \mathfrak{A}_4)} \left( \frac{j \zeta}{2} \pdv{W_4}{\mathcal{A}^{\mathring{n}}} \right)^*_{\tilde{A}_t = 0}
    \, .
  \end{aligned}
\end{equation}

By setting $d_{\tau} C^{\mathring{m}} = 0 = d_{\tau} \mathcal{A}^{\mathring{m}}$ in \eqref{eq:cy2 x r3:2d lg:m-soliton:eqn}, we get the LG $\mathcal{M}^{\theta}(\R_t, \mathfrak{A}_4)$-soliton endpoint equations corresponding to $\sigma^{IJ, KL}(\pm \infty, \theta, \mathfrak{A}_4)$ (that re-expresses \eqref{eq:cy2 x r3:m-soliton:endpts}) as
\begin{equation}
  \label{eq:cy2 x r3:2d lg:m-soliton:endpts}
  \begin{aligned}
    k g^{\mathring{m}\bar{\mathring{n}}}_{\mathcal{M}(\R_{t}, \mathfrak{A}_4)} \left( \frac{j \zeta}{2} \pdv{W_4}{C^{\mathring{n}}} \right)^*_{\tilde{A}_t = 0}
    + g^{\mathring{m}\bar{\mathring{n}}}_{\mathcal{M}(\R_{t}, \mathfrak{A}_4)} \left( \frac{j \zeta}{2} \pdv{W_4}{(\tilde{A}_t)^{\mathring{n}}} \right)^*_{\tilde{A}_t = 0}
    &= 0
    \, ,
    \\
    k_{\theta} g^{\mathring{m}\bar{\mathring{n}}}_{\mathcal{M}(\R_{t}, \mathfrak{A}_4)} \left( \frac{j \zeta}{2} \pdv{W_4}{\mathcal{A}^{\mathring{n}}} \right)^*_{\tilde{A}_t = 0}
    &= 0
    \, .
  \end{aligned}
\end{equation}

Recall from the end of \autoref{sec:cy2 x r3:vw} that we are only considering certain $CY_2$ such that the $\theta$-deformed VW configurations are isolated and non-degenerate.
Next, recall also that such configurations will correspond to the endpoints of the $\mathfrak{A}_4^{\theta}$-solitons; therefore, just like their endpoints, these $\mathfrak{A}_4^{\theta}$-solitons would be isolated and non-degenerate.
As these $\mathfrak{A}_4^{\theta}$-solitons will correspond, in the 2d gauged sigma model, to the endpoints of the $\mathcal{M}^{\theta}(\R_t, \mathfrak{A}_4)$-solitons, i.e., $\sigma^{IJ, KL}(\pm \infty, \theta, \mathfrak{A}_4)$, this means the latter would also be isolated and non-degenerate.
Thus, from their definition in~\eqref{eq:cy2 x r3:2d lg:m-soliton:endpts} which tells us that they are critical points of $W_4(\mathcal{A}, C, \tilde{A}_t)$, we conclude that $W_4(\mathcal{A}, C, \tilde{A}_t)$ can be regarded as a holomorphic Morse function in $\mathcal{M}(\R_t, \mathfrak{A}_4)$.

Just like in \autoref{sec:m2 x r3:fs-cat}, an LG $\mathcal{M}^{\theta}(\R_t, \mathfrak{A}_4)$-soliton defined in \eqref{eq:cy2 x r3:2d lg:m-soliton:eqn} maps to a straight line in the complex $W_4$-plane that starts and ends at the critical values $W_4^{IJ}(\theta) \equiv W_4(\sigma^{IJ}(-\infty, \theta, \mathfrak{A}_4))$ and $W_4^{KL}(\theta) \equiv W_4(\sigma^{KL}(+\infty, \theta, \mathfrak{A}_4))$, respectively, where its slope depends on $\theta$ (via $\zeta$).
We shall also assume that $\Re(W_4^{IJ}(\theta)) < \Re(W_4^{KL}(\theta))$ and that $I \neq J$, $J \neq K$, and $K \neq L$.

\subtitle{The 2d Gauged LG Model as an LG SQM}

With suitable rescalings, we can recast \eqref{eq:cy2 x r3:2d lg:action} as a 1d LG SQM (that re-expresses \eqref{eq:cy2 x r3:sqm action}), where its action will be given by\footnote{%
  In the following expression, we have integrated out $A_\xi$ and omitted the fields corresponding to the finite-energy gauge fields $A_{\{t, \tau\}}$ (as explained in~\autoref{ft:stokes theorem for m2 x r3:sqm}).
  \label{ft:stokes theorem for cy2 x r3:2d-lg sqm}
}
\begin{equation}
  \label{eq:cy2 x r3:2d lg:sqm:action}
  \begin{aligned}
    S_{\text{2d-LG SQM}, \mathcal{M}(\R_{\tau}, \mathcal{M}(\R_t, \mathfrak{A}_4))}
    = \int d\xi \Bigg(
    & \left| \left(
      \dv{C^{\mathring{u}}}{\xi}
      - k \dv{\breve{A}^{\mathring{u}}}{\xi}
      \right)
      + g_{\mathcal{M}(\R_{\tau}, \mathcal{M}(\R_{t}, \mathfrak{A}_4))}^{\mathring{u}\mathring{v}} \left(
      \pdv{H_4}{C^{\mathring{v}}}
      - k \pdv{H_4}{\breve{A}^{\mathring{v}}}
      \right)
      \right|^2
    \\
    & + \left| \dv{\mathcal{A}^{\mathring{u}}}{\xi}
      + g_{\mathcal{M}(\R_{\tau}, \mathcal{M}(\R_t, \mathfrak{A}_4))}^{\mathring{u}\mathring{v}} \pdv{H_4}{\mathcal{A}^{\mathring{v}}}
      \right|^2
      + \dots
      \Bigg)
      \, ,
  \end{aligned}
\end{equation}
where $H_4(\mathcal{A}, C, \breve{A})$ is the \emph{real-valued} potential in $\mathcal{M}(\R_{\tau}, \mathcal{M}(\R_t, \mathfrak{A}_4))$, and the subscript ``2d-LG SQM, $\mathcal{M}(\R_{\tau}, \mathcal{M}(\R_t, \mathfrak{A}_4))$'' is to specify that it is a 1d SQM in $\mathcal{M}(\R_{\tau}, \mathcal{M}(\R_t, \mathfrak{A}_4))$ obtained from the equivalent 2d LG model.
We will also refer to this \emph{1d} LG SQM as ``2d-LG SQM'' in the rest of this subsection.

The 2d-LG SQM will localize onto configurations that \emph{simultaneously} set to zero the LHS and RHS of the expression within the squared terms in~\eqref{eq:cy2 x r3:2d lg:sqm:action}.
In other words, it will localize onto $\xi$-invariant critical points of $H_4(\mathcal{A}, C, \breve{A})$ that will correspond, when $A_{\xi}, A_{\tau}, (\tilde{A}_t)^{\mathring{m}} \rightarrow 0$, to the LG $\mathcal{M}^{\theta}(\R_t, \mathfrak{A}_4)$-solitons defined by~\eqref{eq:cy2 x r3:2d lg:m-soliton:eqn}.
For our choice of $CY_2$, the LG $\mathcal{M}^{\theta}(\R_t, \mathfrak{A}_4)$-solitons, just like their endpoints, will be isolated and non-degenerate.
Thus, $H_4(\mathcal{A}, C, \breve{A})$ can be regarded as a real-valued Morse functional in $\mathcal{M}(\R_{\tau}, \mathcal{M}(\R_t, \mathfrak{A}_4))$.

\subtitle{Morphisms from $\gamma^{IJ}(\theta, \mathfrak{A}_4)$ to $\gamma^{KL}(\theta, \mathfrak{A}_4)$ as Floer Homology Classes of Intersecting Thimbles}

Repeating here the analysis in \autoref{sec:m2 x r3:fs-cat} with \eqref{eq:cy2 x r3:2d lg:sqm:action} as the action of the 2d-LG SQM, we find that we can interpret the LG $\mathcal{M}^{\theta}(\R_t, \mathfrak{A}_4)$-soliton solution $\sigma^{IJ, KL}_{\pm}(\tau, \theta, \mathfrak{A}_4)$ as intersections of thimbles.
Specifically, a $\sigma^{IJ, KL}_{\pm}(\tau, \theta, \mathfrak{A}_4)$-soliton pair, whose left and right endpoints correspond to $\gamma^{IJ}(\theta, \mathfrak{A}_4)$ and $\gamma^{KL}(\theta, \mathfrak{A}_4)$, respectively, can be identified as a pair of intersection points $p^{IJ, KL}_{\text{VW}, \pm}(\theta) \in S^{IJ}_{\text{VW}} \bigcap S^{KL}_{\text{VW}}$ of a left and right thimble in the fiber space over the line segment $[W^{IJ}_4(\theta), W^{KL}_4(\theta)]$.

This means that the 2d-LG SQM in $\mathcal{M}(\R_{\tau}, \mathcal{M}(\R_t, \mathfrak{A}_4))$ with action \eqref{eq:cy2 x r3:2d lg:sqm:action}, will physically realize a Floer homology that we shall name an $\mathcal{M}(\R_t, \mathfrak{A}_4)$-LG Floer homology.
The chains of the $\mathcal{M}(\R_t, \mathfrak{A}_4)$-LG Floer complex will be generated by LG $\mathcal{M}^{\theta}(\R_t, \mathfrak{A}_4)$-solitons which we can identify with $p^{**, **}_{\text{VW}, \pm}(\theta)$, and the $\mathcal{M}(\R_t, \mathfrak{A}_4)$-LG Floer differential will be realized by the flow lines governed by the gradient flow equations satisfied by the $\xi$-varying configurations which set the expression within the squared terms in \eqref{eq:cy2 x r3:2d lg:sqm:action} to zero.
The partition function of the 2d-LG SQM in $\mathcal{M}(\R_\tau, \mathcal{M}(\R_t, \mathfrak{A}_4))$ will then be given by\footnote{%
  The `$\theta$' label is omitted in the LHS of the following expression (as explained in \autoref{ft:theta omission in m2-2d lg partition fn}).
  \label{ft:theta omission in cy2 2d-lg partition fn}
}
\begin{equation}
  \label{eq:cy2 x r3:2d lg:partition function}
  \mathcal{Z}_{\text{2d-LG SQM}, \mathcal{M}(\R_\tau, \mathcal{M}(\R_t, \mathfrak{A}_4))}(G)
  = \sum^n_{I \neq J \neq K \neq L = 1}
  \,
  \sum_{%
    p^{IJ, KL}_{\text{VW}, \pm}
    \in S^{IJ}_{\text{VW}} \cap S^{KL}_{\text{VW}}
  }
  \text{HF}^G_{d_v} \left(
    p^{IJ, KL}_{\text{VW}, \pm}(\theta)
  \right)
  \, .
\end{equation}
Here, the contribution $\text{HF}^G_{d_v} (p^{IJ, KL}_{\text{VW}, \pm}(\theta))$ can be identified with a homology class in an $\mathcal{M}(\R_t, \mathfrak{A}_4)$-LG Floer homology generated by intersection points of thimbles.
These intersection points represent LG $\mathcal{M}^{\theta}(\R_t, \mathfrak{A}_4)$-solitons defined by \eqref{eq:cy2 x r3:2d lg:m-soliton:eqn}, whose endpoints correspond to $\theta$-deformed, non-constant paths in $\mathcal{M}(\R_t, \mathfrak{A}_4)$ defined by \eqref{eq:cy2 x r3:2d lg:m-soliton:endpts}.
The degree of each chain in the complex is $d_v$, and is counted by the number of outgoing flow lines from the fixed critical points of $H_4(\mathcal{A}, C, \breve{A})$ in $\mathcal{M}(\R_{\tau}, \mathcal{M}(\R_t, \mathfrak{A}_4))$ which can also be identified as $p^{IJ, KL}_{\text{VW}, \pm}(\theta)$.

Therefore, $\mathcal{Z}_{\text{2d-LG SQM}, \mathcal{M}(\R_\tau, \mathcal{M}(\R_t, \mathfrak{A}_4))}(G)$ in \eqref{eq:cy2 x r3:2d lg:partition function} is a sum of LG $\mathcal{M}^{\theta}(\R_t, \mathfrak{A}_4)$-solitons defined by \eqref{eq:cy2 x r3:2d lg:m-soliton:eqn} with endpoints \eqref{eq:cy2 x r3:2d lg:m-soliton:endpts}, or equivalently, $\sigma^{IJ, KL}_{\pm}(\tau, \theta, \mathfrak{A}_4)$-solitons defined by \eqref{eq:cy2 x r3:m-soliton eqns} (with $A_{\xi}, A_{\tau}, (\tilde{A}_t)^{\mathring{m}} \rightarrow 0$) with endpoints \eqref{eq:cy2 x r3:m-soliton:endpts}, whose start and end correspond to the non-constant paths $\gamma^{IJ}(\theta, \mathfrak{A}_4)$ and $\gamma^{KL}(\theta, \mathfrak{A}_4)$, respectively.
In other words, we can write
\begin{equation}
  \label{eq:cy2 x r3:2d lg:floer homology as vector}
  \text{CF}_{\mathcal{M}(\R_t, \mathfrak{A}_4)} \left(
    \gamma^{IJ}(\theta, \mathfrak{A}_4), \gamma^{KL}(\theta, \mathfrak{A}_4)
  \right)_{\pm}
  =
  \text{HF}^G_{d_v} \left(
    p^{IJ, KL}_{\text{VW}, \pm}(\theta)
  \right)
\end{equation}
where $\text{CF}_{\mathcal{M}(\R_t, \mathfrak{A}_4)} ( \gamma^{IJ}(\theta, \mathfrak{A}_4), \gamma^{KL}(\theta, \mathfrak{A}_4) )_{\pm}$ is a vector representing a $\sigma^{IJ, KL}_{\pm}(\tau, \theta, \mathfrak{A}_4)$-soliton, such that $\Re(W^{IJ}_4(\theta)) < \Re(W^{KL}_4(\theta))$.
This will lead us to the following one-to-one identification\footnote{%
  The `$\theta$' label is omitted in the following expression (as explained in \autoref{ft:omission of theta in m2 2d-lg}).
  \label{ft:omission of theta in cy2 2d-lg}
}
\begin{equation}
  \label{eq:cy2 x r3:2d lg:floer hom as morphism}
  \boxed{
    \text{Hom} \left(
      \gamma^{IJ}(\mathfrak{A}_4), \gamma^{KL}(\mathfrak{A}_4)
    \right)_{\pm}
    \Longleftrightarrow
    \text{HF}^G_{d_v} \left(
      p^{IJ, KL}_{\text{VW}, \pm}
    \right)
  }
\end{equation}
where the RHS is proportional to the identity class when $I = K$ and $J = L$, and zero when $I \leftrightarrow K$ and $J \leftrightarrow L$ (since the $\sigma^{IJ, KL}_{\pm}(\tau, \theta, \mathfrak{A}_4)$-soliton only moves in one direction from $\gamma^{IJ}(\theta, \mathfrak{A}_4)$ to $\gamma^{KL}(\theta, \mathfrak{A}_4)$).

\subtitle{Soliton String Theory from the 2d LG Model}

Just like in \autoref{sec:m2 x r3:fs-cat}, the 2d gauged LG model in $\mathcal{M}(\R_t, \mathfrak{A}_4)$ with action \eqref{eq:cy2 x r3:2d lg:action} can be interpreted as a soliton string theory in $\mathcal{M}(\R_t, \mathfrak{A}_4)$.
The dynamics of this soliton string theory in $\mathcal{M}(\R_t, \mathfrak{A}_4)$ will be governed by the BPS worldsheet equations of \eqref{eq:cy2 x r3:2d lg:worldsheet:eqn}, where $(\mathcal{A}^{\mathring{m}}, C^{\mathring{m}}, (\tilde{A}_t)^{\mathring{m}})$ are scalars on the worldsheet corresponding to the holomorphic coordinates of $\mathcal{M}(\R_t, \mathfrak{A}_4)$.
At an arbitrary instant in time whence $d_{\xi} \mathcal{A}^{\mathring{m}} = d_{\xi} C^{\mathring{m}} = 0 = d_{\xi} (\tilde{A}_t)^{\mathring{m}} = d_{\xi} A_\tau$ in~\eqref{eq:cy2 x r3:2d lg:worldsheet:eqn}, the dynamics of $(\mathcal{A}^{\mathring{m}}, C^{\mathring{m}}, (\tilde{A}_{\tau})^{\mathring{m}})$ and the 2d gauge fields $(A_{\tau}, A_{\xi})$ along $\tau$ will be governed by the soliton equations
\begin{equation}
  \label{eq:cy2 x r3:2d lg:string m-soliton}
  \begin{aligned}
    \dv{(\tilde{A}_t)^{\mathring{m}}}{\tau}
    + k \dv{C^{\mathring{m}}}{\tau}
    - j \dv{A_\xi}{\tau}
    =& [A_{\xi} - k A_{\tau}, C^{\mathring{m}} - k (\tilde{A}_t)^{\mathring{m}}]
      + j [A_\tau, A_\xi]
      \\
    & + g^{\mathring{m}\bar{\mathring{n}}}_{\mathcal{M}(\R_{t}, \mathfrak{A}_4)} \left( \frac{j \zeta}{2} \pdv{W_4}{C^{\mathring{n}}} \right)^*
      - k g^{\mathring{m}\bar{\mathring{n}}}_{\mathcal{M}(\R_{t}, \mathfrak{A}_4)} \left( \frac{j \zeta}{2} \pdv{W_4}{(\tilde{A}_t)^{\mathring{n}}} \right)^*
      \, ,
    \\
    k_{\theta} \dv{\mathcal{A}^{\mathring{m}}}{\tau}
    =& [A_{\xi} - k_{\theta} A_{\tau}, \mathcal{A}^{\mathring{m}}]
      + g^{\mathring{m}\bar{\mathring{n}}}_{\mathcal{M}(\R_{t}, \mathfrak{A}_4)} \left( \frac{j \zeta}{2} \pdv{W_4}{\mathcal{A}^{\mathring{n}}} \right)^*
      \, .
  \end{aligned}
\end{equation}

\subtitle{The Normalized 7d-Spin$(7)$ Partition Function, LG $\mathcal{M}^{\theta}(\R_t, \mathfrak{A}_4)$-soliton String Scattering, and Maps of an $A_{\infty}$-structure}

Since our 7d-Spin$(7)$ theory, which was derived from Spin$(7)$ theory, is also semi-classical, its normalized 7d partition function will also be a sum over tree-level scattering amplitudes of the LG $\mathcal{M}^{\theta}(\R_t, \mathfrak{A}_4)$-soliton strings defined by \eqref{eq:cy2 x r3:2d lg:m-soliton:eqn}.
The BPS worldsheet underlying such a tree-level scattering is similar to \autoref{fig:m2 x r3:mu-d maps}, where instead of the endpoints of each string being labeled $\gamma^{**}(\mathfrak{A}_2)$, they will now be labeled $\gamma^{**}(\mathfrak{A}_4)$.

In other words, we can, like in~\eqref{eq:m2 x r3:2d lg:normalized partition fn}, express the normalized 7d-Spin$(7)$ partition function as
\begin{equation}
  \label{eq:cy2 x r3:2d lg:normalized partition fn}
  \tilde{\mathcal{Z}}_{\text{7d-Spin}(7), CY_2 \times \R^3}(G) = \sum_{\mathfrak{N}_n} \mu^{\mathfrak{N}_n}_{\mathfrak{A}_4}
  \, ,
  \qquad
  \mathfrak{N}_n = 1, 2, \dots, \left\lfloor \frac{n - 2}{2} \right\rfloor
\end{equation}
where each
\begin{equation}
  \label{eq:cy2 x r3:2d lg:composition maps}
  \boxed{
    \mu^{\mathfrak{N}_n}_{\mathfrak{A}_4}: \bigotimes_{i = 1}^{\mathfrak{N}_n}
    \text{Hom} \left(
      \gamma^{I_{2i - 1} I_{2i}}(\mathfrak{A}_4), \gamma^{I_{2(i + 1) - 1} I_{2(i + 1)}}(\mathfrak{A}_4)
    \right)_-
    \longto
    \text{Hom} \left(
      \gamma^{I_1 I_2}(\mathfrak{A}_4), \gamma^{I_{2\mathfrak{N}_n + 1} I_{2\mathfrak{N}_n + 2}}(\mathfrak{A}_4)
    \right)_+
  }
\end{equation}
is a scattering amplitude of $\mathfrak{N}_n$ incoming LG $\mathcal{M}^{\theta}(\R_t, \mathfrak{A}_4)$-soliton strings $\text{Hom} (\gamma^{I_1 I_2}(\mathfrak{A}_4), \gamma^{I_3 I_4}(\mathfrak{A}_4) )_-$, $\dots$, $\text{Hom} (\gamma^{I_{2\mathfrak{N}_n - 1} I_{2\mathfrak{N}_n}}(\mathfrak{A}_4), \gamma^{I_{2\mathfrak{N}_n + 1} I_{2\mathfrak{N}_n + 2}}(\mathfrak{A}_4) )_-$, and a single outgoing LG $\mathcal{M}^{\theta}(\R_t, \mathfrak{A}_4)$-soliton string $\text{Hom} (\gamma^{I_1 I_2}(\mathfrak{A}_4), \gamma^{I_{2\mathfrak{N}_n + 1} I_{2\mathfrak{N}_n + 2}}(\mathfrak{A}_4) )_+$, with left and right boundaries as labeled, whose underlying worldsheet can be regarded as a disc with $\mathfrak{N}_n + 1$ vertex operators at the boundary.
In short, $\mu^{\mathfrak{N}_n}_{\mathfrak{A}_4}$ counts pseudoholomorphic discs with $\mathfrak{N}_n + 1$ punctures at the boundary that are mapped to $\mathcal{M}(\R_t, \mathfrak{A}_4)$ according to the BPS worldsheet equations \eqref{eq:cy2 x r3:2d lg:worldsheet:eqn}.

Just as in \autoref{sec:m2 x r3:fs-cat}, the collection of $\mu^{\mathfrak{N}_n}_{\mathfrak{A}_4}$ maps in \eqref{eq:cy2 x r3:2d lg:composition maps} can be regarded as composition maps defining an $A_{\infty}$-structure.

\subtitle{An FS type $A_{\infty}$-category of $\mathfrak{A}_4^{\theta}$-solitons}

Altogether, this means that the normalized partition function of 7d-Spin$(7)$ theory on $CY_2 \times \R^3$ as expressed in \eqref{eq:cy2 x r3:2d lg:normalized partition fn}, manifests a \emph{novel} FS type $A_{\infty}$-category defined by the $\mu^{\mathfrak{N}_n}_{\mathfrak{A}_4}$ maps \eqref{eq:cy2 x r3:2d lg:composition maps} and the one-to-one identification \eqref{eq:cy2 x r3:2d lg:floer hom as morphism}, where the $\mathfrak{N}_n + 1$ number of objects $\big\{\gamma^{I_1 I_2}(\mathfrak{A}_4), \gamma^{I_3 I_4}(\mathfrak{A}_4), \dots,$ $\gamma^{I_{2\mathfrak{N}_n + 1} I_{2\mathfrak{N}_n + 2}}(\mathfrak{A}_4) \big\}$ correspond to $\mathfrak{A}_4^{\theta}$-solitons with endpoints themselves corresponding to ($\theta$-deformed) VW configurations on $CY_2$!

\subsection{Soliton Membrane Theory, the 7d-\texorpdfstring{Spin$(7)$}{Spin(7)} Partition Function, and a Fueter type \texorpdfstring{$A_{\infty}$}{A-infinity}-2-category 2-categorifying the HW Floer Homology of \texorpdfstring{$CY_2$}{CY2}}
\label{sec:cy2 x r3:fueter-cat}

Note that we can also express \eqref{eq:cy2 x r3:worldvolume:eqn} (corresponding to \eqref{eq:cy2 x r3:worldsheet:eqn} in the 2d gauged sigma model) as
\begin{equation}
  \label{eq:cy2 x r3:3d-lg:worldvolume:eqn}
  \begin{aligned}
    i \Dv{C^{\mathring{a}}}{\xi} + j \Dv{C^{\mathring{a}}}{\tau} + k \Dv{C^{\mathring{a}}}{t}
    + i F_{t \tau} + j F_{\xi t} + k F_{\tau \xi}
    &= - j q^{\mathring{a}}
      \, ,
    \\
    i e^{-i\theta} \Dv{\mathcal{A}^{\mathring{a}}}{\xi}
    + j \Dv{\mathcal{A}^{\mathring{a}}}{\tau}
    + k \Dv{\mathcal{A}^{\mathring{a}}}{t}
    &=
      - i e^{-i\theta} r^{\mathring{a}}
      \, ,
  \end{aligned}
\end{equation}
which are non-constant, $\theta$-deformed, gauged Fueter equations for the $(\mathcal{A}^{\mathring{a}}, C^{\mathring{a}})$ fields (that will correspond to \eqref{eq:cy2 x r3:2d lg:worldsheet:eqn} in the 2d gauged LG model).
In turn, this means that we can express the action of the 3d gauged sigma model with target space $\mathfrak{A}_4$ in \eqref{eq:cy2 x r3:3d model action} as
\begin{equation}
  \label{eq:cy2 x r3:3d lg:action}
  \begin{aligned}
    S_{\text{3d-LG}, \mathfrak{A}_4} = \frac{1}{e^2} \int dt d\tau d\xi \,
    \bigg(
    & \left| i D_{\xi} C^{\mathring{a}} + j D_{\tau} C^{\mathring{a}} + k D_{t} C^{\mathring{a}}
      + i F_{t \tau} + j F_{\xi t} + k F_{\tau \xi}
      + j q^{\mathring{a}}
      \right|^2
    \\
    & + \left| i e^{-i\theta} D_{\xi} \mathcal{A}^{\mathring{a}} + j D_{\tau} \mathcal{A}^{\mathring{a}} + k D_t \mathcal{A}^{\mathring{a}}
      + i e^{-i\theta} r^{\mathring{a}}
      \right|^2
      + \dots
      \bigg)
    \\
    = \frac{1}{e^2} \int dt d\tau d\xi \,
    \bigg(
    & \left| i D_{\xi} C^{\mathring{a}} + j D_{\tau} C^{\mathring{a}} + k D_{t} C^{\mathring{a}}
      + i F_{t \tau} + j F_{\xi t} + k F_{\tau \xi}
      - g_{\mathfrak{A}_4}^{\mathring{a}\bar{\mathring{b}}} \left(
      \frac{j\zeta}{2} \pdv{\mathcal{W}_4}{C^{\mathring{b}}}
      \right)^{*}
      \right|^2
    \\
    & + \left| i e^{-i\theta} D_{\xi} \mathcal{A}^{\mathring{a}} + j D_{\tau} \mathcal{A}^{\mathring{a}} + k D_t \mathcal{A}^{\mathring{a}}
      - g_{\mathfrak{A}_4}^{\mathring{a}\bar{\mathring{b}}} \left(
      \frac{i\zeta}{2} \pdv{\mathcal{W}_4}{\mathcal{A}^{\mathring{b}}}
      \right)^{*}
      \right|^2
      + \dots
      \bigg)
    \\
    = \frac{1}{e^2} \int dt d\tau d\xi \,
    \bigg(
    & \left| D_\varrho \mathcal{A}^{\mathring{a}} \right|^2
      + \left| D_\varrho C^{\mathring{a}} \right|^2
      + \left| \pdv{\mathcal{W}_4}{\mathcal{A}^{\mathring{a}}} \right|^2
      + \left| \pdv{\mathcal{W}_4}{C^{\mathring{a}}} \right|^2
      + \left| F_{t \tau} \right|^2
      + \left| F_{\xi t} \right|^2
      + \left| F_{\tau \xi} \right|^2
      + \dots
      \bigg)
      \, ,
  \end{aligned}
\end{equation}
where $g_{\mathfrak{A}_4}$ is the metric on $\mathfrak{A}_4$.
In other words, our 3d gauged sigma model can also be interpreted as a 3d gauged LG model in $\mathfrak{A}_4$ with holomorphic superpotential $\mathcal{W}_4(\mathcal{A}, C)$.

By setting $d_{\xi} C^{\mathring{a}} = 0 = d_{\xi} \mathcal{A}^{\mathring{a}}$ and $A_\xi, A_\tau, A_t \rightarrow 0$ in the expression within the squared terms in~\eqref{eq:cy2 x r3:3d lg:action}, we can read off the LG $\mathfrak{A}_4^{\theta}$-sheet equations corresponding to $\Sigma^{IJ, KL}_\pm(\tau, t, \theta, \mathfrak{A}_4)$ (that re-expresses \eqref{eq:cy2 x r3:sheet eqns} with $A_\xi, A_\tau, A_t \rightarrow 0$) as
\begin{equation}
  \label{eq:cy2 x r3:3d lg:sheet:eqns}
  \begin{aligned}
    j \dv{C^{\mathring{a}}}{\tau} + k \dv{C^{\mathring{a}}}{t}
    &= g_{\mathfrak{A}_4}^{\mathring{a}\bar{\mathring{b}}} \left(
      \frac{j\zeta}{2} \pdv{\mathcal{W}_4}{C^{\mathring{a}}}
      \right)^{*}
      \, ,
    \\
    j \dv{\mathcal{A}^{\mathring{a}}}{\tau} + k \dv{\mathcal{A}^{\mathring{a}}}{t}
    &= g_{\mathfrak{A}_4}^{\mathring{a}\bar{\mathring{b}}} \left(
      \frac{i\zeta}{2} \pdv{\mathcal{W}_4}{\mathcal{A}^{\mathring{a}}}
      \right)^{*}
      \, .
  \end{aligned}
\end{equation}

By setting $d_{\tau} C^{\mathring{a}} = 0 = d_{\tau} \mathcal{A}^{\mathring{a}}$ in~\eqref{eq:cy2 x r3:3d lg:sheet:eqns}, we can read off the LG $\mathfrak{A}_4^{\theta}$-soliton equations corresponding to $\Gamma^{IJ}(t, \theta, \mathfrak{A}_4)$ and $\Gamma^{KL}(t, \theta, \mathfrak{A}_4)$, or equivalently, the LG $\mathfrak{A}_4^{\theta}$-sheet edge equations corresponding to $\Sigma^{IJ, KL}(\pm \infty, t, \theta, \mathfrak{A}_4)$, (that re-expresses~\eqref{eq:cy2 x r3:soliton}) as
\begin{equation}
  \label{eq:cy2 x r3:3d lg:soliton:eqns}
  \begin{aligned}
    k \dv{C^{\mathring{a}}}{t}
    &= g_{\mathfrak{A}_4}^{\mathring{a}\bar{\mathring{b}}} \left(
      \frac{j\zeta}{2} \pdv{\mathcal{W}_4}{C^{\mathring{a}}}
      \right)^{*}
      \, ,
    \\
    k \dv{\mathcal{A}^{\mathring{a}}}{t}
    &= g_{\mathfrak{A}_4}^{\mathring{a}\bar{\mathring{b}}} \left(
      \frac{i\zeta}{2} \pdv{\mathcal{W}_4}{\mathcal{A}^{\mathring{a}}}
      \right)^{*}
      \, .
  \end{aligned}
\end{equation}

By setting $d_t C^{\mathring{a}} = 0 = d_t \mathcal{A}^{\mathring{a}}$ in~\eqref{eq:cy2 x r3:3d lg:soliton:eqns}, we can read off the LG $\mathfrak{A}_4^{\theta}$-soliton endpoint equations corresponding to $\Gamma^{**}(\pm \infty, \theta, \mathfrak{A}_4)$, or equivalently, the LG $\mathfrak{A}_4^{\theta}$-sheet vertex equations corresponding to $\Sigma^{IJ, KL}(\pm \infty, \pm \infty, \theta, \mathfrak{A}_4)$ and $\Sigma^{IJ, KL}(\pm \infty, \mp \infty, \theta, \mathfrak{A}_4)$, (that re-expresses~\eqref{eq:cy2 x r3:soliton} with $d_t C^{\mathring{a}} = 0 = d_t \mathcal{A}^{\mathring{a}}$) as
\begin{equation}
  \label{eq:cy2 x r3:3d lg:sheet:endpts}
  g_{\mathfrak{A}_4}^{\mathring{a}\bar{\mathring{b}}} \left(
    \frac{j\zeta}{2} \pdv{\mathcal{W}_4}{C^{\mathring{a}}}
  \right)^{*}
  = 0
  \, ,
  \qquad
  g_{\mathfrak{A}_4}^{\mathring{a}\bar{\mathring{b}}} \left(
    \frac{i\zeta}{2} \pdv{\mathcal{W}_4}{\mathcal{A}^{\mathring{a}}}
  \right)^{*}
  = 0
  \, .
\end{equation}

Recall from the end of \autoref{sec:cy2 x r3:vw} that we are only considering certain $CY_2$ such that (the endpoints $\Gamma^{**}(\pm \infty, \theta, \mathfrak{A}_4)$ and thus) the LG $\mathfrak{A}_4^{\theta}$-solitons, and effectively, (the vertices $\Sigma^{IJ, KL}(\pm \infty, \pm \infty, \theta, \mathfrak{A}_4)$, $\Sigma^{IJ, KL}(\pm \infty, \mp \infty, \theta, \mathfrak{A}_4)$ and thus) the LG $\mathfrak{A}_4^{\theta}$-sheets, are isolated and non-degenerate.
Therefore, from their definition in~\eqref{eq:cy2 x r3:3d lg:sheet:endpts} which tells us that they correspond to critical points of $\mathcal{W}_4(\mathcal{A}, C)$, we conclude that $\mathcal{W}_4(\mathcal{A}, C)$ can be regarded as a holomorphic Morse function in $\mathfrak{A}_4$.

Just like in~\autoref{sec:m2 x r3:fueter-cat}, this means that an LG $\mathfrak{A}_4^{\theta}$-soliton $\Gamma^{IJ}(t, \theta, \mathfrak{A}_4)$ defined in \eqref{eq:cy2 x r3:3d lg:soliton:eqns} maps to a straight line segment $[\mathcal{W}_4^I(\theta), \mathcal{W}_4^J(\theta)]$ in the complex $\mathcal{W}_4$-plane that starts and ends at critical values $\mathcal{W}^I_4(\theta) \equiv \mathcal{W}_4(\Gamma^I(- \infty, \theta, \mathfrak{A}_4))$ and $\mathcal{W}^J_4(\theta) \equiv \mathcal{W}_4(\Gamma^J(+ \infty, \theta, \mathfrak{A}_4))$, respectively, where its slope depends on $\theta$ (via $\zeta$).
Therefore, an LG $\mathfrak{A}_4^{\theta}$-sheet defined in \eqref{eq:cy2 x r3:3d lg:sheet:eqns} maps to a quadrilateral in the complex $\mathcal{W}_4$-plane, whose edges are the straight line segments that the LG $\mathfrak{A}_4^{\theta}$-solitons map to, and whose bottom-left, top-left, bottom-right, and top-right vertices are the critical points $\mathcal{W}^I_4(\theta) \equiv \mathcal{W}_4(\Sigma^I(- \infty, - \infty, \theta, \mathfrak{A}_4))$, $\mathcal{W}^J_4(\theta) \equiv \mathcal{W}_4(\Sigma^J(- \infty, + \infty, \theta, \mathfrak{A}_4))$, $\mathcal{W}^K_4(\theta) \equiv \mathcal{W}_4(\Sigma^K(+ \infty, - \infty, \theta, \mathfrak{A}_4))$, and $\mathcal{W}^L_4(\theta) \equiv $ $\mathcal{W}_4(\Sigma^L(+ \infty, + \infty, \theta, \mathfrak{A}_4))$, respectively, where the slope of the straight line segments between each left-right vertex pair depends on $\theta$ (via $\zeta$).

We shall also assume that $\text{Re}(\mathcal{W}_4^I(\theta)) < \{ \text{Re}(\mathcal{W}_4^J(\theta)), \text{Re}(\mathcal{W}_4^K(\theta)) \} < \text{Re}(\mathcal{W}_4^L(\theta))$.

\subtitle{The 3d Gauged LG Model as an LG SQM}

Last but not least, after suitable rescalings, we can recast~\eqref{eq:cy2 x r3:3d lg:action} as a 1d LG SQM (that re-expresses~\eqref{eq:cy2 x r3:sqm action}), where its action will be given by\footnote{%
  In the following expression, we have integrated out $A_\xi$ and omitted the fields corresponding to the finite-energy gauge fields $A_{\{t, \tau\}}$ (as explained in~\autoref{ft:stokes theorem for m2 x r3:sqm}).
  \label{ft:stokes theorem for cy2 x r3:3d-lg sqm}
}
\begin{equation}
  \label{eq:cy2 x r3:3d lg:sqm action}
  \begin{aligned}
    S_{\text{3d-LG SQM}, \mathcal{M}(\R^2, \mathfrak{A}_4)}
    = \frac{1}{e^2} \int d\xi \Bigg(
    &\left| i \left(
      \dv{C^{\mathring{u}}}{\xi}
      - k \dv{\breve{A}^{\mathring{u}}}{\xi}
      \right)
      + g^{\mathring{u}\mathring{v}}_{\mathcal{M}(\R^2, \mathfrak{A}_4)} \left(
      \pdv{\mathfrak{F}_4}{C^{\mathring{v}}}
      - k \pdv{\mathfrak{F}_4}{\breve{A}^{\mathring{v}}}
      \right)
      \right|^2
    \\
    &
      + \left|
      i e^{-i\theta} \dv{\mathcal{A}^{\mathring{u}}}{\xi}
      + g^{\mathring{u}\mathring{v}}_{\mathcal{M}(\R^2, \mathfrak{A}_4)} \pdv{\mathfrak{F}_4}{\mathcal{A}^{\mathring{v}}}
      \right|^2
      + \dots
      \Bigg)
      \, .
  \end{aligned}
\end{equation}
Here, $\mathfrak{F}_4(\mathcal{A}, C, \breve{A})$ is the \emph{real-valued} potential in $\mathcal{M}(\R^2, \mathfrak{A}_4)$, and the subscript ``3d-LG SQM, $\mathcal{M}(\R^2, \mathfrak{A}_4)$'' is to specify that it is a 1d SQM with target space $\mathcal{M}(\R^2, \mathfrak{A}_4)$ obtained from the equivalent 3d LG model.
We will also refer to this \emph{1d} LG SQM as ``3d-LG SQM'' in the rest of this subsection.

The 3d-LG SQM will localize onto configurations that \emph{simultaneously} set to zero the LHS and RHS of the expression within the squared terms in~\eqref{eq:cy2 x r3:3d lg:sqm action}.
In other words, it will localize onto $\xi$-invariant critical points of $\mathfrak{F}_4(\mathcal{A}, C, \breve{A})$ that will correspond, when $A_{\xi}, A_{\tau}, A_t \rightarrow 0$, to the LG $\mathfrak{A}_4^{\theta}$-sheets defined by~\eqref{eq:cy2 x r3:3d lg:sheet:eqns}.
For our choice of $CY_2$, the LG $\mathfrak{A}_4^{\theta}$-sheets, just like their vertices, will be isolated and non-degenerate.
Thus, $\mathfrak{F}_4(\mathcal{A}, C, \breve{A})$ can be regarded as a \emph{real-valued} Morse functional in $\mathcal{M}(\R^2, \mathfrak{A}_4)$.

\subtitle{Morphisms between $\mathfrak{A}_4^{\theta}$-solitons as Intersection Floer Homology Classes}

Repeating here the analysis in \autoref{sec:m2 x r3:fueter-cat} with \eqref{eq:cy2 x r3:3d lg:sqm action} as the action of the 3d-LG SQM, we find that we can interpret the LG $\mathfrak{A}_4^{\theta}$-soliton solution $\Gamma^{IJ}(t, \theta, \mathfrak{A}_4)$ as a thimble-intersection, and the LG $\mathfrak{A}_4^{\theta}$-sheet solution $\Sigma^{IJ,KL}_{\pm}(\tau, t, \theta, \mathfrak{A}_4)$ as an intersection of thimble-intersections.

Specifically, a $\Gamma^{IJ}(t, \theta, \mathfrak{A}_4)$-soliton, whose bottom and top endpoints correspond to $\mathcal{E}^I_{\text{VW}}(\theta)$ and $\mathcal{E}^J_{\text{VW}}(\theta)$, respectively, can be identified as an intersection point $q^{IJ}_{\text{VW}, \pm}(\theta) \in S^{IJ}_{\text{VW}}(\theta)$ of a bottom and top thimble in the fiber space over the line segment $[\mathcal{W}_4^I(\theta), \mathcal{W}_4^J(\theta)]$.
As a result, a $\Sigma^{IJ,KL}_{\pm}(\tau, t, \theta, \mathfrak{A}_4)$-sheet pair, whose left and right edges correspond to $\Gamma^{IJ}(t, \theta, \mathfrak{A}_4)$ and $\Gamma^{KL}(t, \theta, \mathfrak{A}_4)$, respectively, can be identified as a pair of intersection points $\{q^{IJ}_{\text{VW}, \pm}(\theta), q^{KL}_{\text{VW}, \pm}(\theta) \} \eqqcolon \mathfrak{P}^{IJ, KL}_{\text{VW}, \pm}(\theta) \in S^{IJ}_{\text{VW}}(\theta) \bigcap S^{KL}_{\text{VW}}(\theta)$ of a left and right thimble-intersection in the fiber space over the quadrilateral with vertices $(\mathcal{W}_4^I(\theta), \mathcal{W}_4^J(\theta), \mathcal{W}_4^K(\theta), \mathcal{W}_4^L(\theta))$.

At any rate, the 3d-LG SQM in $\mathcal{M}(\R^2, \mathfrak{A}_4)$ with action \eqref{eq:cy2 x r3:3d lg:sqm action} will physically realize a Floer homology that we shall name an $\mathfrak{A}_4$-3d-LG Floer homology.
The chains of the $\mathfrak{A}_4$-3d-LG Floer complex are generated by LG $\mathfrak{A}_4^{\theta}$-sheets which we can thus identify with $\mathfrak{P}^{**, **}_{\text{VW}, \pm}(\theta)$, and the $\mathfrak{A}_4$-3d-LG Floer differential will be realized by the flow lines governed by the gradient flow equations satisfied by $\xi$-varying configurations which set the expression within the squared terms of \eqref{eq:cy2 x r3:3d lg:sqm action} to zero.
The partition function of the 3d-LG SQM in $\mathcal{M}(\R^2, \mathfrak{A}_4)$ will be given by\footnote{%
  The `$\theta$' label is omitted in the LHS of the following expression (as explained in \autoref{ft:theta omission in m2-2d lg partition fn}).
  \label{ft:theta omission in 3d lg partition fn:cy2 x r3}
}
\begin{equation}
  \label{eq:cy2 x r3:3d lg:sqm:partition fn}
  \mathcal{Z}_{\text{3d-LG SQM}, \mathcal{M}(\R^2, \mathfrak{A}_4)}(G)
  = \sum_{I \neq J \neq K \neq L = 1}^n \,
  \sum_{\substack{\mathfrak{P}^{IJ, KL}_{\text{VW}, \pm} \\ \in S^{IJ}_{\text{VW}} \cap S^{KL}_{\text{VW}}}}
  \text{HF}^G_{d_v} \left(
    \mathfrak{P}^{IJ, KL}_{\text{VW}, \pm}(\theta)
  \right)
  \, ,
\end{equation}
where the contribution $\text{HF}^G_{d_v}(\mathfrak{P}^{IJ, KL}_{\text{VW}, \pm}(\theta))$ can be identified with a homology class in an $\mathfrak{A}_4$-3d-LG Floer homology generated by intersection points of thimble-intersections.
These intersection points represent LG $\mathfrak{A}_4^{\theta}$-sheets defined by \eqref{eq:cy2 x r3:3d lg:sheet:eqns}, whose edges correspond to LG $\mathfrak{A}_4^{\theta}$-solitons defined by \eqref{eq:cy2 x r3:3d lg:soliton:eqns}, and whose vertices defined by \eqref{eq:cy2 x r3:3d lg:sheet:endpts} will correspond to $\theta$-deformed VW configurations on $CY_2$.
The degree of each chain in the complex is $d_v$, and is counted by the number of outgoing flow lines from the fixed critical points of $\mathfrak{F}_4(\mathcal{A}, C, \breve{A})$ in $\mathcal{M}(\R^2, \mathfrak{A}_4)$ which can also be identified as $\mathfrak{P}^{IJ, KL}_{\text{VW},\pm}(\theta)$.

Therefore, $\mathcal{Z}_{\text{3d-LG SQM}, \mathcal{M}(\R^2, \mathfrak{A}_4)}(G)$ in \eqref{eq:cy2 x r3:3d lg:sqm:partition fn} is a sum of LG $\mathfrak{A}_4^{\theta}$-sheets defined by (i) \eqref{eq:cy2 x r3:3d lg:sheet:eqns} with (ii) edges \eqref{eq:cy2 x r3:3d lg:soliton:eqns} and (iii) vertices \eqref{eq:cy2 x r3:3d lg:sheet:endpts}, or equivalently, $\Sigma_{\pm}^{IJ, KL}(\tau, t, \theta, \mathfrak{A}_4)$-sheets defined by (i) \eqref{eq:cy2 x r3:sheet eqns} and~\eqref{eq:cy2 x r3:sheet eqns:aux cond} (with $A_{\xi}, A_{\tau}, A_t \rightarrow 0$) with (ii) edges \eqref{eq:cy2 x r3:soliton} and~\eqref{eq:cy2 x r3:sheet eqns:aux cond}, and (iii) vertices \eqref{eq:cy2 x r3:soliton} and~\eqref{eq:cy2 x r3:sheet eqns:aux cond} (with $d_t C^{\mathring{a}} = 0 = d_t \mathcal{A}^{\mathring{a}}$), respectively.
In other words, we can write
\begin{equation}
  \label{eq:cy2 x r3:3d lg:floer-hom as vector}
  \text{CF}_{\mathcal{M}(\R^2, \mathfrak{A}_4)} \left(
    \Gamma^{IJ}(t, \theta, \mathfrak{A}_4),
    \Gamma^{KL}(t, \theta, \mathfrak{A}_4)
  \right)_\pm
  =
  \text{HF}^G_{d_v} \left(
    \mathfrak{P}^{IJ, KL}_{\text{VW}, \pm}(\theta)
  \right)
  \, ,
\end{equation}
where $\text{CF}_{\mathcal{M}(\R^2, \mathfrak{A}_4)} (\Gamma^{IJ}(t, \theta, \mathfrak{A}_4), \Gamma^{KL}(t, \theta, \mathfrak{A}_4) )_\pm$ is a vector representing a $\Sigma^{IJ, KL}_{\pm}(\tau, t, \theta, \mathfrak{A}_4)$-sheet, whose left and right edges correspond to $\Gamma^{IJ}(t, \theta, \mathfrak{A}_4)$ and $\Gamma^{KL}(t, \theta, \mathfrak{A}_4)$, respectively, and whose bottom-left, top-left, bottom-right, and top-right vertices correspond to $\mathcal{E}^I_{\text{VW}}(\theta)$, $\mathcal{E}^J_{\text{VW}}(\theta)$, $\mathcal{E}^K_{\text{VW}}(\theta)$, and $\mathcal{E}^L_{\text{VW}}(\theta)$, respectively, such that $\text{Re}(\mathcal{W}_4^I(\theta)) < \{ \text{Re}(\mathcal{W}_4^J(\theta)), \text{Re}(\mathcal{W}_4^K(\theta)) \} < \text{Re}(\mathcal{W}_4^L(\theta))$.
This will lead us to the following one-to-one identifications\footnote{%
  The `$\theta$' label is omitted in the following expression (as explained in \autoref{ft:omission of theta in m2 2d-lg}).
  \label{ft:omission of theta in m2 3d-lg:cy2 x r3}
}
\begin{equation}
  \label{eq:cy2 x r3:3d lg:2-morphism}
  \boxed{
    \text{Hom} \left(
      \Gamma^{IJ}(t, \mathfrak{A}_4),
      \Gamma^{KL}(t, \mathfrak{A}_4)
    \right)_\pm
    \Longleftrightarrow
    \text{Hom} \left(
      \text{Hom}(\mathcal{E}^I_{\text{VW}}, \mathcal{E}^J_{\text{VW}}),
      \text{Hom}(\mathcal{E}^K_{\text{VW}}, \mathcal{E}^L_{\text{VW}})
    \right)_\pm
    \Longleftrightarrow
    \text{HF}^G_{d_v} \left(
      \mathfrak{P}^{IJ, KL}_{\text{VW}, \pm}
    \right)
  }
\end{equation}
where the RHS is proportional to the identity class when $I = K$ and $J = L$, and zero when (i) $I \leftrightarrow K$ and $J \leftrightarrow L$ (since the $\Sigma^{IJ, KL}(\tau, t, \theta, \mathfrak{A}_4)$-sheet only moves in one direction from $\Gamma^{IJ}(t, \theta, \mathfrak{A}_4)$ to $\Gamma^{KL}(t, \theta, \mathfrak{A}_4)$), and (ii) $I \leftrightarrow J$ or $K \leftrightarrow L$ (since the $\Gamma^{**}(t, \theta, \mathfrak{A}_4)$-solitons only move in one direction from $\mathcal{E}^I_{\text{VW}}(\theta)$ to $\mathcal{E}^J_{\text{VW}}(\theta)$ or $\mathcal{E}^K_{\text{VW}}(\theta)$ to $\mathcal{E}^L_{\text{VW}}(\theta)$).

\subtitle{Soliton Membrane Theory from the 3d LG Model}

Just like in \autoref{sec:m2 x r3:fueter-cat}, the 3d gauged LG model in $\mathfrak{A}_4$ with action \eqref{eq:cy2 x r3:3d lg:action} can be interpreted as a soliton membrane theory in $\mathfrak{A}_4$.
The dynamics of this soliton membrane theory in $\mathfrak{A}_4$ will be governed by the BPS worldvolume equations of \eqref{eq:cy2 x r3:3d-lg:worldvolume:eqn}, where $(\mathcal{A}^{\mathring{a}}, C^{\mathring{a}})$ are scalars on the worldvolume corresponding to the holomorphic coordinates of $\mathfrak{A}_4$.
At an arbitrary instant in time whence $d_{\xi} \mathcal{A}^{\mathring{a}} = d_{\xi} C^{\mathring{a}} = 0 = d_{\xi} A_{\{t, \tau\}}$ in \eqref{eq:cy2 x r3:3d-lg:worldvolume:eqn}, the dynamics of $(\mathcal{A}^{\mathring{a}}, C^{\mathring{a}})$ and the 3d gauge fields $(A_t, A_{\tau}, A_{\xi})$ along $(\tau, t)$ will be governed by the membrane equations
\begin{equation}
  \label{eq:cy2 x r3:soliton sheet:eqn}
  \begin{aligned}
    j \dv{C^{\mathring{a}}}{\tau} + k \dv{C^{\mathring{a}}}{t}
    + i \dv{A_\tau}{t} - i \dv{A_t}{\tau}
    - j \dv{A_{\xi}}{t} + k \dv{A_{\xi}}{\tau}
    &= - [ i A_{\xi} + j A_{\tau} + k A_t, C^{\mathring{a}}]
      - i [A_t, A_\tau]
    \\
    & \qquad
      + [j A_t - k A_{\tau} , A_{\xi}]
      + g^{\mathring{a} \bar{\mathring{b}}}_{\mathfrak{A}_4} \left(
      \frac{j \zeta}{2} \pdv{\mathcal{W}_4}{C^{\mathring{b}}}
      \right)^{*}
      \, ,
    \\
    j \dv{\mathcal{A}^{\mathring{a}}}{\tau} + k \dv{\mathcal{A}^{\mathring{a}}}{t}
    &= - [ i e^{-i\theta} A_{\xi} + j A_{\tau} + k A_t, \mathcal{A}^{\mathring{a}}]
      + g^{\mathring{a} \bar{\mathring{b}}}_{\mathfrak{A}_4} \left(
      \frac{i \zeta}{2} \pdv{\mathcal{W}_4}{\mathcal{A}^{\mathring{a}}}
      \right)^{*}
      \, .
  \end{aligned}
\end{equation}

\subtitle{The Normalized 7d-Spin$(7)$ Partition Function, Soliton Membrane Scattering, and Maps of an $A_{\infty}$-structure}

The normalized 7d-Spin$(7)$ partition function can be regarded as a sum over tree-level scattering amplitudes of the LG $\mathfrak{A}_4^{\theta}$-sheets defined by \eqref{eq:cy2 x r3:3d lg:sheet:eqns}.
The BPS worldvolume underlying such a tree-level scattering amplitude is similar to \autoref{fig:m2 x r3:fueter composition maps}, where instead of the vertices of each sheet being labeled $\mathcal{E}^{*}_{\text{BF}}$, they will now be labeled $\mathcal{E}^{*}_{\text{VW}}$.

In other words, we can, like in \eqref{eq:m2 x r3:3d lg:partition fn}, express the normalized 7d-Spin$(7)$ partition function as
\begin{equation}
  \label{eq:cy2 x r3:3d lg:partition fn}
  \mathcal{\tilde{Z}}_{\text{7d-Spin}(7), CY_2 \times \R^3}(G)
  = \sum_{\mathfrak{N}_n} \varPi_{\mathfrak{A}_4}^{\mathfrak{N}_n}
  \, ,
  \qquad
  \mathfrak{N}_n = 1, 2, \dots, \left\lfloor \frac{n - 2}{2} \right\rfloor
\end{equation}
where each
\begin{equation}
  \label{eq:cy2 x r3:fueter composition maps}
  \boxed{
    \begin{aligned}
      \varPi^{\mathfrak{N}_n}_{\mathfrak{A}_4}: \bigotimes_{i = 1}^{\mathfrak{N}_n}
      & \text{Hom} \left(
        \text{Hom} \left( \mathcal{E}^{I_{2i - 1}}_{\text{VW}}, \mathcal{E}^{I_{2i}}_{\text{VW}} \right),
        \text{Hom} \left( \mathcal{E}^{I_{2(i + 1) - 1}}_{\text{VW}}, \mathcal{E}^{I_{2(i + 1)}}_{\text{VW}} \right)
        \right)_-
      \\
      &\longto
        \text{Hom} \left(
        \text{Hom} \left( \mathcal{E}^{I_1}_{\text{VW}}, \mathcal{E}^{I_2}_{\text{VW}} \right),
        \text{Hom} \left( \mathcal{E}^{I_{2 \mathfrak{N}_n + 1}}_{\text{VW}}, \mathcal{E}^{I_{2 \mathfrak{N}_n + 2}}_{\text{VW}} \right)
        \right)_+
    \end{aligned}
  }
\end{equation}
is a scattering amplitude of $\mathfrak{N}_n$ incoming LG $\mathfrak{A}_4^{\theta}$-soliton membranes $\text{Hom} \Big( \text{Hom} \big(\mathcal{E}^{I_1}_{\text{VW}}, \mathcal{E}^{I_2}_{\text{VW}} \big) , \text{Hom} \big(\mathcal{E}^{I_3}_{\text{VW}}, \mathcal{E}^{I_4}_{\text{VW}}\big) \Big)_-$, $\dots$, $\text{Hom} \Big( \text{Hom}\big(\mathcal{E}^{I_{2\mathfrak{N}_n - 1}}_{\text{VW}}, \mathcal{E}^{I_{2 \mathfrak{N}_n}}_{\text{VW}}\big) , \text{Hom}\big(\mathcal{E}^{I_{2\mathfrak{N}_n + 1}}_{\text{VW}}, \mathcal{E}^{I_{2\mathfrak{N}_n + 2}}_{\text{VW}}\big) \Big)_-$, and a single outgoing LG $\mathfrak{A}_4^{\theta}$-soliton membrane $\text{Hom} \Big( \text{Hom}(\mathcal{E}^{I_1}_{\text{VW}}, \mathcal{E}^{I_2}_{\text{VW}}) , \text{Hom}(\mathcal{E}^{I_{2\mathfrak{N}_n + 1}}_{\text{VW}}, \mathcal{E}^{I_{2\mathfrak{N}_n + 2}}_{\text{VW}}) \Big)_+$, with vertices as labeled.

Just as in \autoref{sec:m2 x r3:fueter-cat}, the collection of $\varPi^{\mathfrak{N}_n}_{\mathfrak{A}_4}$ maps in~\eqref{eq:cy2 x r3:fueter composition maps} which involve 2-morphisms, can also be regarded as composition maps defining an $A_{\infty}$-structure of a 2-category whose $n$ objects $\{\mathcal{E}^1_{\text{VW}}, \mathcal{E}^2_{\text{VW}}, \dots, \mathcal{E}^n_{\text{VW}}\}$ correspond to ($\theta$-deformed) VW configurations on $CY_2$.

\subtitle{A Fueter type $A_\infty$-2-category 2-categorifying the HW Floer Homology of $CY_2$}

As VW configurations on $CY_2$ are known to generate the HW Floer homology of $CY_2$ which is itself a 0-category, this 2-category is a 2-categorification of the said Floer homology.

Altogether, this means that the normalized partition function of 7d-Spin$(7)$ theory on $CY_2 \times \R^3$, as expressed in~\eqref{eq:cy2 x r3:3d lg:partition fn}, manifests a \emph{novel} Fueter type $A_\infty$-2-category, defined by the maps~\eqref{eq:cy2 x r3:fueter composition maps} and the identifications~\eqref{eq:cy2 x r3:3d lg:2-morphism}, which 2-categorifies the HW Floer homology of $CY_2$!

\subtitle{An Equivalence Between a Fueter type $A_{\infty}$-2-category and an FS type $A_{\infty}$-category}

Recall from \autoref{sec:cy2 x r3:fs-cat} that the normalized partition function of 7d-Spin$(7)$ theory on $CY_2 \times \R^3$ also manifests the FS type $A_{\infty}$-category of $\mathfrak{A}_4^{\theta}$-solitons.
This means that we have a \emph{novel} equivalence between the Fueter type $A_\infty$-2-category 2-categorifying the HW Floer homology of $CY_2$ and the FS type $A_\infty$-category of $\mathfrak{A}_4^\theta$-solitons!

\section{A Cauchy-Riemann-Fueter type \texorpdfstring{$A_{\infty}$}{A-infinity}-3-category of Four-Manifolds}
\label{sec:cy2 x r4}

In this section, we will study Spin$(7)$ theory on $\text{Spin}(7) = M_4 \times \R^4$, where $M_4 = CY_2$.
We will recast it as a 4d gauged LG model on $\R^4$, a 2d gauged LG model on $\R^2$, or a 1d LG SQM.
Following the approach in \autoref{sec:cy2 x s x r3}, we will, via the 8d Spin$(7)$ partition function and its equivalent 2d gauged LG model, be able to physically realize a novel FS type $A_{\infty}$-category of sheets whose vertices correspond to VW configurations on $CY_2$ that generate a HW Floer homology.
Similarly, via the 8d Spin$(7)$ partition function and its equivalent 4d gauged LG model, we will be able to also physically realize a novel Cauchy-Riemann-Fueter $A_{\infty}$-3-category that 3-categorifies the HW Floer homology of $CY_2$.

\subsection{\texorpdfstring{Spin$(7)$}{Spin(7)} Theory on \texorpdfstring{$CY_2 \times \R^4$}{CY2 x R4} as a 4d Model on \texorpdfstring{$\R^4$}{R4}, 2d Model on \texorpdfstring{$\R^2$}{R2}, or 1d SQM}
\label{sec:cy2 x r4:theory}

Let us now specialize our Spin$(7)$-manifold to one which is actually a product manifold containing an $\R^4$ submanifold within.
One such possibility is a Spin$(7)$-manifold of the form $\text{Spin}(7) = M_4 \times \R^4$, where $M_4$ is a closed and compact hyperkähler four-manifold, i.e., $M_4 = CY_2$ \cite{donaldson-1996-gauge, cao-2016-gauge-theor}.
We will consider this case, and study Spin$(7)$ theory on $\text{Spin}(7) = CY_2 \times \R^4$.

\subtitle{Spin$(7)$ Theory on $CY_2 \times \R^4$}

We will re-use the $(t, \tau, \xi, y)$ labels from \autoref{sec:cy2 x s x r3} for the $(x^0, x^1, x^2, x^3)$ coordinates of $\R^4$.
Exploiting the self-duality of the $F^+$'s in the Spin$(7)$ instanton equation, we can express the action of Spin$(7)$ theory in \eqref{eq:spin7 action} on $CY_2 \times \R^4$ as
\begin{equation}
  \label{eq:cy2 x r4:action:raw}
  S_{\text{Spin}(7), CY_2 \times \R^4}
  = \frac{1}{e^2} \int_{CY_2 \times \R^4} dt d\tau d\xi dy d^4x \, \Tr \left(
  |F_{y\alpha}^+|^2
  + |F_{yt}^+|^2
  + |F_{y\tau}^+|^2
  + |F_{y\xi}^+|^2
  + \dots
  \right)
  \, ,
\end{equation}
where $x^{\alpha}$ for $\alpha \in \{4, 5, 6, 7\}$ are the coordinates of $CY_2$.

The conditions the bosons that minimize the action \eqref{eq:cy2 x r4:action:raw}, i.e., the BPS equations of Spin$(7)$ theory on $CY_2 \times \R^4$, are easily identified by setting to zero the expression within the squared terms therein, i.e.,
\begin{equation}
  \label{eq:cy2 x r4:bps eqns:raw}
  \begin{aligned}
    F_{y \alpha} - I F_{\xi \alpha} - J F_{t \alpha} - K F_{\tau \alpha}
    &= 0
      \, ,
    \\
    F_{y \xi} + F_{\tau t} + \omega^{\alpha\beta}_{(I)} F_{\alpha\beta}
    &= 0
      \, ,
    \\
    F_{y t} + F_{\xi \tau} + \omega^{\alpha\beta}_{(J)} F_{\alpha\beta}
    &= 0
      \, ,
    \\
    F_{y \tau} + F_{t \xi} + \omega^{\alpha\beta}_{(K)} F_{\alpha\beta}
    &= 0
      \, ,
  \end{aligned}
\end{equation}
where $(I, J, K)$ are the three complex structures of the hyperkähler $CY_2$ whose actions on the cotangent bases of $CY_2$, i.e., $e_{\alpha}$, are defined as
\begin{equation}
  \label{eq:cy2 x r4:complex structure:action}
  \begin{aligned}
    I e_4
    &= - e_5
      \, ,
    &\qquad
      I e_6
    &= - e_7
      \, ,
    \\
    J e_4
    &= e_7
      \, ,
    &\qquad
      J e_5
    &= e_6
      \, ,
    \\
    K e_4
    &= e_6
      \, ,
    &\qquad
      K e_5
    &= - e_7
      \, ,
  \end{aligned}
\end{equation}
and $\omega_{(I/J/K)}$ is the Kähler two-form w.r.t. $(I/J/K)$.

Choosing the $I$-complex structure, i.e., using complex coordinates $z^\kappa$ defined as $z^1 = x^5 + i x^4$ and $z^2 = x^7 +  ix^6$ that are holomorphic w.r.t. $I$, we can express \eqref{eq:cy2 x r4:bps eqns:raw} as
\begin{equation}
  \label{eq:cy2 x r4:bps eqns}
  \begin{aligned}
    &(D_y - I D_{\xi} - J D_t - K D_{\tau}) \mathcal{A}^{(I)}_\kappa
      + I (F_{y \xi} + F_{\tau t})
      + J (F_{y t} + F_{\xi \tau})
      + K (F_{y \tau} + F_{t \xi})
    \\
    & = \partial^{(I)}_\kappa (A_y - I A_{\xi} - J A_t - K A_{\tau})
      - 2 I \omega^{\kappa\bar{\lambda}}_{(I)} \mathcal{F}^{(I)}_{\kappa\bar{\lambda}}
      + 4 K \varepsilon^{\kappa\lambda}_{(I)} \mathcal{F}^{(I)}_{\kappa\lambda}
      \, ,
  \end{aligned}
\end{equation}
where $\mathcal{A}^{(I)}_\kappa \in \Omega^{(1, 0)}(CY_2, \text{ad}(G)) \otimes \Omega^0(\R^4, \text{ad}(G))$ is a holomorphic gauge connection w.r.t. $I$;
$\mathcal{F}^{(I)}_{\kappa\lambda}$ are components of the $(2, 0)$-form field strength in $\mathcal{A}^{(I)}$;
$\mathcal{F}^{(I)}_{\kappa\bar{\lambda}}$ are components of the $(1, 1)$-form field strength in $\mathcal{A}^{(I)}$ and $\bar{\mathcal{A}}^{(I)}$ (its complex conjugate that is anti-holomorphic w.r.t $I$);
$\partial^{(I)}_\kappa$ is the holomorphic derivative w.r.t $I$;
and $\varepsilon^{\kappa\lambda}_{(I)}$ are components of the holomorphic symplectic form w.r.t. $I$.

Next, note that we are physically free to rotate the $(t, \tau)$ and $(\xi, y)$-planes of $\R^4$ by an angle $\theta$, whence \eqref{eq:cy2 x r4:bps eqns} becomes\footnote{%
  If the complex structure chosen was, instead, $J$ or $K$,~\eqref{eq:cy2 x r4:bps eqns} would still have the same expression (with the ``$(I)$'' label and the complex structures multiplying $\omega^{\kappa\bar{\lambda}}$ and $\varepsilon^{\kappa\lambda}$ replaced appropriately (up to a sign)).
  Since the form of the expression is the same no matter the choice of the complex structure, we will choose to work in the $I$-complex structure in the rest of this section and omit the ``$(I)$'' label in the expressions hereafter.
  \label{ft:omit I label in bps eqns}
}
\begin{equation}
  \label{eq:cy2 x r4:bps eqns:rotated}
  \begin{aligned}
    &(D_y - I D_{\xi} - J D_t - K D_{\tau}) \mathcal{A}_\kappa
      + I (F_{y \xi} + F_{\tau t})
      + J (F_{y t} + F_{\xi \tau})
      + K (F_{y \tau} + F_{t \xi})
    \\
    & = \partial_\kappa (A_y - I A_{\xi} - J A_t - K A_{\tau})
      - 2 I e^{-I\theta} \omega^{\kappa\bar{\lambda}} \mathcal{F}_{\kappa\bar{\lambda}}
      + 4 K_{\theta} \varepsilon^{\kappa\lambda} \mathcal{F}_{\kappa\lambda}
      \, ,
  \end{aligned}
\end{equation}
where $K_{\theta} \coloneqq e^{I\theta/2} K e^{-I\theta/2}$.
This allows us to write the action for Spin$(7)$ theory on $CY_2 \times \R^4$ as
\begin{equation}
  \label{eq:cy2 x r4:action}
  \begin{aligned}
    & S_{\text{Spin}(7), CY_2 \times \R^4}
    \\
    &= \frac{1}{e^2} \int_{\R^4} dt d\tau d\xi dy \int_{CY_2} |dz|^4 \, \Tr \Bigg(
    \bigg|
      ( D_y - I D_{\xi} - J D_t - K D_{\tau} ) \mathcal{A}_\kappa
      + I (F_{y \xi} + F_{\tau t})
      + J (F_{y t} + F_{\xi \tau})
    \\
    & \qquad \qquad \qquad \qquad \qquad \qquad \qquad \quad
      + K (F_{y \tau} + F_{t \xi})
      + r_{\kappa}
      + p
      + q
      \bigg|^2
      + \dots
      \Bigg)
      \, ,
  \end{aligned}
\end{equation}
where
\begin{equation}
  \label{eq:cy2 x r4:action:components}
  r_{\kappa}
  = - \partial_{\kappa} (A_y - I A_{\xi} - J A_t - K A_{\tau})
  \, ,
  \qquad
  p = 2 I e^{-I\theta} \omega^{\kappa\bar{\lambda}} \mathcal{F}_{\kappa\bar{\lambda}}
  \, ,
  \qquad
  q = - 4 K_{\theta} \varepsilon^{\kappa\lambda} \mathcal{F}_{\kappa\lambda}
  \, .
\end{equation}

\subtitle{Spin$(7)$ Theory as a 4d Model}

After suitable rescalings, we can recast \eqref{eq:cy2 x r4:action} as a 4d model on $\R^4$, where its action now reads\footnote{%
  Just like in \autoref{ft:stokes theorem for m2 x r3:3d model}, to arrive at the following expression, we have (i) employed Stokes' theorem and the fact that $CY_2$ has no boundary to omit terms with $\partial_{\kappa} A_{\{t, \tau, \xi, y\}}$ as they will vanish when integrated over $CY_2$, and (ii) integrated out the scalar field $\mathfrak{g}_4(p) = 2 I e^{-I\theta} \omega^{a\bar{b}} \mathcal{F}_{a\bar{b}}$, whose contribution to the action is $|\mathfrak{g}_4(p)|^2$.
  \label{ft:stokes theorem for cy2 x r4:4d model}
}
\begin{equation}
  \label{eq:cy2 x r4:4d model action}
  \begin{aligned}
    S_{\text{4d}, \mathfrak{A}_4}
    &= \frac{1}{e^2} \int_{\R^4} dt d\tau d\xi dy \Bigg(
    \bigg|
    ( D_y - I D_{\xi} - J D_t - K D_{\tau} ) \mathcal{A}^a
      + I (F_{y \xi} + F_{\tau t})
      + J (F_{y t} + F_{\xi \tau})
    \\
    & \qquad \qquad \qquad \qquad \quad \,
      + K (F_{y \tau} + F_{t \xi})
      + q^a
      \bigg|^2
      + \dots
      \Bigg)
    \\
    &= \frac{1}{e^2} \int_{\R_y \times \R_{\xi}} d\xi dy \int_{\R_\tau \times \R_t} dt d\tau \Bigg(
      \bigg|
      (D_y -  I D_{\xi}) \mathcal{A}^a + P^a
      + I F_{y\xi}
      + J (D_y A_t + D_{\xi} A_{\tau})
    \\
    & \qquad \qquad \qquad \qquad \qquad \qquad \qquad
      + K (D_y A_{\tau} - D_{\xi} A_t)
      + Q
      + q^a
      \Bigg)
      \, .
  \end{aligned}
\end{equation}
Here $\mathcal{A}^a$ and $a$ are coordinates and indices on the space $\mathfrak{A}_4$ or irreducible $\mathcal{A}_w$ fields on $CY_2$, and
\begin{equation}
  \label{eq:cy2 x r4:4d model action:components}
  P^a
  = - J D_t \mathcal{A}^a
  - K D_{\tau} \mathcal{A}^a
  \, ,
  \qquad
  Q
  = I F_{\tau t}
  - J (\partial_t A_y + \partial_{\tau} A_{\xi})
  - K (\partial_{\tau} A_y - \partial_t A_{\xi})
  \, ,
\end{equation}
with $q^a$ corresponding to $q$ in \eqref{eq:cy2 x r4:action:components}.

In other words, Spin$(7)$ theory on $CY_2 \times \R^4$ can be regarded as a 4d gauged sigma model along the $(t, \tau, \xi, y)$-directions with target space $\mathfrak{A}_4$ and action \eqref{eq:cy2 x r4:4d model action}.

\subtitle{Spin$(7)$ Theory as a 2d Model}

From \eqref{eq:cy2 x r4:4d model action}, one can see that we can, after suitable rescalings, also recast the 4d model action as an equivalent 2d model action\footnote{%
  Just like in \autoref{ft:stokes theorem for m2 x r3:2d model}, to arrive at the following expression, we have employed Stokes' theorem and the fact that the finite-energy gauge fields $A_{\{t, \tau, \xi, y\}}$ would vanish at $t, \tau \rightarrow \pm \infty$.
  \label{ft:stokes theorem for cy2 x r4:2d model}
}
\begin{equation}
  \label{eq:cy2 x r4:2d model action}
  \begin{aligned}
    S_{\text{2d}, \mathcal{M}(\R_t \times \R_{\tau}, \mathfrak{A}_4)}
    = \frac{1}{e^2} \int_{\R^2} d\xi dy \Bigg(
    \bigg|
    & (D_y - I D_{\xi}) \mathcal{A}^m + P^m
      + I F_{y \xi}
      + J \left( D_y (\tilde{A}_t)^m + D_{\xi} (\tilde{A}_{\tau})^m \right)
    \\
    & + K \left( D_y (\tilde{A}_{\tau})^m - D_{\xi} (\tilde{A}_t)^m \right)
      + q^m
      \bigg|^2
      + \dots
      \Bigg) \, .
  \end{aligned}
\end{equation}
Here, $(\mathcal{A}^m, (\tilde{A}_{\tau})^m, (\tilde{A}_t)^m)$ and $m$ are coordinates and indices on the double path space $\mathcal{M}(\R_t \times \R_{\tau}, \mathfrak{A}_4)$ of smooth double paths from $\R_t \times \R_{\tau}$ to $\mathfrak{A}_4$, with
\begin{equation}
  \label{eq:cy2 x r4:2d model action:components}
  P^m = - J (\tilde{D}_t \mathcal{A})^m - K (\tilde{D}_\tau \mathcal{A})^m
\end{equation}
corresponding to $P^a$ in \eqref{eq:cy2 x r4:4d model action:components}, $q^m$ corresponding to $q^a$, and $(\tilde{A}_t, \tilde{A}_{\tau}, \tilde{D}_t, \tilde{D}_{\tau})$ corresponding to $(A_t, A_{\tau}, D_t, D_{\tau})$, in the underlying 4d model.

In other words, Spin$(7)$ theory on $CY_2 \times \R^4$ can also be regarded as a 2d gauged sigma model along the $(\xi, y)$-directions with target space $\mathcal{M}(\R_t \times \R_{\tau}, \mathfrak{A}_4)$ and action \eqref{eq:cy2 x r4:2d model action}.

\subtitle{Spin$(7)$ Theory as a 1d SQM}

Singling out $y$ as the direction in ``time'', the equivalent SQM action can be obtained from \eqref{eq:cy2 x r4:2d model action} after suitable rescalings as\footnote{%
  In the resulting SQM, we have integrated out $A_y$ (for a similar reason explained in \autoref{ft:stokes theorem for m2 x r3:sqm}) and applied Stokes' theorem and the fact that the terms containing the fields corresponding to $A_{\{t, \tau, \xi\}}$ would vanish at $\xi \rightarrow \pm \infty$.
  \label{ft:stokes theorem for cy2 x r4:sqm}
}
\begin{equation}
  \label{eq:cy2 x r4:sqm action}
  S_{\text{SQM}, \mathcal{M}(\R_{\xi}, \mathcal{M}(\R_t \times \R_{\tau}, \mathfrak{A}_4))}
  = \frac{1}{e^2 } \int dy \left(
    \left|
      (\partial_y \mathcal{A}^u + \partial_y \widehat{A}^u)
      + g^{uv}_{\mathcal{M}(\R_{\xi}, \mathcal{M}(\R_t \times \R_{\tau}, \mathfrak{A}_4))} \left(
        \pdv{\mathcal{H}_4}{\mathcal{A}^u}
        + \pdv{\mathcal{H}_4}{\widehat{A}^u}
      \right)
    \right|^2
    + \dots
  \right)
  \, ,
\end{equation}
where $(\mathcal{A}^u, \widehat{A}^u)$ and $(u, v)$ are coordinates and indices on the path space $\mathcal{M}(\R_{\xi}, \mathcal{M}(\R_t \times \R_{\tau}, \mathfrak{A}_4))$ of smooth maps from $\R_{\xi}$ to $\mathcal{M}(\R_t \times \R_{\tau}, \mathfrak{A}_4)$ with $\widehat{A}^u \coloneqq - (I \widehat{A}_{\xi} + J \widehat{A}_t + K \widehat{A}_{\tau})^u$ in $\mathcal{M}(\R_{\xi}, \mathcal{M}(\R_{\tau} \times \R_{\tau}, \mathfrak{A}_4))$ corresponding to $-(I A_{\xi} + J (\tilde{A}_t)^m + K (\widetilde{A}_{\tau})^m)$ in the underlying 2d model, and to $- (I A_{\xi} + J A_t + K A_{\tau})$ in the underlying 4d model;
$g_{\mathcal{M}(\R_{\xi}, \mathcal{M}(\R_t \times \R_{\tau}, \mathfrak{A}_4))}$ is the metric of $\mathcal{M}(\R_{\xi}, \mathcal{M}(\R_t \times \R_{\tau}, \mathfrak{A}_4))$;
and $\mathcal{H}_4(\mathcal{A}, \widehat{A})$ is the SQM potential function.
Note that we can also interpret $\mathcal{M}(\R_{\xi}, \mathcal{M}(\R_t \times \R_{\tau}, \mathfrak{A}_4))$ as the \emph{triple} path space $\mathcal{M}(\R^3, \mathfrak{A}_4)$ of smooth maps from $\R^3$ to $\mathfrak{A}_4$.

In short, Spin$(7)$ theory on $CY_2 \times \R^4$ can also be regarded as a 1d SQM along the $y$-direction in $\mathcal{M}(\R^3, \mathfrak{A}_4)$ whose action is \eqref{eq:cy2 x r4:sqm action}.

\subsection{Non-constant Paths, Solitons, Threebranes, and the HW Floer Homology of \texorpdfstring{$CY_2$}{CY2}}
\label{sec:cy2 x r4:vw}

By a similar analysis to that which was done in \autoref{sec:m2 x r3:gc-bf}, we find that the equivalent 1d SQM of Spin$(7)$ theory on $CY_2 \times \R^4$ will localize onto \emph{$y$-invariant, $\theta$-deformed, non-constant} triple paths in $\mathcal{M}(\R_{\xi}, \mathcal{M}(\R_t \times \R_{\tau}, \mathfrak{A}_4))$ which will correspond, in the 2d gauged sigma model with target space $\mathcal{M}(\R_t \times \R_{\tau}, \mathfrak{A}_4)$, to \emph{$y$-invariant, $\theta$-deformed} solitons along the $\xi$-direction that we shall refer to as $\mathcal{M}^{\theta}(\R_t \times \R_{\tau}, \mathfrak{A}_4)$-solitons.

\subtitle{$\mathcal{M}^{\theta}(\R_t \times \R_{\tau}, \mathfrak{A}_4)$-solitons in the 2d Gauged Model}

Specifically, such $\mathcal{M}^{\theta}(\R_t \times \R_{\tau}, \mathfrak{A}_4)$-solitons are defined by
\begin{equation}
  \label{eq:cy2 x r4:m-soliton eqns}
  \begin{aligned}
    I \partial_{\xi} \left(
    \mathcal{A}^m + A_y + J (\tilde{A}_t)^m + K (\tilde{A}_{\tau})^m
    \right)
    =& - I [A_{\xi}, \mathcal{A}^m + A_y + J (\tilde{A}_t)^m + K (\tilde{A}_{\tau})^m]
    \\
     & + [A_y, \mathcal{A}^m + J (\tilde{A}_t)^m + K (\tilde{A}_{\tau})^m]
    \\
     & - J (\tilde{D}_t \mathcal{A})^m - K (\tilde{D}_\tau \mathcal{A})^m
       + q^m
       \, .
  \end{aligned}
\end{equation}

\subtitle{$\mathfrak{A}_4^{\theta}$-threebranes in the 4d Gauged Model}

In turn, they will correspond, in the 4d gauged sigma model with target space $\mathfrak{A}_4$, to $y$-invariant, $\theta$-deformed threebranes along the $(\xi, \tau, t)$-directions that are defined by
\begin{equation}
  \label{eq:cy2 x r4:threebrane eqn}
  \begin{aligned}
    & (I \partial_{\xi} + J \partial_t + K \partial_{\tau}) (\mathcal{A}^a + A_y)
      - I (\partial_{\tau} A_{t} - \partial_t A_{\tau})
      - J (\partial_{\xi} A_{\tau} - \partial_{\tau} A_{\xi})
      - K (\partial_t A_{\xi} - \partial_{\xi} A_t)
    \\
    &= - [I A_{\xi} + J A_t + K A_{\tau}, \mathcal{A}^a + A_y]
      + [A_y, \mathcal{A}^a]
      + I [A_{\tau}, A_t]
      + J [A_{\xi}, A_{\tau}]
      + K [A_t, A_{\xi}]
      + q^a
      \, ,
  \end{aligned}
\end{equation}
and the condition
\begin{equation}
  \label{eq:cy2 x r4:threebrane eqn:aux cond}
  \mathfrak{g}_4(p) = 0
  \, ,
\end{equation}
where $\mathfrak{g}_4(p)$ is the auxiliary scalar field defined in \autoref{ft:stokes theorem for cy2 x r4:4d model}.
We shall refer to such threebranes defined by \eqref{eq:cy2 x r4:threebrane eqn} and \eqref{eq:cy2 x r4:threebrane eqn:aux cond} as $\mathfrak{A}_4^{\theta}$-threebranes.

\subtitle{$y$-independent, $\theta$-deformed Spin$(7)$ Configurations on Spin$(7)$ Theory}

In turn, the 4d configurations defined by \eqref{eq:cy2 x r4:threebrane eqn} and \eqref{eq:cy2 x r4:threebrane eqn:aux cond} will correspond, in Spin$(7)$ theory, to $y$-independent, $\theta$-deformed Spin$(7)$ configurations on $CY_2 \times \R^4$ that are defined, via \eqref{eq:cy2 x r4:action:components}, by
\begin{equation}
  \label{eq:cy2 x r4:spin7 configs}
  \begin{aligned}
    & (I \partial_{\xi} + J \partial_t + K \partial_{\tau}) (\mathcal{A}_{\kappa} + A_y)
      - I (\partial_{\tau} A_{t} - \partial_t A_{\tau})
      - J (\partial_{\xi} A_{\tau} - \partial_{\tau} A_{\xi})
      - K (\partial_t A_{\xi} - \partial_{\xi} A_t)
    \\
    &= - \mathcal{D}_{\kappa} (A_y - I A_{\xi} - J A_t - K A_{\tau})
      - [I A_{\xi} + J A_t + K A_{\tau}, A_y]
      + I [A_{\tau}, A_t]
      + J [A_{\xi}, A_{\tau}]
    \\
    & \quad
      + K [A_t, A_{\xi}]
      - 4 K_{\theta} \varepsilon^{\kappa\lambda} \mathcal{F}_{\kappa\lambda}
      \, ,
  \end{aligned}
\end{equation}
and the condition
\begin{equation}
  \label{eq:cy2 x r4:spin7 configs:aux cond}
  2Ie^{-I\theta} \omega^{\kappa\bar{\lambda}} \mathcal{F}_{\kappa\bar{\lambda}}
  = 0
  \, .
\end{equation}

\subtitle{Spin$(7)$ Configurations, $\mathfrak{A}_4^{\theta}$-threebranes, $\mathcal{M}^{\theta}(\R_t \times \R_{\tau}, \mathfrak{A}_4)$-solitons, and Non-constant Triple Paths}

In short, these \emph{$y$-independent, $\theta$-deformed} Spin$(7)$ configurations on $CY_2 \times \R^4$ that are defined by \eqref{eq:cy2 x r4:spin7 configs} and \eqref{eq:cy2 x r4:spin7 configs:aux cond}, will correspond to the $\mathfrak{A}_4^{\theta}$-threebranes defined by \eqref{eq:cy2 x r4:threebrane eqn} and \eqref{eq:cy2 x r4:threebrane eqn:aux cond}, which, in turn, will correspond to the $\mathcal{M}^{\theta}(\R_t \times \R_{\tau}, \mathfrak{A}_4)$-solitons defined by \eqref{eq:cy2 x r4:m-soliton eqns}, which, in turn, will correspond to the $y$-invariant, $\theta$-deformed, non-constant triple paths in $\mathcal{M}(\R^3, \mathfrak{A}_4)$ defined by setting both the LHS and RHS of the expression within the squared term of \eqref{eq:cy2 x r4:sqm action} \emph{simultaneously} to zero.

\subtitle{$\mathcal{M}^{\theta}(\R_t \times \R_{\tau}, \mathfrak{A}_4)$-soliton Endpoints Corresponding to Non-constant Double Paths}

Consider now the fixed endpoints of the $\mathcal{M}^{\theta}(\R_t \times \R_{\tau}, \mathfrak{A}_4)$-solitons at $\xi = \pm \infty$, where we also expect the fields in the 2d gauged sigma model corresponding to the finite-energy 4d gauge fields $A_y, A_{\xi}, A_{\tau}, A_t$ to decay to zero.
They are given by \eqref{eq:cy2 x r4:m-soliton eqns} with $\partial_{\xi} \mathcal{A}^m = 0$ and $A_y, A_{\xi}, (\tilde{A}_{\tau})^m, (\tilde{A}_t)^m \rightarrow 0$, i.e.,
\begin{equation}
  \label{eq:cy2 x r4:m-soliton:endpts}
  J (\tilde{\partial}_t \mathcal{A})^m
  + K (\tilde{\partial}_{\tau} \mathcal{A})^m
  = q^m
  \, .
\end{equation}
These are $(y, \xi)$-invariant, $\theta$-deformed, non-constant \emph{double} paths in $\mathcal{M}(\R_t \times \R_{\tau}, \mathfrak{A}_4)$.

\subtitle{$\mathfrak{A}_4^{\theta}$-threebrane Faces Corresponding to $\mathfrak{A}_4^{\theta}$-sheets in the 4d Gauged Model}

In turn, \eqref{eq:cy2 x r4:m-soliton:endpts} will correspond, in the 4d gauged sigma model, to the fixed faces of the $\mathfrak{A}_4^{\theta}$-threebranes at $\xi = \pm \infty$, i.e., $(y, \xi)$-invariant, $\theta$-deformed sheets along the $(\tau, t)$-direction that are defined by
\begin{equation}
  \label{eq:cy2 x r4:sheet}
  J \partial_t \mathcal{A}^a + K \partial_{\tau} \mathcal{A}^a
  = q^a
  \, .
\end{equation}
Notice that these equations can also be obtained from \eqref{eq:cy2 x r4:threebrane eqn} with $\partial_{\xi} \mathcal{A}^a = 0$ and $A_y, A_{\xi}, A_{\tau}, A_t \rightarrow 0$.
Such sheets defined by \eqref{eq:cy2 x r4:sheet} and \eqref{eq:cy2 x r4:threebrane eqn:aux cond} are $\mathfrak{A}_4^{\theta}$-sheets.

\subtitle{$\mathfrak{A}_4^{\theta}$-threebrane or $\mathfrak{A}_4^{\theta}$-sheet Edges Corresponding to  $\mathfrak{A}_4^{\theta}$-solitons in the 4d Gauged Model}

Consider now the fixed edges of (i) the $\mathfrak{A}_4^{\theta}$-threebranes at $\xi, \tau = \pm \infty$, or equivalently (ii) the $\mathfrak{A}_4^{\theta}$-sheets at $\tau = \pm \infty$.
These are $(y, \xi, \tau)$-invariant, $\theta$-deformed solitons along the $t$-direction and are given by (i) \eqref{eq:cy2 x r4:threebrane eqn} with $\partial_{\{\tau, \xi\}} \mathcal{A}^a = 0$ and $A_y, A_{\xi}, A_{\tau}, A_t \rightarrow 0$, or equivalently (ii) \eqref{eq:cy2 x r4:sheet} with $\partial_{\tau} \mathcal{A}^a = 0$, i.e.,
\begin{equation}
  \label{eq:cy2 x r4:soliton}
  J \partial_t \mathcal{A}^a = q^a
  \, .
\end{equation}
Such solitons defined by \eqref{eq:cy2 x r4:soliton} and \eqref{eq:cy2 x r4:threebrane eqn:aux cond} are $\mathfrak{A}_4^{\theta}$-solitons.

\subtitle{$\mathfrak{A}_4^{\theta}$-soliton Endpoints, and $\mathfrak{A}_4^{\theta}$-sheet or $\mathfrak{A}_4^{\theta}$-threebrane Vertices Corresponding to $\theta$-deformed VW Configurations on $CY_2$}

Consider now (i) the fixed endpoints of the $\mathfrak{A}_4^{\theta}$-solitons at $t = \pm \infty$, (ii) the vertices of the $\mathfrak{A}_4^{\theta}$-sheets at $\tau, t = \pm \infty$, or equivalently (iii) the vertices of the $\mathfrak{A}_4^{\theta}$-threebranes at $\xi, \tau, t = \pm \infty$.
They are given by (i) \eqref{eq:cy2 x r4:soliton} and \eqref{eq:cy2 x r4:threebrane eqn:aux cond} with $\partial_t \mathcal{A}^a = 0$, (ii) \eqref{eq:cy2 x r4:sheet} and~\eqref{eq:cy2 x r4:threebrane eqn:aux cond} with $\partial_{\{t, \tau\}} \mathcal{A}^a = 0$, or equivalently (iii) \eqref{eq:cy2 x r4:threebrane eqn} and~\eqref{eq:cy2 x r4:threebrane eqn:aux cond} with $\partial_{\{t, \tau, \xi\}} \mathcal{A}^a = 0$ and $A_y, A_{\xi}, A_{\tau}, A_t \rightarrow 0$.
In turn, they will correspond, in Spin$(7)$ theory, to $(y, \xi, \tau, t)$-independent, $\theta$-deformed configurations that obey \eqref{eq:cy2 x r4:spin7 configs} and~\eqref{eq:cy2 x r4:spin7 configs:aux cond} with $\partial_{\{t, \tau, \xi\}} \mathcal{A}^a = 0$ and $A_y, A_{\xi}, A_{\tau}, A_t \rightarrow 0$, i.e.,
\begin{equation}
  \label{eq:cy2 x r4:soliton:endpts:spin7}
  K_{\theta} \varepsilon^{\kappa\lambda} \mathcal{F}_{\kappa\lambda}
  = 0
  \, ,
  \qquad
  I e^{-I\theta} \omega^{\kappa\bar{\lambda}} \mathcal{F}_{\kappa\bar{\lambda}}
  = 0
  \, .
\end{equation}

At $\theta = 0, \pi$, \eqref{eq:cy2 x r4:soliton:endpts:spin7} can be written, in the real coordinates of $CY_2$, as
\begin{equation}
  \label{eq:cy2 x r4:vw eqns}
  F_{\alpha\beta}^+ = 0
  \, .
\end{equation}
This is the ASD instanton equation on $CY_2$, which can be interpreted as VW equations on $CY_2$ with the self-dual two-form field being zero (i.e., \eqref{eq:cy2 x r3:vw eqns} with $B = 0$).
We shall, in the rest of this section, refer to configurations that span the space of solutions to this equation as VW configurations on $CY_2$.
Recall also, from \autoref{sec:cy2 x r3:vw}, that VW configurations on $CY_2$ are known to generate the HW Floer homology of $CY_2$.

In other words, the $(y, \xi, \tau, t)$-independent, $\theta$-deformed Spin$(7)$ configurations corresponding to the endpoints of the $\mathfrak{A}_4^{\theta}$-solitons (or equivalently, the vertices of the $\mathfrak{A}_4^{\theta}$-sheets and $\mathfrak{A}_4^{\theta}$-threebranes) are $\theta$-deformed VW configurations on $CY_2$.
We will also assume choices of $CY_2$ satisfying \autoref{ft:isolation of vw} whereby such configurations are isolated and non-degenerate.

\subtitle{Non-constant Triple Paths, $\mathcal{M}^{\theta}(\R_t \times \R_{\tau}, \mathfrak{A}_4)$-solitons, $\mathfrak{A}_4^{\theta}$-solitons, $\mathfrak{A}_4^{\theta}$-sheets, and $\mathfrak{A}_4^{\theta}$-threebranes.}

In short, from the equivalent 1d SQM of Spin$(7)$ theory on $CY_2 \times \R^4$, the theory localizes onto $y$-invariant, $\theta$-deformed, non-constant triple paths in $\mathcal{M}(\R^3, \mathfrak{A}_4)$, which, in turn, will correspond to $\mathcal{M}^{\theta}(\R_t \times \R_{\tau}, \mathfrak{A}_4)$-solitons in the 2d gauged sigma model whose endpoints are $(y, \xi)$-invariant, $\theta$-deformed, non-constant double paths in $\mathcal{M}(\R_t \times \R_{\tau}, \mathfrak{A}_4)$.
In the 4d gauged sigma model, these $\mathcal{M}^{\theta}(\R_t \times \R_{\tau}, \mathfrak{A}_4)$-solitons will correspond to $\mathfrak{A}_4^{\theta}$-threebranes, whose faces are $\mathfrak{A}_4^{\theta}$-sheets, whose edges are $\mathfrak{A}_4^{\theta}$-solitons, and whose vertices will correspond to $\theta$-deformed VW configurations on $CY_2$ that generate the HW Floer homology of $CY_2$.

\subsection{The 2d Model and Open Strings, the 4d Model and Open Threebranes}
\label{sec:cy2 x r4:2d-4d model}

By following a similar analysis to that which was done in \autoref{sec:m2 x r3:2d-3d model}, we find that the 2d gauged sigma model with target space $\mathcal{M}(\R_t \times \R_{\tau}, \mathfrak{A}_4)$ whose action is \eqref{eq:cy2 x r4:2d model action}, will define an open string theory in $\mathcal{M}(\R_t \times \R_{\tau}, \mathfrak{A}_4)$.
Similarly, we find that the 4d gauged sigma model with target space $\mathfrak{A}_4$ whose action is \eqref{eq:cy2 x r4:4d model action} will define an open \emph{threebrane} theory in $\mathfrak{A}_4$.
We will now work out the details pertaining to the BPS worldsheets and BPS worldvolumes (including their boundaries) that are necessary to define this open string theory and open threebrane theory, respectively.

\subtitle{BPS Worldsheets of the 2d Model}

The BPS worldsheets of the 2d gauged sigma model with target space $\mathcal{M}(\R_t \times \R_{\tau}, \mathfrak{A}_4)$ correspond to its classical trajectories.
Specifically, these are defined by setting to zero the expression within the squared term in \eqref{eq:cy2 x r4:2d model action}, i.e.,
\begin{equation}
  \label{eq:cy2 x r4:worldsheet eqn}
  \Dv{\mathcal{A}^m}{y} - I \Dv{\mathcal{A}^m}{\xi}
  + I F_{y \xi}
  + J \left(
    \Dv{(\tilde{A}_t)^m}{y} + \Dv{(\tilde{A}_{\tau})^m}{\xi}
  \right)
  + K \left(
    \Dv{(\tilde{A}_{\tau})^m}{y} - \Dv{(\tilde{A}_t)^m}{\xi}
  \right)
  = - P^m - q^m
  \, .
\end{equation}

\subtitle{BPS Worldsheets with Boundaries Labeled by Non-constant Double Paths in $\mathcal{M}(\R_t \times \R_{\tau}, \mathfrak{A}_4)$}

The boundaries of the BPS worldsheets are traced out by the endpoints of the $\mathcal{M}^{\theta}(\R_t \times \R_{\tau}, \mathfrak{A}_4)$-solitons as they propagate in $y$.
As we have seen at the end of \autoref{sec:cy2 x r4:vw}, these endpoints correspond to $(y, \xi)$-invariant, $\theta$-deformed, non-constant double paths in $\mathcal{M}(\R_t \times \R_{\tau}, \mathfrak{A}_4)$ that we shall, at $y = \pm \infty$, denote as $\varsigma_{\pm}(\theta, \mathfrak{A}_4)$.
In turn, they will correspond, in the 4d gauged sigma model with target space $\mathfrak{A_4}$, to $\mathfrak{A}_4^{\theta}$-sheets that we shall, at $y = \pm \infty$, denote as $\Sigma_{\pm}(\tau, t, \theta, \mathfrak{A}_4)$.
The fixed edges of the $\Sigma_{\pm}(\tau, t, \theta, \mathfrak{A}_4)$-sheets at $\tau = \pm \infty$ are $\mathfrak{A}_4^{\theta}$-solitons that we shall, at $y = \pm \infty$, denote as $\Gamma_{\pm}(t, \theta, \mathfrak{A}_4)$, whose endpoints will correspond to VW configurations on $CY_2$.

Suppose there are $n \geq 8$ such configurations $\{ \mathcal{E}^1_{\text{VW}}(\theta), \mathcal{E}^2_{\text{VW}}(\theta), \dots, \mathcal{E}^n_{\text{VW}}(\theta) \}$.
Just as in \autoref{sec:cy2 x r3:2d-3d model}, we can further specify any $\Gamma_{\pm}(t, \theta, \mathfrak{A}_4)$ as $\Gamma_{\pm}^{IJ}(t, \theta, \mathfrak{A}_4)$, where its endpoints would correspond to $\mathcal{E}^I_{\text{VW}}(\theta)$ and $\mathcal{E}^J_{\text{VW}}(\theta)$.
We can also further specify any $\Sigma_{\pm}(\tau, t, \theta, \mathfrak{A}_4)$ as $\Sigma^{IJ, KL}_{\pm}(\tau, t, \theta, \mathfrak{A}_4)$, where (i) its left and right edges correspond to the $\mathfrak{A}_4^{\theta}$-solitons $\Gamma^{IJ}(t, \theta, \mathfrak{A}_4)$ and $\Gamma^{KL}(t, \theta, \mathfrak{A}_4)$, and (ii) its four vertices would correspond to $\mathcal{E}^I_{\text{VW}}(\theta)$, $\mathcal{E}^J_{\text{VW}}(\theta)$, $\mathcal{E}^K_{\text{VW}}(\theta)$, and $\mathcal{E}^L_{\text{VW}}(\theta)$, similar to the kind shown in \autoref{fig:m2 x r3:frakA-sheet}.
However, instead of the face being labeled $\Sigma^{IJ, KL}(t, \tau, \theta, \mathfrak{A}_2)$, they will be labeled $\Sigma^{IJ, KL}(\tau, t, \theta, \mathfrak{A}_4)$.
Instead of the edges being labeled $\Gamma^{**}(\tau, \theta, \mathfrak{A}_2)$, they will be labeled $\Gamma^{**}(t, \theta, \mathfrak{A}_4)$.
And, instead of the vertices being labeled $\mathcal{E}^*_{\text{BF}}(\theta)$, they will be labeled $\mathcal{E}^*_{\text{VW}}(\theta)$.

Consequently, in the 2d gauged sigma model, we can further specify any $\varsigma_{\pm}(\theta, \mathfrak{A}_4)$ as $\varsigma^{IJ, KL}_{\pm}(\theta, \mathfrak{A}_4)$, where the latter will correspond to a $\Sigma^{IJ, KL}_{\pm}(\tau, t, \theta, \mathfrak{A}_4)$ $\mathfrak{A}_4^{\theta}$-sheet in the equivalent 4d gauged sigma model.

Since the endpoints of an $\mathcal{M}^{\theta}(\R_t \times \R_{\tau}, \mathfrak{A}_4)$-soliton are now denoted as $\varsigma^{**, **}_{\pm}(\theta, \mathfrak{A}_4)$, we can also denote and specify the former at $y = \pm \infty$ as $\Xi^{\{IJ, KL\}, \{MN, PQ\}}_{\pm}(\xi, \theta, \mathfrak{A}_4)$,\footnote{%
  Just like in \autoref{ft:fixing m-A2-soliton centre of mass dof}, the $y$-invariant $\mathcal{M}^{\theta}(\R_t \times \R_{\tau}, \mathfrak{A}_4)$-solitons can be fixed at $y = \pm \infty$ by adding physically-inconsequential $\mathcal{Q}$-exact terms to the SQM action.
  \label{ft:fixing m-A4-sheet centre of mass dof}
}~where its left and right endpoints would be $\varsigma^{IJ, KL}_{\pm}(\theta, \mathfrak{A}_4)$ and $\varsigma^{MN, PQ}_{\pm}(\theta, \mathfrak{A}_4)$, respectively.

As the $\varsigma^{**, **}(\theta, \mathfrak{A}_4)$'s are $y$-invariant and therefore, have the same value for all $y$, we have BPS worldsheets of the kind similar to~\autoref{fig:m2 x r3:bps worldsheet}. This time, however, instead of $\xi$ and $\tau$ being the temporal and spatial directions, we will have $y$ and $\xi$, respectively.
Instead of the boundaries being labeled $\gamma^{**}(\theta, \mathfrak{A}_2)$, they will be labeled $\varsigma^{**, **}(\theta, \mathfrak{A}_4)$.
And, at $y = \pm \infty$, instead of the $\mathcal{M}^{\theta}(\R_{\tau}, \mathfrak{A}_2)$-solitons $\sigma^{IJ, KL}_{\pm}(t, \theta, \mathfrak{A}_2)$, we will have $\mathcal{M}^{\theta}(\R_t \times \R_{\tau}, \mathfrak{A}_4)$-solitons $\Xi^{\{IJ, KL\}, \{MN, PQ\}}_{\pm}(\xi, \theta, \mathfrak{A}_4)$.

\subtitle{The 2d Model on $\R^2$ and an Open String Theory in $\mathcal{M}(\R_t \times \R_{\tau}, \mathfrak{A}_4)$}

Thus, like in \autoref{sec:m2 x r3:2d-3d model}, one can understand the 2d gauged sigma model with target space $\mathcal{M}(\R_t \times \R_{\tau}, \mathfrak{A}_4)$ to define an open string theory in $\mathcal{M}(\R_t \times \R_{\tau}, \mathfrak{A}_4)$, whose \emph{effective} worldsheet and boundaries are similar to \autoref{fig:m2 x r3:bps worldsheet}, where the coordinates are as described above.

\subtitle{BPS Worldvolumes of the 4d Model}

The BPS worldvolumes of the 4d gauged sigma model with target space $\mathfrak{A}_4$ correspond to its classical trajectories.
Specifically, these are defined by setting to zero the expression within the squared term in \eqref{eq:cy2 x r4:4d model action}, i.e.,
\begin{equation}
  \label{eq:cy2 x r4:worldvolume eqn}
  \Dv{\mathcal{A}^a}{y} - I \Dv{\mathcal{A}^a}{\xi} - J \Dv{\mathcal{A}^a}{t} - K \Dv{\mathcal{A}^a}{\tau}
  + I (F_{y \xi} + F_{\tau t})
  + J (F_{y t} + F_{\xi \tau})
  + K (F_{y \tau} + F_{t \xi})
  = - q^a
  \, .
\end{equation}

\subtitle{BPS Worldvolumes with Boundaries Labeled by $\mathfrak{A}_4^{\theta}$-sheets and $\mathfrak{A}_4^{\theta}$-solitons, and Edges Labeled by VW Configurations on $CY_2$}

The (i) boundaries and (ii) edges of the \emph{4d} BPS worldvolumes are traced out by the (i) (open) faces and edges, and (ii) vertices of the $\mathfrak{A}_4^{\theta}$-threebranes as they propagate in $y$.
As we have seen at the end of \autoref{sec:cy2 x r4:vw}, these (open) faces, edges, and vertices would correspond to $\mathfrak{A}_4^{\theta}$-sheets, $\mathfrak{A}_4^{\theta}$-solitons, and VW configurations on $CY_2$, respectively.

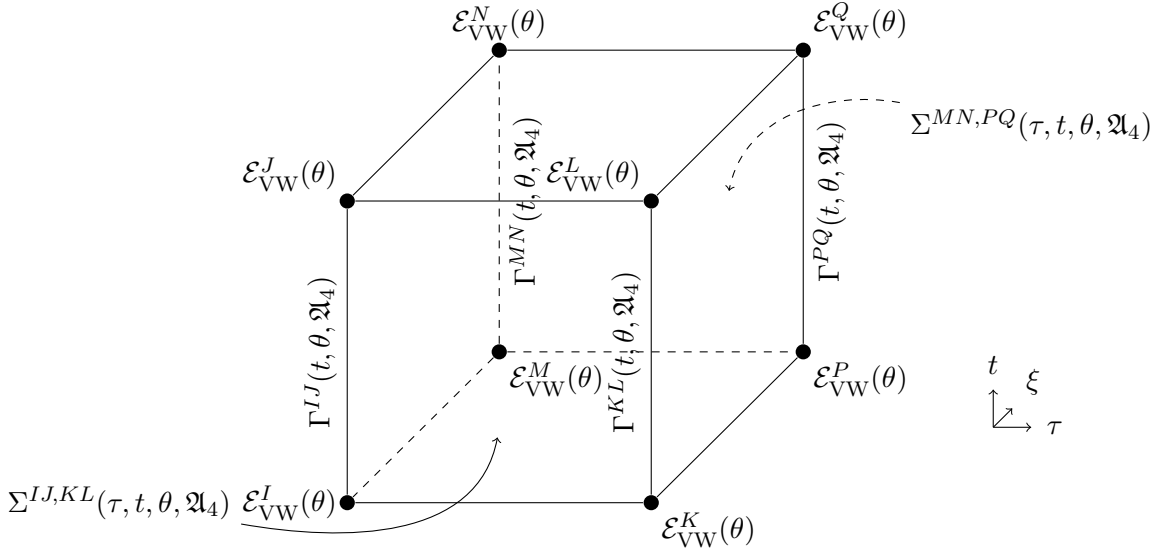
\begin{figure}
  \centering
  \begin{tikzpicture}[%
    auto,%
    every edge/.style={draw},%
    relation/.style={scale=1, sloped, anchor=center, align=center,%
      color=black},%
    vertRelation/.style={scale=1, anchor=center, align=center},%
    dot/.style={circle, fill, minimum size=2*\radius, node contents={},%
      inner sep=0pt},%
    ]
    \let\radius\undefined
    \newlength{\radius}
    \setlength{\radius}{1mm}
    \node (ftl) at (-4,4) [dot];
    \node at (ftl) [above left]  {$\mathcal{E}^J_{\text{VW}}(\theta)$};
    \node (ftr) at (0,4) [dot];
    \node at (ftr) [above left]  {$\mathcal{E}^L_{\text{VW}}(\theta)$};
    \node (fbl) at (-4,0) [dot];
    \node at (fbl) [left]  {$\mathcal{E}^I_{\text{VW}}(\theta)$};
    \node (fbr) at (0,0) [dot];
    \node at (fbr) [below right]  {$\mathcal{E}^K_{\text{VW}}(\theta)$};
    \node (btl) at (-2,6) [dot];
    \node at (btl) [above]  {$\mathcal{E}^N_{\text{VW}}(\theta)$};
    \node (bbl) at (-2,2) [dot];
    \node at (bbl) [below right]  {$\mathcal{E}^M_{\text{VW}}(\theta)$};
    \node (btr) at (2,6) [dot];
    \node at (btr) [above right]  {$\mathcal{E}^Q_{\text{VW}}(\theta)$};
    \node (bbr) at (2,2) [dot];
    \node at (bbr) [below right]  {$\mathcal{E}^P_{\text{VW}}(\theta)$};
    \draw
    (fbl) edge node[relation, above] {$\Gamma^{IJ}(t, \theta, \mathfrak{A}_4)$} (ftl)
    (ftl) edge (btl)
    (bbl) edge[dashed] node[relation, below] {$\Gamma^{MN}(t, \theta, \mathfrak{A}_4)$} (btl)
    (bbl) edge[dashed] (fbl)
    (fbr) edge node[relation, above] {$\Gamma^{KL}(t, \theta, \mathfrak{A}_4)$} (ftr)
    (ftr) edge (btr)
    (bbr) edge node[relation, below] {$\Gamma^{PQ}(t, \theta, \mathfrak{A}_4)$} (btr)
    (bbr) edge (fbr)
    (fbl) edge (fbr)
    (ftl) edge (ftr)
    (bbl) edge[dashed] (bbr)
    (btl) edge (btr);
    \node (fn) at (-7, 0) {$\Sigma^{IJ, KL}(\tau, t, \theta, \mathfrak{A}_4)$};
    \node (ff) at (-2, 1) {};
    \draw [->]
    (fn) edge [out=350, in=260] (ff);
    \node (bn) at (5, 5) {$\Sigma^{MN, PQ}(\tau, t, \theta, \mathfrak{A}_4)$};
    \node (bf) at (1, 4) {};
    \draw[->]
    (bn) edge [out=170, in=80 , dashed] (bf);
    \coordinate (cco) at (4.5, 1);
    \coordinate (ccx) at (5, 1);
    \coordinate (ccy) at (4.75,1.25);
    \coordinate (ccz) at (4.5, 1.5);
    \node at (ccx) [right=2pt of ccx] {$\tau$};
    \node at (ccy) [above right=2pt of ccy] {$\xi$};
    \node at (ccz) [above=2pt of ccz] {$t$};
    \draw[->] (cco) -- (ccx);
    \draw[->] (cco) -- (ccy);
    \draw[->] (cco) -- (ccz);
  \end{tikzpicture}
  \caption{%
    $\mathfrak{A}_4^{\theta}$-threebrane $\Upsilon^{\{IJ, KL\}, \{MN, PQ\}}$ with
    (i) faces being $\mathfrak{A}_4^{\theta}$-sheets $\Sigma^{IJ, KL}_\pm(\tau, t, \theta, \mathfrak{A}_4)$ and $\Sigma^{MN, PQ}_\pm(\tau, t, \theta, \mathfrak{A}_4)$,
    (ii) edges being $\mathfrak{A}_4^{\theta}$-solitons $\Gamma^{IJ}(t, \theta, \mathfrak{A}_4)$, $\Gamma^{KL}(t, \theta, \mathfrak{A}_4)$, $\Gamma^{MN}(t, \theta, \mathfrak{A}_4)$, $\Gamma^{PQ}(t, \theta, \mathfrak{A}_4)$,
    and (iii) vertices corresponding to $\mathcal{E}^I_{\text{VW}}(\theta)$, $\mathcal{E}^J_{\text{VW}}(\theta)$, $\mathcal{E}^K_{\text{VW}}(\theta)$, $\mathcal{E}^L_{\text{VW}}(\theta)$, $\mathcal{E}^M_{\text{VW}}(\theta)$, $\mathcal{E}^N_{\text{VW}}(\theta)$, $\mathcal{E}^P_{\text{VW}}(\theta)$, $\mathcal{E}^Q_{\text{VW}}(\theta)$.
  }
  \label{fig:cy2 x r4:frakA-threebrane}
\end{figure}

This means that we can denote and specify any $\mathfrak{A}_4^{\theta}$-threebrane at $y = \pm \infty$ as \\
$\Upsilon^{\{IJ, KL\}, \{MN, PQ\}}_{\pm}(\xi, \tau, t, \theta, \mathfrak{A}_4)$,\footnote{%
  Just like in \autoref{ft:fixing A2-sheet centre of mass dof}, the $y$-invariant $\mathfrak{A}_4^{\theta}$-threebranes can be fixed at $y = \pm \infty$ by adding physically-inconsequential $\mathcal{Q}$-exact terms to the SQM action.
  \label{ft:fixing A4-threebrane centre of mass dof}
}~where (i) its front and back faces correspond to the $\mathfrak{A}_4^{\theta}$-sheets $\Sigma^{IJ, KL}(\tau, t, \theta, \mathfrak{A}_4)$ and $\Sigma^{MN, PQ}(\tau, t, \theta, \mathfrak{A}_4)$, respectively, (ii) its four vertical edges correspond to the $\mathfrak{A}_4^{\theta}$-solitons $\Gamma^{IJ}(t, \theta, \mathfrak{A}_4)$, $\Gamma^{KL}(t, \theta, \mathfrak{A}_4)$, $\Gamma^{MN}(t, \theta, \mathfrak{A}_4)$, and $\Gamma^{PQ}(t, \theta, \mathfrak{A}_4)$, and (iii) its eight vertices correspond to $\mathcal{E}^I_{\text{VW}}(\theta)$, $\mathcal{E}^J_{\text{VW}}(\theta)$, $\mathcal{E}^K_{\text{VW}}(\theta)$, $\mathcal{E}^L_{\text{VW}}(\theta)$, $\mathcal{E}^M_{\text{VW}}(\theta)$, $\mathcal{E}^N_{\text{VW}}(\theta)$, $\mathcal{E}^P_{\text{VW}}(\theta)$, and $\mathcal{E}^Q_{\text{VW}}(\theta)$, similar to the kind shown in \autoref{fig:cy2 x r4:frakA-threebrane}.

Since the $\mathcal{E}^{*}_{\text{VW}}(\theta)$'s, $\Gamma^{**}(t, \theta, \mathfrak{A}_4)$, and $\Sigma^{**, **}(\tau, t, \theta, \mathfrak{A}_4)$'s are $y$-invariant and therefore, have the same value for all $y$, we have 4d BPS worldvolumes, that can be interpreted via \autoref{fig:m2 x r3:bps worldvolume}.
This time, however, instead of $\xi$ being the temporal direction and $(t, \tau)$ being the spatial directions, we will have $y$ and $(t \times \tau, \xi)$, respectively, where ``$t \times \tau$'' represents the $(t, \tau)$-plane.
Instead of the faces being labeled $\Sigma^{**, **}_{\pm}(t, \tau, \theta, \mathfrak{A}_2)$ and $\Gamma^{**}(\tau, \theta, \mathfrak{A}_2)$, they will be labeled $\Upsilon^{\{**, **\}, \{**, **\}}_{\pm}(\xi, \tau, t, \theta, \mathfrak{A}_4)$ and $\Sigma^{**, **}(\tau, t, \theta, \mathfrak{A}_4)$, respectively.
And instead of the edges being labeled $\mathcal{E}^{*}_{\text{BF}}(\theta)$, they will be labeled $\Gamma^{**}(t, \theta, \mathfrak{A}_4)$, i.e., a sheet in the 4d BPS worldvolumes will be labeled by $\Gamma^{**}(t, \theta, \mathfrak{A}_4 )$.
Consequently, the edges of the sheets will be labeled by the endpoints of $\Gamma^{**}(t, \theta, \mathfrak{A}_4)$, i.e., $\mathcal{E}^{*}_{\text{VW}}(\theta)$.

\subtitle{The 4d Model on $\R^4$ and an Open Threebrane Theory in $\mathfrak{A}_4$}

Hence, one can understand the 4d gauged sigma model on $\R^4$ with target space $\mathfrak{A}_4$ to define an open threebrane theory in $\mathfrak{A}_4$, whose \emph{effective} worldvolume, boundaries, and edges, as described above.

\subsection{Soliton String Theory, the \texorpdfstring{Spin$(7)$}{Spin(7)} Partition Function, and an FS type \texorpdfstring{$A_{\infty}$}{A-infinity}-category of \texorpdfstring{$\mathfrak{A}_4^{\theta}$}{A4-theta}-sheets}
\label{sec:cy2 x r4:fs-cat}

\subtitle{The 2d Model as a 2d Gauged LG Model}

Notice that we can also express \eqref{eq:cy2 x r4:worldsheet eqn} as
\begin{equation}
  \label{eq:cy2 x r4:2d lg:worldsheet eqn}
  \Dv{\mathcal{A}^m}{y} - I \Dv{\mathcal{A}^m}{\xi}
  + \Dv{\breve{A}^m}{y} - I \Dv{\breve{A}^m}{\xi}
  + I F_{y \xi}
  = - P^m - q^m
  \, .
\end{equation}
Here $\mathcal{A}^m$ and $\breve{A}^m \coloneq J (\tilde{A}_t)^m + K (\tilde{A}_{\tau})^m$ can be interpreted as holomorphic coordinates on the complex path space $\mathcal{M}(\C, \mathfrak{A}_4)$ of paths from the $I$-complexified $\R_t \times \R_{\tau}$ to $\mathfrak{A}_4$.
In turn, this means that we can express the action of the 2d gauged sigma model with target space $\mathcal{M}(\R_t \times \R_{\tau}, \mathfrak{A}_4)$ in \eqref{eq:cy2 x r4:2d model action} as
\begin{equation}
  \label{eq:cy2 x r4:2d lg:action}
  \begin{aligned}
    S_{\text{2d-LG}, \mathcal{M}(\C, \mathfrak{A}_4)}
    &= \frac{1}{e^2} \int d\xi dy \Bigg(
      \bigg|
      (D_y - I D_{\xi}) \left(
      \mathcal{A}^m + \breve{A}^m
      \right)
      + I F_{y \xi}
      + P^m
      + q^m
      \bigg|^2
      + \dots
      \Bigg)
    \\
    & = \frac{1}{e^2} \int d\xi dy \Bigg(
      \bigg|
      (D_y - I D_{\xi}) \left(
      \mathcal{A}^m + \breve{A}^m
      \right)
      + I F_{y \xi}
      + g^{m\bar{n}}_{\mathcal{M}(\C, \mathfrak{A}_4)} \left(
      \frac{K \zeta}{2} \pdv{\mathfrak{w}_4}{\mathcal{A}^n}
      \right)^{*}
    \\
    & \qquad \qquad \qquad \quad
      + g^{m\bar{n}}_{\mathcal{M}(\C, \mathfrak{A}_4)} \left(
      \frac{K \zeta}{2} \pdv{\mathfrak{w}_4}{\breve{A}^n}
      \right)^{*}
      \bigg|^2
      + \dots
      \Bigg)
    \\
    & = \frac{1}{e^2} \int d\xi dy \Bigg(
      |D_{\rho} \mathcal{A}^m|^2
      + |D_{\rho} \breve{A}^m|^2
      + \left| \pdv{\mathfrak{w}_4}{\mathcal{A}^m} \right|^2
      + \left| \pdv{\mathfrak{w}_4}{\breve{A}^m} \right|^2
      + |F_{y\xi}|^2
      + \dots
      \Bigg)
      \, ,
  \end{aligned}
\end{equation}
where $g_{\mathcal{M}(\C, \mathfrak{A}_4)}$ is the metric on $\mathcal{M}(\C, \mathfrak{A}_4)$ and $\zeta \coloneqq e^{I\theta}$.
In other words, the 2d gauged sigma model with target space $\mathcal{M}(\R_t \times \R_{\tau}, \mathfrak{A}_4)$ can also be interpreted as a 2d gauged LG model with target space $\mathcal{M}(\C, \mathfrak{A}_4)$ and a holomorphic superpotential $\mathfrak{w}_4(\mathcal{A}, \breve{A})$.

By setting $d_y \mathcal{A}^m = 0$ and $A_y, A_{\xi}, \breve{A}^m \rightarrow 0$ in the expression within the squared term in \eqref{eq:cy2 x r4:2d lg:action}, we can read off the LG $\mathcal{M}^{\theta}(\C, \mathfrak{A}_4)$-soliton equation corresponding to $\Xi^{\{IJ, KL\}, \{MN, PQ\}}_{\pm}(\xi, \theta, \mathfrak{A}_4)$ (that re-expresses \eqref{eq:cy2 x r4:m-soliton eqns} with $A_y, A_{\xi}, (\tilde{A}_t)^m, (\tilde{A}_{\tau})^m \rightarrow 0$) as
\begin{equation}
  \label{eq:cy2 x r4:2d lg:m-soliton:eqn}
  I \dv{\mathcal{A}^m}{\xi}
  = g^{m\bar{n}}_{\mathcal{M}(\C, \mathfrak{A}_4)} \left(
    \frac{K \zeta}{2} \pdv{\mathfrak{w}_4}{\mathcal{A}^n}
  \right)^{*}_{\breve{A} = 0}
  + g^{m\bar{n}}_{\mathcal{M}(\C, \mathfrak{A}_4)} \left(
    \frac{K \zeta}{2} \pdv{\mathfrak{w}_4}{(\breve{A})^n}
  \right)^{*}_{\breve{A} = 0}
  \, .
\end{equation}
By setting $d_{\xi} \mathcal{A}^m = 0$ in \eqref{eq:cy2 x r4:2d lg:m-soliton:eqn}, we get the LG $\mathcal{M}^{\theta}(\C, \mathfrak{A}_4)$-soliton endpoint equation corresponding to $\Xi^{\{IJ, KL\}, \{MN, PQ\}}(\pm \infty, \theta, \mathfrak{A}_4)$ (that re-expresses \eqref{eq:cy2 x r4:m-soliton:endpts}) as
\begin{equation}
  \label{eq:cy2 x r4:2d lg:m-soliton:endpt eqns}
  g^{m\bar{n}}_{\mathcal{M}(\C, \mathfrak{A}_4)} \left(
    \frac{K \zeta}{2} \pdv{\mathfrak{w}_4}{\mathcal{A}^n}
  \right)^{*}_{\breve{A} = 0}
  + g^{m\bar{n}}_{\mathcal{M}(\C, \mathfrak{A}_4)} \left(
    \frac{K \zeta}{2} \pdv{\mathfrak{w}_4}{\breve{A}^n}
  \right)^{*}_{\breve{A} = 0}
  = 0
  \, .
\end{equation}

Recall from the end of \autoref{sec:cy2 x r4:vw} that we are only considering certain $CY_2$ such that the $\theta$-deformed VW configurations are isolated and non-degenerate.
Next, recall also that such configurations will correspond to the vertices of the $\mathfrak{A}_4^{\theta}$-sheets; therefore, just like their vertices, these $\mathfrak{A}_4^{\theta}$-sheets would be isolated and non-degenerate.
As these $\mathfrak{A}_4^{\theta}$-sheets will correspond, in the 2d gauged sigma model, to the endpoints of the $\mathcal{M}^{\theta}(\R_t \times \R_{\tau}, \mathfrak{A}_4)$-solitons (and thus the endpoints of the LG $\mathcal{M}^{\theta}(\C, \mathfrak{A}_4)$-solitons), i.e., $\Xi^{\{IJ, KL\}, \{MN, PQ\}}(\pm \infty, \theta, \mathfrak{A}_4)$, this means that the latter would also be isolated and non-degenerate.
Thus, from their definition in \eqref{eq:cy2 x r4:2d lg:m-soliton:endpt eqns} which tells us that they are critical points of $\mathfrak{w}_4(\mathcal{A}, \breve{A})$, we conclude that $\mathfrak{w}_4(\mathcal{A}, \breve{A})$ can be regarded as a holomorphic Morse function in $\mathcal{M}(\C, \mathfrak{A}_4)$.

Similar to \autoref{sec:m2 x r3:fs-cat}, an LG $\mathcal{M}^{\theta}(\C, \mathfrak{A}_4)$-soliton defined in \eqref{eq:cy2 x r4:2d lg:m-soliton:eqn} maps to a straight line in the complex $\mathfrak{w}_4$-plane that starts and ends at the critical points $\mathfrak{w}_4^{IJ, KL}(\theta) \equiv \mathfrak{w}_4\left(\Xi^{\{IJ, KL\}, \{MN, PQ\}}(- \infty, \theta, \mathfrak{A}_4)\right)$ and $\mathfrak{w}_4^{MN, PQ}(\theta) \equiv \mathfrak{w}_4\left(\Xi^{\{IJ, KL\}, \{MN, PQ\}}(+ \infty, \theta, \mathfrak{A}_4)\right)$, respectively, where its slope depends on $\theta$ (via $\zeta$).
We shall also assume that $\Re ( \mathfrak{w}_4^{IJ, KL}(\theta) ) < \Re ( \mathfrak{w}_4^{MN, PQ}(\theta) )$ and that $I \neq J$, $K \neq L$, $J \neq K$, $M \neq N$, $P \neq Q$, and $N \neq P$.

\subtitle{The 2d Gauged LG Model as an LG SQM}

With suitable rescalings, we can recast \eqref{eq:cy2 x r4:2d lg:action} as a 1d LG SQM (that re-expresses \eqref{eq:cy2 x r4:sqm action}), where its action will be given by\footnote{%
  In the following expression, we have integrated out $A_y$ and omitted the fields corresponding to the finite-energy gauge fields $A_{\{t, \tau, \xi\}}$ (as explained in \autoref{ft:stokes theorem for cy2 x r4:sqm}).
  \label{ft:stokes theorem for cy2 x r4:lg sqm}
}
\begin{equation}
  \label{eq:cy2 x r4:2d lg:sqm action}
  S_{\text{2d-LG SQM}, \mathcal{M}(\R_{\xi}, \mathcal{M}(\C, \mathfrak{A}_4))}
  = \frac{1}{e^2} \int dy \left(
    \left|
      \left(
        \dv{\mathcal{A}^u}{y} + \dv{\widehat{A}^u}{y}
      \right)
      + g^{uv}_{\mathcal{M}(\R_{\xi}, \mathcal{M}(\C, \mathfrak{A}_4))} \left(
        \pdv{\mathfrak{H}_4}{\mathcal{A}^u}
        + \pdv{\mathfrak{H}_4}{\widehat{A}^u}
      \right)
    \right|^2
    + \dots
  \right)
  \, ,
\end{equation}
where $\mathfrak{H}_4(\mathcal{A}, \widehat{A})$ is the \emph{real-valued} potential in $\mathcal{M}(\R_{\xi}, \mathcal{M}(\C, \mathfrak{A}_4))$, and the subscript ``$\text{2d-LG SQM}$, $\mathcal{M}(\R_{\xi}, \mathcal{M}(\C, \mathfrak{A}_4))$'' is to specify that it is a 1d SQM in $\mathcal{M}(\R_{\xi}, \mathcal{M}(\C, \mathfrak{A}_4))$ obtained from the equivalent 2d LG model.
We will also refer to this \emph{1d} LG SQM as ``2d-LG SQM'' in the rest of this subsection.

The 2d-LG SQM will localize onto configurations that \emph{simultaneously} set to zero the LHS and RHS of the expression within the squared term in \eqref{eq:cy2 x r4:2d lg:sqm action}.
In other words, it will localize onto $y$-invariant critical points of $\mathfrak{H}_4(\mathcal{A}, \widehat{A})$ that will correspond, when $A_y, A_{\xi}, \breve{A}^m \rightarrow 0$, to the LG $\mathcal{M}^{\theta}(\C, \mathfrak{A}_4)$-solitons defined by \eqref{eq:cy2 x r4:2d lg:m-soliton:eqn}.
For our choice of $CY_2$, the LG $\mathcal{M}^{\theta}(\C, \mathfrak{A}_4)$-solitons, just like their endpoints, will be isolated and non-degenerate.
Thus, $\mathfrak{H}_4(\mathcal{A}, \widehat{A})$ can be regarded as a real-valued Morse functional in $\mathcal{M}(\R_{\xi}, \mathcal{M}(\C, \mathfrak{A}_4))$.

\subtitle{Morphisms from $\varsigma^{IJ, KL}(\theta, \mathfrak{A}_4)$ to $\varsigma^{MN, PQ}(\theta, \mathfrak{A}_4)$ as Floer Homology Classes of Intersecting Thimbles}

Repeating here the analysis in \autoref{sec:m2 x r3:fs-cat} with \eqref{eq:cy2 x r4:2d lg:sqm action} as the action of the 2d-LG SQM, we find that we can interpret the LG $\mathcal{M}^{\theta}(\C, \mathfrak{A}_4)$-soliton solution $\Xi^{\{IJ, KL\}, \{MN, PQ\}}_{\pm}(\xi, \theta, \mathfrak{A}_4)$ as intersections of thimbles.
Specifically, a $\Xi^{\{IJ, KL\}, \{MN, PQ\}}_{\pm}(\xi, \theta, \mathfrak{A}_4)$-soliton pair, whose left and right endpoints correspond to $\varsigma^{IJ, KL}(\theta, \mathfrak{A}_4)$ and $\varsigma^{MN, PQ}(\theta, \mathfrak{A}_4)$, respectively, can be identified as a pair of intersection points $\mathfrak{p}^{\{IJ, KL\}, \{MN, PQ\}}_{\text{VW}, \pm}(\theta) \in \big( S^{IJ}_{\text{VW}} \cap S^{KL}_{\text{VW}} \big) \bigcap \big( S^{MN}_{\text{VW}} \cap S^{PQ}_{\text{VW}} \big)$ of a left and right thimble in the fiber space over the line segment $\big[ \mathfrak{w}_4^{IJ, KL}(\theta), \mathfrak{w}_4^{MN, PQ}(\theta) \big]$.

This means that the 2d-LG SQM in $\mathcal{M}(\R_{\xi}, \mathcal{M}(\C, \mathfrak{A}_4))$ with action \eqref{eq:cy2 x r4:2d lg:sqm action}, will physically realize a Floer homology that we shall name an $\mathcal{M}(\C, \mathfrak{A}_4)$-LG Floer homology.
The chains of the $\mathcal{M}(\C, \mathfrak{A}_4)$-LG Floer complex will be generated by LG $\mathcal{M}^{\theta}(\C, \mathfrak{A}_4)$-solitons which we can identify with $\mathfrak{p}^{\{**, **\}, \{**, **\}}_{\text{VW}, \pm}(\theta)$, and the $\mathcal{M}(\C, \mathfrak{A}_4)$-LG Floer differential will be realized by the flow lines governed by the gradient flow equations satisfied by the $y$-varying configurations that set the expression within the squared term in \eqref{eq:cy2 x r4:2d lg:sqm action} to zero.
The partition function of the 2d-LG SQM in $\mathcal{M}(\R_{\xi}, \mathcal{M}(\C, \mathfrak{A}_4))$ will then be given by\footnote{%
  The `$\theta$' label is omitted in the LHS of the following expression (as explained in \autoref{ft:theta omission in m2-2d lg partition fn}).
  \label{ft:theta omission in cy2xr4:2d lg parition fn}
}
\begin{equation}
  \label{eq:cy2 x r4:2d lg:partition function}
  \mathcal{Z}_{\text{2d-LG SQM}, \mathcal{M}(\R_{\xi}, \mathcal{M}(\C, \mathfrak{A}_4))}(G)
  = \sum_{\substack{I \neq J \neq K \neq L \neq M \\ \neq N \neq P \neq Q = 1}}
  \sum_{\substack{\mathfrak{p}^{\{IJ, KL\}, \{MN, PQ\}}_{\text{VW}, \pm} \\ \in \left( S^{IJ}_{\text{VW}} \cap S^{KL}_{\text{VW}} \right) \bigcap \left( S^{MN}_{\text{VW}} \cap S^{PQ}_{\text{VW}} \right)}}
  \text{HF}^G_{d_v} \left(
    \mathfrak{p}^{\{IJ, KL\}, \{MN, PQ\}}_{\text{VW}, \pm} (\theta)
  \right)
  \, .
\end{equation}
Here, the contribution $\text{HF}^G_{d_v} (\mathfrak{p}^{\{IJ, KL\}, \{MN, PQ\}}_{\text{VW}, \pm} (\theta))$ can be identified with a homology class in an $\mathcal{M}(\C, \mathfrak{A}_4)$-LG Floer homology generated by intersection points of thimbles.
These intersection points represent LG $\mathcal{M}^{\theta}(\C, \mathfrak{A}_4)$-solitons defined by \eqref{eq:cy2 x r4:2d lg:m-soliton:eqn}, whose endpoints correspond to $\theta$-deformed, non-constant complex paths in $\mathcal{M}(\C, \mathfrak{A}_4)$ defined by \eqref{eq:cy2 x r4:2d lg:m-soliton:endpt eqns}.
The degree of each chain in the complex is $d_v$, and is counted by the number of outgoing flow lines from the fixed critical points of $\mathfrak{H}_4(\mathcal{A}, \widehat{A})$ in $\mathcal{M}(\R_{\xi}, \mathcal{M}(\C, \mathfrak{A}_4))$ which can also be identified as $\mathfrak{p}^{\{IJ, KL\}, \{MN, PQ\}}_{\text{VW}, \pm} (\theta)$.

Therefore, $\mathcal{Z}_{\text{2d-LG SQM}, \mathcal{M}(\R_{\xi}, \mathcal{M}(\C, \mathfrak{A}_4))}(G)$ in \eqref{eq:cy2 x r4:2d lg:partition function} is a sum of LG $\mathcal{M}^{\theta}(\C, \mathfrak{A}_4)$-solitons defined by \eqref{eq:cy2 x r4:2d lg:m-soliton:eqn} with endpoints \eqref{eq:cy2 x r4:2d lg:m-soliton:endpt eqns}, or equivalently, $\Xi^{\{IJ, KL\}, \{MN, PQ\}}(\xi, \theta, \mathfrak{A}_4)$-solitons defined by \eqref{eq:cy2 x r4:m-soliton eqns} (with $A_y, A_{\xi}, (\tilde{A}_t)^m, (\tilde{A}_{\tau})^m \rightarrow 0$) with endpoints \eqref{eq:cy2 x r4:m-soliton:endpts}, whose start and end correspond to the non-constant double paths $\varsigma^{IJ, KL}(\theta, \mathfrak{A}_4)$ and $\varsigma^{MN, PQ}(\theta, \mathfrak{A}_4)$, respectively.
In other words, we can write
\begin{equation}
  \label{eq:cy2 x r4:2d lg:floer homology as vector}
  \text{CF}_{\mathcal{M}(\R_t \times \R_{\tau}, \mathfrak{A}_4)} \left(
    \varsigma^{IJ, KL}(\theta, \mathfrak{A}_4),
    \varsigma^{MN, PQ}(\theta, \mathfrak{A}_4)
  \right)_{\pm}
  = \text{HF}^G_{d_v} \left(
    \mathfrak{p}^{\{IJ, KL\}, \{MN, PQ\}}_{\text{VW}, \pm} (\theta)
  \right)
  \, ,
\end{equation}
where $\text{CF}_{\mathcal{M}(\R_t \times \R_{\tau}, \mathfrak{A}_4)} (\varsigma^{IJ, KL}(\theta, \mathfrak{A}_4), \varsigma^{MN, PQ}(\theta, \mathfrak{A}_4) )_{\pm}$ is a vector representing a $\Xi^{\{IJ, KL\}, \{MN, PQ\}}(\xi, \theta, \mathfrak{A}_4)$-soliton, such that $\Re ( \mathfrak{w}_4^{IJ, KL}(\theta) ) < \Re ( \mathfrak{w}_4^{MN, PQ}(\theta) )$.
This will lead us to the following one-to-one identification\footnote{%
  The `$\theta$' label is omitted in the following expression (as explained in \autoref{ft:omission of theta in m2 2d-lg}).
  \label{ft:omission of theta in cy2 x r4 2d-lg}
}
\begin{equation}
  \label{eq:cy2 x r4:2d lg:floer hom as morphism}
  \boxed{
    \text{Hom} \left(
      \varsigma^{IJ, KL}(\mathfrak{A}_4),
      \varsigma^{MN, PQ}(\mathfrak{A}_4)
    \right)_{\pm}
    \Longleftrightarrow
    \text{HF}^G_{d_v} \left(
      \mathfrak{p}^{\{IJ, KL\}, \{MN, PQ\}}_{\text{VW}, \pm}
    \right)
  }
\end{equation}
where the RHS is (i) proportional to the identity class when $I = M$, $J = N$, $K = P$, and $L = Q$, and (ii) zero when $I \leftrightarrow M$, $J \leftrightarrow N$, $K \leftrightarrow P$, and $L \leftrightarrow Q$ (since the $\Xi^{\{IJ, KL\}, \{MN, PQ\}}(\xi, \theta, \mathfrak{A}_4)$-soliton only moves in one direction from $\varsigma^{IJ, KL}(\theta, \mathfrak{A}_4)$ to $\varsigma^{MN, PQ}(\theta, \mathfrak{A}_4)$).

\subtitle{Soliton String Theory from the 2d LG Model}

Just like in~\autoref{sec:m2 x r3:fs-cat}, the 2d gauged LG model in $\mathcal{M}(\C, \mathfrak{A}_4)$ with action \eqref{eq:cy2 x r4:2d lg:action} can be interpreted as a soliton string theory in $\mathcal{M}(\C, \mathfrak{A}_4)$.
The dynamics of this soliton string theory in $\mathcal{M}(\C, \mathfrak{A}_4)$ will be governed by the BPS worldsheet equation of \eqref{eq:cy2 x r4:2d lg:worldsheet eqn}, where $(\mathcal{A}^m, \breve{A}^m)$ are scalars on the worldsheet corresponding to the holomorphic coordinates of $\mathcal{M}(\C, \mathfrak{A}_4)$.
At an arbitrary instant in time whence $d_y \mathcal{A}^m = d_y \breve{A}^m = 0 = d_y A_{\xi}$ in \eqref{eq:cy2 x r4:2d lg:worldsheet eqn}, the dynamics of  $(\mathcal{A}^m, \breve{A}^m)$ and the 2d gauge fields $(A_y, A_{\xi})$ along $\xi$ will be governed by the soliton equation
\begin{equation}
  \label{eq:cy2 x r4:2d lg:string m-soliton}
  \begin{aligned}
    I \dv{\mathcal{A}^m}{\xi}
    + I \dv{\breve{A}^m}{\xi}
    + I \dv{A_y}{\xi}
    =& - I [A_{\xi}, \mathcal{A}^m + \breve{A}^m + A_y]
       + [A_y, \mathcal{A}^m + \breve{A}^m]
    \\
     & + g^{m\bar{n}}_{\mathcal{M}(\C, \mathfrak{A}_4)} \left(
       \frac{K \zeta}{2} \pdv{\mathfrak{w}_4}{\mathcal{A}^n}
       \right)^{*}
       + g^{m\bar{n}}_{\mathcal{M}(\C, \mathfrak{A}_4)} \left(
       \frac{K \zeta}{2} \pdv{\mathfrak{w}_4}{\breve{A}^n}
       \right)^{*}
       \, .
  \end{aligned}
\end{equation}

\subtitle{The Normalized Spin$(7)$ Partition Function, LG $\mathcal{M}^{\theta}(\C, \mathfrak{A}_4)$-soliton String Scattering, and Maps of an $A_{\infty}$-structure}

Since our Spin$(7)$ theory is semi-classical, its normalized 8d partition function will be a sum over tree-level scattering amplitudes of the $\mathcal{M}^{\theta}(\C, \mathfrak{A}_4)$-soliton strings defined by \eqref{eq:cy2 x r4:2d lg:m-soliton:eqn}.
The BPS worldsheet underlying such a tree-level scattering is similar to \autoref{fig:m2 x r3:mu-d maps}, where instead of the endpoints of each string being labeled $\gamma^{**}(\mathfrak{A}_2)$, they are now labeled $\varsigma^{**, **}(\mathfrak{A}_4)$.

In other words, we can express the normalized Spin$(7)$ partition function as
\begin{equation}
  \label{eq:cy2 x r4:2d lg:normalized partition fn}
  \tilde{\mathcal{Z}}_{\text{Spin}(7), CY_2 \times \R^4}(G) = \sum_{\mathfrak{K}_n} \mho^{\mathfrak{K}_n}_{\mathfrak{A}_2}
  \, ,
  \qquad
  \mathfrak{K}_n = 1, 2, \dots \left\lfloor {\frac{n - 4}{4}} \right\rfloor
\end{equation}
where each
\begin{equation}
  \label{eq:cy2 x r4:2d lg:composition maps}
  \boxed{
    \begin{aligned}
      \mho^{\mathfrak{K}_n}_{\mathfrak{A}_4} : \bigotimes_{i = 1}^{\mathfrak{K}_n}
      & \text{Hom} \left(
        \varsigma^{\{I_{4i - 3} I_{4i - 2}\}, \{I_{4i - 1} I_{4i}\}}(\mathfrak{A}_4),
        \varsigma^{\{I_{4(i + 1) - 3} I_{4(i + 1) - 2}\}, \{I_{4(i + 1) - 1} I_{4(i + 1)}\}}(\mathfrak{A}_4)
      \right)_-
      \\
      & \longto
      \text{Hom} \left(
        \varsigma^{\{I_1 I_2\}, \{I_3 I_4 \}}(\mathfrak{A}_4),
        \varsigma^{\{I_{4\mathfrak{K}_n + 1} I_{4\mathfrak{K}_n + 2}\}, \{I_{4\mathfrak{K}_n + 3} I_{4\mathfrak{K}_n + 4}\}}(\mathfrak{A}_4)
      \right)_+
    \end{aligned}
  }
\end{equation}
is a scattering amplitude of $\mathfrak{K}_n$ incoming LG $\mathcal{M}^{\theta}(\C, \mathfrak{A}_4)$-soliton strings $\text{Hom}(\varsigma^{**, **}(\mathfrak{A}_4), \varsigma^{**, **}(\mathfrak{A}_4))_-$, and a single outgoing $\text{Hom}(\varsigma^{**, **}(\mathfrak{A}_4), \varsigma^{**, **}(\mathfrak{A}_4))_+$, with left and right boundaries as labeled, whose underlying worldsheet can be regarded as a disc with $\mathfrak{K}_n + 1$ vertex operators at the boundary.
In short, $\mho^{\mathfrak{K}_n}_{\mathfrak{A}_4}$ counts pseudoholomorphic discs with $\mathfrak{K}_n + 1$ punctures at the boundary that are mapped to $\mathcal{M}(\C, \mathfrak{A}_4)$ according to the BPS worldsheet equation \eqref{eq:cy2 x r4:2d lg:worldsheet eqn}.

Just as in \autoref{sec:m2 x r3:fs-cat}, the collection of $\mho^{\mathfrak{K}_n}_{\mathfrak{A}_4}$ maps in \eqref{eq:cy2 x r4:2d lg:composition maps} can be regarded as composition maps defining an $A_{\infty}$-structure.

\subtitle{An FS type $A_{\infty}$-category of $\mathfrak{A}_4^{\theta}$-sheets}

Altogether, this means that the normalized partition function of Spin$(7)$ theory on $CY_2 \times \R^4$ as expressed in \eqref{eq:cy2 x r4:2d lg:normalized partition fn}, manifests a \emph{novel} FS type $A_{\infty}$-category defined by the $\mho^{\mathfrak{K}_n}_{\mathfrak{A}_4}$ maps \eqref{eq:cy2 x r4:2d lg:composition maps} and the one-to-one identification \eqref{eq:cy2 x r4:2d lg:floer hom as morphism}, where the $\mathfrak{K}_n + 1$ number of objects $\{\varsigma^{I_1 I_2, I_3 I_4}(\mathfrak{A}_4), \varsigma^{I_5 I_6, I_7 I_8}(\mathfrak{A}_4), \dots$, $\varsigma^{I_{4\mathfrak{K}_n + 1} I_{4\mathfrak{K}_n + 2}, I_{4\mathfrak{K}_n + 3} I_{4\mathfrak{K}_n + 4}}(\mathfrak{A}_4)\}$ correspond to $\mathfrak{A}_4^{\theta}$-sheets with vertices themselves corresponding to ($\theta$-deformed) VW configurations on $CY_2$!

\subsection{Soliton Threebrane Theory, the \texorpdfstring{Spin$(7)$}{Spin(7)} Partition Function, and a Cauchy-Riemann-Fueter type \texorpdfstring{$A_{\infty}$}{A-infinity}-3-category 3-categorifying the HW Floer Homology of \texorpdfstring{$CY_2$}{CY2}}
\label{sec:cy2 x r4:crf-cat}

Notice that the action of the 4d gauged sigma model with target space $\mathfrak{A}_4$ in \eqref{eq:cy2 x r4:4d model action} can also be expressed as
\begin{equation}
  \label{eq:cy2 x r4:4d lg:action}
  \begin{aligned}
    S_{\text{4d-LG}, \mathfrak{A}_4}
    &= \frac{1}{e^2} \int dt d\tau d\xi dy \Bigg(
      \bigg|
      ( D_y - I D_{\xi} - J D_t - K D_{\tau} ) \mathcal{A}^a
      + I (F_{y \xi} + F_{\tau t})
      + J (F_{y t} + F_{\xi \tau})
    \\
    & \qquad \qquad \qquad \qquad \quad \,
      + K (F_{y \tau} + F_{t \xi})
      + q^a
      \bigg|^2
      + \dots
      \Bigg)
    \\
    &= \frac{1}{e^2} \int dt d\tau d\xi dy \Bigg(
      \bigg|
      ( D_y - I D_{\xi} - J D_t - K D_{\tau} ) \mathcal{A}^a
      + I (F_{y \xi} + F_{\tau t})
      + J (F_{y t} + F_{\xi \tau})
    \\
    & \qquad \qquad \qquad \qquad \quad \,
      + K (F_{y \tau} + F_{t \xi})
      + g^{a\bar{b}}_{\mathfrak{A}_4} \left(
      \frac{K\zeta}{2} \pdv{\mathfrak{W}_4}{\mathcal{A}^b}
      \right)^*
      \bigg|^2
      + \dots
      \Bigg)
    \\
    &= \frac{1}{e^2} \int dt d\tau d\xi dy \Bigg(
      \left| D_{\varpi} \mathcal{A}^a \right|^2
      + \left| \pdv{\mathfrak{W}_4}{\mathcal{A}^b} \right|^2
      + \left| F_{y \xi} \right|^2
      + \left| F_{\tau t} \right|^2
      + \left| F_{y t} \right|^2
      + \left| F_{\xi \tau} \right|^2
    \\
    & \qquad \qquad \qquad \qquad \quad \,
      + \left| F_{y \tau} \right|^2
      + \left| F_{t \xi} \right|^2
      + \dots
      \Bigg)
      \, ,
  \end{aligned}
\end{equation}
where $\varpi$ is the index on the $\R^4$ worldvolume.
In other words, our 4d gauged sigma model can also be interpreted as a 4d gauged LG model in $\mathfrak{A}_4$ with holomorphic superpotential $\mathfrak{W}_4(\mathcal{A})$.
Noting that the gradient vector field of $\mathfrak{W}_4(\mathcal{A})$ is the $\theta$-independent part of $q^a$, i.e., $\mathcal{F}$, we find that the holomorphic superpotential $\mathfrak{W}_4(\mathcal{A})$ must therefore be $CS(\mathcal{A})$, a \emph{Chern-Simons function of $\mathcal{A}$}.
Furthermore, $\mathfrak{W}_4(\mathcal{A})$ is \emph{complex-valued} (since $\mathcal{A}$ is complex).

Another important fact to note at this point is that the BPS worldvolume equation \eqref{eq:cy2 x r4:worldvolume eqn} (i.e., the BPS equation of the 4d gauged sigma model, and thus the 4d gauged LG model) is a non-constant, $\theta$-deformed, gauged Cauchy-Riemann-Fueter equation for the $\mathcal{A}^a$ fields (that will correspond to \eqref{eq:cy2 x r4:2d lg:worldsheet eqn} in the 2d gauged LG model).

By setting $d_y \mathcal{A}^a = 0$ and $A_y, A_{\xi}, A_{\tau}, A_t \rightarrow 0$ in the expression within the squared term in \eqref{eq:cy2 x r4:4d lg:action}, we can read off the LG $\mathfrak{A}_4^{\theta}$-threebrane equation corresponding to $\Upsilon^{\{IJ, KL\}, \{MN, PQ\}}_{\pm}(\xi, \tau, t, \theta, \mathfrak{A}_4)$ (that re-expresses \eqref{eq:cy2 x r4:threebrane eqn} with $A_y, A_{\xi}, A_{\tau}, A_t \rightarrow 0$) as
\begin{equation}
  \label{eq:cy2 x r4:4d lg:threebrane eqn}
  I \dv{\mathcal{A}^a}{\xi} + J \dv{\mathcal{A}^a}{t} + K \dv{\mathcal{A}^a}{\tau}
  = g^{a\bar{b}}_{\mathfrak{A}_4} \left(
    \frac{K\zeta}{2} \pdv{\mathfrak{W}_4}{\mathcal{A}^b}
  \right)^*
  \, .
\end{equation}
By setting $d_{\xi} \mathcal{A}^a = 0$ in \eqref{eq:cy2 x r4:4d lg:threebrane eqn}, we can read off
(i) the LG $\mathfrak{A}_4^{\theta}$-threebrane face equation corresponding to $\Upsilon^{\{IJ, KL\}, \{MN, PQ\}}_{\pm}(\pm \infty, \tau, t, \theta, \mathfrak{A}_4)$, or equivalently,
(ii) the LG $\mathfrak{A}_4^{\theta}$-sheet equation corresponding to $\Sigma^{IJ, KL}(\tau, t, \theta, \mathfrak{A}_4)$ and $\Sigma^{MN, PQ}(\tau, t, \theta, \mathfrak{A}_4)$,
(that re-expresses \eqref{eq:cy2 x r4:sheet}) as
\begin{equation}
  \label{eq:cy2 x r4:4d lg:sheet eqn}
  J \dv{\mathcal{A}^a}{t} + K \dv{\mathcal{A}^a}{\tau}
  = g^{a\bar{b}}_{\mathfrak{A}_4} \left(
    \frac{K\zeta}{2} \pdv{\mathfrak{W}_4}{\mathcal{A}^b}
  \right)^*
  \, .
\end{equation}
By setting $d_{\tau} \mathcal{A}^a = 0$ in \eqref{eq:cy2 x r4:4d lg:sheet eqn}, we can read off
(i) the LG $\mathfrak{A}_4^{\theta}$-threebrane edge equation corresponding to $\Upsilon^{\{IJ, KL\}, \{MN, PQ\}}_{\pm}(\pm \infty, \pm \infty, t, \theta, \mathfrak{A}_4)$ and $\Upsilon^{\{IJ, KL\}, \{MN, PQ\}}_{\pm}(\pm \infty, \mp \infty, t, \theta, \mathfrak{A}_4)$
(ii) the LG $\mathfrak{A}_4^{\theta}$-sheet edge equation corresponding to $\Sigma^{IJ, KL}(\pm \infty, t, \theta, \mathfrak{A}_4)$ and $\Sigma^{MN, PQ}(\pm \infty, t, \theta, \mathfrak{A}_4)$, or equivalently,
(iii) the LG $\mathfrak{A}_4^{\theta}$-soliton equation corresponding to $\Gamma^{IJ}(t, \theta, \mathfrak{A}_4)$, $\Gamma^{KL}(t, \theta, \mathfrak{A}_4)$, $\Gamma^{MN}(t, \theta, \mathfrak{A}_4)$, and $\Gamma^{PQ}(t, \theta, \mathfrak{A}_4)$,
(that re-expresses \eqref{eq:cy2 x r4:soliton}) as
\begin{equation}
  \label{eq:cy2 x r4:4d lg:soliton eqn}
  J \dv{\mathcal{A}^a}{t}
  = g^{a\bar{b}}_{\mathfrak{A}_4} \left(
    \frac{K\zeta}{2} \pdv{\mathfrak{W}_4}{\mathcal{A}^b}
  \right)^*
  \, .
\end{equation}
By setting $d_t \mathcal{A}^a = 0$, we can read off
(i) the LG $\mathfrak{A}_4^{\theta}$-threebrane vertex equation corresponding to $\Upsilon^{\{IJ, KL\}, \{MN, PQ\}}_{\pm}(\pm \infty, \pm \infty, \pm \infty, \theta, \mathfrak{A}_4)$, $\Upsilon^{\{IJ, KL\}, \{MN, PQ\}}_{\pm}(\pm \infty, \pm \infty, \mp \infty, \theta, \mathfrak{A}_4)$, \\ $\Upsilon^{\{IJ, KL\}, \{MN, PQ\}}_{\pm}(\pm \infty, \mp \infty, \pm \infty, \theta, \mathfrak{A}_4)$, and $\Upsilon^{\{IJ, KL\}, \{MN, PQ\}}_{\pm}(\pm \infty, \mp \infty, \mp \infty, \theta, \mathfrak{A}_4)$,
(ii) the LG $\mathfrak{A}_4^{\theta}$-sheet vertex equation corresponding to $\Sigma^{**, **}(\pm \infty, \pm \infty, \theta, \mathfrak{A}_4)$ and $\Sigma^{**, **}(\pm \infty, \mp \infty, \theta, \mathfrak{A}_4)$, or equivalently,
(iii) the LG $\mathfrak{A}_4^{\theta}$-soliton endpoint equation corresponding to $\Gamma^{**}(\pm \infty, \theta, \mathfrak{A}_4)$,
(that re-expresses \eqref{eq:cy2 x r4:soliton} with $d_t \mathcal{A}^a = 0$) as
\begin{equation}
  \label{eq:cy2 x r4:4d lg:soliton:endpt eqn}
  g^{a\bar{b}}_{\mathfrak{A}_4} \left(
    \frac{K\zeta}{2} \pdv{\mathfrak{W}_4}{\mathcal{A}^b}
  \right)^*
  = 0
  \, .
\end{equation}

Recall from the end of \autoref{sec:cy2 x r4:vw} that we are only considering certain $CY_2$ such that (the endpoints $\Gamma^{**}(\pm \infty, \theta, \mathfrak{A}_4)$ and thus) the LG $\mathfrak{A}_4^{\theta}$-solitons, and effectively, (the vertices at $\xi, \tau, t = \pm \infty$ of $\Upsilon^{\{IJ, KL\}, \{MN, PQ\}}_{\pm}(\xi, \tau, t, \theta, \mathfrak{A}_4)$ and thus) the LG $\mathfrak{A}_4^{\theta}$-threebranes, are isolated and non-degenerate.
Therefore, from their definition in \eqref{eq:cy2 x r4:4d lg:soliton:endpt eqn} which tells us that they correspond to critical points of $\mathfrak{W}_4(\mathcal{A})$, we conclude that $\mathfrak{W}_4(\mathcal{A})$ can be regarded as a holomorphic Morse function in $\mathfrak{A}_4$.

Just like in \autoref{sec:m2 x r3:fueter-cat}, this means an LG $\mathfrak{A}_4^{\theta}$-soliton $\Gamma^{IJ}(t, \theta, \mathfrak{A}_4)$ defined in \eqref{eq:cy2 x r4:4d lg:soliton eqn} maps to a straight line segment $[\mathfrak{W}^I_4(\theta), \mathfrak{W}^J_4(\theta)]$ in the complex $\mathfrak{W}_4$-plane that starts and ends at critical values $\mathfrak{W}^I_4(\theta) \equiv \mathfrak{W}_4(\Gamma^I(-\infty, \theta, \mathfrak{A}_4))$ and $\mathfrak{W}^J_4(\theta) \equiv \mathfrak{W}_4(\Gamma^J(+\infty, \theta, \mathfrak{A}_4))$, respectively, where its slope depends on $\theta$ (via $\zeta$).
Therefore, an LG $\mathfrak{A}_4^{\theta}$-sheet $\Sigma^{IJ, KL}(\tau, t, \theta, \mathfrak{A}_4)$ defined in \eqref{eq:cy2 x r4:4d lg:sheet eqn} maps to a quadrilateral in the complex $\mathfrak{W}_4^{\theta}$-plane, whose edges are the straight line segments that the LG $\mathfrak{A}_4^{\theta}$-solitons map to, and whose bottom-left, top-left, bottom-right, top-right vertices are the critical points $\mathfrak{W}^I_4(\theta) \equiv \mathfrak{W}_4(\Sigma^I(-\infty, -\infty, \theta, \mathfrak{A}_4))$, $\mathfrak{W}^J_4(\theta) \equiv \mathfrak{W}_4(\Sigma^J(-\infty, +\infty, \theta, \mathfrak{A}_4))$, $\mathfrak{W}^K_4(\theta) \equiv \mathfrak{W}_4(\Sigma^K(+\infty, -\infty, \theta, \mathfrak{A}_4))$, and $\mathfrak{W}^L_4(\theta) \equiv \mathfrak{W}_4(\Sigma^L(+\infty, +\infty, \theta, \mathfrak{A}_4))$, respectively, where the slopes between each left-right vertex pair depends on $\theta$ (via $\zeta$).

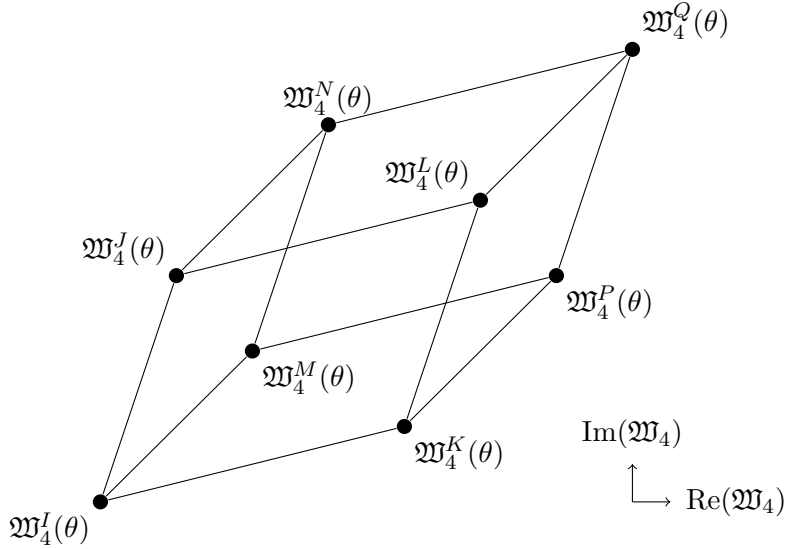
\begin{figure}
  \centering
  \begin{tikzpicture}[%
    auto,%
    every edge/.style={draw},%
    relation/.style={scale=1, sloped, anchor=center, align=center,%
      color=black},%
    vertRelation/.style={scale=1, anchor=center, align=center},%
    dot/.style={circle, fill, minimum size=2*\radius, node contents={},%
      inner sep=0pt},%
    ]
    \let\radius\undefined
    \newlength{\radius}
    \setlength{\radius}{1mm}
    \node (ftl) at (-2,3) [dot];
    \node at (ftl) [above left] {$\mathfrak{W}^J_4(\theta)$};
    \node (ftr) at (2,4) [dot];
    \node at (ftr) [above left]  {$\mathfrak{W}^L_4(\theta)$};
    \node (fbl) at (-3,0) [dot];
    \node at (fbl) [below left]  {$\mathfrak{W}^I_4(\theta)$};
    \node (fbr) at (1,1) [dot];
    \node at (fbr) [below right]  {$\mathfrak{W}^K_4(\theta)$};
    \node (btl) at (0,5) [dot];
    \node at (btl) [above]  {$\mathfrak{W}^N_4(\theta)$};
    \node (btr) at (4,6) [dot];
    \node at (btr) [above right]  {$\mathfrak{W}^Q_4(\theta)$};
    \node (brl) at (-1,2) [dot];
    \node at (brl) [below right]  {$\mathfrak{W}^M_4(\theta)$};
    \node (bbr) at (3,3) [dot];
    \node at (bbr) [below right]  {$\mathfrak{W}^P_4(\theta)$};
    \draw
    (fbl) edge (ftl)
    (ftl) edge (btl)
    (brl) edge (btl)
    (brl) edge (fbl)
    (fbr) edge (ftr)
    (ftr) edge (btr)
    (bbr) edge (btr)
    (bbr) edge (fbr)
    (fbl) edge (fbr)
    (ftl) edge (ftr)
    (brl) edge (bbr)
    (btl) edge (btr)
    ;
    \coordinate (cco) at (4, 0);
    \coordinate (ccx) at (4.5, 0);
    \coordinate (ccz) at (4, 0.5);
    \node at (ccx) [right=2pt of ccx] {$\Re(\mathfrak{W}_4)$};
    \node at (ccz) [above=2pt of ccz] {$\Im(\mathfrak{W}_4)$};
    \draw[->] (cco) -- (ccx);
    \draw[->] (cco) -- (ccz);
  \end{tikzpicture}
  \caption{An LG $\mathfrak{A}_4^\theta$-threebrane $\Upsilon^{\{IJ, KL\}, \{MN, PQ\}}(\xi, \tau, t, \theta, \mathfrak{A}_4)$ that is mapped to the complex $\mathfrak{W}_4$-plane.
  }
  \label{fig:cy2 x r4:frakA-threebrane projection}
\end{figure}

Since the faces of the LG $\mathfrak{A}_4^{\theta}$-threebrane are LG $\mathfrak{A}_4^{\theta}$-sheets, the LG $\mathfrak{A}_4^{\theta}$-threebrane can be mapped to the complex $\mathfrak{W}_4$-plane as shown in \autoref{fig:cy2 x r4:frakA-threebrane projection}.
The vertices are critical points $\mathfrak{W}_4^I \equiv \mathfrak{W}_4(\Upsilon^I(- \infty, - \infty, - \infty, \theta, \mathfrak{A}_4))$, \dots, $\mathfrak{W}_4^Q \equiv \mathfrak{W}_4(\Upsilon^Q(+ \infty, + \infty, + \infty, \theta, \mathfrak{A}_4))$, and the slope of the line segments between each front-back vertex pair depends on $\theta$ (via $\zeta$).

We shall also assume that
(i) $\Re(\mathfrak{W}_4^I(\theta)) < \{ \Re(\mathfrak{W}_4^J(\theta)), \Re(\mathfrak{W}_4^K(\theta)), \Re(\mathfrak{W}_4^M(\theta)) \}$,
(ii) $\Re(\mathfrak{W}_4^J(\theta)) < \{ \Re(\mathfrak{W}_4^L(\theta)), \Re(\mathfrak{W}_4^N(\theta)) \}$,
(iii) $\Re(\mathfrak{W}_4^K(\theta)) < \{ \Re(\mathfrak{W}_4^L(\theta)), \Re(\mathfrak{W}_4^P(\theta)) \}$,
(iv) $\Re(\mathfrak{W}_4^M(\theta)) < \{ \Re(\mathfrak{W}_4^N(\theta)),$ $\Re(\mathfrak{W}_4^P(\theta)) \}$,
and (v) $\{ \Re(\mathfrak{W}_4^L(\theta)), \Re(\mathfrak{W}_4^N(\theta)), \Re(\mathfrak{W}_4^P(\theta)) \} < \Re(\mathfrak{W}_4^Q(\theta))$,
as depicted in~\autoref{fig:cy2 x r4:frakA-threebrane projection}.

\subtitle{The 4d Gauged LG Model as an LG SQM}

Last but not least, after suitable rescalings, we can recast \eqref{eq:cy2 x r4:4d lg:action} as a 1d LG SQM (that re-expresses \eqref{eq:cy2 x r4:sqm action}), where its action will be given by\footnote{%
  In the following expression, we have integrated out $A_y$ and omitted the fields corresponding to the finite-energy gauge fields $A_{\{t, \tau, \xi, y\}}$ (as explained in \autoref{ft:stokes theorem for cy2 x r4:sqm}).
  \label{ft:stokes theorem for cy2 x r4:4d lg sqm}
}
\begin{equation}
  \label{eq:cy2 x r4:4d lg:sqm action}
  S_{\text{4d-LG SQM}, \mathcal{M}(\R^3, \mathfrak{A}_4)}
  = \frac{1}{e^2} \int dy
  \left(
    \left|
      \left( \dv{\mathcal{A}^u}{y}
        + \dv{\widehat{A}^u}{y}
      \right)
      + g^{uv}_{\mathcal{M}(\R^3, \mathfrak{A}_4)}
      \left( \pdv{\mathfrak{G}_4}{\mathcal{A}^v}
        + \pdv{\mathfrak{G}_4}{\widehat{A}^v}
      \right)
    \right|^2
    + \dots
  \right)
  \, .
\end{equation}
Here $\mathfrak{G}_4(\mathcal{A}, \widehat{A})$ is the \emph{real-valued} potential in $\mathcal{M}(\R^3, \mathfrak{A}_4)$, and the subscript ``$\text{4d-LG SQM}, \mathcal{M}(\R^3, \mathfrak{A}_4)$'' is to specify that it is a 1d SQM with target space $\mathcal{M}(\R^3, \mathfrak{A}_4)$ obtained from the equivalent 4d LG model.
We will also refer to this \emph{1d} LG SQM as ``4d-LG SQM'' in the rest of this subsection.

The 4d-LG SQM will localize onto configurations that \emph{simultaneously} set to zero the LHS and RHS of the expression within the squared term in \eqref{eq:cy2 x r4:4d lg:sqm action}.
In other words, it will localize onto $y$-invariant critical points of $\mathfrak{G}_4(\mathcal{A}, \widehat{A})$ that will correspond, when $A_y, A_{\xi}, A_{\tau}, A_t \rightarrow 0$, to the LG $\mathfrak{A}_4^{\theta}$-threebranes defined by \eqref{eq:cy2 x r4:4d lg:threebrane eqn}.
For our choice of $CY_2$, the LG $\mathfrak{A}_4^{\theta}$-threebranes, just like their vertices, will be isolated and non-degenerate.
Thus, $\mathfrak{G}_4(\mathcal{A}, \widehat{A})$ can be regarded as a \emph{real-valued} Morse functional in $\mathcal{M}(\R^3, \mathfrak{A}_4)$.

\subtitle{Morphisms between $\mathfrak{A}_4^{\theta}$-sheets as Intersection Floer Homology Classes}

Applying a similar analysis to that in \autoref{sec:m2 x r3:fueter-cat} with \eqref{eq:cy2 x r4:4d lg:sqm action} as the action of the 4d-LG SQM, we find that we can interpret the LG $\mathfrak{A}_4^{\theta}$-soliton solution $\Gamma^{IJ}(t, \theta, \mathfrak{A}_4)$ as a thimble-intersection, the LG $\mathfrak{A}_4^{\theta}$-sheet solution $\Sigma^{IJ, KL}(\tau, t, \theta, \mathfrak{A}_4)$ as an intersection of thimble-intersections for which we shall call an ``intersecting pair of thimble-intersections'', and finally, the LG $\mathfrak{A}_4^{\theta}$-threebrane solution $\Upsilon^{\{IJ, KL\}, \{MN, PQ\}}(\xi, \tau, t, \theta, \mathfrak{A}_4)$ as an intersection of ``intersecting pairs of thimble-intersections''.

Specifically, a $\Gamma^{IJ}(t, \theta, \mathfrak{A}_4)$-soliton, whose bottom and top endpoints correspond to $\mathcal{E}^I_{\text{VW}}(\theta)$ and $\mathcal{E}^J_{\text{VW}}(\theta)$, respectively, can be identified as an intersection point $q^{IJ}_{\text{VW}, \pm}(\theta) \in S^{IJ}_{\text{VW}}(\theta)$ of a bottom and top thimble in the fiber space over the line segment $[\mathfrak{W}_4^I(\theta), \mathfrak{W}_4^J(\theta))]$.
In turn, a $\Sigma^{IJ, KL}(\tau, t, \theta, \mathfrak{A}_4)$-sheet pair, whose left and right edges correspond to $\Gamma^{IJ}(t, \theta, \mathfrak{A}_4)$ and $\Gamma^{KL}(t, \theta, \mathfrak{A}_4)$, respectively, can be identified as a pair of intersection points $\{ q^{IJ}_{\text{VW}}(\theta), q^{KL}_{\text{VW}}(\theta) \} \eqqcolon \mathfrak{P}^{IJ, KL}_{\text{VW}}(\theta) \in S^{IJ}_{\text{VW}}(\theta) \bigcap S^{KL}_{\text{VW}}(\theta)$ of a left and right thimble-intersection in the fiber space over the quadrilateral with vertices $(\mathfrak{W}_4^I(\theta), \mathfrak{W}_4^J(\theta), \mathfrak{W}_4^K(\theta), \mathfrak{W}_4^L(\theta))$.

As a result, a $\Upsilon^{\{IJ, KL\}, \{MN, PQ\}}_{\pm}(\xi, \tau, t, \theta, \mathfrak{A}_4)$-threebrane pair, whose front and back faces correspond to $\Sigma^{IJ, KL}(\tau, t, \theta, \mathfrak{A}_4)$ and $\Sigma^{MN, PQ}(\tau, t, \theta, \mathfrak{A}_4)$, respectively, can be identified as a pair of intersection points $\{\mathfrak{P}^{IJ, KL}_{\text{VW}, \pm}(\theta), \mathfrak{P}^{MN, PQ}_{\text{VW}, \pm}(\theta) \} \eqqcolon \mathfrak{Q}^{\{IJ, KL\}, \{MN, PQ\}}_{\text{VW}, \pm}(\theta) \in (S^{IJ}_{\text{VW}}(\theta) \cap S^{KL}_{\text{VW}}(\theta)) \bigcap (S^{MN}_{\text{VW}}(\theta) \cap S^{PQ}_{\text{VW}}(\theta))$ of a front and back ``intersecting pair of thimble-intersections'' in the fiber space over complex $\mathfrak{W}_4$-plane.

At any rate, the 4d-LG SQM in $\mathcal{M}(\R^3, \mathfrak{A}_4)$ with action \eqref{eq:cy2 x r4:4d lg:sqm action} will physically realize a Floer homology that we shall name an $\mathfrak{A}_4$-4d-LG Floer homology.
The chains of the $\mathfrak{A}_4$-4d-LG Floer complex are generated by LG $\mathfrak{A}_4^{\theta}$-threebranes which we can thus identify with $\mathfrak{Q}^{\{**, **\}, \{**, **\}}_{\text{VW}, \pm}(\theta)$, and the $\mathfrak{A}_4$-4d-LG Floer differential will be realized by the flow lines governed by the gradient flow equation satisfied by $y$-varying configurations which set the expression within the squared term of \eqref{eq:cy2 x r4:4d lg:sqm action} to zero.
The partition function of the 4d-LG SQM in $\mathcal{M}(\R^3, \mathfrak{A}_4)$ will be given by\footnote{%
  The `$\theta$' label is omitted in the LHS of the following expression (as explained in \autoref{ft:theta omission in m2-2d lg partition fn}).
}
\begin{equation}
  \label{eq:cy2 x r4:4d lg:partition fn}
  \mathcal{Z}_{\text{4d-LG SQM}, \mathcal{M}(\R^3, \mathfrak{A}_4)}(G)
  = \sum_{\substack{I \neq J \neq K \neq L \neq M \\ \neq N \neq P \neq Q = 1}}
  \sum_{\substack{\mathfrak{Q}^{\{IJ, KL\}, \{MN, PQ\}}_{\text{VW}, \pm} \\ \in \left( S^{IJ}_{\text{VW}} \cap S^{KL}_{\text{VW}} \right) \bigcap \left( S^{MN}_{\text{VW}} \cap S^{PQ}_{\text{VW}} \right)}}
  \text{HF}^G_{d_v} \left(
    \mathfrak{Q}^{\{IJ, KL\}, \{MN, PQ\}}_{\text{VW}, \pm} (\theta)
  \right)
  \, ,
\end{equation}
where the contribution $\text{HF}^G_{d_v} (\mathfrak{Q}^{\{IJ, KL\}, \{MN, PQ\}}_{\text{VW}, \pm} (\theta))$ can be identified with a homology class in an $\mathfrak{A}_4$-4d-LG Floer homology generated by intersection points of ``intersecting pairs of thimble-intersections''.
These intersection points represent LG $\mathfrak{A}_4^{\theta}$-threebranes defined by \eqref{eq:cy2 x r4:4d lg:threebrane eqn}, whose faces correspond to LG $\mathfrak{A}_4^{\theta}$-sheets defined by \eqref{eq:cy2 x r4:4d lg:sheet eqn}, whose edges correspond to LG $\mathfrak{A}_4^{\theta}$-solitons defined by \eqref{eq:cy2 x r4:4d lg:soliton eqn}, and whose vertices defined by \eqref{eq:cy2 x r4:4d lg:soliton:endpt eqn} will correspond to $\theta$-deformed VW configurations on $CY_2$.
The degree of each chain in the complex is $d_v$, and is counted by the number of outgoing flow lines from the fixed critical points of $\mathfrak{G}_4(\mathcal{A}, \widehat{A})$ in $\mathcal{M}(\R^3, \mathfrak{A}_4)$ which can also be identified as $\mathfrak{Q}^{\{IJ, KL\}, \{MN, PQ\}}_{\text{VW}, \pm}(\theta)$.

Therefore, $\mathcal{Z}_{\text{4d-LG SQM}, \mathcal{M}(\R^3, \mathfrak{A}_4)}(G)$ in \eqref{eq:cy2 x r4:4d lg:partition fn} is a sum of LG $\mathfrak{A}_4^{\theta}$-sheets defined by (i)~\eqref{eq:cy2 x r4:4d lg:threebrane eqn} with (ii) faces \eqref{eq:cy2 x r4:4d lg:sheet eqn}, (iii) edges \eqref{eq:cy2 x r4:4d lg:soliton eqn}, and (iv) vertices \eqref{eq:cy2 x r4:4d lg:soliton:endpt eqn}, or equivalently, $\Upsilon^{\{IJ, KL\}, \{MN, PQ\}}_{\pm}(\xi, \tau, t, \theta, \mathfrak{A}_4)$-threebranes defined by (i) \eqref{eq:cy2 x r4:threebrane eqn} and \eqref{eq:cy2 x r4:threebrane eqn:aux cond} (with $A_y, A_{\xi}, A_{\tau}, A_t \rightarrow 0$) with (ii) faces \eqref{eq:cy2 x r4:sheet} and \eqref{eq:cy2 x r4:threebrane eqn:aux cond}, (iii) edges \eqref{eq:cy2 x r4:soliton} and \eqref{eq:cy2 x r4:threebrane eqn:aux cond}, and (iv) vertices \eqref{eq:cy2 x r4:soliton} and \eqref{eq:cy2 x r4:threebrane eqn:aux cond} (with $d_t \mathcal{A}^a = 0$), respectively.
In other words, we can write
\begin{equation}
  \label{eq:cy2 x r4:4d lg:floer-hom as vector}
  \text{CF}_{\mathcal{M}(\R^3, \mathfrak{A}_4)} \left(
    \Sigma^{IJ, KL}(\tau, t, \theta, \mathfrak{A}_4),
    \Sigma^{MN, PQ}(\tau, t, \theta, \mathfrak{A}_4)
  \right)_{\pm}
  =
  \text{HF}^G_{d_v} \left(
    \mathfrak{Q}^{\{IJ, KL\}, \{MN, PQ\}}_{\text{VW}, \pm} (\theta)
  \right)
  \, ,
\end{equation}
where the LHS is a vector representing a $\Upsilon^{\{IJ, KL\}, \{MN, PQ\}}_{\pm}(\xi, \tau, t, \theta, \mathfrak{A}_4)$-threebrane, as depicted in \autoref{fig:cy2 x r4:frakA-threebrane}.
Also, recall that
(i) $\Re(\mathfrak{W}_4^I(\theta)) < \{ \Re(\mathfrak{W}_4^J(\theta)), \Re(\mathfrak{W}_4^K(\theta)), \Re(\mathfrak{W}_4^M(\theta)) \}$,
(ii) $\Re(\mathfrak{W}_4^J(\theta)) < \{ \Re(\mathfrak{W}_4^L(\theta)), \Re(\mathfrak{W}_4^N(\theta)) \}$,
(iii) $\Re(\mathfrak{W}_4^K(\theta)) < \{ \Re(\mathfrak{W}_4^L(\theta)), \Re(\mathfrak{W}_4^P(\theta)) \}$,
(iv) $\Re(\mathfrak{W}_4^M(\theta)) < \{ \Re(\mathfrak{W}_4^N(\theta)),$ $\Re(\mathfrak{W}_4^P(\theta)) \}$,
and (v) $\{ \Re(\mathfrak{W}_4^L(\theta)), \Re(\mathfrak{W}_4^N(\theta)), \Re(\mathfrak{W}_4^P(\theta)) \} < \Re(\mathfrak{W}_4^Q(\theta))$.

Just like in \autoref{sec:m2 x r3:fueter-cat}, we can regard a $\Sigma^{IJ, KL}(\tau, t, \theta, \mathfrak{A}_4)$-sheet as a 1-morphism $\text{Hom}(\Gamma^{IJ}(t, \theta, \mathfrak{A}_4),$ $\Gamma^{KL}(t, \theta, \mathfrak{A}_4))$, from its left edge to its right edge corresponding to $\Gamma^{IJ}(t, \theta, \mathfrak{A}_4)$ and $\Gamma^{KL}(t, \theta, \mathfrak{A}_4)$, and a 2-morphism $\text{Hom}( \text{Hom}(\mathcal{E}^I_{\text{VW}}(\theta), \mathcal{E}^J_{\text{VW}}(\theta)), \text{Hom}(\mathcal{E}^K_{\text{VW}}(\theta), \mathcal{E}^L_{\text{VW}}(\theta)) )$, amongst its vertices $\{ \mathcal{E}^I_{\text{VW}}(\theta),$ $\mathcal{E}^J_{\text{VW}}(\theta), \mathcal{E}^K_{\text{VW}}(\theta), \mathcal{E}^L_{\text{VW}}(\theta) \}$.
Thus, we have the following one-to-one identifications\footnote{%
  The `$\theta$' label is omitted in the following expression (as explained in \autoref{ft:omission of theta in m2 2d-lg}).
  \label{ft:omission of theta in cy2 x r4 4d-lg}
}
\begin{equation}
  \label{eq:cy2 x r4:4d-lg:3-morphism}
  \boxed{
    \begin{gathered}
      \text{Hom} \left[
        \Sigma^{IJ, KL}(\tau, t, \mathfrak{A}_4),
        \Sigma^{MN, PQ}(\tau, t, \mathfrak{A}_4)
      \right]_{\pm}
      \\
      \Updownarrow
      \\
      \text{Hom} \left[
        \text{Hom} \left(
          \Gamma^{IJ}(t, \mathfrak{A}_4),
          \Gamma^{KL}(t, \mathfrak{A}_4)
        \right),
        \text{Hom} \left(
          \Gamma^{MN}(t, \mathfrak{A}_4),
          \Gamma^{PQ}(t, \mathfrak{A}_4)
        \right)
      \right]_{\pm}
      \\
      \Updownarrow
      \\
      \text{Hom} \left[
        \text{Hom} \Big(
          \text{Hom} \left(
            \mathcal{E}^I_{\text{VW}},
            \mathcal{E}^J_{\text{VW}}
          \right),
          \text{Hom} \left(
            \mathcal{E}^K_{\text{VW}},
            \mathcal{E}^L_{\text{VW}}
          \right)
        \Big),
        \text{Hom} \Big(
          \text{Hom} \left(
            \mathcal{E}^M_{\text{VW}},
            \mathcal{E}^N_{\text{VW}}
          \right),
          \text{Hom} \big(
            \mathcal{E}^P_{\text{VW}},
            \mathcal{E}^Q_{\text{VW}}
          \big)
        \Big)
      \right]_{\pm}
      \\
      \Updownarrow
      \\
      \text{HF}^G_{d_v} \left(
        \mathfrak{Q}^{\{IJ, KL\}, \{MN, PQ\}}_{\text{VW}, \pm}
      \right)
    \end{gathered}
  }
\end{equation}
where the bottom-most entry is proportional to the identity class when $I = M$, $J = N$, $K = P$, and $L = Q$, and zero when (i) $I \leftrightarrow M$, $J \leftrightarrow N$, $K \leftrightarrow P$, and $L \leftrightarrow Q$ (since the $\Upsilon^{\{IJ, KL\}, \{MN, PQ\}}(\xi, \tau, t, \theta, \mathfrak{A}_4)$-threebrane only moves in one direction from $\Sigma^{IJ, KL}(\tau, t, \theta, \mathfrak{A}_4)$ to $\Sigma^{MN, PQ}(\tau, t, \theta, \mathfrak{A}_4)$, as depicted in \autoref{fig:cy2 x r4:frakA-threebrane}), (ii) $I \leftrightarrow K$ and $J \leftrightarrow L$ (since the $\Sigma^{IJ, KL}(\tau, t, \theta, \mathfrak{A}_4)$-sheet only moves in one direction from $\Gamma^{IJ}(t, \theta, \mathfrak{A}_4)$ to $\Gamma^{KL}(t, \theta, \mathfrak{A}_4)$) and likewise for $M \leftrightarrow P$ and $N \leftrightarrow Q$, and (iii) $I \leftrightarrow J$ (since the $\Gamma^{IJ}(t, \theta, \mathfrak{A}_4)$-soliton only moves in one direction from $\mathcal{E}^I_{\text{VW}}(\theta)$ to $\mathcal{E}^J_{\text{VW}}(\theta)$) and likewise for $K \leftrightarrow L$, $M \leftrightarrow N$, or $P \leftrightarrow Q$.

\subtitle{Soliton Threebrane Theory from the 4d LG Model}

Just like the 4d gauged sigma model, the equivalent 4d gauged LG model with target space $\mathfrak{A}_4$ will define an open threebrane theory in $\mathfrak{A}_4$ with effective worldvolumes and boundaries described at the end of \autoref{sec:cy2 x r4:2d-4d model}.

The dynamics of this open threebrane theory in $\mathfrak{A}_4$ will be governed by the BPS worldvolume equation of \eqref{eq:cy2 x r4:worldvolume eqn}, where $\mathcal{A}^a$ are scalars on the worldvolume corresponding to the holomorphic coordinates of $\mathfrak{A}_4$.
At an arbitrary instant in time whence $d_y \mathcal{A}^a = 0 = d_y A_{\{t, \tau, \xi\}}$ in \eqref{eq:cy2 x r4:worldvolume eqn}, the dynamics of $\mathcal{A}^a$ and the 4d gauge fields $(A_t, A_{\tau}, A_{\xi}, A_y)$ along $(\xi, \tau, t)$ will be governed by the threebrane equation
\begin{equation}
  \label{eq:cy2 x r4:4d lg:soliton threebrane eqn}
  \begin{aligned}
    & I \dv{\mathcal{A}^a}{\xi} + J \dv{\mathcal{A}^a}{t} + K \dv{\mathcal{A}^a}{\tau}
      + I \left( \dv{A_y}{\xi} - \dv{A_t}{\tau} + \dv{A_{\tau}}{t} \right)
      + J \left( \dv{A_y}{t} - \dv{A_{\tau}}{\xi} + \dv{A_{\xi}}{\tau} \right)
      + K \left( \dv{A_y}{\tau} - \dv{A_{\xi}}{t} + \dv{A_t}{\xi} \right)
    \\
    &= [A_y, \mathcal{A}^a]
      - [I A_{\xi} + J A_t + K A_{\tau}, \mathcal{A}^a + A_y]
      + I [A_{\tau}, A_t]
      + J [A_{\xi}, A_{\tau}]
      + K [A_t, A_{\xi}]
      + g^{a\bar{b}}_{\mathfrak{A}_4} \left(
      \frac{K\zeta}{2} \pdv{\mathfrak{W}_4}{\mathcal{A}^b}
      \right)^*
      \, .
  \end{aligned}
\end{equation}
Just like how the membrane equations of \eqref{eq:m2 x r3:soliton sheet:eqn} can be seen as 2d soliton membrane equations, we can see \eqref{eq:cy2 x r4:4d lg:soliton threebrane eqn} as a 3d soliton threebrane equation.

Hence, just as how the 3d gauged LG model in \autoref{sec:m2 x r3:fueter-cat} can be interpreted as a \emph{soliton membrane theory} whose dynamics are governed by the soliton membrane equations of \eqref{eq:m2 x r3:soliton sheet:eqn}, we can interpret our 4d gauged LG model as a \emph{soliton threebrane theory} whose dynamics are governed by the soliton threebrane equation of~\eqref{eq:cy2 x r4:4d lg:soliton threebrane eqn}.

\subtitle{The Normalized Spin$(7)$ Partition Function, Soliton Threebrane Scattering, and Maps of an $A_{\infty}$-structure}

The normalized Spin$(7)$ partition function can be regarded as a sum over tree-level scattering amplitudes of LG $\mathfrak{A}_4^{\theta}$-threebranes defined by \eqref{eq:cy2 x r4:4d lg:threebrane eqn}.
In other words, we can express the normalized Spin$(7)$ partition function as
\begin{equation}
  \label{eq:cy2 x r4:4d lg:partition function}
  \mathcal{Z}_{\text{Spin}(7), CY_2 \times \R^4}(G)
  = \sum_{\mathfrak{K}_n} \varDelta^{\mathfrak{K}_n}_{\mathfrak{A}_4}
  \, ,
  \quad
  \mathfrak{K}_n = 1, 2, \dots, \left\lfloor \frac{n - 4}{4} \right\rfloor
\end{equation}
where each
\begin{equation}
  \label{eq:cy2 x r4:4d lg:crf composition maps}
  \boxed{
    \begin{aligned}
      \varDelta^{\mathfrak{K}_n}_{\mathfrak{A}_4}
      : \bigotimes_{i = 1}^{\mathfrak{K}_n}
      & \text{Hom} \bigg[
        \text{Hom} \left(
        \text{Hom} \left(
        \mathcal{E}^{I_{4i-3}}_{\text{VW}},
        \mathcal{E}^{I_{4i-2}}_{\text{VW}}
        \right),
        \text{Hom} \left(
        \mathcal{E}^{I_{4i-1}}_{\text{VW}},
        \mathcal{E}^{I_{4i}}_{\text{VW}}
        \right)
        \right),
      \\
      & \qquad \quad
        \text{Hom} \left(
        \mathcal{E}^{I_{4(i + 1)-3}}_{\text{VW}},
        \mathcal{E}^{I_{4(i + 1)-2}}_{\text{VW}}
        \right),
        \text{Hom} \left(
        \mathcal{E}^{I_{4(i + 1)-1}}_{\text{VW}},
        \mathcal{E}^{I_{4(i + 1)}}_{\text{VW}}
        \right)
        \bigg]_-
      \\
      & \longrightarrow
        \text{Hom} \bigg[
        \text{Hom} \left(
        \text{Hom} \left(
        \mathcal{E}^{I_1}_{\text{VW}},
        \mathcal{E}^{I_2}_{\text{VW}}
        \right),
        \text{Hom} \left(
        \mathcal{E}^{I_3}_{\text{VW}},
        \mathcal{E}^{I_4}_{\text{VW}}
        \right)
        \right),
      \\
      & \qquad \qquad \quad
        \text{Hom} \left(
        \mathcal{E}^{I_{4\mathfrak{K}_n + 1}}_{\text{VW}},
        \mathcal{E}^{I_{4\mathfrak{K}_n + 2}}_{\text{VW}}
        \right),
        \text{Hom} \left(
        \mathcal{E}^{I_{4\mathfrak{K}_n + 3}}_{\text{VW}},
        \mathcal{E}^{I_{4\mathfrak{K}_n + 4}}_{\text{VW}}
        \right)
        \bigg]_+
    \end{aligned}
  }
\end{equation}
is a scattering amplitude of $\mathfrak{K}_n$ incoming LG $\mathfrak{A}_4^{\theta}$-soliton threebranes $\text{Hom} [ \text{Hom}\big( \text{Hom}(\mathcal{E}^{*}_{\text{VW}}, \mathcal{E}^{*}_{\text{VW}}),$ $\text{Hom}(\mathcal{E}^{*}_{\text{VW}}, \mathcal{E}^{*}_{\text{VW}}) \big), \text{Hom}\big( \text{Hom}(\mathcal{E}^{*}_{\text{VW}}, \mathcal{E}^{*}_{\text{VW}}), \text{Hom}(\mathcal{E}^{*}_{\text{VW}}, \mathcal{E}^{*}_{\text{VW}}) \big)]_-$, and a single outgoing LG $\mathfrak{A}_4^{\theta}$-threebrane $\text{Hom} [ \text{Hom}\big( \text{Hom}(\mathcal{E}^{*}_{\text{VW}}, \mathcal{E}^{*}_{\text{VW}}), \text{Hom}(\mathcal{E}^{*}_{\text{VW}}, \mathcal{E}^{*}_{\text{VW}}) \big), \text{Hom}\big( \text{Hom}(\mathcal{E}^{*}_{\text{VW}}, \mathcal{E}^{*}_{\text{VW}}), \text{Hom}(\mathcal{E}^{*}_{\text{VW}}, \mathcal{E}^{*}_{\text{VW}}) \big)]_+$, with vertices labeled.

Notice that the $\varDelta^{\mathfrak{K}_n}_{\mathfrak{A}_4}$ maps in \eqref{eq:cy2 x r4:4d lg:crf composition maps} which involve 3-morphisms, can also be regarded as composition maps defining an $A_{\infty}$-structure -- in particular, that of a 3-category whose $n$ objects $\mathcal{E}^1_{\text{VW}}, \mathcal{E}^2_{\text{VW}}, \dots, \mathcal{E}^n_{\text{VW}} \}$ correspond to ($\theta$-deformed) VW configurations on $CY_2$.

\subtitle{A Cauchy-Riemann-Fueter type $A_{\infty}$-3-category 3-categorifying the HW Floer Homology of $CY_2$}

As VW configurations on $CY_2$ are known to generate the HW Floer homology of $CY_2$ which is itself a 0-category, this 3-category is a 3-categorification of the said Floer homology.

Since this 3-category is determined by the gauged Cauchy-Riemann-Fueter equation in \eqref{eq:cy2 x r4:worldvolume eqn}, we shall name it a Cauchy-Riemann-Fueter type 3-category.

Altogether, this means that the normalized partition function of Spin$(7)$ theory on $CY_2 \times \R^4$, as expressed in \eqref{eq:cy2 x r4:4d lg:partition function}, manifests a \emph{novel} Cauchy-Riemann-Fueter type $A_{\infty}$-3-category, defined by the maps \eqref{eq:cy2 x r4:4d lg:crf composition maps} and the identifications \eqref{eq:cy2 x r4:4d-lg:3-morphism}, which 3-categorifies the HW Floer homology of $CY_2$!

\subtitle{An Equivalence Between a Cauchy-Riemann-Fueter type $A_{\infty}$-3-category and an FS type $A_{\infty}$-category}

Recall from \autoref{sec:cy2 x r4:fs-cat} that the normalized partition function of Spin$(7)$ theory on $CY_2 \times \R^4$ also manifests the FS type $A_{\infty}$-category of $\mathfrak{A}_4^{\theta}$-sheets.
This means that we have a \emph{novel} equivalence between the Cauchy-Riemann-Fueter type $A_{\infty}$-3-category 3-categorifying the HW Floer homology of $CY_2$ and the FS type $A_{\infty}$-category of $\mathfrak{A}_4^{\theta}$-sheets!

\section{Physical Proofs and Generalizations of Mathematical Conjectures}
\label{sec:proofs}

In this section, using the results that we have obtained thus far, and some relevant results from our previous works in \cite{er-2024-topol-gauge-theor, er-2023-topol-n}, we will furnish physical proofs and generalizations of the mathematical conjectures by Bousseau in \cite{bousseau-2024-holom-floer}, Doan-Rezchikov in \cite{doan-2022-holom-floer}, and Cao in \cite{cao-2016-gauge-theor}.

\subsection{Proving and Generalizing Bousseau's Atiyah-Floer type Correspondences}
\label{sec:proofs:atiyah-floer}

\subtitle{An $A_{\infty}$-2-category from \cite{er-2023-topol-n}}

In \cite[eqn.~(9.43)]{er-2023-topol-n}, on the RHS, the Hom-category is physically realized by a 3d sigma model on $\R^2 \times I$ with target space $\mathcal{M}^{G, \theta}_{\text{H}}(\Sigma)$, the hyperkähler moduli space of $\theta$-deformed $G$-Hitchin equations on a Riemann surface $\Sigma$, whose BPS equations would be $\theta$-deformed, Fueter equations on $\R^2 \times I$ that depend on the complex structure of $\mathcal{M}^{G, \theta}_{\text{H}}(\Sigma)$.
In addition, the $L$'s are Lagrangian branes in $\mathcal{M}^{G, \theta}_{\text{H}}(\Sigma)$.

Furthermore, since the LHS of \cite[eqn.~(9.43)]{er-2023-topol-n} obeys an $A_\infty$-structure map, that would mean likewise for the RHS.

Altogether, from the RHS of \cite[eqn.~(9.43)]{er-2023-topol-n}, this means that we would get an $A_{\infty}$-2-category, determined by a Fueter equation, and whose objects are Lagrangian branes in $\mathcal{M}^{G, \theta}_{\text{H}}(\Sigma)$.

\subtitle{An $A_{\infty}$-2-category from \cite{er-2024-topol-gauge-theor}}

In \cite[eqn.~(12.15)]{er-2024-topol-gauge-theor}, on the top entry, the Hom-category is physically realized by a 3d sigma model on $\R^2 \times I$ with target space $\mathcal{M}^{G, \theta}_{\text{inst}}(CY_2)$, the hyperkähler moduli space of $\theta$-deformed instantons on $CY_2$, whose BPS equations would be $\theta$-deformed, Fueter equations on $\R^2 \times I$ that depend on the complex structure of $\mathcal{M}^{G, \theta}_{\text{inst}}(CY_2)$.
In addition, the $L$'s are Lagrangian branes in $\mathcal{M}^{G, \theta}_{\text{inst}}(CY_2)$.

Furthermore, since the center entry of \cite[eqn.~(12.15)]{er-2024-topol-gauge-theor} obeys an $A_\infty$-structure map, that would mean likewise for the top entry.

Altogether, from the top entry of \cite[eqn.~(12.15)]{er-2024-topol-gauge-theor}, this means that we would get an $A_{\infty}$-2-category, determined by a Fueter equation, and whose objects are Lagrangian branes in $\mathcal{M}^{G, \theta}_{\text{inst}}(CY_2)$.

\subtitle{Bousseau-Doan-Rezchikov's Fueter 2-category}

A 2-category with an $A_{\infty}$-structure, determined by a Fueter equation, whose objects are Lagrangian branes of a hyperkähler manifold, were recently constructed and studied by Bousseau \cite{bousseau-2024-holom-floer} and Doan-Rezchikov \cite{doan-2022-holom-floer}.
It was named a Fueter 2-category.

Therefore, the $A_{\infty}$-2-categories that we have above, realize their construction of a Fueter 2-category.

\subtitle{Physical Proofs and Generalizations of Bousseau's Mathematical Conjectures}

With this latest interpretation of \cite[eqn.~(9.43)]{er-2023-topol-n} in terms of a Fueter 2-category, our results in \cite[$\S$9.5]{er-2023-topol-n} will mean that we have an \emph{equivalence} between (i) an FS type $A_{\infty}$-category of $\theta$-deformed $G$-Hitchin configurations on $M_3$ and (ii) a Fueter 2-category of Lagrangian branes in $\mathcal{M}^{G, \theta}_{\text{H}}(\Sigma)$.
At $\theta = \pi /2 $, this will amount to an equivalence between (i) an FS type $A_{\infty}$-category of flat $G_{\C}$ connections on $M_3$ and (ii) a Fueter 2-category of Lagrangian branes in $\mathcal{M}^{G_{\C}}_{\text{flat}}(\Sigma)$ (which spans the space of flat $G_{\C}$ connections on $\Sigma$ that can be extended to all of $M_3 = M_3' \bigcup_{\Sigma} M_3''$).
This equivalence was conjectured by Bousseau in \cite[Conjecture 2.14]{bousseau-2024-holom-floer}.

Likewise, with this latest interpretation of \cite[eqn.~(12.15)]{er-2024-topol-gauge-theor} in terms of a Fueter 2-category, our results in \cite[$\S$12.2]{er-2024-topol-gauge-theor} will mean that we have an \emph{equivalence} between (i) an FS type $A_{\infty}$-category of $\theta$-deformed holomorphic vector bundles on $CY_3$ and (ii) a Fueter 2-category of Lagrangian branes in $\mathcal{M}^{G, \theta}_{\text{inst}}(CY_2)$.
At $\theta = 0$, this will amount to an equivalence between (i) an FS type $A_{\infty}$-category of holomorphic vector bundles on $CY_3$ and (ii) a Fueter 2-category of Lagrangian branes in $\mathcal{M}^{G}_{\text{inst}}(CY_2)$ (which, for an algebraic surface $CY_2$, spans the space of holomorphic vector bundles on $CY_2$ that can be extended to all of $CY_3 = CY_3' \bigcup_{CY_2} CY_3''$).
This equivalence was conjectured by Bousseau in \cite[Conjecture 2.15]{bousseau-2024-holom-floer}.

Thus, we have physically proved and generalized (for general $\theta$) his mathematical conjectures.

\subsection{Proving Bousseau-Doan-Rezchikov's Conjecture, and Generalizing Bousseau-Doan-Rezchikov's and Cao's Conjectures}
\label{sec:proofs:equivalences}

\subtitle{An FS $A_{\infty}$-category of Lagrangian Branes in Path Space from \cite{er-2023-topol-n}}

In \cite[eqn.~(9.38)]{er-2023-topol-n}, we found a correspondence between (i) an FS type $A_{\infty}$-category of $\theta$-deformed $G$-Hitchin configurations on $M_3$ and (ii) an FS $A_{\infty}$-category of Lagrangian branes in the path space $\mathcal{M}(\R, \mathcal{M}^{G, \theta}_{\text{H}}(\Sigma))$ of smooth maps from $\R$ to $\mathcal{M}^{G, \theta}_{\text{H}}(\Sigma)$.
The result in \autoref{sec:proofs:atiyah-floer} thus implies an \emph{equivalence} between (i) a Fueter 2-category of Lagrangian branes in $\mathcal{M}^{G, \theta}_{\text{H}}(\Sigma)$ and (ii) an FS $A_{\infty}$-category of Lagrangian branes in $\mathcal{M}(\R, \mathcal{M}^{G, \theta}_{\text{H}}(\Sigma))$.

\subtitle{An FS $A_{\infty}$-category of Lagrangian Branes in Path Space from \cite{er-2024-topol-gauge-theor}}

In \cite[eqn.~(9.41) and (12.9)]{er-2024-topol-gauge-theor}, we found a correspondence between (i) an FS type $A_{\infty}$-category of $\theta$-deformed holomorphic vector bundles on $CY_3$ and (ii) an FS $A_{\infty}$-category of Lagrangian branes in the path space $\mathcal{M}(\R, \mathcal{M}^{G, \theta}_{\text{inst}}(CY_2))$ of smooth maps from $\R$ to $\mathcal{M}^{G, \theta}_{\text{inst}}(CY_2)$.
The result in \autoref{sec:proofs:atiyah-floer} thus implies an \emph{equivalence} between (i) a Fueter 2-category of Lagrangian branes in $\mathcal{M}^{G, \theta}_{\text{inst}}(CY_2)$ and (ii) an FS $A_{\infty}$-category of Lagrangian branes in $\mathcal{M}(\R, \mathcal{M}^{G, \theta}_{\text{inst}}(CY_2))$.

\subtitle{A Physical Proof of Bousseau-Doan-Rezchikov's Mathematical Conjecture}

An equivalence between a Fueter 2-category of Lagrangian branes and an FS $A_{\infty}$-category of Lagrangian branes in path space, was conjectured by Bousseau in \cite[Conjecture 2.10]{bousseau-2024-holom-floer} and Doan-Rezchikov in \cite[Conjecture 1.5]{doan-2022-holom-floer}.

Thus, we have physically proved their mathematical conjecture.

\subtitle{A Gauge-theoretic Generalization of Bousseau-Doan-Rezchikov's Mathematical Conjecture}

Moreover, in \autoref{sec:m2 x r3}, \autoref{sec:cy2 x s x r3}, and \autoref{sec:cy2 x r3}, we found that to the space $\mathfrak{A}_D$ of irreducible fields on a $D$-manifold, we have equivalences between (i) a Fueter type $A_{\infty}$-2-category whose objects are in $\mathfrak{A}_D$, and (ii) an FS type $A_{\infty}$-category whose objects are non-constant paths in the path space $\mathcal{M}(\R_{\tau}, \mathfrak{A}_D)$.
Since the objects of our categories in \autoref{sec:m2 x r3}, \autoref{sec:cy2 x s x r3}, and \autoref{sec:cy2 x r3}, are gauge-theoretic, the aforementioned equivalences can be understood as a gauge-theoretic generalization of our statements above.

Thus, we have a gauge-theoretic generalization of Bousseau-Doan-Rezchikov's mathematical conjecture!

\subtitle{A Gauge-theoretic Generalization of Cao's Mathematical Conjecture}

In \cite[Proposal 1.2.2 and Remark 1.2.3]{cao-2016-gauge-theor}, Cao conjectured the existence of a 3-category with an $A_{\infty}$-structure, whose objects are Lagrangian branes, that is determined by the Cauchy-Riemann-Fueter equation, which is related to an FS $A_{\infty}$-category.
Recall that in \autoref{sec:cy2 x r4}, our Cauchy-Riemann-Fueter type $A_{\infty}$-3-category is a 3-category with an $A_{\infty}$-structure, whose objects are \emph{gauge-theoretic} configurations in $\mathfrak{A}_4$, that is determined by the \emph{gauged} Cauchy-Riemann-Fueter equation, which is \emph{equivalent} to an FS type $A_{\infty}$-category.

Thus, we have a gauge-theoretic generalization of Cao's mathematical conjecture!

\glsaddall
\setglossarystyle{altlist}
\printglossary[title={Glossary of Notation}]

\printbibliography

\end{document}